\documentclass[apjl,ip,twocolumn]{aastex61}
\usepackage{comment}
\usepackage{ifthen}
\usepackage[switch]{lineno}

\newcommand{\forloop}[5][1]%
{%
\setcounter{#2}{#3}%
\ifthenelse{#4}%
	{%
	#5%
	\addtocounter{#2}{#1}%
	\forloop[#1]{#2}{\value{#2}}{#4}{#5}%
	}%
	{%
	}%
}%

\newcommand{\ctbd}[1]{}

\newcommand{\lc}{light curve}
\newcommand{\lcs}{light curves}
\newcommand{\Lc}{Light curve}

\newcommand{\masy}{\ensuremath{\rm mas\,yr^{-1}}}
\newcommand{\kms}{\ensuremath{\rm km\,s^{-1}}}
\newcommand{\ms}{\ensuremath{\rm m\,s^{-1}}}

\newcommand{\gcmc}{\ensuremath{\rm g\,cm^{-3}}}

\newcommand{\vsini}{\ensuremath{v \sin{i}}}
\newcommand{\feh}{\ensuremath{\rm [Fe/H]}}

\newcommand{\vmac}{\ensuremath{v_{\rm mac}}}
\newcommand{\vmic}{\ensuremath{v_{\rm mic}}}

\newcommand{\rsun}{\ensuremath{R_\sun}}
\newcommand{\msun}{\ensuremath{M_\sun}}
\newcommand{\lsun}{\ensuremath{L_\sun}}

\newcommand{\rstar}{\ensuremath{R_\star}}
\newcommand{\mstar}{\ensuremath{M_\star}}
\newcommand{\lstar}{\ensuremath{L_\star}}

\newcommand{\teffstar}{\ensuremath{T_{\rm eff\star}}}
\newcommand{\rhostar}{\ensuremath{\rho_\star}}
\newcommand{\loggstar}{\ensuremath{\log{g_{\star}}}}

\newcommand{\rpl}{\ensuremath{R_{p}}}
\newcommand{\mpl}{\ensuremath{M_{p}}}

\newcommand{\rhopl}{\ensuremath{\rho_{p}}}

\newcommand{\arstar}{\ensuremath{a/\rstar}}
\newcommand{\zrstar}{\ensuremath{\zeta/\rstar}}

\newcommand{\rjup}{\ensuremath{R_{\rm J}}}
\newcommand{\mjup}{\ensuremath{M_{\rm J}}}

\newcommand{\reffigl}[1]{Figure~\ref{fig:#1}}
\newcommand{\refsecl}[1]{\mbox{Section \ref{sec:#1}}}

\newcommand{\reftabl}[1]{Table~\ref{tab:#1}}

\newcommand{\loopand}{\ifnum\value{planetcounter}=2 and \else\fi}
\newcommand{\loopcomma}{\ifnum\value{planetcounter}<2 ,\else. \fi}
\newcommand{\loopcommanoperiod}{\ifnum\value{planetcounter}<2 ,\else \space\fi}
\newcommand{\loopcommanospace}{\ifnum\value{planetcounter}<2 ,\else \fi}

\newcommand{\hatcurhtrxxxxA}{HATS580-019}                      
\newcommand{\hatcurfieldxxxxA}{\ensuremath{string}}            
\newcommand{\hatcurCCraxxxxA}{\ensuremath{20^{\mathrm h}01^{\mathrm m}42.60{\mathrm s}}}                     
\newcommand{\hatcurCCdecxxxxA}{\ensuremath{-26{\arcdeg}04{\arcmin}39.3{\arcsec}}}                    
\newcommand{\hatcurCCmagxxxxA}{14.033}                         
\newcommand{\hatcurCCtwomassxxxxA}{2MASS~20014273-2604392}     
\newcommand{\hatcurCCgscxxxxA}{GSC~6896-01012}                 
\newcommand{\hatcurCCtassmvxxxxA}{\ensuremath{14.033\pm0.050}} 
\newcommand{\hatcurCCtassmvshortxxxxA}{\ensuremath{14.0}}      
\newcommand{\hatcurCCtassmBxxxxA}{\ensuremath{14.718\pm0.010}} 
\newcommand{\hatcurCCtassmBshortxxxxA}{\ensuremath{14.7}}      
\newcommand{\hatcurCCtassmIxxxxA}{\ensuremath{nff\pmnff}}      
\newcommand{\hatcurCCtassmIshortxxxxA}{\ensuremath{0.0}}       
\newcommand{\hatcurCCtassmgxxxxA}{\ensuremath{nff\pmnff}}      
\newcommand{\hatcurCCtassmgshortxxxxA}{\ensuremath{0.0}}       
\newcommand{\hatcurCCtassmrxxxxA}{\ensuremath{nff\pmnff}}      
\newcommand{\hatcurCCtassmrshortxxxxA}{\ensuremath{0.0}}       
\newcommand{\hatcurCCtassmixxxxA}{\ensuremath{13.535\pm0.010}} 
\newcommand{\hatcurCCtassmishortxxxxA}{\ensuremath{13.5}}      
\newcommand{\hatcurCCtwomassJmagxxxxA}{\ensuremath{12.643\pm0.025}} 
\newcommand{\hatcurCCtwomassHmagxxxxA}{\ensuremath{12.373\pm0.034}} 
\newcommand{\hatcurCCtwomassKmagxxxxA}{\ensuremath{12.289\pm0.027}} 
\newcommand{\hatcurCCcitJmagxxxxA}{\ensuremath{12.661\pm0.025}} 
\newcommand{\hatcurCCcitHmagxxxxA}{\ensuremath{12.367\pm0.034}} 
\newcommand{\hatcurCCcitKmagxxxxA}{\ensuremath{12.313\pm0.027}} 
\newcommand{\hatcurCCbbJmagxxxxA}{\ensuremath{12.709\pm0.027}} 
\newcommand{\hatcurCCbbHmagxxxxA}{\ensuremath{12.389\pm0.035}} 
\newcommand{\hatcurCCbbKmagxxxxA}{\ensuremath{12.333\pm0.027}} 
\newcommand{\hatcurCCesoJmagxxxxA}{\ensuremath{12.711\pm0.028}} 
\newcommand{\hatcurCCesoHmagxxxxA}{\ensuremath{12.385\pm0.039}} 
\newcommand{\hatcurCCesoKmagxxxxA}{\ensuremath{12.332\pm0.028}} 
\newcommand{\hatcurCCesoJHmagxxxxA}{\ensuremath{0.325\pm0.046}} 
\newcommand{\hatcurCCesoJKmagxxxxA}{\ensuremath{0.379\pm0.039}} 
\newcommand{\hatcurCCesoHKmagxxxxA}{\ensuremath{0.053\pm0.048}} 
\newcommand{\hatcurLCdipxxxxA}{\ensuremath{11.3}}              
\newcommand{\hatcurLCrprstarxxxxA}{\ensuremath{0.1038\pm0.0025}} 
\newcommand{\hatcurLCbsqxxxxA}{\ensuremath{0.177_{-0.073}^{+0.073}}} 
\newcommand{\hatcurLCimpxxxxA}{\ensuremath{0.421_{-0.099}^{+0.079}}} 
\newcommand{\hatcurLCzetaxxxxA}{\ensuremath{17.55\pm0.24}}     
\newcommand{\hatcurLCdurxxxxA}{\ensuremath{0.1283\pm0.0021}}   
\newcommand{\hatcurLCdurshortxxxxA}{\ensuremath{0.1283}}       
\newcommand{\hatcurLCdurhrxxxxA}{\ensuremath{3.079\pm0.051}}   
\newcommand{\hatcurLCdurhrshortxxxxA}{\ensuremath{3.079}}      
\newcommand{\hatcurLCqxxxxA}{\ensuremath{0.03350\pm0.00056}}   
\newcommand{\hatcurLCqshortxxxxA}{\ensuremath{0.034}}          
\newcommand{\hatcurLCingdurxxxxA}{\ensuremath{0.0144\pm0.0015}} 
\newcommand{\hatcurLCPxxxxA}{\ensuremath{3.8297015\pm0.0000046}} 
\newcommand{\hatcurLCPprecxxxxA}{\ensuremath{3.8297015}}       
\newcommand{\hatcurLCPshortxxxxA}{\ensuremath{3.8297}}         
\newcommand{\hatcurLCTxxxxA}{\ensuremath{2456870.34792\pm0.00068}} 
\newcommand{\hatcurLCTAxxxxA}{\ensuremath{2455277.1922\pm0.0019}} 
\newcommand{\hatcurLCTBxxxxA}{\ensuremath{2457180.55375\pm0.00084}} 
\newcommand{\hatcurLChatnetmAxxxxA}{\ensuremath{13.65037\pm0.00011}} 
\newcommand{\hatcurLCiblendAxxxxA}{\ensuremath{0.816\pm0.063}} 
\newcommand{\hatcurLChatnetmBxxxxA}{\ensuremath{13.65040\pm0.00011}} 
\newcommand{\hatcurLCiblendBxxxxA}{\ensuremath{0.880\pm0.062}} 
\newcommand{\hatcurLCrhoxxxxA}{\ensuremath{1.38_{-0.23}^{+0.16}}} 
\newcommand{\hatcurSMEiteffxxxxA}{\ensuremath{5990\pm110}}     
\newcommand{\hatcurSMEizfehxxxxA}{\ensuremath{0.300\pm0.056}}  
\newcommand{\hatcurSMEizfehshortxxxxA}{\ensuremath{0.30}}      
\newcommand{\hatcurSMEiloggxxxxA}{\ensuremath{4.51\pm0.17}}    
\newcommand{\hatcurSMEivsinxxxxA}{\ensuremath{3.76\pm0.54}}    
\newcommand{\hatcurSMEivmacxxxxA}{\ensuremath{4.32\pm0.17}}    
\newcommand{\hatcurSMEivmicxxxxA}{\ensuremath{1.225\pm0.085}}  
\newcommand{\hatcurLBizxxxxA}{\ensuremath{0.1881}}             
\newcommand{\hatcurLBiizxxxxA}{\ensuremath{0.3457}}            
\newcommand{\hatcurLBiixxxxA}{\ensuremath{0.2495}}             
\newcommand{\hatcurLBiiixxxxA}{\ensuremath{0.3488}}            
\newcommand{\hatcurLBiIxxxxA}{\ensuremath{0.2285}}             
\newcommand{\hatcurLBiiIxxxxA}{\ensuremath{0.3488}}            
\newcommand{\hatcurLBigxxxxA}{\ensuremath{0.5282}}             
\newcommand{\hatcurLBiigxxxxA}{\ensuremath{0.2582}}            
\newcommand{\hatcurLBirxxxxA}{\ensuremath{0.3362}}             
\newcommand{\hatcurLBiirxxxxA}{\ensuremath{0.3446}}            
\newcommand{\hatcurLBiRxxxxA}{\ensuremath{0.3120}}             
\newcommand{\hatcurLBiiRxxxxA}{\ensuremath{0.3470}}            
\newcommand{\hatcurLBikepxxxxA}{\ensuremath{0.1000}}           
\newcommand{\hatcurLBiikepxxxxA}{\ensuremath{0.1000}}          
\newcommand{\hatcurISOmxxxxA}{\ensuremath{1.168\pm0.042}}      
\newcommand{\hatcurISOmshortxxxxA}{\ensuremath{1.17}}          
\newcommand{\hatcurISOmlongxxxxA}{\ensuremath{1.168\pm0.042}}  
\newcommand{\hatcurISOrxxxxA}{\ensuremath{1.117\pm0.060}}      
\newcommand{\hatcurISOrshortxxxxA}{\ensuremath{1.12}}          
\newcommand{\hatcurISOrlongxxxxA}{\ensuremath{1.117\pm0.060}}  
\newcommand{\hatcurISOrhoxxxxA}{\ensuremath{1.19\pm0.16}}      
\newcommand{\hatcurISOrholongxxxxA}{\ensuremath{1.19\pm0.16}}  
\newcommand{\hatcurISOloggxxxxA}{\ensuremath{4.411\pm0.038}}   
\newcommand{\hatcurISOlumxxxxA}{\ensuremath{1.39\pm0.23}}      
\newcommand{\hatcurISOlumshortxxxxA}{\ensuremath{1.39}}        
\newcommand{\hatcurISOmvxxxxA}{\ensuremath{4.44\pm0.19}}       
\newcommand{\hatcurISOvixxxxA}{\ensuremath{0.655\pm0.031}}     
\newcommand{\hatcurISOagexxxxA}{\ensuremath{1.2\pm1.1}}        
\newcommand{\hatcurISOsigmaxxxxA}{\ensuremath{0.00050\pm0.00014}} 
\newcommand{\hatcurISOMJxxxxA}{\ensuremath{3.36\pm0.15}}       
\newcommand{\hatcurISOMHxxxxA}{\ensuremath{3.06\pm0.13}}       
\newcommand{\hatcurISOMKxxxxA}{\ensuremath{3.01\pm0.13}}       
\newcommand{\hatcurISOJKxxxxA}{\ensuremath{0.360\pm0.020}}     
\newcommand{\hatcurISOspecxxxxA}{G}                            
\newcommand{\hatcurRVKxxxxA}{\ensuremath{45\pm12}}             
\newcommand{\hatcurRVrkxxxxA}{\ensuremath{0\pm0}}              
\newcommand{\hatcurRVrhxxxxA}{\ensuremath{0\pm0}}              
\newcommand{\hatcurRVkxxxxA}{\ensuremath{0\pm0}}               
\newcommand{\hatcurRVhxxxxA}{\ensuremath{0\pm0}}               
\newcommand{\hatcurRVtronexxxxA}{\ensuremath{0\pm0}}           
\newcommand{\hatcurRVtrtwoxxxxA}{\ensuremath{0\pm0}}           
\newcommand{\hatcurRVgammaAxxxxA}{\ensuremath{-20250\pm13}}    
\newcommand{\hatcurRVjitterAxxxxA}{\ensuremath{69\pm11}}       
\newcommand{\hatcurRVjittertwosiglimAxxxxA}{\ensuremath{<90.7}} 
\newcommand{\hatcurRVfitrmsAxxxxA}{\ensuremath{0.0}}           
\newcommand{\hatcurRVgammaBxxxxA}{\ensuremath{-20176\pm94}}    
\newcommand{\hatcurRVjitterBxxxxA}{\ensuremath{60\pm130}}      
\newcommand{\hatcurRVjittertwosiglimBxxxxA}{\ensuremath{<286.0}} 
\newcommand{\hatcurRVfitrmsBxxxxA}{\ensuremath{0.0}}           
\newcommand{\hatcurRVgammaCxxxxA}{\ensuremath{5\pm12}}         
\newcommand{\hatcurRVjitterCxxxxA}{\ensuremath{28\pm10}}       
\newcommand{\hatcurRVjittertwosiglimCxxxxA}{\ensuremath{<48.2}} 
\newcommand{\hatcurRVfitrmsCxxxxA}{\ensuremath{0.0}}           
\newcommand{\hatcurRVeccenxxxxA}{\ensuremath{0\pm0}}           
\newcommand{\hatcurRVeccentwosiglimxxxxA}{\ensuremath{<0.000}} 
\newcommand{\hatcurRVomegaxxxxA}{\ensuremath{0\pm0}}           
\newcommand{\hatcurPPixxxxA}{\ensuremath{87.54\pm0.66}}        
\newcommand{\hatcurPPgxxxxA}{\ensuremath{7.5\pm2.2}}           
\newcommand{\hatcurPPloggxxxxA}{\ensuremath{2.87_{-0.14}^{+0.11}}} 
\newcommand{\hatcurPParxxxxA}{\ensuremath{9.72\pm0.44}}        
\newcommand{\hatcurPParelxxxxA}{\ensuremath{0.05046\pm0.00060}} 
\newcommand{\hatcurPPrhoxxxxA}{\ensuremath{0.33\pm0.11}}       
\newcommand{\hatcurPPmxxxxA}{\ensuremath{0.39\pm0.10}}         
\newcommand{\hatcurPPmshortxxxxA}{\ensuremath{0.39}}           
\newcommand{\hatcurPPmlongxxxxA}{\ensuremath{0.39\pm0.10}}     
\newcommand{\hatcurPPmexxxxA}{\ensuremath{123\pm33}}           
\newcommand{\hatcurPPmeshortxxxxA}{\ensuremath{123.0}}         
\newcommand{\hatcurPPmelongxxxxA}{\ensuremath{123\pm33}}       
\newcommand{\hatcurPPrxxxxA}{\ensuremath{1.130\pm0.075}}       
\newcommand{\hatcurPPrshortxxxxA}{\ensuremath{1.13}}           
\newcommand{\hatcurPPrlongxxxxA}{\ensuremath{1.130\pm0.075}}   
\newcommand{\hatcurPPrexxxxA}{\ensuremath{12.67\pm0.84}}       
\newcommand{\hatcurPPreshortxxxxA}{\ensuremath{12.7}}          
\newcommand{\hatcurPPrelongxxxxA}{\ensuremath{12.67\pm0.84}}   
\newcommand{\hatcurPPmrcorrxxxxA}{\ensuremath{0.09}}           
\newcommand{\hatcurPPteffxxxxA}{\ensuremath{1348\pm47}}        
\newcommand{\hatcurPPthetaxxxxA}{\ensuremath{0.0296\pm0.0080}} 
\newcommand{\hatcurPPfluxperixxxxA}{\ensuremath{7.4\pm1.1}}    
\newcommand{\hatcurPPfluxperidimxxxxA}{\ensuremath{8}}         
\newcommand{\hatcurPPfluxapxxxxA}{\ensuremath{7.4\pm1.1}}      
\newcommand{\hatcurPPfluxapdimxxxxA}{\ensuremath{8}}           
\newcommand{\hatcurPPfluxavgxxxxA}{\ensuremath{7.4\pm1.1}}     
\newcommand{\hatcurPPfluxavgdimxxxxA}{\ensuremath{8}}          
\newcommand{\hatcurPPfluxavglogxxxxA}{\ensuremath{8.872\pm0.060}} 
\newcommand{\hatcurXsecphasexxxxA}{\ensuremath{0\pm0}}         
\newcommand{\hatcurXsecondaryxxxxA}{\ensuremath{2456872.26277\pm0.00068}} 
\newcommand{\hatcurXsecdurxxxxA}{\ensuremath{0.1283\pm0.0021}} 
\newcommand{\hatcurXsecingdurxxxxA}{\ensuremath{0.0144\pm0.0015}} 
\newcommand{\hatcurPPphiconjxxxxA}{\ensuremath{0\pm0}}         
\newcommand{\hatcurPPperixxxxA}{\ensuremath{2456869.39049\pm0.00068}} 
\newcommand{\hatcurPPaequivxxxxA}{\ensuremath{0.0428\pm0.0030}} 
\newcommand{\hatcurPPtcircxxxxA}{\ensuremath{231_{-79}^{+109}}} 
\newcommand{\hatcurPPtinfallxxxxA}{\ensuremath{9500_{-2600}^{+4300}}} 
\newcommand{\hatcurXdistxxxxA}{\ensuremath{734\pm46}}          
\newcommand{\hatcurXAvxxxxA}{\ensuremath{0.305\pm0.098}}       
\newcommand{\hatcurXdistredxxxxA}{\ensuremath{717\pm43}}       
\newcommand{\hatcurXEBVxxxxA}{\ensuremath{0.099\pm0.031}}      
\newcommand{\hatcurXmvisoredxxxxA}{\ensuremath{14.022\pm0.048}} 
\newcommand{\hatcurXmiisoredxxxxA}{\ensuremath{13.207\pm0.022}} 
\newcommand{\hatcurXmjisoredxxxxA}{\ensuremath{12.727\pm0.017}} 
\newcommand{\hatcurXmhisoredxxxxA}{\ensuremath{12.390\pm0.019}} 
\newcommand{\hatcurXmkisoredxxxxA}{\ensuremath{12.316\pm0.020}} 
\newcommand{\hatcurXviisoredxxxxA}{\ensuremath{0.815\pm0.034}} 
\newcommand{\hatcurXvkisoredxxxxA}{\ensuremath{1.706\pm0.056}} 
\newcommand{\hatcurXjhisoredxxxxA}{\ensuremath{0.337\pm0.011}} 
\newcommand{\hatcurXjkisoredxxxxA}{\ensuremath{0.411\pm0.012}} 
\newcommand{\hatcurCCpmraxxxxA}{\ensuremath{3.2\pm1.6}}        
\newcommand{\hatcurCCpmdecxxxxA}{\ensuremath{1.6\pm1.6}}       
\newcommand{\hatcurCCpmxxxxA}{\ensuremath{3.6\pm2.3}}          

\newcommand{\hatcurhtrxxxxB}{HATS601-050}                      
\newcommand{\hatcurfieldxxxxB}{\ensuremath{string}}            
\newcommand{\hatcurCCraxxxxB}{\ensuremath{06^{\mathrm h}51^{\mathrm m}23.40{\mathrm s}}}                     
\newcommand{\hatcurCCdecxxxxB}{\ensuremath{-29{\arcdeg}03{\arcmin}31.0{\arcsec}}}                    
\newcommand{\hatcurCCmagxxxxB}{12.471}                         
\newcommand{\hatcurCCtwomassxxxxB}{2MASS~06512340-2903309}     
\newcommand{\hatcurCCgscxxxxB}{GSC~6534-00607}                 
\newcommand{\hatcurCCtassmvxxxxB}{\ensuremath{12.471\pm0.030}} 
\newcommand{\hatcurCCtassmvshortxxxxB}{\ensuremath{12.5}}      
\newcommand{\hatcurCCtassmBxxxxB}{\ensuremath{13.190\pm0.030}} 
\newcommand{\hatcurCCtassmBshortxxxxB}{\ensuremath{13.2}}      
\newcommand{\hatcurCCtassmIxxxxB}{\ensuremath{nff\pmnff}}      
\newcommand{\hatcurCCtassmIshortxxxxB}{\ensuremath{0.0}}       
\newcommand{\hatcurCCtassmgxxxxB}{\ensuremath{12.766\pm0.030}} 
\newcommand{\hatcurCCtassmgshortxxxxB}{\ensuremath{12.8}}      
\newcommand{\hatcurCCtassmrxxxxB}{\ensuremath{12.269\pm0.040}} 
\newcommand{\hatcurCCtassmrshortxxxxB}{\ensuremath{12.3}}      
\newcommand{\hatcurCCtassmixxxxB}{\ensuremath{12.115\pm0.040}} 
\newcommand{\hatcurCCtassmishortxxxxB}{\ensuremath{12.1}}      
\newcommand{\hatcurCCtwomassJmagxxxxB}{\ensuremath{11.241\pm0.023}} 
\newcommand{\hatcurCCtwomassHmagxxxxB}{\ensuremath{10.955\pm0.024}} 
\newcommand{\hatcurCCtwomassKmagxxxxB}{\ensuremath{10.867\pm0.021}} 
\newcommand{\hatcurCCcitJmagxxxxB}{\ensuremath{11.258\pm0.023}} 
\newcommand{\hatcurCCcitHmagxxxxB}{\ensuremath{10.949\pm0.024}} 
\newcommand{\hatcurCCcitKmagxxxxB}{\ensuremath{10.891\pm0.021}} 
\newcommand{\hatcurCCbbJmagxxxxB}{\ensuremath{11.307\pm0.025}} 
\newcommand{\hatcurCCbbHmagxxxxB}{\ensuremath{10.971\pm0.025}} 
\newcommand{\hatcurCCbbKmagxxxxB}{\ensuremath{10.911\pm0.021}} 
\newcommand{\hatcurCCesoJmagxxxxB}{\ensuremath{11.309\pm0.026}} 
\newcommand{\hatcurCCesoHmagxxxxB}{\ensuremath{10.967\pm0.029}} 
\newcommand{\hatcurCCesoKmagxxxxB}{\ensuremath{10.910\pm0.022}} 
\newcommand{\hatcurCCesoJHmagxxxxB}{\ensuremath{0.342\pm0.037}} 
\newcommand{\hatcurCCesoJKmagxxxxB}{\ensuremath{0.400\pm0.034}} 
\newcommand{\hatcurCCesoHKmagxxxxB}{\ensuremath{0.057\pm0.036}} 
\newcommand{\hatcurLCdipxxxxB}{\ensuremath{11.1}}              
\newcommand{\hatcurLCrprstarxxxxB}{\ensuremath{0.1010\pm0.0038}} 
\newcommand{\hatcurLCbsqxxxxB}{\ensuremath{0.093_{-0.066}^{+0.095}}} 
\newcommand{\hatcurLCimpxxxxB}{\ensuremath{0.30_{-0.14}^{+0.13}}} 
\newcommand{\hatcurLCzetaxxxxB}{\ensuremath{16.08\pm0.16}}     
\newcommand{\hatcurLCdurxxxxB}{\ensuremath{0.1384\pm0.0020}}   
\newcommand{\hatcurLCdurshortxxxxB}{\ensuremath{0.1384}}       
\newcommand{\hatcurLCdurhrxxxxB}{\ensuremath{3.321\pm0.047}}   
\newcommand{\hatcurLCdurhrshortxxxxB}{\ensuremath{3.321}}      
\newcommand{\hatcurLCqxxxxB}{\ensuremath{0.04130\pm0.00059}}   
\newcommand{\hatcurLCqshortxxxxB}{\ensuremath{0.041}}          
\newcommand{\hatcurLCingdurxxxxB}{\ensuremath{0.0138\pm0.0017}} 
\newcommand{\hatcurLCPxxxxB}{\ensuremath{3.3488702\pm0.0000039}} 
\newcommand{\hatcurLCPprecxxxxB}{\ensuremath{3.3488702}}       
\newcommand{\hatcurLCPshortxxxxB}{\ensuremath{3.3489}}         
\newcommand{\hatcurLCTxxxxB}{\ensuremath{2457042.00405\pm0.00058}} 
\newcommand{\hatcurLCTAxxxxB}{\ensuremath{2455796.2243\pm0.0015}} 
\newcommand{\hatcurLCTBxxxxB}{\ensuremath{2457299.86705\pm0.00070}} 
\newcommand{\hatcurLChatnetmxxxxB}{\ensuremath{12.321600\pm0.000061}} 
\newcommand{\hatcurLCiblendxxxxB}{\ensuremath{0.827\pm0.063}}  
\newcommand{\hatcurLCrhoxxxxB}{\ensuremath{0.56_{-0.16}^{+0.21}}} 
\newcommand{\hatcurSMEiteffxxxxB}{\ensuremath{5895\pm71}}      
\newcommand{\hatcurSMEizfehxxxxB}{\ensuremath{0.340\pm0.030}}  
\newcommand{\hatcurSMEizfehshortxxxxB}{\ensuremath{0.34}}      
\newcommand{\hatcurSMEiloggxxxxB}{\ensuremath{4.51\pm0.13}}    
\newcommand{\hatcurSMEivsinxxxxB}{\ensuremath{3.76\pm0.37}}    
\newcommand{\hatcurSMEivmacxxxxB}{\ensuremath{4.17\pm0.11}}    
\newcommand{\hatcurSMEivmicxxxxB}{\ensuremath{1.156\pm0.048}}  
\newcommand{\hatcurSMEiiteffxxxxB}{\ensuremath{5758\pm58}}     
\newcommand{\hatcurSMEiizfehxxxxB}{\ensuremath{0.300\pm0.030}} 
\newcommand{\hatcurSMEiizfehshortxxxxB}{\ensuremath{0.30}}     
\newcommand{\hatcurSMEiiloggxxxxB}{\ensuremath{4.324\pm0.042}} 
\newcommand{\hatcurSMEiivsinxxxxB}{\ensuremath{3.98\pm0.26}}   
\newcommand{\hatcurSMEiivmacxxxxB}{\ensuremath{3.962\pm0.088}} 
\newcommand{\hatcurSMEiivmicxxxxB}{\ensuremath{1.070\pm0.034}} 
\newcommand{\hatcurLBizxxxxB}{\ensuremath{0.2162}}             
\newcommand{\hatcurLBiizxxxxB}{\ensuremath{0.3324}}            
\newcommand{\hatcurLBiixxxxB}{\ensuremath{0.2833}}             
\newcommand{\hatcurLBiiixxxxB}{\ensuremath{0.3306}}            
\newcommand{\hatcurLBiIxxxxB}{\ensuremath{0.2606}}             
\newcommand{\hatcurLBiiIxxxxB}{\ensuremath{0.3322}}            
\newcommand{\hatcurLBigxxxxB}{\ensuremath{0.5877}}             
\newcommand{\hatcurLBiigxxxxB}{\ensuremath{0.2128}}            
\newcommand{\hatcurLBirxxxxB}{\ensuremath{0.3801}}             
\newcommand{\hatcurLBiirxxxxB}{\ensuremath{0.3173}}            
\newcommand{\hatcurLBiRxxxxB}{\ensuremath{0.3531}}             
\newcommand{\hatcurLBiiRxxxxB}{\ensuremath{0.3221}}            
\newcommand{\hatcurLBikepxxxxB}{\ensuremath{0.1000}}           
\newcommand{\hatcurLBiikepxxxxB}{\ensuremath{0.1000}}          
\newcommand{\hatcurISOmxxxxB}{\ensuremath{1.187\pm0.060}}      
\newcommand{\hatcurISOmshortxxxxB}{\ensuremath{1.19}}          
\newcommand{\hatcurISOmlongxxxxB}{\ensuremath{1.187\pm0.060}}  
\newcommand{\hatcurISOrxxxxB}{\ensuremath{1.44\pm0.18}}        
\newcommand{\hatcurISOrshortxxxxB}{\ensuremath{1.44}}          
\newcommand{\hatcurISOrlongxxxxB}{\ensuremath{1.44\pm0.18}}    
\newcommand{\hatcurISOrhoxxxxB}{\ensuremath{0.56_{-0.16}^{+0.22}}} 
\newcommand{\hatcurISOrholongxxxxB}{\ensuremath{0.56_{-0.16}^{+0.22}}} 
\newcommand{\hatcurISOloggxxxxB}{\ensuremath{4.198\pm0.088}}   
\newcommand{\hatcurISOlumxxxxB}{\ensuremath{2.04\pm0.54}}      
\newcommand{\hatcurISOlumshortxxxxB}{\ensuremath{2.04}}        
\newcommand{\hatcurISOmvxxxxB}{\ensuremath{4.05\pm0.27}}       
\newcommand{\hatcurISOvixxxxB}{\ensuremath{0.709\pm0.019}}     
\newcommand{\hatcurISOagexxxxB}{\ensuremath{4.74_{-0.51}^{+0.70}}} 
\newcommand{\hatcurISOsigmaxxxxB}{\ensuremath{0.00050\pm0.00012}} 
\newcommand{\hatcurISOMJxxxxB}{\ensuremath{2.90\pm0.26}}       
\newcommand{\hatcurISOMHxxxxB}{\ensuremath{2.56\pm0.26}}       
\newcommand{\hatcurISOMKxxxxB}{\ensuremath{2.51\pm0.26}}       
\newcommand{\hatcurISOJKxxxxB}{\ensuremath{0.390\pm0.010}}     
\newcommand{\hatcurISOspecxxxxB}{G}                            
\newcommand{\hatcurRVKxxxxB}{\ensuremath{94.9\pm5.1}}          
\newcommand{\hatcurRVrkxxxxB}{\ensuremath{0.24\pm0.13}}        
\newcommand{\hatcurRVrhxxxxB}{\ensuremath{0.34_{-0.22}^{+0.14}}} 
\newcommand{\hatcurRVkxxxxB}{\ensuremath{0.100\pm0.060}}       
\newcommand{\hatcurRVhxxxxB}{\ensuremath{0.144\pm0.100}}       
\newcommand{\hatcurRVtronexxxxB}{\ensuremath{0\pm0}}           
\newcommand{\hatcurRVtrtwoxxxxB}{\ensuremath{0\pm0}}           
\newcommand{\hatcurRVgammaAxxxxB}{\ensuremath{3093\pm15}}      
\newcommand{\hatcurRVjitterAxxxxB}{\ensuremath{49\pm11}}       
\newcommand{\hatcurRVjittertwosiglimAxxxxB}{\ensuremath{<69.9}} 
\newcommand{\hatcurRVfitrmsAxxxxB}{\ensuremath{0.0}}           
\newcommand{\hatcurRVgammaBxxxxB}{\ensuremath{3086\pm14}}      
\newcommand{\hatcurRVjitterBxxxxB}{\ensuremath{58\pm12}}       
\newcommand{\hatcurRVjittertwosiglimBxxxxB}{\ensuremath{<82.0}} 
\newcommand{\hatcurRVfitrmsBxxxxB}{\ensuremath{0.0}}           
\newcommand{\hatcurRVgammaCxxxxB}{\ensuremath{3097.8\pm6.6}}   
\newcommand{\hatcurRVjitterCxxxxB}{\ensuremath{25.2\pm7.8}}    
\newcommand{\hatcurRVjittertwosiglimCxxxxB}{\ensuremath{<41.5}} 
\newcommand{\hatcurRVfitrmsCxxxxB}{\ensuremath{0.0}}           
\newcommand{\hatcurRVeccenxxxxB}{\ensuremath{0.190\pm0.080}}   
\newcommand{\hatcurRVeccentwosiglimxxxxB}{\ensuremath{<0.330}} 
\newcommand{\hatcurRVomegaxxxxB}{\ensuremath{60\pm76}}         
\newcommand{\hatcurPPixxxxB}{\ensuremath{87.1\pm1.6}}          
\newcommand{\hatcurPPgxxxxB}{\ensuremath{9.5\pm2.4}}           
\newcommand{\hatcurPPloggxxxxB}{\ensuremath{2.98\pm0.11}}      
\newcommand{\hatcurPParxxxxB}{\ensuremath{6.94\pm0.74}}        
\newcommand{\hatcurPParelxxxxB}{\ensuremath{0.04639\pm0.00077}} 
\newcommand{\hatcurPPrhoxxxxB}{\ensuremath{0.34_{-0.11}^{+0.16}}} 
\newcommand{\hatcurPPmxxxxB}{\ensuremath{0.768\pm0.045}}       
\newcommand{\hatcurPPmshortxxxxB}{\ensuremath{0.77}}           
\newcommand{\hatcurPPmlongxxxxB}{\ensuremath{0.768\pm0.045}}   
\newcommand{\hatcurPPmexxxxB}{\ensuremath{244\pm14}}           
\newcommand{\hatcurPPmeshortxxxxB}{\ensuremath{244.2}}         
\newcommand{\hatcurPPmelongxxxxB}{\ensuremath{244\pm14}}       
\newcommand{\hatcurPPrxxxxB}{\ensuremath{1.41\pm0.19}}         
\newcommand{\hatcurPPrshortxxxxB}{\ensuremath{1.41}}           
\newcommand{\hatcurPPrlongxxxxB}{\ensuremath{1.41\pm0.19}}     
\newcommand{\hatcurPPrexxxxB}{\ensuremath{15.8\pm2.1}}         
\newcommand{\hatcurPPreshortxxxxB}{\ensuremath{15.8}}          
\newcommand{\hatcurPPrelongxxxxB}{\ensuremath{15.8\pm2.1}}     
\newcommand{\hatcurPPmrcorrxxxxB}{\ensuremath{0.25}}           
\newcommand{\hatcurPPteffxxxxB}{\ensuremath{1553\pm92}}        
\newcommand{\hatcurPPthetaxxxxB}{\ensuremath{0.0421\pm0.0064}} 
\newcommand{\hatcurPPfluxperixxxxB}{\ensuremath{1.95_{-0.62}^{+1.06}}} 
\newcommand{\hatcurPPfluxperidimxxxxB}{\ensuremath{9}}         
\newcommand{\hatcurPPfluxapxxxxB}{\ensuremath{9.1\pm1.3}}      
\newcommand{\hatcurPPfluxapdimxxxxB}{\ensuremath{8}}           
\newcommand{\hatcurPPfluxavgxxxxB}{\ensuremath{1.31_{-0.27}^{+0.36}}} 
\newcommand{\hatcurPPfluxavgdimxxxxB}{\ensuremath{9}}          
\newcommand{\hatcurPPfluxavglogxxxxB}{\ensuremath{9.12\pm0.10}} 
\newcommand{\hatcurXsecphasexxxxB}{\ensuremath{0.564\pm0.039}} 
\newcommand{\hatcurXsecondaryxxxxB}{\ensuremath{2457043.89\pm0.13}} 
\newcommand{\hatcurXsecdurxxxxB}{\ensuremath{0.175\pm0.031}}   
\newcommand{\hatcurXsecingdurxxxxB}{\ensuremath{0.020\pm0.011}} 
\newcommand{\hatcurPPphiconjxxxxB}{\ensuremath{0.062_{-0.042}^{+0.087}}} 
\newcommand{\hatcurPPperixxxxB}{\ensuremath{2457041.80\pm0.28}} 
\newcommand{\hatcurPPaequivxxxxB}{\ensuremath{0.0326\pm0.0035}} 
\newcommand{\hatcurPPtcircxxxxB}{\ensuremath{69_{-40}^{+76}}}  
\newcommand{\hatcurPPtinfallxxxxB}{\ensuremath{790_{-330}^{+550}}} 
\newcommand{\hatcurXdistxxxxB}{\ensuremath{479\pm59}}          
\newcommand{\hatcurXAvxxxxB}{\ensuremath{0.024_{-0.024}^{+0.059}}} 
\newcommand{\hatcurXdistredxxxxB}{\ensuremath{478\pm59}}       
\newcommand{\hatcurXEBVxxxxB}{\ensuremath{0.0080_{-0.0080}^{+0.0190}}} 
\newcommand{\hatcurXmvisoredxxxxB}{\ensuremath{12.481\pm0.027}} 
\newcommand{\hatcurXmiisoredxxxxB}{\ensuremath{11.753\pm0.015}} 
\newcommand{\hatcurXmjisoredxxxxB}{\ensuremath{11.303\pm0.013}} 
\newcommand{\hatcurXmhisoredxxxxB}{\ensuremath{10.967\pm0.015}} 
\newcommand{\hatcurXmkisoredxxxxB}{\ensuremath{10.909\pm0.016}} 
\newcommand{\hatcurXviisoredxxxxB}{\ensuremath{0.728\pm0.017}} 
\newcommand{\hatcurXvkisoredxxxxB}{\ensuremath{1.572\pm0.032}} 
\newcommand{\hatcurXjhisoredxxxxB}{\ensuremath{0.3350\pm0.0080}} 
\newcommand{\hatcurXjkisoredxxxxB}{\ensuremath{0.3930\pm0.0089}} 
\newcommand{\hatcurCCpmraxxxxB}{\ensuremath{-15.5\pm1.2}}      
\newcommand{\hatcurCCpmdecxxxxB}{\ensuremath{-6.9\pm1.1}}      
\newcommand{\hatcurCCpmxxxxB}{\ensuremath{17.0\pm1.6}}         

\newcommand{\hatcurhtrxxxxC}{HATS606-028}                      
\newcommand{\hatcurfieldxxxxC}{\ensuremath{string}}            
\newcommand{\hatcurCCraxxxxC}{\ensuremath{09^{\mathrm h}20^{\mathrm m}21.05{\mathrm s}}}                     
\newcommand{\hatcurCCdecxxxxC}{\ensuremath{-31{\arcdeg}16{\arcmin}09.6{\arcsec}}}                    
\newcommand{\hatcurCCmagxxxxC}{13.669}                         
\newcommand{\hatcurCCtwomassxxxxC}{2MASS~09202105-3116095}     
\newcommand{\hatcurCCgscxxxxC}{GSC~7153-01785}                 
\newcommand{\hatcurCCtassmvxxxxC}{\ensuremath{13.669\pm0.040}} 
\newcommand{\hatcurCCtassmvshortxxxxC}{\ensuremath{13.7}}      
\newcommand{\hatcurCCtassmBxxxxC}{\ensuremath{14.316\pm0.030}} 
\newcommand{\hatcurCCtassmBshortxxxxC}{\ensuremath{14.3}}      
\newcommand{\hatcurCCtassmIxxxxC}{\ensuremath{nff\pmnff}}      
\newcommand{\hatcurCCtassmIshortxxxxC}{\ensuremath{0.0}}       
\newcommand{\hatcurCCtassmgxxxxC}{\ensuremath{13.962\pm0.020}} 
\newcommand{\hatcurCCtassmgshortxxxxC}{\ensuremath{14.0}}      
\newcommand{\hatcurCCtassmrxxxxC}{\ensuremath{13.490\pm0.060}} 
\newcommand{\hatcurCCtassmrshortxxxxC}{\ensuremath{13.5}}      
\newcommand{\hatcurCCtassmixxxxC}{\ensuremath{13.409\pm0.070}} 
\newcommand{\hatcurCCtassmishortxxxxC}{\ensuremath{13.4}}      
\newcommand{\hatcurCCtwomassJmagxxxxC}{\ensuremath{12.523\pm0.034}} 
\newcommand{\hatcurCCtwomassHmagxxxxC}{\ensuremath{12.218\pm0.035}} 
\newcommand{\hatcurCCtwomassKmagxxxxC}{\ensuremath{12.114\pm0.030}} 
\newcommand{\hatcurCCcitJmagxxxxC}{\ensuremath{12.538\pm0.033}} 
\newcommand{\hatcurCCcitHmagxxxxC}{\ensuremath{12.212\pm0.035}} 
\newcommand{\hatcurCCcitKmagxxxxC}{\ensuremath{12.138\pm0.030}} 
\newcommand{\hatcurCCbbJmagxxxxC}{\ensuremath{12.590\pm0.036}} 
\newcommand{\hatcurCCbbHmagxxxxC}{\ensuremath{12.234\pm0.036}} 
\newcommand{\hatcurCCbbKmagxxxxC}{\ensuremath{12.158\pm0.030}} 
\newcommand{\hatcurCCesoJmagxxxxC}{\ensuremath{12.593\pm0.037}} 
\newcommand{\hatcurCCesoHmagxxxxC}{\ensuremath{12.230\pm0.041}} 
\newcommand{\hatcurCCesoKmagxxxxC}{\ensuremath{12.157\pm0.031}} 
\newcommand{\hatcurCCesoJHmagxxxxC}{\ensuremath{0.362\pm0.053}} 
\newcommand{\hatcurCCesoJKmagxxxxC}{\ensuremath{0.436\pm0.048}} 
\newcommand{\hatcurCCesoHKmagxxxxC}{\ensuremath{0.075\pm0.052}} 
\newcommand{\hatcurLCdipxxxxC}{\ensuremath{20.0}}              
\newcommand{\hatcurLCrprstarxxxxC}{\ensuremath{0.1352\pm0.0028}} 
\newcommand{\hatcurLCbsqxxxxC}{\ensuremath{0.231_{-0.074}^{+0.073}}} 
\newcommand{\hatcurLCimpxxxxC}{\ensuremath{0.481_{-0.084}^{+0.071}}} 
\newcommand{\hatcurLCzetaxxxxC}{\ensuremath{26.90\pm0.27}}     
\newcommand{\hatcurLCdurxxxxC}{\ensuremath{0.0871\pm0.0013}}   
\newcommand{\hatcurLCdurshortxxxxC}{\ensuremath{0.0871}}       
\newcommand{\hatcurLCdurhrxxxxC}{\ensuremath{2.091\pm0.032}}   
\newcommand{\hatcurLCdurhrshortxxxxC}{\ensuremath{2.091}}      
\newcommand{\hatcurLCqxxxxC}{\ensuremath{0.06380\pm0.00097}}   
\newcommand{\hatcurLCqshortxxxxC}{\ensuremath{0.064}}          
\newcommand{\hatcurLCingdurxxxxC}{\ensuremath{0.0131\pm0.0013}} 
\newcommand{\hatcurLCPxxxxC}{\ensuremath{1.36665436\pm0.00000094}} 
\newcommand{\hatcurLCPprecxxxxC}{\ensuremath{1.3666544}}       
\newcommand{\hatcurLCPshortxxxxC}{\ensuremath{1.3667}}         
\newcommand{\hatcurLCTxxxxC}{\ensuremath{2456929.03039\pm0.00033}} 
\newcommand{\hatcurLCTAxxxxC}{\ensuremath{2455972.37225\pm0.00075}} 
\newcommand{\hatcurLCTBxxxxC}{\ensuremath{2457318.52686\pm0.00041}} 
\newcommand{\hatcurLChatnetmxxxxC}{\ensuremath{13.62605\pm0.00013}} 
\newcommand{\hatcurLCiblendxxxxC}{\ensuremath{0.916\pm0.048}}  
\newcommand{\hatcurLCrhoxxxxC}{\ensuremath{1.70_{-0.24}^{+0.17}}} 
\newcommand{\hatcurSMEiteffxxxxC}{\ensuremath{5760\pm130}}     
\newcommand{\hatcurSMEizfehxxxxC}{\ensuremath{0.000\pm0.076}}  
\newcommand{\hatcurSMEizfehshortxxxxC}{\ensuremath{0.00}}      
\newcommand{\hatcurSMEiloggxxxxC}{\ensuremath{4.04\pm0.24}}    
\newcommand{\hatcurSMEivsinxxxxC}{\ensuremath{5.02\pm0.94}}    
\newcommand{\hatcurSMEivmacxxxxC}{\ensuremath{3.97\pm0.19}}    
\newcommand{\hatcurSMEivmicxxxxC}{\ensuremath{1.071\pm0.075}}  
\newcommand{\hatcurSMEiiteffxxxxC}{\ensuremath{6010\pm150}}    
\newcommand{\hatcurSMEiizfehxxxxC}{\ensuremath{0.22\pm0.10}}   
\newcommand{\hatcurSMEiizfehshortxxxxC}{\ensuremath{0.22}}     
\newcommand{\hatcurSMEiiloggxxxxC}{\ensuremath{4.448\pm0.066}} 
\newcommand{\hatcurSMEiivsinxxxxC}{\ensuremath{4.59\pm0.64}}   
\newcommand{\hatcurSMEiivmacxxxxC}{\ensuremath{4.35\pm0.23}}   
\newcommand{\hatcurSMEiivmicxxxxC}{\ensuremath{1.24\pm0.12}}   
\newcommand{\hatcurLBizxxxxC}{\ensuremath{0.1828}}             
\newcommand{\hatcurLBiizxxxxC}{\ensuremath{0.3458}}            
\newcommand{\hatcurLBiixxxxC}{\ensuremath{0.2419}}             
\newcommand{\hatcurLBiiixxxxC}{\ensuremath{0.3502}}            
\newcommand{\hatcurLBiIxxxxC}{\ensuremath{0.2216}}             
\newcommand{\hatcurLBiiIxxxxC}{\ensuremath{0.3497}}            
\newcommand{\hatcurLBigxxxxC}{\ensuremath{0.5139}}             
\newcommand{\hatcurLBiigxxxxC}{\ensuremath{0.2679}}            
\newcommand{\hatcurLBirxxxxC}{\ensuremath{0.3262}}             
\newcommand{\hatcurLBiirxxxxC}{\ensuremath{0.3487}}            
\newcommand{\hatcurLBiRxxxxC}{\ensuremath{0.3027}}             
\newcommand{\hatcurLBiiRxxxxC}{\ensuremath{0.3503}}            
\newcommand{\hatcurLBikepxxxxC}{\ensuremath{0.1000}}           
\newcommand{\hatcurLBiikepxxxxC}{\ensuremath{0.1000}}          
\newcommand{\hatcurISOmxxxxC}{\ensuremath{1.111\pm0.054}}      
\newcommand{\hatcurISOmshortxxxxC}{\ensuremath{1.11}}          
\newcommand{\hatcurISOmlongxxxxC}{\ensuremath{1.111\pm0.054}}  
\newcommand{\hatcurISOrxxxxC}{\ensuremath{1.046\pm0.058}}      
\newcommand{\hatcurISOrshortxxxxC}{\ensuremath{1.05}}          
\newcommand{\hatcurISOrlongxxxxC}{\ensuremath{1.046\pm0.058}}  
\newcommand{\hatcurISOrhoxxxxC}{\ensuremath{1.37\pm0.18}}      
\newcommand{\hatcurISOrholongxxxxC}{\ensuremath{1.37\pm0.18}}  
\newcommand{\hatcurISOloggxxxxC}{\ensuremath{4.445\pm0.034}}   
\newcommand{\hatcurISOlumxxxxC}{\ensuremath{1.17\pm0.22}}      
\newcommand{\hatcurISOlumshortxxxxC}{\ensuremath{1.17}}        
\newcommand{\hatcurISOmvxxxxC}{\ensuremath{4.64\pm0.22}}       
\newcommand{\hatcurISOvixxxxC}{\ensuremath{0.672\pm0.040}}     
\newcommand{\hatcurISOagexxxxC}{\ensuremath{1.2_{-1.1}^{+1.5}}} 
\newcommand{\hatcurISOsigmaxxxxC}{\ensuremath{0.00320\pm0.00030}} 
\newcommand{\hatcurISOMJxxxxC}{\ensuremath{3.54\pm0.17}}       
\newcommand{\hatcurISOMHxxxxC}{\ensuremath{3.22\pm0.15}}       
\newcommand{\hatcurISOMKxxxxC}{\ensuremath{3.17\pm0.15}}       
\newcommand{\hatcurISOJKxxxxC}{\ensuremath{0.380\pm0.020}}     
\newcommand{\hatcurISOspecxxxxC}{G}                            
\newcommand{\hatcurRVKxxxxC}{\ensuremath{380\pm23}}            
\newcommand{\hatcurRVrkxxxxC}{\ensuremath{0\pm0}}              
\newcommand{\hatcurRVrhxxxxC}{\ensuremath{0\pm0}}              
\newcommand{\hatcurRVkxxxxC}{\ensuremath{0\pm0}}               
\newcommand{\hatcurRVhxxxxC}{\ensuremath{0\pm0}}               
\newcommand{\hatcurRVtronexxxxC}{\ensuremath{0\pm0}}           
\newcommand{\hatcurRVtrtwoxxxxC}{\ensuremath{0\pm0}}           
\newcommand{\hatcurRVgammaAxxxxC}{\ensuremath{13456\pm43}}     
\newcommand{\hatcurRVjitterAxxxxC}{\ensuremath{124\pm39}}      
\newcommand{\hatcurRVjittertwosiglimAxxxxC}{\ensuremath{<185.6}} 
\newcommand{\hatcurRVfitrmsAxxxxC}{\ensuremath{0.0}}           
\newcommand{\hatcurRVgammaBxxxxC}{\ensuremath{13624\pm0}}      
\newcommand{\hatcurRVjitterBxxxxC}{\ensuremath{0\pm3400}}      
\newcommand{\hatcurRVjittertwosiglimBxxxxC}{\ensuremath{<4138.0}} 
\newcommand{\hatcurRVfitrmsBxxxxC}{\ensuremath{0.0}}           
\newcommand{\hatcurRVgammaCxxxxC}{\ensuremath{13384\pm21}}     
\newcommand{\hatcurRVjitterCxxxxC}{\ensuremath{0\pm32}}        
\newcommand{\hatcurRVjittertwosiglimCxxxxC}{\ensuremath{<114.5}} 
\newcommand{\hatcurRVfitrmsCxxxxC}{\ensuremath{0.0}}           
\newcommand{\hatcurRVeccenxxxxC}{\ensuremath{0\pm0}}           
\newcommand{\hatcurRVeccentwosiglimxxxxC}{\ensuremath{<0.000}} 
\newcommand{\hatcurRVomegaxxxxC}{\ensuremath{0\pm0}}           
\newcommand{\hatcurPPixxxxC}{\ensuremath{84.7\pm1.1}}          
\newcommand{\hatcurPPgxxxxC}{\ensuremath{29.4\pm3.6}}          
\newcommand{\hatcurPPloggxxxxC}{\ensuremath{3.468\pm0.052}}    
\newcommand{\hatcurPParxxxxC}{\ensuremath{5.14\pm0.22}}        
\newcommand{\hatcurPParelxxxxC}{\ensuremath{0.02498\pm0.00040}} 
\newcommand{\hatcurPPrhoxxxxC}{\ensuremath{1.06\pm0.19}}       
\newcommand{\hatcurPPmxxxxC}{\ensuremath{2.24\pm0.15}}         
\newcommand{\hatcurPPmshortxxxxC}{\ensuremath{2.24}}           
\newcommand{\hatcurPPmlongxxxxC}{\ensuremath{2.24\pm0.15}}     
\newcommand{\hatcurPPmexxxxC}{\ensuremath{713\pm49}}           
\newcommand{\hatcurPPmeshortxxxxC}{\ensuremath{713.1}}         
\newcommand{\hatcurPPmelongxxxxC}{\ensuremath{713\pm49}}       
\newcommand{\hatcurPPrxxxxC}{\ensuremath{1.382\pm0.086}}       
\newcommand{\hatcurPPrshortxxxxC}{\ensuremath{1.38}}           
\newcommand{\hatcurPPrlongxxxxC}{\ensuremath{1.382\pm0.086}}   
\newcommand{\hatcurPPrexxxxC}{\ensuremath{15.49\pm0.96}}       
\newcommand{\hatcurPPreshortxxxxC}{\ensuremath{15.5}}          
\newcommand{\hatcurPPrelongxxxxC}{\ensuremath{15.49\pm0.96}}   
\newcommand{\hatcurPPmrcorrxxxxC}{\ensuremath{0.35}}           
\newcommand{\hatcurPPteffxxxxC}{\ensuremath{1834\pm73}}        
\newcommand{\hatcurPPthetaxxxxC}{\ensuremath{0.0725\pm0.0064}} 
\newcommand{\hatcurPPfluxperixxxxC}{\ensuremath{2.55\pm0.41}}  
\newcommand{\hatcurPPfluxperidimxxxxC}{\ensuremath{9}}         
\newcommand{\hatcurPPfluxapxxxxC}{\ensuremath{2.55\pm0.41}}    
\newcommand{\hatcurPPfluxapdimxxxxC}{\ensuremath{9}}           
\newcommand{\hatcurPPfluxavgxxxxC}{\ensuremath{2.55\pm0.41}}   
\newcommand{\hatcurPPfluxavgdimxxxxC}{\ensuremath{9}}          
\newcommand{\hatcurPPfluxavglogxxxxC}{\ensuremath{9.407\pm0.069}} 
\newcommand{\hatcurXsecphasexxxxC}{\ensuremath{0\pm0}}         
\newcommand{\hatcurXsecondaryxxxxC}{\ensuremath{2456929.71371\pm0.00033}} 
\newcommand{\hatcurXsecdurxxxxC}{\ensuremath{0.0871\pm0.0013}} 
\newcommand{\hatcurXsecingdurxxxxC}{\ensuremath{0.0131\pm0.0013}} 
\newcommand{\hatcurPPphiconjxxxxC}{\ensuremath{0\pm0}}         
\newcommand{\hatcurPPperixxxxC}{\ensuremath{2456928.68872\pm0.00033}} 
\newcommand{\hatcurPPaequivxxxxC}{\ensuremath{0.0231\pm0.0018}} 
\newcommand{\hatcurPPtcircxxxxC}{\ensuremath{5.7\pm1.7}}       
\newcommand{\hatcurPPtinfallxxxxC}{\ensuremath{22.9\pm5.0}}    
\newcommand{\hatcurXdistxxxxC}{\ensuremath{628\pm44}}          
\newcommand{\hatcurXAvxxxxC}{\ensuremath{0.025_{-0.025}^{+0.108}}} 
\newcommand{\hatcurXdistredxxxxC}{\ensuremath{631\pm44}}       
\newcommand{\hatcurXEBVxxxxC}{\ensuremath{0.0080_{-0.0080}^{+0.0350}}} 
\newcommand{\hatcurXmvisoredxxxxC}{\ensuremath{13.695\pm0.044}} 
\newcommand{\hatcurXmiisoredxxxxC}{\ensuremath{12.995\pm0.024}} 
\newcommand{\hatcurXmjisoredxxxxC}{\ensuremath{12.554\pm0.019}} 
\newcommand{\hatcurXmhisoredxxxxC}{\ensuremath{12.226\pm0.023}} 
\newcommand{\hatcurXmkisoredxxxxC}{\ensuremath{12.170\pm0.024}} 
\newcommand{\hatcurXviisoredxxxxC}{\ensuremath{0.701\pm0.027}} 
\newcommand{\hatcurXvkisoredxxxxC}{\ensuremath{1.524\pm0.055}} 
\newcommand{\hatcurXjhisoredxxxxC}{\ensuremath{0.327\pm0.017}} 
\newcommand{\hatcurXjkisoredxxxxC}{\ensuremath{0.383\pm0.018}} 
\newcommand{\hatcurCCpmraxxxxC}{\ensuremath{-27.7\pm3.7}}      
\newcommand{\hatcurCCpmdecxxxxC}{\ensuremath{16.0\pm3.7}}      
\newcommand{\hatcurCCpmxxxxC}{\ensuremath{32.0\pm5.2}}         

\newcommand{\hatcurhtrxxxxD}{HATS655-001}                      
\newcommand{\hatcurfieldxxxxD}{\ensuremath{string}}            
\newcommand{\hatcurCCraxxxxD}{\ensuremath{11^{\mathrm h}46^{\mathrm m}30.72{\mathrm s}}}                     
\newcommand{\hatcurCCdecxxxxD}{\ensuremath{-33{\arcdeg}51{\arcmin}36.2{\arcsec}}}                    
\newcommand{\hatcurCCmagxxxxD}{13.790}                         
\newcommand{\hatcurCCtwomassxxxxD}{2MASS~11463084-3351361}     
\newcommand{\hatcurCCgscxxxxD}{GSC~7225-00413}                 
\newcommand{\hatcurCCtassmvxxxxD}{\ensuremath{13.790\pm0.030}} 
\newcommand{\hatcurCCtassmvshortxxxxD}{\ensuremath{13.8}}      
\newcommand{\hatcurCCtassmBxxxxD}{\ensuremath{14.548\pm0.040}} 
\newcommand{\hatcurCCtassmBshortxxxxD}{\ensuremath{14.5}}      
\newcommand{\hatcurCCtassmIxxxxD}{\ensuremath{nff\pmnff}}      
\newcommand{\hatcurCCtassmIshortxxxxD}{\ensuremath{0.0}}       
\newcommand{\hatcurCCtassmgxxxxD}{\ensuremath{14.137\pm0.020}} 
\newcommand{\hatcurCCtassmgshortxxxxD}{\ensuremath{14.1}}      
\newcommand{\hatcurCCtassmrxxxxD}{\ensuremath{13.579\pm0.030}} 
\newcommand{\hatcurCCtassmrshortxxxxD}{\ensuremath{13.6}}      
\newcommand{\hatcurCCtassmixxxxD}{\ensuremath{13.30\pm0.28}}   
\newcommand{\hatcurCCtassmishortxxxxD}{\ensuremath{13.3}}      
\newcommand{\hatcurCCtwomassJmagxxxxD}{\ensuremath{12.458\pm0.025}} 
\newcommand{\hatcurCCtwomassHmagxxxxD}{\ensuremath{12.088\pm0.027}} 
\newcommand{\hatcurCCtwomassKmagxxxxD}{\ensuremath{12.046\pm0.027}} 
\newcommand{\hatcurCCcitJmagxxxxD}{\ensuremath{12.473\pm0.025}} 
\newcommand{\hatcurCCcitHmagxxxxD}{\ensuremath{12.083\pm0.028}} 
\newcommand{\hatcurCCcitKmagxxxxD}{\ensuremath{12.070\pm0.027}} 
\newcommand{\hatcurCCbbJmagxxxxD}{\ensuremath{12.525\pm0.027}} 
\newcommand{\hatcurCCbbHmagxxxxD}{\ensuremath{12.104\pm0.028}} 
\newcommand{\hatcurCCbbKmagxxxxD}{\ensuremath{12.090\pm0.027}} 
\newcommand{\hatcurCCesoJmagxxxxD}{\ensuremath{12.528\pm0.028}} 
\newcommand{\hatcurCCesoHmagxxxxD}{\ensuremath{12.097\pm0.029}} 
\newcommand{\hatcurCCesoKmagxxxxD}{\ensuremath{12.089\pm0.028}} 
\newcommand{\hatcurCCesoJHmagxxxxD}{\ensuremath{0.431\pm0.039}} 
\newcommand{\hatcurCCesoJKmagxxxxD}{\ensuremath{0.439\pm0.040}} 
\newcommand{\hatcurCCesoHKmagxxxxD}{\ensuremath{0.0080\pm0.0070}} 
\newcommand{\hatcurLCdipxxxxD}{\ensuremath{19.0}}              
\newcommand{\hatcurLCrprstarxxxxD}{\ensuremath{0.1250\pm0.0028}} 
\newcommand{\hatcurLCbsqxxxxD}{\ensuremath{0.039_{-0.029}^{+0.039}}} 
\newcommand{\hatcurLCimpxxxxD}{\ensuremath{0.198_{-0.096}^{+0.082}}} 
\newcommand{\hatcurLCzetaxxxxD}{\ensuremath{15.47\pm0.11}}     
\newcommand{\hatcurLCdurxxxxD}{\ensuremath{0.1461\pm0.0016}}   
\newcommand{\hatcurLCdurshortxxxxD}{\ensuremath{0.1461}}       
\newcommand{\hatcurLCdurhrxxxxD}{\ensuremath{3.507\pm0.037}}   
\newcommand{\hatcurLCdurhrshortxxxxD}{\ensuremath{3.507}}      
\newcommand{\hatcurLCqxxxxD}{\ensuremath{0.03790\pm0.00041}}   
\newcommand{\hatcurLCqshortxxxxD}{\ensuremath{0.038}}          
\newcommand{\hatcurLCingdurxxxxD}{\ensuremath{0.01679\pm0.00095}} 
\newcommand{\hatcurLCPxxxxD}{\ensuremath{3.8537768\pm0.0000038}} 
\newcommand{\hatcurLCPprecxxxxD}{\ensuremath{3.8537768}}       
\newcommand{\hatcurLCPshortxxxxD}{\ensuremath{3.8538}}         
\newcommand{\hatcurLCTxxxxD}{\ensuremath{2457236.75653\pm0.00049}} 
\newcommand{\hatcurLCTAxxxxD}{\ensuremath{2455672.1233\pm0.0015}} 
\newcommand{\hatcurLCTBxxxxD}{\ensuremath{2457433.29917\pm0.00055}} 
\newcommand{\hatcurLChatnetmxxxxD}{\ensuremath{13.623140\pm0.000089}} 
\newcommand{\hatcurLCiblendxxxxD}{\ensuremath{0.919\pm0.044}}  
\newcommand{\hatcurLCrhoxxxxD}{\ensuremath{1.026_{-0.071}^{+0.048}}} 
\newcommand{\hatcurSMEiteffxxxxD}{\ensuremath{5700\pm110}}     
\newcommand{\hatcurSMEizfehxxxxD}{\ensuremath{-0.040\pm0.063}} 
\newcommand{\hatcurSMEizfehshortxxxxD}{\ensuremath{-0.04}}     
\newcommand{\hatcurSMEiloggxxxxD}{\ensuremath{4.49\pm0.14}}    
\newcommand{\hatcurSMEivsinxxxxD}{\ensuremath{2.04\pm0.98}}    
\newcommand{\hatcurSMEivmacxxxxD}{\ensuremath{3.87\pm0.17}}    
\newcommand{\hatcurSMEivmicxxxxD}{\ensuremath{1.037\pm0.063}}  
\newcommand{\hatcurSMEiiteffxxxxD}{\ensuremath{5644\pm94}}     
\newcommand{\hatcurSMEiizfehxxxxD}{\ensuremath{0.010\pm0.066}} 
\newcommand{\hatcurSMEiizfehshortxxxxD}{\ensuremath{0.01}}     
\newcommand{\hatcurSMEiiloggxxxxD}{\ensuremath{4.338\pm0.022}} 
\newcommand{\hatcurSMEiivsinxxxxD}{\ensuremath{2.50\pm0.76}}   
\newcommand{\hatcurSMEiivmacxxxxD}{\ensuremath{3.79\pm0.14}}   
\newcommand{\hatcurSMEiivmicxxxxD}{\ensuremath{1.006\pm0.049}} 
\newcommand{\hatcurLBizxxxxD}{\ensuremath{0.2257}}             
\newcommand{\hatcurLBiizxxxxD}{\ensuremath{0.3193}}            
\newcommand{\hatcurLBiixxxxD}{\ensuremath{0.2888}}             
\newcommand{\hatcurLBiiixxxxD}{\ensuremath{0.3179}}            
\newcommand{\hatcurLBiIxxxxD}{\ensuremath{0.2676}}             
\newcommand{\hatcurLBiiIxxxxD}{\ensuremath{0.3192}}            
\newcommand{\hatcurLBigxxxxD}{\ensuremath{0.5830}}             
\newcommand{\hatcurLBiigxxxxD}{\ensuremath{0.2134}}            
\newcommand{\hatcurLBirxxxxD}{\ensuremath{0.3816}}             
\newcommand{\hatcurLBiirxxxxD}{\ensuremath{0.3101}}            
\newcommand{\hatcurLBiRxxxxD}{\ensuremath{0.3559}}             
\newcommand{\hatcurLBiiRxxxxD}{\ensuremath{0.3132}}            
\newcommand{\hatcurLBikepxxxxD}{\ensuremath{0.1000}}           
\newcommand{\hatcurLBiikepxxxxD}{\ensuremath{0.1000}}          
\newcommand{\hatcurISOmxxxxD}{\ensuremath{0.964\pm0.040}}      
\newcommand{\hatcurISOmshortxxxxD}{\ensuremath{0.96}}          
\newcommand{\hatcurISOmlongxxxxD}{\ensuremath{0.964\pm0.040}}  
\newcommand{\hatcurISOrxxxxD}{\ensuremath{1.101_{-0.024}^{+0.031}}} 
\newcommand{\hatcurISOrshortxxxxD}{\ensuremath{1.10}}          
\newcommand{\hatcurISOrlongxxxxD}{\ensuremath{1.101_{-0.024}^{+0.031}}} 
\newcommand{\hatcurISOrhoxxxxD}{\ensuremath{1.027_{-0.071}^{+0.050}}} 
\newcommand{\hatcurISOrholongxxxxD}{\ensuremath{1.027_{-0.071}^{+0.050}}} 
\newcommand{\hatcurISOloggxxxxD}{\ensuremath{4.340\pm0.019}}   
\newcommand{\hatcurISOlumxxxxD}{\ensuremath{1.11\pm0.11}}      
\newcommand{\hatcurISOlumshortxxxxD}{\ensuremath{1.11}}        
\newcommand{\hatcurISOmvxxxxD}{\ensuremath{4.74\pm0.12}}       
\newcommand{\hatcurISOvixxxxD}{\ensuremath{0.736\pm0.027}}     
\newcommand{\hatcurISOagexxxxD}{\ensuremath{9.0\pm1.9}}        
\newcommand{\hatcurISOsigmaxxxxD}{\ensuremath{0.00070\pm0.00012}} 
\newcommand{\hatcurISOMJxxxxD}{\ensuremath{3.543\pm0.081}}     
\newcommand{\hatcurISOMHxxxxD}{\ensuremath{3.179\pm0.068}}     
\newcommand{\hatcurISOMKxxxxD}{\ensuremath{3.121\pm0.066}}     
\newcommand{\hatcurISOJKxxxxD}{\ensuremath{0.420\pm0.020}}     
\newcommand{\hatcurISOspecxxxxD}{G}                            
\newcommand{\hatcurRVKxxxxD}{\ensuremath{79\pm12}}             
\newcommand{\hatcurRVrkxxxxD}{\ensuremath{0\pm0}}              
\newcommand{\hatcurRVrhxxxxD}{\ensuremath{0\pm0}}              
\newcommand{\hatcurRVkxxxxD}{\ensuremath{0\pm0}}               
\newcommand{\hatcurRVhxxxxD}{\ensuremath{0\pm0}}               
\newcommand{\hatcurRVtronexxxxD}{\ensuremath{0\pm0}}           
\newcommand{\hatcurRVtrtwoxxxxD}{\ensuremath{0\pm0}}           
\newcommand{\hatcurRVgammaAxxxxD}{\ensuremath{71949.5\pm7.1}}  
\newcommand{\hatcurRVjitterAxxxxD}{\ensuremath{18\pm12}}       
\newcommand{\hatcurRVjittertwosiglimAxxxxD}{\ensuremath{<34.4}} 
\newcommand{\hatcurRVfitrmsAxxxxD}{\ensuremath{0.0}}           
\newcommand{\hatcurRVgammaBxxxxD}{\ensuremath{71945\pm24}}     
\newcommand{\hatcurRVjitterBxxxxD}{\ensuremath{0\pm15}}        
\newcommand{\hatcurRVjittertwosiglimBxxxxD}{\ensuremath{<31.0}} 
\newcommand{\hatcurRVfitrmsBxxxxD}{\ensuremath{0.0}}           
\newcommand{\hatcurRVeccenxxxxD}{\ensuremath{0\pm0}}           
\newcommand{\hatcurRVeccentwosiglimxxxxD}{\ensuremath{<0.000}} 
\newcommand{\hatcurRVomegaxxxxD}{\ensuremath{0\pm0}}           
\newcommand{\hatcurPPixxxxD}{\ensuremath{88.78\pm0.55}}        
\newcommand{\hatcurPPgxxxxD}{\ensuremath{8.2\pm1.3}}           
\newcommand{\hatcurPPloggxxxxD}{\ensuremath{2.912_{-0.087}^{+0.060}}} 
\newcommand{\hatcurPParxxxxD}{\ensuremath{9.30_{-0.22}^{+0.15}}} 
\newcommand{\hatcurPParelxxxxD}{\ensuremath{0.04753\pm0.00066}} 
\newcommand{\hatcurPPrhoxxxxD}{\ensuremath{0.303\pm0.055}}     
\newcommand{\hatcurPPmxxxxD}{\ensuremath{0.595\pm0.089}}       
\newcommand{\hatcurPPmshortxxxxD}{\ensuremath{0.60}}           
\newcommand{\hatcurPPmlongxxxxD}{\ensuremath{0.595\pm0.089}}   
\newcommand{\hatcurPPmexxxxD}{\ensuremath{189\pm28}}           
\newcommand{\hatcurPPmeshortxxxxD}{\ensuremath{189.2}}         
\newcommand{\hatcurPPmelongxxxxD}{\ensuremath{189\pm28}}       
\newcommand{\hatcurPPrxxxxD}{\ensuremath{1.340\pm0.056}}       
\newcommand{\hatcurPPrshortxxxxD}{\ensuremath{1.34}}           
\newcommand{\hatcurPPrlongxxxxD}{\ensuremath{1.340\pm0.056}}   
\newcommand{\hatcurPPrexxxxD}{\ensuremath{15.02\pm0.62}}       
\newcommand{\hatcurPPreshortxxxxD}{\ensuremath{15.0}}          
\newcommand{\hatcurPPrelongxxxxD}{\ensuremath{15.02\pm0.62}}   
\newcommand{\hatcurPPmrcorrxxxxD}{\ensuremath{0.11}}           
\newcommand{\hatcurPPteffxxxxD}{\ensuremath{1312\pm25}}        
\newcommand{\hatcurPPthetaxxxxD}{\ensuremath{0.0436_{-0.0073}^{+0.0051}}} 
\newcommand{\hatcurPPfluxperixxxxD}{\ensuremath{6.69\pm0.52}}  
\newcommand{\hatcurPPfluxperidimxxxxD}{\ensuremath{8}}         
\newcommand{\hatcurPPfluxapxxxxD}{\ensuremath{6.69\pm0.52}}    
\newcommand{\hatcurPPfluxapdimxxxxD}{\ensuremath{8}}           
\newcommand{\hatcurPPfluxavgxxxxD}{\ensuremath{6.69\pm0.52}}   
\newcommand{\hatcurPPfluxavgdimxxxxD}{\ensuremath{8}}          
\newcommand{\hatcurPPfluxavglogxxxxD}{\ensuremath{8.825\pm0.033}} 
\newcommand{\hatcurXsecphasexxxxD}{\ensuremath{0\pm0}}         
\newcommand{\hatcurXsecondaryxxxxD}{\ensuremath{2457238.68341\pm0.00049}} 
\newcommand{\hatcurXsecdurxxxxD}{\ensuremath{0.1461\pm0.0016}} 
\newcommand{\hatcurXsecingdurxxxxD}{\ensuremath{0.01679\pm0.00095}} 
\newcommand{\hatcurPPphiconjxxxxD}{\ensuremath{0\pm0}}         
\newcommand{\hatcurPPperixxxxD}{\ensuremath{2457235.79308\pm0.00049}} 
\newcommand{\hatcurPPaequivxxxxD}{\ensuremath{0.0452\pm0.0017}} 
\newcommand{\hatcurPPtcircxxxxD}{\ensuremath{139\pm33}}        
\newcommand{\hatcurPPtinfallxxxxD}{\ensuremath{4140\pm870}}    
\newcommand{\hatcurXdistxxxxD}{\ensuremath{622\pm21}}          
\newcommand{\hatcurXAvxxxxD}{\ensuremath{0.112\pm0.078}}       
\newcommand{\hatcurXdistredxxxxD}{\ensuremath{613\pm19}}       
\newcommand{\hatcurXEBVxxxxD}{\ensuremath{0.036\pm0.025}}      
\newcommand{\hatcurXmvisoredxxxxD}{\ensuremath{13.792\pm0.029}} 
\newcommand{\hatcurXmiisoredxxxxD}{\ensuremath{12.997\pm0.019}} 
\newcommand{\hatcurXmjisoredxxxxD}{\ensuremath{12.511\pm0.016}} 
\newcommand{\hatcurXmhisoredxxxxD}{\ensuremath{12.136\pm0.017}} 
\newcommand{\hatcurXmkisoredxxxxD}{\ensuremath{12.068\pm0.018}} 
\newcommand{\hatcurXviisoredxxxxD}{\ensuremath{0.794\pm0.022}} 
\newcommand{\hatcurXvkisoredxxxxD}{\ensuremath{1.724\pm0.036}} 
\newcommand{\hatcurXjhisoredxxxxD}{\ensuremath{0.375\pm0.012}} 
\newcommand{\hatcurXjkisoredxxxxD}{\ensuremath{0.442\pm0.012}} 
\newcommand{\hatcurCCpmraxxxxD}{\ensuremath{-35.0\pm2.0}}      
\newcommand{\hatcurCCpmdecxxxxD}{\ensuremath{-5.0\pm2.1}}      
\newcommand{\hatcurCCpmxxxxD}{\ensuremath{35.4\pm2.9}}         

\newcommand{\hatcurCCbbHmag}[1]{\ifnum#1=50 %
\hatcurCCbbHmagxxxxA
\else
\ifnum#1=51 %
\hatcurCCbbHmagxxxxB
\else
\ifnum#1=52 %
\hatcurCCbbHmagxxxxC
\else
\ifnum#1=53 %
\hatcurCCbbHmagxxxxD
\else
??????\fi
\fi
\fi
\fi
}
\newcommand{\hatcurCCbbJmag}[1]{\ifnum#1=50 %
\hatcurCCbbJmagxxxxA
\else
\ifnum#1=51 %
\hatcurCCbbJmagxxxxB
\else
\ifnum#1=52 %
\hatcurCCbbJmagxxxxC
\else
\ifnum#1=53 %
\hatcurCCbbJmagxxxxD
\else
??????\fi
\fi
\fi
\fi
}
\newcommand{\hatcurCCbbKmag}[1]{\ifnum#1=50 %
\hatcurCCbbKmagxxxxA
\else
\ifnum#1=51 %
\hatcurCCbbKmagxxxxB
\else
\ifnum#1=52 %
\hatcurCCbbKmagxxxxC
\else
\ifnum#1=53 %
\hatcurCCbbKmagxxxxD
\else
??????\fi
\fi
\fi
\fi
}
\newcommand{\hatcurCCcitHmag}[1]{\ifnum#1=50 %
\hatcurCCcitHmagxxxxA
\else
\ifnum#1=51 %
\hatcurCCcitHmagxxxxB
\else
\ifnum#1=52 %
\hatcurCCcitHmagxxxxC
\else
\ifnum#1=53 %
\hatcurCCcitHmagxxxxD
\else
??????\fi
\fi
\fi
\fi
}
\newcommand{\hatcurCCcitJmag}[1]{\ifnum#1=50 %
\hatcurCCcitJmagxxxxA
\else
\ifnum#1=51 %
\hatcurCCcitJmagxxxxB
\else
\ifnum#1=52 %
\hatcurCCcitJmagxxxxC
\else
\ifnum#1=53 %
\hatcurCCcitJmagxxxxD
\else
??????\fi
\fi
\fi
\fi
}
\newcommand{\hatcurCCcitKmag}[1]{\ifnum#1=50 %
\hatcurCCcitKmagxxxxA
\else
\ifnum#1=51 %
\hatcurCCcitKmagxxxxB
\else
\ifnum#1=52 %
\hatcurCCcitKmagxxxxC
\else
\ifnum#1=53 %
\hatcurCCcitKmagxxxxD
\else
??????\fi
\fi
\fi
\fi
}
\newcommand{\hatcurCCdec}[1]{\ifnum#1=50 %
\hatcurCCdecxxxxA
\else
\ifnum#1=51 %
\hatcurCCdecxxxxB
\else
\ifnum#1=52 %
\hatcurCCdecxxxxC
\else
\ifnum#1=53 %
\hatcurCCdecxxxxD
\else
??????\fi
\fi
\fi
\fi
}
\newcommand{\hatcurCCesoHKmag}[1]{\ifnum#1=50 %
\hatcurCCesoHKmagxxxxA
\else
\ifnum#1=51 %
\hatcurCCesoHKmagxxxxB
\else
\ifnum#1=52 %
\hatcurCCesoHKmagxxxxC
\else
\ifnum#1=53 %
\hatcurCCesoHKmagxxxxD
\else
??????\fi
\fi
\fi
\fi
}
\newcommand{\hatcurCCesoHmag}[1]{\ifnum#1=50 %
\hatcurCCesoHmagxxxxA
\else
\ifnum#1=51 %
\hatcurCCesoHmagxxxxB
\else
\ifnum#1=52 %
\hatcurCCesoHmagxxxxC
\else
\ifnum#1=53 %
\hatcurCCesoHmagxxxxD
\else
??????\fi
\fi
\fi
\fi
}
\newcommand{\hatcurCCesoJHmag}[1]{\ifnum#1=50 %
\hatcurCCesoJHmagxxxxA
\else
\ifnum#1=51 %
\hatcurCCesoJHmagxxxxB
\else
\ifnum#1=52 %
\hatcurCCesoJHmagxxxxC
\else
\ifnum#1=53 %
\hatcurCCesoJHmagxxxxD
\else
??????\fi
\fi
\fi
\fi
}
\newcommand{\hatcurCCesoJKmag}[1]{\ifnum#1=50 %
\hatcurCCesoJKmagxxxxA
\else
\ifnum#1=51 %
\hatcurCCesoJKmagxxxxB
\else
\ifnum#1=52 %
\hatcurCCesoJKmagxxxxC
\else
\ifnum#1=53 %
\hatcurCCesoJKmagxxxxD
\else
??????\fi
\fi
\fi
\fi
}
\newcommand{\hatcurCCesoJmag}[1]{\ifnum#1=50 %
\hatcurCCesoJmagxxxxA
\else
\ifnum#1=51 %
\hatcurCCesoJmagxxxxB
\else
\ifnum#1=52 %
\hatcurCCesoJmagxxxxC
\else
\ifnum#1=53 %
\hatcurCCesoJmagxxxxD
\else
??????\fi
\fi
\fi
\fi
}
\newcommand{\hatcurCCesoKmag}[1]{\ifnum#1=50 %
\hatcurCCesoKmagxxxxA
\else
\ifnum#1=51 %
\hatcurCCesoKmagxxxxB
\else
\ifnum#1=52 %
\hatcurCCesoKmagxxxxC
\else
\ifnum#1=53 %
\hatcurCCesoKmagxxxxD
\else
??????\fi
\fi
\fi
\fi
}
\newcommand{\hatcurCCgsc}[1]{\ifnum#1=50 %
\hatcurCCgscxxxxA
\else
\ifnum#1=51 %
\hatcurCCgscxxxxB
\else
\ifnum#1=52 %
\hatcurCCgscxxxxC
\else
\ifnum#1=53 %
\hatcurCCgscxxxxD
\else
??????\fi
\fi
\fi
\fi
}
\newcommand{\hatcurCCmag}[1]{\ifnum#1=50 %
\hatcurCCmagxxxxA
\else
\ifnum#1=51 %
\hatcurCCmagxxxxB
\else
\ifnum#1=52 %
\hatcurCCmagxxxxC
\else
\ifnum#1=53 %
\hatcurCCmagxxxxD
\else
??????\fi
\fi
\fi
\fi
}
\newcommand{\hatcurCCpm}[1]{\ifnum#1=50 %
\hatcurCCpmxxxxA
\else
\ifnum#1=51 %
\hatcurCCpmxxxxB
\else
\ifnum#1=52 %
\hatcurCCpmxxxxC
\else
\ifnum#1=53 %
\hatcurCCpmxxxxD
\else
??????\fi
\fi
\fi
\fi
}
\newcommand{\hatcurCCpmdec}[1]{\ifnum#1=50 %
\hatcurCCpmdecxxxxA
\else
\ifnum#1=51 %
\hatcurCCpmdecxxxxB
\else
\ifnum#1=52 %
\hatcurCCpmdecxxxxC
\else
\ifnum#1=53 %
\hatcurCCpmdecxxxxD
\else
??????\fi
\fi
\fi
\fi
}
\newcommand{\hatcurCCpmra}[1]{\ifnum#1=50 %
\hatcurCCpmraxxxxA
\else
\ifnum#1=51 %
\hatcurCCpmraxxxxB
\else
\ifnum#1=52 %
\hatcurCCpmraxxxxC
\else
\ifnum#1=53 %
\hatcurCCpmraxxxxD
\else
??????\fi
\fi
\fi
\fi
}
\newcommand{\hatcurCCra}[1]{\ifnum#1=50 %
\hatcurCCraxxxxA
\else
\ifnum#1=51 %
\hatcurCCraxxxxB
\else
\ifnum#1=52 %
\hatcurCCraxxxxC
\else
\ifnum#1=53 %
\hatcurCCraxxxxD
\else
??????\fi
\fi
\fi
\fi
}
\newcommand{\hatcurCCtassmB}[1]{\ifnum#1=50 %
\hatcurCCtassmBxxxxA
\else
\ifnum#1=51 %
\hatcurCCtassmBxxxxB
\else
\ifnum#1=52 %
\hatcurCCtassmBxxxxC
\else
\ifnum#1=53 %
\hatcurCCtassmBxxxxD
\else
??????\fi
\fi
\fi
\fi
}
\newcommand{\hatcurCCtassmBshort}[1]{\ifnum#1=50 %
\hatcurCCtassmBshortxxxxA
\else
\ifnum#1=51 %
\hatcurCCtassmBshortxxxxB
\else
\ifnum#1=52 %
\hatcurCCtassmBshortxxxxC
\else
\ifnum#1=53 %
\hatcurCCtassmBshortxxxxD
\else
??????\fi
\fi
\fi
\fi
}
\newcommand{\hatcurCCtassmg}[1]{\ifnum#1=50 %
\hatcurCCtassmgxxxxA
\else
\ifnum#1=51 %
\hatcurCCtassmgxxxxB
\else
\ifnum#1=52 %
\hatcurCCtassmgxxxxC
\else
\ifnum#1=53 %
\hatcurCCtassmgxxxxD
\else
??????\fi
\fi
\fi
\fi
}
\newcommand{\hatcurCCtassmgshort}[1]{\ifnum#1=50 %
\hatcurCCtassmgshortxxxxA
\else
\ifnum#1=51 %
\hatcurCCtassmgshortxxxxB
\else
\ifnum#1=52 %
\hatcurCCtassmgshortxxxxC
\else
\ifnum#1=53 %
\hatcurCCtassmgshortxxxxD
\else
??????\fi
\fi
\fi
\fi
}
\newcommand{\hatcurCCtassmi}[1]{\ifnum#1=50 %
\hatcurCCtassmixxxxA
\else
\ifnum#1=51 %
\hatcurCCtassmixxxxB
\else
\ifnum#1=52 %
\hatcurCCtassmixxxxC
\else
\ifnum#1=53 %
\hatcurCCtassmixxxxD
\else
??????\fi
\fi
\fi
\fi
}
\newcommand{\hatcurCCtassmI}[1]{\ifnum#1=50 %
\hatcurCCtassmIxxxxA
\else
\ifnum#1=51 %
\hatcurCCtassmIxxxxB
\else
\ifnum#1=52 %
\hatcurCCtassmIxxxxC
\else
\ifnum#1=53 %
\hatcurCCtassmIxxxxD
\else
??????\fi
\fi
\fi
\fi
}
\newcommand{\hatcurCCtassmishort}[1]{\ifnum#1=50 %
\hatcurCCtassmishortxxxxA
\else
\ifnum#1=51 %
\hatcurCCtassmishortxxxxB
\else
\ifnum#1=52 %
\hatcurCCtassmishortxxxxC
\else
\ifnum#1=53 %
\hatcurCCtassmishortxxxxD
\else
??????\fi
\fi
\fi
\fi
}
\newcommand{\hatcurCCtassmIshort}[1]{\ifnum#1=50 %
\hatcurCCtassmIshortxxxxA
\else
\ifnum#1=51 %
\hatcurCCtassmIshortxxxxB
\else
\ifnum#1=52 %
\hatcurCCtassmIshortxxxxC
\else
\ifnum#1=53 %
\hatcurCCtassmIshortxxxxD
\else
??????\fi
\fi
\fi
\fi
}
\newcommand{\hatcurCCtassmr}[1]{\ifnum#1=50 %
\hatcurCCtassmrxxxxA
\else
\ifnum#1=51 %
\hatcurCCtassmrxxxxB
\else
\ifnum#1=52 %
\hatcurCCtassmrxxxxC
\else
\ifnum#1=53 %
\hatcurCCtassmrxxxxD
\else
??????\fi
\fi
\fi
\fi
}
\newcommand{\hatcurCCtassmrshort}[1]{\ifnum#1=50 %
\hatcurCCtassmrshortxxxxA
\else
\ifnum#1=51 %
\hatcurCCtassmrshortxxxxB
\else
\ifnum#1=52 %
\hatcurCCtassmrshortxxxxC
\else
\ifnum#1=53 %
\hatcurCCtassmrshortxxxxD
\else
??????\fi
\fi
\fi
\fi
}
\newcommand{\hatcurCCtassmv}[1]{\ifnum#1=50 %
\hatcurCCtassmvxxxxA
\else
\ifnum#1=51 %
\hatcurCCtassmvxxxxB
\else
\ifnum#1=52 %
\hatcurCCtassmvxxxxC
\else
\ifnum#1=53 %
\hatcurCCtassmvxxxxD
\else
??????\fi
\fi
\fi
\fi
}
\newcommand{\hatcurCCtassmvshort}[1]{\ifnum#1=50 %
\hatcurCCtassmvshortxxxxA
\else
\ifnum#1=51 %
\hatcurCCtassmvshortxxxxB
\else
\ifnum#1=52 %
\hatcurCCtassmvshortxxxxC
\else
\ifnum#1=53 %
\hatcurCCtassmvshortxxxxD
\else
??????\fi
\fi
\fi
\fi
}
\newcommand{\hatcurCCtwomass}[1]{\ifnum#1=50 %
\hatcurCCtwomassxxxxA
\else
\ifnum#1=51 %
\hatcurCCtwomassxxxxB
\else
\ifnum#1=52 %
\hatcurCCtwomassxxxxC
\else
\ifnum#1=53 %
\hatcurCCtwomassxxxxD
\else
??????\fi
\fi
\fi
\fi
}
\newcommand{\hatcurCCtwomassHmag}[1]{\ifnum#1=50 %
\hatcurCCtwomassHmagxxxxA
\else
\ifnum#1=51 %
\hatcurCCtwomassHmagxxxxB
\else
\ifnum#1=52 %
\hatcurCCtwomassHmagxxxxC
\else
\ifnum#1=53 %
\hatcurCCtwomassHmagxxxxD
\else
??????\fi
\fi
\fi
\fi
}
\newcommand{\hatcurCCtwomassJmag}[1]{\ifnum#1=50 %
\hatcurCCtwomassJmagxxxxA
\else
\ifnum#1=51 %
\hatcurCCtwomassJmagxxxxB
\else
\ifnum#1=52 %
\hatcurCCtwomassJmagxxxxC
\else
\ifnum#1=53 %
\hatcurCCtwomassJmagxxxxD
\else
??????\fi
\fi
\fi
\fi
}
\newcommand{\hatcurCCtwomassKmag}[1]{\ifnum#1=50 %
\hatcurCCtwomassKmagxxxxA
\else
\ifnum#1=51 %
\hatcurCCtwomassKmagxxxxB
\else
\ifnum#1=52 %
\hatcurCCtwomassKmagxxxxC
\else
\ifnum#1=53 %
\hatcurCCtwomassKmagxxxxD
\else
??????\fi
\fi
\fi
\fi
}
\newcommand{\hatcurfield}[1]{\ifnum#1=50 %
\hatcurfieldxxxxA
\else
\ifnum#1=51 %
\hatcurfieldxxxxB
\else
\ifnum#1=52 %
\hatcurfieldxxxxC
\else
\ifnum#1=53 %
\hatcurfieldxxxxD
\else
??????\fi
\fi
\fi
\fi
}
\newcommand{\hatcurhtr}[1]{\ifnum#1=50 %
\hatcurhtrxxxxA
\else
\ifnum#1=51 %
\hatcurhtrxxxxB
\else
\ifnum#1=52 %
\hatcurhtrxxxxC
\else
\ifnum#1=53 %
\hatcurhtrxxxxD
\else
??????\fi
\fi
\fi
\fi
}
\newcommand{\hatcurISOage}[1]{\ifnum#1=50 %
\hatcurISOagexxxxA
\else
\ifnum#1=51 %
\hatcurISOagexxxxB
\else
\ifnum#1=52 %
\hatcurISOagexxxxC
\else
\ifnum#1=53 %
\hatcurISOagexxxxD
\else
??????\fi
\fi
\fi
\fi
}
\newcommand{\hatcurISOJK}[1]{\ifnum#1=50 %
\hatcurISOJKxxxxA
\else
\ifnum#1=51 %
\hatcurISOJKxxxxB
\else
\ifnum#1=52 %
\hatcurISOJKxxxxC
\else
\ifnum#1=53 %
\hatcurISOJKxxxxD
\else
??????\fi
\fi
\fi
\fi
}
\newcommand{\hatcurISOlogg}[1]{\ifnum#1=50 %
\hatcurISOloggxxxxA
\else
\ifnum#1=51 %
\hatcurISOloggxxxxB
\else
\ifnum#1=52 %
\hatcurISOloggxxxxC
\else
\ifnum#1=53 %
\hatcurISOloggxxxxD
\else
??????\fi
\fi
\fi
\fi
}
\newcommand{\hatcurISOlum}[1]{\ifnum#1=50 %
\hatcurISOlumxxxxA
\else
\ifnum#1=51 %
\hatcurISOlumxxxxB
\else
\ifnum#1=52 %
\hatcurISOlumxxxxC
\else
\ifnum#1=53 %
\hatcurISOlumxxxxD
\else
??????\fi
\fi
\fi
\fi
}
\newcommand{\hatcurISOlumshort}[1]{\ifnum#1=50 %
\hatcurISOlumshortxxxxA
\else
\ifnum#1=51 %
\hatcurISOlumshortxxxxB
\else
\ifnum#1=52 %
\hatcurISOlumshortxxxxC
\else
\ifnum#1=53 %
\hatcurISOlumshortxxxxD
\else
??????\fi
\fi
\fi
\fi
}
\newcommand{\hatcurISOm}[1]{\ifnum#1=50 %
\hatcurISOmxxxxA
\else
\ifnum#1=51 %
\hatcurISOmxxxxB
\else
\ifnum#1=52 %
\hatcurISOmxxxxC
\else
\ifnum#1=53 %
\hatcurISOmxxxxD
\else
??????\fi
\fi
\fi
\fi
}
\newcommand{\hatcurISOMH}[1]{\ifnum#1=50 %
\hatcurISOMHxxxxA
\else
\ifnum#1=51 %
\hatcurISOMHxxxxB
\else
\ifnum#1=52 %
\hatcurISOMHxxxxC
\else
\ifnum#1=53 %
\hatcurISOMHxxxxD
\else
??????\fi
\fi
\fi
\fi
}
\newcommand{\hatcurISOMJ}[1]{\ifnum#1=50 %
\hatcurISOMJxxxxA
\else
\ifnum#1=51 %
\hatcurISOMJxxxxB
\else
\ifnum#1=52 %
\hatcurISOMJxxxxC
\else
\ifnum#1=53 %
\hatcurISOMJxxxxD
\else
??????\fi
\fi
\fi
\fi
}
\newcommand{\hatcurISOMK}[1]{\ifnum#1=50 %
\hatcurISOMKxxxxA
\else
\ifnum#1=51 %
\hatcurISOMKxxxxB
\else
\ifnum#1=52 %
\hatcurISOMKxxxxC
\else
\ifnum#1=53 %
\hatcurISOMKxxxxD
\else
??????\fi
\fi
\fi
\fi
}
\newcommand{\hatcurISOmlong}[1]{\ifnum#1=50 %
\hatcurISOmlongxxxxA
\else
\ifnum#1=51 %
\hatcurISOmlongxxxxB
\else
\ifnum#1=52 %
\hatcurISOmlongxxxxC
\else
\ifnum#1=53 %
\hatcurISOmlongxxxxD
\else
??????\fi
\fi
\fi
\fi
}
\newcommand{\hatcurISOmshort}[1]{\ifnum#1=50 %
\hatcurISOmshortxxxxA
\else
\ifnum#1=51 %
\hatcurISOmshortxxxxB
\else
\ifnum#1=52 %
\hatcurISOmshortxxxxC
\else
\ifnum#1=53 %
\hatcurISOmshortxxxxD
\else
??????\fi
\fi
\fi
\fi
}
\newcommand{\hatcurISOmv}[1]{\ifnum#1=50 %
\hatcurISOmvxxxxA
\else
\ifnum#1=51 %
\hatcurISOmvxxxxB
\else
\ifnum#1=52 %
\hatcurISOmvxxxxC
\else
\ifnum#1=53 %
\hatcurISOmvxxxxD
\else
??????\fi
\fi
\fi
\fi
}
\newcommand{\hatcurISOr}[1]{\ifnum#1=50 %
\hatcurISOrxxxxA
\else
\ifnum#1=51 %
\hatcurISOrxxxxB
\else
\ifnum#1=52 %
\hatcurISOrxxxxC
\else
\ifnum#1=53 %
\hatcurISOrxxxxD
\else
??????\fi
\fi
\fi
\fi
}
\newcommand{\hatcurISOrho}[1]{\ifnum#1=50 %
\hatcurISOrhoxxxxA
\else
\ifnum#1=51 %
\hatcurISOrhoxxxxB
\else
\ifnum#1=52 %
\hatcurISOrhoxxxxC
\else
\ifnum#1=53 %
\hatcurISOrhoxxxxD
\else
??????\fi
\fi
\fi
\fi
}
\newcommand{\hatcurISOrholong}[1]{\ifnum#1=50 %
\hatcurISOrholongxxxxA
\else
\ifnum#1=51 %
\hatcurISOrholongxxxxB
\else
\ifnum#1=52 %
\hatcurISOrholongxxxxC
\else
\ifnum#1=53 %
\hatcurISOrholongxxxxD
\else
??????\fi
\fi
\fi
\fi
}
\newcommand{\hatcurISOrlong}[1]{\ifnum#1=50 %
\hatcurISOrlongxxxxA
\else
\ifnum#1=51 %
\hatcurISOrlongxxxxB
\else
\ifnum#1=52 %
\hatcurISOrlongxxxxC
\else
\ifnum#1=53 %
\hatcurISOrlongxxxxD
\else
??????\fi
\fi
\fi
\fi
}
\newcommand{\hatcurISOrshort}[1]{\ifnum#1=50 %
\hatcurISOrshortxxxxA
\else
\ifnum#1=51 %
\hatcurISOrshortxxxxB
\else
\ifnum#1=52 %
\hatcurISOrshortxxxxC
\else
\ifnum#1=53 %
\hatcurISOrshortxxxxD
\else
??????\fi
\fi
\fi
\fi
}
\newcommand{\hatcurISOsigma}[1]{\ifnum#1=50 %
\hatcurISOsigmaxxxxA
\else
\ifnum#1=51 %
\hatcurISOsigmaxxxxB
\else
\ifnum#1=52 %
\hatcurISOsigmaxxxxC
\else
\ifnum#1=53 %
\hatcurISOsigmaxxxxD
\else
??????\fi
\fi
\fi
\fi
}
\newcommand{\hatcurISOspec}[1]{\ifnum#1=50 %
\hatcurISOspecxxxxA
\else
\ifnum#1=51 %
\hatcurISOspecxxxxB
\else
\ifnum#1=52 %
\hatcurISOspecxxxxC
\else
\ifnum#1=53 %
\hatcurISOspecxxxxD
\else
??????\fi
\fi
\fi
\fi
}
\newcommand{\hatcurISOvi}[1]{\ifnum#1=50 %
\hatcurISOvixxxxA
\else
\ifnum#1=51 %
\hatcurISOvixxxxB
\else
\ifnum#1=52 %
\hatcurISOvixxxxC
\else
\ifnum#1=53 %
\hatcurISOvixxxxD
\else
??????\fi
\fi
\fi
\fi
}
\newcommand{\hatcurLBig}[1]{\ifnum#1=50 %
\hatcurLBigxxxxA
\else
\ifnum#1=51 %
\hatcurLBigxxxxB
\else
\ifnum#1=52 %
\hatcurLBigxxxxC
\else
\ifnum#1=53 %
\hatcurLBigxxxxD
\else
??????\fi
\fi
\fi
\fi
}
\newcommand{\hatcurLBii}[1]{\ifnum#1=50 %
\hatcurLBiixxxxA
\else
\ifnum#1=51 %
\hatcurLBiixxxxB
\else
\ifnum#1=52 %
\hatcurLBiixxxxC
\else
\ifnum#1=53 %
\hatcurLBiixxxxD
\else
??????\fi
\fi
\fi
\fi
}
\newcommand{\hatcurLBiI}[1]{\ifnum#1=50 %
\hatcurLBiIxxxxA
\else
\ifnum#1=51 %
\hatcurLBiIxxxxB
\else
\ifnum#1=52 %
\hatcurLBiIxxxxC
\else
\ifnum#1=53 %
\hatcurLBiIxxxxD
\else
??????\fi
\fi
\fi
\fi
}
\newcommand{\hatcurLBiig}[1]{\ifnum#1=50 %
\hatcurLBiigxxxxA
\else
\ifnum#1=51 %
\hatcurLBiigxxxxB
\else
\ifnum#1=52 %
\hatcurLBiigxxxxC
\else
\ifnum#1=53 %
\hatcurLBiigxxxxD
\else
??????\fi
\fi
\fi
\fi
}
\newcommand{\hatcurLBiii}[1]{\ifnum#1=50 %
\hatcurLBiiixxxxA
\else
\ifnum#1=51 %
\hatcurLBiiixxxxB
\else
\ifnum#1=52 %
\hatcurLBiiixxxxC
\else
\ifnum#1=53 %
\hatcurLBiiixxxxD
\else
??????\fi
\fi
\fi
\fi
}
\newcommand{\hatcurLBiiI}[1]{\ifnum#1=50 %
\hatcurLBiiIxxxxA
\else
\ifnum#1=51 %
\hatcurLBiiIxxxxB
\else
\ifnum#1=52 %
\hatcurLBiiIxxxxC
\else
\ifnum#1=53 %
\hatcurLBiiIxxxxD
\else
??????\fi
\fi
\fi
\fi
}
\newcommand{\hatcurLBiikep}[1]{\ifnum#1=50 %
\hatcurLBiikepxxxxA
\else
\ifnum#1=51 %
\hatcurLBiikepxxxxB
\else
\ifnum#1=52 %
\hatcurLBiikepxxxxC
\else
\ifnum#1=53 %
\hatcurLBiikepxxxxD
\else
??????\fi
\fi
\fi
\fi
}
\newcommand{\hatcurLBiir}[1]{\ifnum#1=50 %
\hatcurLBiirxxxxA
\else
\ifnum#1=51 %
\hatcurLBiirxxxxB
\else
\ifnum#1=52 %
\hatcurLBiirxxxxC
\else
\ifnum#1=53 %
\hatcurLBiirxxxxD
\else
??????\fi
\fi
\fi
\fi
}
\newcommand{\hatcurLBiiR}[1]{\ifnum#1=50 %
\hatcurLBiiRxxxxA
\else
\ifnum#1=51 %
\hatcurLBiiRxxxxB
\else
\ifnum#1=52 %
\hatcurLBiiRxxxxC
\else
\ifnum#1=53 %
\hatcurLBiiRxxxxD
\else
??????\fi
\fi
\fi
\fi
}
\newcommand{\hatcurLBiiz}[1]{\ifnum#1=50 %
\hatcurLBiizxxxxA
\else
\ifnum#1=51 %
\hatcurLBiizxxxxB
\else
\ifnum#1=52 %
\hatcurLBiizxxxxC
\else
\ifnum#1=53 %
\hatcurLBiizxxxxD
\else
??????\fi
\fi
\fi
\fi
}
\newcommand{\hatcurLBikep}[1]{\ifnum#1=50 %
\hatcurLBikepxxxxA
\else
\ifnum#1=51 %
\hatcurLBikepxxxxB
\else
\ifnum#1=52 %
\hatcurLBikepxxxxC
\else
\ifnum#1=53 %
\hatcurLBikepxxxxD
\else
??????\fi
\fi
\fi
\fi
}
\newcommand{\hatcurLBir}[1]{\ifnum#1=50 %
\hatcurLBirxxxxA
\else
\ifnum#1=51 %
\hatcurLBirxxxxB
\else
\ifnum#1=52 %
\hatcurLBirxxxxC
\else
\ifnum#1=53 %
\hatcurLBirxxxxD
\else
??????\fi
\fi
\fi
\fi
}
\newcommand{\hatcurLBiR}[1]{\ifnum#1=50 %
\hatcurLBiRxxxxA
\else
\ifnum#1=51 %
\hatcurLBiRxxxxB
\else
\ifnum#1=52 %
\hatcurLBiRxxxxC
\else
\ifnum#1=53 %
\hatcurLBiRxxxxD
\else
??????\fi
\fi
\fi
\fi
}
\newcommand{\hatcurLBiz}[1]{\ifnum#1=50 %
\hatcurLBizxxxxA
\else
\ifnum#1=51 %
\hatcurLBizxxxxB
\else
\ifnum#1=52 %
\hatcurLBizxxxxC
\else
\ifnum#1=53 %
\hatcurLBizxxxxD
\else
??????\fi
\fi
\fi
\fi
}
\newcommand{\hatcurLCbsq}[1]{\ifnum#1=50 %
\hatcurLCbsqxxxxA
\else
\ifnum#1=51 %
\hatcurLCbsqxxxxB
\else
\ifnum#1=52 %
\hatcurLCbsqxxxxC
\else
\ifnum#1=53 %
\hatcurLCbsqxxxxD
\else
??????\fi
\fi
\fi
\fi
}
\newcommand{\hatcurLCdip}[1]{\ifnum#1=50 %
\hatcurLCdipxxxxA
\else
\ifnum#1=51 %
\hatcurLCdipxxxxB
\else
\ifnum#1=52 %
\hatcurLCdipxxxxC
\else
\ifnum#1=53 %
\hatcurLCdipxxxxD
\else
??????\fi
\fi
\fi
\fi
}
\newcommand{\hatcurLCdur}[1]{\ifnum#1=50 %
\hatcurLCdurxxxxA
\else
\ifnum#1=51 %
\hatcurLCdurxxxxB
\else
\ifnum#1=52 %
\hatcurLCdurxxxxC
\else
\ifnum#1=53 %
\hatcurLCdurxxxxD
\else
??????\fi
\fi
\fi
\fi
}
\newcommand{\hatcurLCdurhr}[1]{\ifnum#1=50 %
\hatcurLCdurhrxxxxA
\else
\ifnum#1=51 %
\hatcurLCdurhrxxxxB
\else
\ifnum#1=52 %
\hatcurLCdurhrxxxxC
\else
\ifnum#1=53 %
\hatcurLCdurhrxxxxD
\else
??????\fi
\fi
\fi
\fi
}
\newcommand{\hatcurLCdurhrshort}[1]{\ifnum#1=50 %
\hatcurLCdurhrshortxxxxA
\else
\ifnum#1=51 %
\hatcurLCdurhrshortxxxxB
\else
\ifnum#1=52 %
\hatcurLCdurhrshortxxxxC
\else
\ifnum#1=53 %
\hatcurLCdurhrshortxxxxD
\else
??????\fi
\fi
\fi
\fi
}
\newcommand{\hatcurLCdurshort}[1]{\ifnum#1=50 %
\hatcurLCdurshortxxxxA
\else
\ifnum#1=51 %
\hatcurLCdurshortxxxxB
\else
\ifnum#1=52 %
\hatcurLCdurshortxxxxC
\else
\ifnum#1=53 %
\hatcurLCdurshortxxxxD
\else
??????\fi
\fi
\fi
\fi
}
\newcommand{\hatcurLChatnetm}[1]{\ifnum#1=51 %
\hatcurLChatnetmxxxxB
\else
\ifnum#1=52 %
\hatcurLChatnetmxxxxC
\else
\ifnum#1=53 %
\hatcurLChatnetmxxxxD
\else
??????\fi
\fi
\fi
}
\newcommand{\hatcurLChatnetmA}[1]{\ifnum#1=50 %
\hatcurLChatnetmAxxxxA
\else
??????\fi
}
\newcommand{\hatcurLChatnetmB}[1]{\ifnum#1=50 %
\hatcurLChatnetmBxxxxA
\else
??????\fi
}
\newcommand{\hatcurLCiblend}[1]{\ifnum#1=51 %
\hatcurLCiblendxxxxB
\else
\ifnum#1=52 %
\hatcurLCiblendxxxxC
\else
\ifnum#1=53 %
\hatcurLCiblendxxxxD
\else
??????\fi
\fi
\fi
}
\newcommand{\hatcurLCiblendA}[1]{\ifnum#1=50 %
\hatcurLCiblendAxxxxA
\else
??????\fi
}
\newcommand{\hatcurLCiblendB}[1]{\ifnum#1=50 %
\hatcurLCiblendBxxxxA
\else
??????\fi
}
\newcommand{\hatcurLCimp}[1]{\ifnum#1=50 %
\hatcurLCimpxxxxA
\else
\ifnum#1=51 %
\hatcurLCimpxxxxB
\else
\ifnum#1=52 %
\hatcurLCimpxxxxC
\else
\ifnum#1=53 %
\hatcurLCimpxxxxD
\else
??????\fi
\fi
\fi
\fi
}
\newcommand{\hatcurLCingdur}[1]{\ifnum#1=50 %
\hatcurLCingdurxxxxA
\else
\ifnum#1=51 %
\hatcurLCingdurxxxxB
\else
\ifnum#1=52 %
\hatcurLCingdurxxxxC
\else
\ifnum#1=53 %
\hatcurLCingdurxxxxD
\else
??????\fi
\fi
\fi
\fi
}
\newcommand{\hatcurLCP}[1]{\ifnum#1=50 %
\hatcurLCPxxxxA
\else
\ifnum#1=51 %
\hatcurLCPxxxxB
\else
\ifnum#1=52 %
\hatcurLCPxxxxC
\else
\ifnum#1=53 %
\hatcurLCPxxxxD
\else
??????\fi
\fi
\fi
\fi
}
\newcommand{\hatcurLCPprec}[1]{\ifnum#1=50 %
\hatcurLCPprecxxxxA
\else
\ifnum#1=51 %
\hatcurLCPprecxxxxB
\else
\ifnum#1=52 %
\hatcurLCPprecxxxxC
\else
\ifnum#1=53 %
\hatcurLCPprecxxxxD
\else
??????\fi
\fi
\fi
\fi
}
\newcommand{\hatcurLCPshort}[1]{\ifnum#1=50 %
\hatcurLCPshortxxxxA
\else
\ifnum#1=51 %
\hatcurLCPshortxxxxB
\else
\ifnum#1=52 %
\hatcurLCPshortxxxxC
\else
\ifnum#1=53 %
\hatcurLCPshortxxxxD
\else
??????\fi
\fi
\fi
\fi
}
\newcommand{\hatcurLCq}[1]{\ifnum#1=50 %
\hatcurLCqxxxxA
\else
\ifnum#1=51 %
\hatcurLCqxxxxB
\else
\ifnum#1=52 %
\hatcurLCqxxxxC
\else
\ifnum#1=53 %
\hatcurLCqxxxxD
\else
??????\fi
\fi
\fi
\fi
}
\newcommand{\hatcurLCqshort}[1]{\ifnum#1=50 %
\hatcurLCqshortxxxxA
\else
\ifnum#1=51 %
\hatcurLCqshortxxxxB
\else
\ifnum#1=52 %
\hatcurLCqshortxxxxC
\else
\ifnum#1=53 %
\hatcurLCqshortxxxxD
\else
??????\fi
\fi
\fi
\fi
}
\newcommand{\hatcurLCrho}[1]{\ifnum#1=50 %
\hatcurLCrhoxxxxA
\else
\ifnum#1=51 %
\hatcurLCrhoxxxxB
\else
\ifnum#1=52 %
\hatcurLCrhoxxxxC
\else
\ifnum#1=53 %
\hatcurLCrhoxxxxD
\else
??????\fi
\fi
\fi
\fi
}
\newcommand{\hatcurLCrprstar}[1]{\ifnum#1=50 %
\hatcurLCrprstarxxxxA
\else
\ifnum#1=51 %
\hatcurLCrprstarxxxxB
\else
\ifnum#1=52 %
\hatcurLCrprstarxxxxC
\else
\ifnum#1=53 %
\hatcurLCrprstarxxxxD
\else
??????\fi
\fi
\fi
\fi
}
\newcommand{\hatcurLCT}[1]{\ifnum#1=50 %
\hatcurLCTxxxxA
\else
\ifnum#1=51 %
\hatcurLCTxxxxB
\else
\ifnum#1=52 %
\hatcurLCTxxxxC
\else
\ifnum#1=53 %
\hatcurLCTxxxxD
\else
??????\fi
\fi
\fi
\fi
}
\newcommand{\hatcurLCTA}[1]{\ifnum#1=50 %
\hatcurLCTAxxxxA
\else
\ifnum#1=51 %
\hatcurLCTAxxxxB
\else
\ifnum#1=52 %
\hatcurLCTAxxxxC
\else
\ifnum#1=53 %
\hatcurLCTAxxxxD
\else
??????\fi
\fi
\fi
\fi
}
\newcommand{\hatcurLCTB}[1]{\ifnum#1=50 %
\hatcurLCTBxxxxA
\else
\ifnum#1=51 %
\hatcurLCTBxxxxB
\else
\ifnum#1=52 %
\hatcurLCTBxxxxC
\else
\ifnum#1=53 %
\hatcurLCTBxxxxD
\else
??????\fi
\fi
\fi
\fi
}
\newcommand{\hatcurLCzeta}[1]{\ifnum#1=50 %
\hatcurLCzetaxxxxA
\else
\ifnum#1=51 %
\hatcurLCzetaxxxxB
\else
\ifnum#1=52 %
\hatcurLCzetaxxxxC
\else
\ifnum#1=53 %
\hatcurLCzetaxxxxD
\else
??????\fi
\fi
\fi
\fi
}
\newcommand{\hatcurPPaequiv}[1]{\ifnum#1=50 %
\hatcurPPaequivxxxxA
\else
\ifnum#1=51 %
\hatcurPPaequivxxxxB
\else
\ifnum#1=52 %
\hatcurPPaequivxxxxC
\else
\ifnum#1=53 %
\hatcurPPaequivxxxxD
\else
??????\fi
\fi
\fi
\fi
}
\newcommand{\hatcurPPar}[1]{\ifnum#1=50 %
\hatcurPParxxxxA
\else
\ifnum#1=51 %
\hatcurPParxxxxB
\else
\ifnum#1=52 %
\hatcurPParxxxxC
\else
\ifnum#1=53 %
\hatcurPParxxxxD
\else
??????\fi
\fi
\fi
\fi
}
\newcommand{\hatcurPParel}[1]{\ifnum#1=50 %
\hatcurPParelxxxxA
\else
\ifnum#1=51 %
\hatcurPParelxxxxB
\else
\ifnum#1=52 %
\hatcurPParelxxxxC
\else
\ifnum#1=53 %
\hatcurPParelxxxxD
\else
??????\fi
\fi
\fi
\fi
}
\newcommand{\hatcurPPfluxap}[1]{\ifnum#1=50 %
\hatcurPPfluxapxxxxA
\else
\ifnum#1=51 %
\hatcurPPfluxapxxxxB
\else
\ifnum#1=52 %
\hatcurPPfluxapxxxxC
\else
\ifnum#1=53 %
\hatcurPPfluxapxxxxD
\else
??????\fi
\fi
\fi
\fi
}
\newcommand{\hatcurPPfluxapdim}[1]{\ifnum#1=50 %
\hatcurPPfluxapdimxxxxA
\else
\ifnum#1=51 %
\hatcurPPfluxapdimxxxxB
\else
\ifnum#1=52 %
\hatcurPPfluxapdimxxxxC
\else
\ifnum#1=53 %
\hatcurPPfluxapdimxxxxD
\else
??????\fi
\fi
\fi
\fi
}
\newcommand{\hatcurPPfluxavg}[1]{\ifnum#1=50 %
\hatcurPPfluxavgxxxxA
\else
\ifnum#1=51 %
\hatcurPPfluxavgxxxxB
\else
\ifnum#1=52 %
\hatcurPPfluxavgxxxxC
\else
\ifnum#1=53 %
\hatcurPPfluxavgxxxxD
\else
??????\fi
\fi
\fi
\fi
}
\newcommand{\hatcurPPfluxavgdim}[1]{\ifnum#1=50 %
\hatcurPPfluxavgdimxxxxA
\else
\ifnum#1=51 %
\hatcurPPfluxavgdimxxxxB
\else
\ifnum#1=52 %
\hatcurPPfluxavgdimxxxxC
\else
\ifnum#1=53 %
\hatcurPPfluxavgdimxxxxD
\else
??????\fi
\fi
\fi
\fi
}
\newcommand{\hatcurPPfluxavglog}[1]{\ifnum#1=50 %
\hatcurPPfluxavglogxxxxA
\else
\ifnum#1=51 %
\hatcurPPfluxavglogxxxxB
\else
\ifnum#1=52 %
\hatcurPPfluxavglogxxxxC
\else
\ifnum#1=53 %
\hatcurPPfluxavglogxxxxD
\else
??????\fi
\fi
\fi
\fi
}
\newcommand{\hatcurPPfluxperi}[1]{\ifnum#1=50 %
\hatcurPPfluxperixxxxA
\else
\ifnum#1=51 %
\hatcurPPfluxperixxxxB
\else
\ifnum#1=52 %
\hatcurPPfluxperixxxxC
\else
\ifnum#1=53 %
\hatcurPPfluxperixxxxD
\else
??????\fi
\fi
\fi
\fi
}
\newcommand{\hatcurPPfluxperidim}[1]{\ifnum#1=50 %
\hatcurPPfluxperidimxxxxA
\else
\ifnum#1=51 %
\hatcurPPfluxperidimxxxxB
\else
\ifnum#1=52 %
\hatcurPPfluxperidimxxxxC
\else
\ifnum#1=53 %
\hatcurPPfluxperidimxxxxD
\else
??????\fi
\fi
\fi
\fi
}
\newcommand{\hatcurPPg}[1]{\ifnum#1=50 %
\hatcurPPgxxxxA
\else
\ifnum#1=51 %
\hatcurPPgxxxxB
\else
\ifnum#1=52 %
\hatcurPPgxxxxC
\else
\ifnum#1=53 %
\hatcurPPgxxxxD
\else
??????\fi
\fi
\fi
\fi
}
\newcommand{\hatcurPPi}[1]{\ifnum#1=50 %
\hatcurPPixxxxA
\else
\ifnum#1=51 %
\hatcurPPixxxxB
\else
\ifnum#1=52 %
\hatcurPPixxxxC
\else
\ifnum#1=53 %
\hatcurPPixxxxD
\else
??????\fi
\fi
\fi
\fi
}
\newcommand{\hatcurPPlogg}[1]{\ifnum#1=50 %
\hatcurPPloggxxxxA
\else
\ifnum#1=51 %
\hatcurPPloggxxxxB
\else
\ifnum#1=52 %
\hatcurPPloggxxxxC
\else
\ifnum#1=53 %
\hatcurPPloggxxxxD
\else
??????\fi
\fi
\fi
\fi
}
\newcommand{\hatcurPPm}[1]{\ifnum#1=50 %
\hatcurPPmxxxxA
\else
\ifnum#1=51 %
\hatcurPPmxxxxB
\else
\ifnum#1=52 %
\hatcurPPmxxxxC
\else
\ifnum#1=53 %
\hatcurPPmxxxxD
\else
??????\fi
\fi
\fi
\fi
}
\newcommand{\hatcurPPme}[1]{\ifnum#1=50 %
\hatcurPPmexxxxA
\else
\ifnum#1=51 %
\hatcurPPmexxxxB
\else
\ifnum#1=52 %
\hatcurPPmexxxxC
\else
\ifnum#1=53 %
\hatcurPPmexxxxD
\else
??????\fi
\fi
\fi
\fi
}
\newcommand{\hatcurPPmelong}[1]{\ifnum#1=50 %
\hatcurPPmelongxxxxA
\else
\ifnum#1=51 %
\hatcurPPmelongxxxxB
\else
\ifnum#1=52 %
\hatcurPPmelongxxxxC
\else
\ifnum#1=53 %
\hatcurPPmelongxxxxD
\else
??????\fi
\fi
\fi
\fi
}
\newcommand{\hatcurPPmeshort}[1]{\ifnum#1=50 %
\hatcurPPmeshortxxxxA
\else
\ifnum#1=51 %
\hatcurPPmeshortxxxxB
\else
\ifnum#1=52 %
\hatcurPPmeshortxxxxC
\else
\ifnum#1=53 %
\hatcurPPmeshortxxxxD
\else
??????\fi
\fi
\fi
\fi
}
\newcommand{\hatcurPPmlong}[1]{\ifnum#1=50 %
\hatcurPPmlongxxxxA
\else
\ifnum#1=51 %
\hatcurPPmlongxxxxB
\else
\ifnum#1=52 %
\hatcurPPmlongxxxxC
\else
\ifnum#1=53 %
\hatcurPPmlongxxxxD
\else
??????\fi
\fi
\fi
\fi
}
\newcommand{\hatcurPPmrcorr}[1]{\ifnum#1=50 %
\hatcurPPmrcorrxxxxA
\else
\ifnum#1=51 %
\hatcurPPmrcorrxxxxB
\else
\ifnum#1=52 %
\hatcurPPmrcorrxxxxC
\else
\ifnum#1=53 %
\hatcurPPmrcorrxxxxD
\else
??????\fi
\fi
\fi
\fi
}
\newcommand{\hatcurPPmshort}[1]{\ifnum#1=50 %
\hatcurPPmshortxxxxA
\else
\ifnum#1=51 %
\hatcurPPmshortxxxxB
\else
\ifnum#1=52 %
\hatcurPPmshortxxxxC
\else
\ifnum#1=53 %
\hatcurPPmshortxxxxD
\else
??????\fi
\fi
\fi
\fi
}
\newcommand{\hatcurPPperi}[1]{\ifnum#1=50 %
\hatcurPPperixxxxA
\else
\ifnum#1=51 %
\hatcurPPperixxxxB
\else
\ifnum#1=52 %
\hatcurPPperixxxxC
\else
\ifnum#1=53 %
\hatcurPPperixxxxD
\else
??????\fi
\fi
\fi
\fi
}
\newcommand{\hatcurPPphiconj}[1]{\ifnum#1=50 %
\hatcurPPphiconjxxxxA
\else
\ifnum#1=51 %
\hatcurPPphiconjxxxxB
\else
\ifnum#1=52 %
\hatcurPPphiconjxxxxC
\else
\ifnum#1=53 %
\hatcurPPphiconjxxxxD
\else
??????\fi
\fi
\fi
\fi
}
\newcommand{\hatcurPPr}[1]{\ifnum#1=50 %
\hatcurPPrxxxxA
\else
\ifnum#1=51 %
\hatcurPPrxxxxB
\else
\ifnum#1=52 %
\hatcurPPrxxxxC
\else
\ifnum#1=53 %
\hatcurPPrxxxxD
\else
??????\fi
\fi
\fi
\fi
}
\newcommand{\hatcurPPre}[1]{\ifnum#1=50 %
\hatcurPPrexxxxA
\else
\ifnum#1=51 %
\hatcurPPrexxxxB
\else
\ifnum#1=52 %
\hatcurPPrexxxxC
\else
\ifnum#1=53 %
\hatcurPPrexxxxD
\else
??????\fi
\fi
\fi
\fi
}
\newcommand{\hatcurPPrelong}[1]{\ifnum#1=50 %
\hatcurPPrelongxxxxA
\else
\ifnum#1=51 %
\hatcurPPrelongxxxxB
\else
\ifnum#1=52 %
\hatcurPPrelongxxxxC
\else
\ifnum#1=53 %
\hatcurPPrelongxxxxD
\else
??????\fi
\fi
\fi
\fi
}
\newcommand{\hatcurPPreshort}[1]{\ifnum#1=50 %
\hatcurPPreshortxxxxA
\else
\ifnum#1=51 %
\hatcurPPreshortxxxxB
\else
\ifnum#1=52 %
\hatcurPPreshortxxxxC
\else
\ifnum#1=53 %
\hatcurPPreshortxxxxD
\else
??????\fi
\fi
\fi
\fi
}
\newcommand{\hatcurPPrho}[1]{\ifnum#1=50 %
\hatcurPPrhoxxxxA
\else
\ifnum#1=51 %
\hatcurPPrhoxxxxB
\else
\ifnum#1=52 %
\hatcurPPrhoxxxxC
\else
\ifnum#1=53 %
\hatcurPPrhoxxxxD
\else
??????\fi
\fi
\fi
\fi
}
\newcommand{\hatcurPPrlong}[1]{\ifnum#1=50 %
\hatcurPPrlongxxxxA
\else
\ifnum#1=51 %
\hatcurPPrlongxxxxB
\else
\ifnum#1=52 %
\hatcurPPrlongxxxxC
\else
\ifnum#1=53 %
\hatcurPPrlongxxxxD
\else
??????\fi
\fi
\fi
\fi
}
\newcommand{\hatcurPPrshort}[1]{\ifnum#1=50 %
\hatcurPPrshortxxxxA
\else
\ifnum#1=51 %
\hatcurPPrshortxxxxB
\else
\ifnum#1=52 %
\hatcurPPrshortxxxxC
\else
\ifnum#1=53 %
\hatcurPPrshortxxxxD
\else
??????\fi
\fi
\fi
\fi
}
\newcommand{\hatcurPPtcirc}[1]{\ifnum#1=50 %
\hatcurPPtcircxxxxA
\else
\ifnum#1=51 %
\hatcurPPtcircxxxxB
\else
\ifnum#1=52 %
\hatcurPPtcircxxxxC
\else
\ifnum#1=53 %
\hatcurPPtcircxxxxD
\else
??????\fi
\fi
\fi
\fi
}
\newcommand{\hatcurPPteff}[1]{\ifnum#1=50 %
\hatcurPPteffxxxxA
\else
\ifnum#1=51 %
\hatcurPPteffxxxxB
\else
\ifnum#1=52 %
\hatcurPPteffxxxxC
\else
\ifnum#1=53 %
\hatcurPPteffxxxxD
\else
??????\fi
\fi
\fi
\fi
}
\newcommand{\hatcurPPtheta}[1]{\ifnum#1=50 %
\hatcurPPthetaxxxxA
\else
\ifnum#1=51 %
\hatcurPPthetaxxxxB
\else
\ifnum#1=52 %
\hatcurPPthetaxxxxC
\else
\ifnum#1=53 %
\hatcurPPthetaxxxxD
\else
??????\fi
\fi
\fi
\fi
}
\newcommand{\hatcurPPtinfall}[1]{\ifnum#1=50 %
\hatcurPPtinfallxxxxA
\else
\ifnum#1=51 %
\hatcurPPtinfallxxxxB
\else
\ifnum#1=52 %
\hatcurPPtinfallxxxxC
\else
\ifnum#1=53 %
\hatcurPPtinfallxxxxD
\else
??????\fi
\fi
\fi
\fi
}
\newcommand{\hatcurRVeccen}[1]{\ifnum#1=50 %
\hatcurRVeccenxxxxA
\else
\ifnum#1=51 %
\hatcurRVeccenxxxxB
\else
\ifnum#1=52 %
\hatcurRVeccenxxxxC
\else
\ifnum#1=53 %
\hatcurRVeccenxxxxD
\else
??????\fi
\fi
\fi
\fi
}
\newcommand{\hatcurRVeccentwosiglim}[1]{\ifnum#1=50 %
\hatcurRVeccentwosiglimxxxxA
\else
\ifnum#1=51 %
\hatcurRVeccentwosiglimxxxxB
\else
\ifnum#1=52 %
\hatcurRVeccentwosiglimxxxxC
\else
\ifnum#1=53 %
\hatcurRVeccentwosiglimxxxxD
\else
??????\fi
\fi
\fi
\fi
}
\newcommand{\hatcurRVfitrmsA}[1]{\ifnum#1=50 %
\hatcurRVfitrmsAxxxxA
\else
\ifnum#1=51 %
\hatcurRVfitrmsAxxxxB
\else
\ifnum#1=52 %
\hatcurRVfitrmsAxxxxC
\else
\ifnum#1=53 %
\hatcurRVfitrmsAxxxxD
\else
??????\fi
\fi
\fi
\fi
}
\newcommand{\hatcurRVfitrmsB}[1]{\ifnum#1=50 %
\hatcurRVfitrmsBxxxxA
\else
\ifnum#1=51 %
\hatcurRVfitrmsBxxxxB
\else
\ifnum#1=52 %
\hatcurRVfitrmsBxxxxC
\else
\ifnum#1=53 %
\hatcurRVfitrmsBxxxxD
\else
??????\fi
\fi
\fi
\fi
}
\newcommand{\hatcurRVfitrmsC}[1]{\ifnum#1=50 %
\hatcurRVfitrmsCxxxxA
\else
\ifnum#1=51 %
\hatcurRVfitrmsCxxxxB
\else
\ifnum#1=52 %
\hatcurRVfitrmsCxxxxC
\else
??????\fi
\fi
\fi
}
\newcommand{\hatcurRVgammaA}[1]{\ifnum#1=50 %
\hatcurRVgammaAxxxxA
\else
\ifnum#1=51 %
\hatcurRVgammaAxxxxB
\else
\ifnum#1=52 %
\hatcurRVgammaAxxxxC
\else
\ifnum#1=53 %
\hatcurRVgammaAxxxxD
\else
??????\fi
\fi
\fi
\fi
}
\newcommand{\hatcurRVgammaB}[1]{\ifnum#1=50 %
\hatcurRVgammaBxxxxA
\else
\ifnum#1=51 %
\hatcurRVgammaBxxxxB
\else
\ifnum#1=52 %
\hatcurRVgammaBxxxxC
\else
\ifnum#1=53 %
\hatcurRVgammaBxxxxD
\else
??????\fi
\fi
\fi
\fi
}
\newcommand{\hatcurRVgammaC}[1]{\ifnum#1=50 %
\hatcurRVgammaCxxxxA
\else
\ifnum#1=51 %
\hatcurRVgammaCxxxxB
\else
\ifnum#1=52 %
\hatcurRVgammaCxxxxC
\else
??????\fi
\fi
\fi
}
\newcommand{\hatcurRVh}[1]{\ifnum#1=50 %
\hatcurRVhxxxxA
\else
\ifnum#1=51 %
\hatcurRVhxxxxB
\else
\ifnum#1=52 %
\hatcurRVhxxxxC
\else
\ifnum#1=53 %
\hatcurRVhxxxxD
\else
??????\fi
\fi
\fi
\fi
}
\newcommand{\hatcurRVjitterA}[1]{\ifnum#1=50 %
\hatcurRVjitterAxxxxA
\else
\ifnum#1=51 %
\hatcurRVjitterAxxxxB
\else
\ifnum#1=52 %
\hatcurRVjitterAxxxxC
\else
\ifnum#1=53 %
\hatcurRVjitterAxxxxD
\else
??????\fi
\fi
\fi
\fi
}
\newcommand{\hatcurRVjitterB}[1]{\ifnum#1=50 %
\hatcurRVjitterBxxxxA
\else
\ifnum#1=51 %
\hatcurRVjitterBxxxxB
\else
\ifnum#1=52 %
\hatcurRVjitterBxxxxC
\else
\ifnum#1=53 %
\hatcurRVjitterBxxxxD
\else
??????\fi
\fi
\fi
\fi
}
\newcommand{\hatcurRVjitterC}[1]{\ifnum#1=50 %
\hatcurRVjitterCxxxxA
\else
\ifnum#1=51 %
\hatcurRVjitterCxxxxB
\else
\ifnum#1=52 %
\hatcurRVjitterCxxxxC
\else
??????\fi
\fi
\fi
}
\newcommand{\hatcurRVjittertwosiglimA}[1]{\ifnum#1=50 %
\hatcurRVjittertwosiglimAxxxxA
\else
\ifnum#1=51 %
\hatcurRVjittertwosiglimAxxxxB
\else
\ifnum#1=52 %
\hatcurRVjittertwosiglimAxxxxC
\else
\ifnum#1=53 %
\hatcurRVjittertwosiglimAxxxxD
\else
??????\fi
\fi
\fi
\fi
}
\newcommand{\hatcurRVjittertwosiglimB}[1]{\ifnum#1=50 %
\hatcurRVjittertwosiglimBxxxxA
\else
\ifnum#1=51 %
\hatcurRVjittertwosiglimBxxxxB
\else
\ifnum#1=52 %
\hatcurRVjittertwosiglimBxxxxC
\else
\ifnum#1=53 %
\hatcurRVjittertwosiglimBxxxxD
\else
??????\fi
\fi
\fi
\fi
}
\newcommand{\hatcurRVjittertwosiglimC}[1]{\ifnum#1=50 %
\hatcurRVjittertwosiglimCxxxxA
\else
\ifnum#1=51 %
\hatcurRVjittertwosiglimCxxxxB
\else
\ifnum#1=52 %
\hatcurRVjittertwosiglimCxxxxC
\else
??????\fi
\fi
\fi
}
\newcommand{\hatcurRVk}[1]{\ifnum#1=50 %
\hatcurRVkxxxxA
\else
\ifnum#1=51 %
\hatcurRVkxxxxB
\else
\ifnum#1=52 %
\hatcurRVkxxxxC
\else
\ifnum#1=53 %
\hatcurRVkxxxxD
\else
??????\fi
\fi
\fi
\fi
}
\newcommand{\hatcurRVK}[1]{\ifnum#1=50 %
\hatcurRVKxxxxA
\else
\ifnum#1=51 %
\hatcurRVKxxxxB
\else
\ifnum#1=52 %
\hatcurRVKxxxxC
\else
\ifnum#1=53 %
\hatcurRVKxxxxD
\else
??????\fi
\fi
\fi
\fi
}
\newcommand{\hatcurRVomega}[1]{\ifnum#1=50 %
\hatcurRVomegaxxxxA
\else
\ifnum#1=51 %
\hatcurRVomegaxxxxB
\else
\ifnum#1=52 %
\hatcurRVomegaxxxxC
\else
\ifnum#1=53 %
\hatcurRVomegaxxxxD
\else
??????\fi
\fi
\fi
\fi
}
\newcommand{\hatcurRVrh}[1]{\ifnum#1=50 %
\hatcurRVrhxxxxA
\else
\ifnum#1=51 %
\hatcurRVrhxxxxB
\else
\ifnum#1=52 %
\hatcurRVrhxxxxC
\else
\ifnum#1=53 %
\hatcurRVrhxxxxD
\else
??????\fi
\fi
\fi
\fi
}
\newcommand{\hatcurRVrk}[1]{\ifnum#1=50 %
\hatcurRVrkxxxxA
\else
\ifnum#1=51 %
\hatcurRVrkxxxxB
\else
\ifnum#1=52 %
\hatcurRVrkxxxxC
\else
\ifnum#1=53 %
\hatcurRVrkxxxxD
\else
??????\fi
\fi
\fi
\fi
}
\newcommand{\hatcurRVtrone}[1]{\ifnum#1=50 %
\hatcurRVtronexxxxA
\else
\ifnum#1=51 %
\hatcurRVtronexxxxB
\else
\ifnum#1=52 %
\hatcurRVtronexxxxC
\else
\ifnum#1=53 %
\hatcurRVtronexxxxD
\else
??????\fi
\fi
\fi
\fi
}
\newcommand{\hatcurRVtrtwo}[1]{\ifnum#1=50 %
\hatcurRVtrtwoxxxxA
\else
\ifnum#1=51 %
\hatcurRVtrtwoxxxxB
\else
\ifnum#1=52 %
\hatcurRVtrtwoxxxxC
\else
\ifnum#1=53 %
\hatcurRVtrtwoxxxxD
\else
??????\fi
\fi
\fi
\fi
}
\newcommand{\hatcurSMEiilogg}[1]{\ifnum#1=51 %
\hatcurSMEiiloggxxxxB
\else
\ifnum#1=52 %
\hatcurSMEiiloggxxxxC
\else
\ifnum#1=53 %
\hatcurSMEiiloggxxxxD
\else
??????\fi
\fi
\fi
}
\newcommand{\hatcurSMEiiteff}[1]{\ifnum#1=51 %
\hatcurSMEiiteffxxxxB
\else
\ifnum#1=52 %
\hatcurSMEiiteffxxxxC
\else
\ifnum#1=53 %
\hatcurSMEiiteffxxxxD
\else
??????\fi
\fi
\fi
}
\newcommand{\hatcurSMEiivmac}[1]{\ifnum#1=51 %
\hatcurSMEiivmacxxxxB
\else
\ifnum#1=52 %
\hatcurSMEiivmacxxxxC
\else
\ifnum#1=53 %
\hatcurSMEiivmacxxxxD
\else
??????\fi
\fi
\fi
}
\newcommand{\hatcurSMEiivmic}[1]{\ifnum#1=51 %
\hatcurSMEiivmicxxxxB
\else
\ifnum#1=52 %
\hatcurSMEiivmicxxxxC
\else
\ifnum#1=53 %
\hatcurSMEiivmicxxxxD
\else
??????\fi
\fi
\fi
}
\newcommand{\hatcurSMEiivsin}[1]{\ifnum#1=51 %
\hatcurSMEiivsinxxxxB
\else
\ifnum#1=52 %
\hatcurSMEiivsinxxxxC
\else
\ifnum#1=53 %
\hatcurSMEiivsinxxxxD
\else
??????\fi
\fi
\fi
}
\newcommand{\hatcurSMEiizfeh}[1]{\ifnum#1=51 %
\hatcurSMEiizfehxxxxB
\else
\ifnum#1=52 %
\hatcurSMEiizfehxxxxC
\else
\ifnum#1=53 %
\hatcurSMEiizfehxxxxD
\else
??????\fi
\fi
\fi
}
\newcommand{\hatcurSMEiizfehshort}[1]{\ifnum#1=51 %
\hatcurSMEiizfehshortxxxxB
\else
\ifnum#1=52 %
\hatcurSMEiizfehshortxxxxC
\else
\ifnum#1=53 %
\hatcurSMEiizfehshortxxxxD
\else
??????\fi
\fi
\fi
}
\newcommand{\hatcurSMEilogg}[1]{\ifnum#1=50 %
\hatcurSMEiloggxxxxA
\else
\ifnum#1=51 %
\hatcurSMEiloggxxxxB
\else
\ifnum#1=52 %
\hatcurSMEiloggxxxxC
\else
\ifnum#1=53 %
\hatcurSMEiloggxxxxD
\else
??????\fi
\fi
\fi
\fi
}
\newcommand{\hatcurSMEiteff}[1]{\ifnum#1=50 %
\hatcurSMEiteffxxxxA
\else
\ifnum#1=51 %
\hatcurSMEiteffxxxxB
\else
\ifnum#1=52 %
\hatcurSMEiteffxxxxC
\else
\ifnum#1=53 %
\hatcurSMEiteffxxxxD
\else
??????\fi
\fi
\fi
\fi
}
\newcommand{\hatcurSMEivmac}[1]{\ifnum#1=50 %
\hatcurSMEivmacxxxxA
\else
\ifnum#1=51 %
\hatcurSMEivmacxxxxB
\else
\ifnum#1=52 %
\hatcurSMEivmacxxxxC
\else
\ifnum#1=53 %
\hatcurSMEivmacxxxxD
\else
??????\fi
\fi
\fi
\fi
}
\newcommand{\hatcurSMEivmic}[1]{\ifnum#1=50 %
\hatcurSMEivmicxxxxA
\else
\ifnum#1=51 %
\hatcurSMEivmicxxxxB
\else
\ifnum#1=52 %
\hatcurSMEivmicxxxxC
\else
\ifnum#1=53 %
\hatcurSMEivmicxxxxD
\else
??????\fi
\fi
\fi
\fi
}
\newcommand{\hatcurSMEivsin}[1]{\ifnum#1=50 %
\hatcurSMEivsinxxxxA
\else
\ifnum#1=51 %
\hatcurSMEivsinxxxxB
\else
\ifnum#1=52 %
\hatcurSMEivsinxxxxC
\else
\ifnum#1=53 %
\hatcurSMEivsinxxxxD
\else
??????\fi
\fi
\fi
\fi
}
\newcommand{\hatcurSMEizfeh}[1]{\ifnum#1=50 %
\hatcurSMEizfehxxxxA
\else
\ifnum#1=51 %
\hatcurSMEizfehxxxxB
\else
\ifnum#1=52 %
\hatcurSMEizfehxxxxC
\else
\ifnum#1=53 %
\hatcurSMEizfehxxxxD
\else
??????\fi
\fi
\fi
\fi
}
\newcommand{\hatcurSMEizfehshort}[1]{\ifnum#1=50 %
\hatcurSMEizfehshortxxxxA
\else
\ifnum#1=51 %
\hatcurSMEizfehshortxxxxB
\else
\ifnum#1=52 %
\hatcurSMEizfehshortxxxxC
\else
\ifnum#1=53 %
\hatcurSMEizfehshortxxxxD
\else
??????\fi
\fi
\fi
\fi
}
\newcommand{\hatcurXAv}[1]{\ifnum#1=50 %
\hatcurXAvxxxxA
\else
\ifnum#1=51 %
\hatcurXAvxxxxB
\else
\ifnum#1=52 %
\hatcurXAvxxxxC
\else
\ifnum#1=53 %
\hatcurXAvxxxxD
\else
??????\fi
\fi
\fi
\fi
}
\newcommand{\hatcurXdist}[1]{\ifnum#1=50 %
\hatcurXdistxxxxA
\else
\ifnum#1=51 %
\hatcurXdistxxxxB
\else
\ifnum#1=52 %
\hatcurXdistxxxxC
\else
\ifnum#1=53 %
\hatcurXdistxxxxD
\else
??????\fi
\fi
\fi
\fi
}
\newcommand{\hatcurXdistred}[1]{\ifnum#1=50 %
\hatcurXdistredxxxxA
\else
\ifnum#1=51 %
\hatcurXdistredxxxxB
\else
\ifnum#1=52 %
\hatcurXdistredxxxxC
\else
\ifnum#1=53 %
\hatcurXdistredxxxxD
\else
??????\fi
\fi
\fi
\fi
}
\newcommand{\hatcurXEBV}[1]{\ifnum#1=50 %
\hatcurXEBVxxxxA
\else
\ifnum#1=51 %
\hatcurXEBVxxxxB
\else
\ifnum#1=52 %
\hatcurXEBVxxxxC
\else
\ifnum#1=53 %
\hatcurXEBVxxxxD
\else
??????\fi
\fi
\fi
\fi
}
\newcommand{\hatcurXjhisored}[1]{\ifnum#1=50 %
\hatcurXjhisoredxxxxA
\else
\ifnum#1=51 %
\hatcurXjhisoredxxxxB
\else
\ifnum#1=52 %
\hatcurXjhisoredxxxxC
\else
\ifnum#1=53 %
\hatcurXjhisoredxxxxD
\else
??????\fi
\fi
\fi
\fi
}
\newcommand{\hatcurXjkisored}[1]{\ifnum#1=50 %
\hatcurXjkisoredxxxxA
\else
\ifnum#1=51 %
\hatcurXjkisoredxxxxB
\else
\ifnum#1=52 %
\hatcurXjkisoredxxxxC
\else
\ifnum#1=53 %
\hatcurXjkisoredxxxxD
\else
??????\fi
\fi
\fi
\fi
}
\newcommand{\hatcurXmhisored}[1]{\ifnum#1=50 %
\hatcurXmhisoredxxxxA
\else
\ifnum#1=51 %
\hatcurXmhisoredxxxxB
\else
\ifnum#1=52 %
\hatcurXmhisoredxxxxC
\else
\ifnum#1=53 %
\hatcurXmhisoredxxxxD
\else
??????\fi
\fi
\fi
\fi
}
\newcommand{\hatcurXmiisored}[1]{\ifnum#1=50 %
\hatcurXmiisoredxxxxA
\else
\ifnum#1=51 %
\hatcurXmiisoredxxxxB
\else
\ifnum#1=52 %
\hatcurXmiisoredxxxxC
\else
\ifnum#1=53 %
\hatcurXmiisoredxxxxD
\else
??????\fi
\fi
\fi
\fi
}
\newcommand{\hatcurXmjisored}[1]{\ifnum#1=50 %
\hatcurXmjisoredxxxxA
\else
\ifnum#1=51 %
\hatcurXmjisoredxxxxB
\else
\ifnum#1=52 %
\hatcurXmjisoredxxxxC
\else
\ifnum#1=53 %
\hatcurXmjisoredxxxxD
\else
??????\fi
\fi
\fi
\fi
}
\newcommand{\hatcurXmkisored}[1]{\ifnum#1=50 %
\hatcurXmkisoredxxxxA
\else
\ifnum#1=51 %
\hatcurXmkisoredxxxxB
\else
\ifnum#1=52 %
\hatcurXmkisoredxxxxC
\else
\ifnum#1=53 %
\hatcurXmkisoredxxxxD
\else
??????\fi
\fi
\fi
\fi
}
\newcommand{\hatcurXmvisored}[1]{\ifnum#1=50 %
\hatcurXmvisoredxxxxA
\else
\ifnum#1=51 %
\hatcurXmvisoredxxxxB
\else
\ifnum#1=52 %
\hatcurXmvisoredxxxxC
\else
\ifnum#1=53 %
\hatcurXmvisoredxxxxD
\else
??????\fi
\fi
\fi
\fi
}
\newcommand{\hatcurXsecdur}[1]{\ifnum#1=50 %
\hatcurXsecdurxxxxA
\else
\ifnum#1=51 %
\hatcurXsecdurxxxxB
\else
\ifnum#1=52 %
\hatcurXsecdurxxxxC
\else
\ifnum#1=53 %
\hatcurXsecdurxxxxD
\else
??????\fi
\fi
\fi
\fi
}
\newcommand{\hatcurXsecingdur}[1]{\ifnum#1=50 %
\hatcurXsecingdurxxxxA
\else
\ifnum#1=51 %
\hatcurXsecingdurxxxxB
\else
\ifnum#1=52 %
\hatcurXsecingdurxxxxC
\else
\ifnum#1=53 %
\hatcurXsecingdurxxxxD
\else
??????\fi
\fi
\fi
\fi
}
\newcommand{\hatcurXsecondary}[1]{\ifnum#1=50 %
\hatcurXsecondaryxxxxA
\else
\ifnum#1=51 %
\hatcurXsecondaryxxxxB
\else
\ifnum#1=52 %
\hatcurXsecondaryxxxxC
\else
\ifnum#1=53 %
\hatcurXsecondaryxxxxD
\else
??????\fi
\fi
\fi
\fi
}
\newcommand{\hatcurXsecphase}[1]{\ifnum#1=50 %
\hatcurXsecphasexxxxA
\else
\ifnum#1=51 %
\hatcurXsecphasexxxxB
\else
\ifnum#1=52 %
\hatcurXsecphasexxxxC
\else
\ifnum#1=53 %
\hatcurXsecphasexxxxD
\else
??????\fi
\fi
\fi
\fi
}
\newcommand{\hatcurXviisored}[1]{\ifnum#1=50 %
\hatcurXviisoredxxxxA
\else
\ifnum#1=51 %
\hatcurXviisoredxxxxB
\else
\ifnum#1=52 %
\hatcurXviisoredxxxxC
\else
\ifnum#1=53 %
\hatcurXviisoredxxxxD
\else
??????\fi
\fi
\fi
\fi
}
\newcommand{\hatcurXvkisored}[1]{\ifnum#1=50 %
\hatcurXvkisoredxxxxA
\else
\ifnum#1=51 %
\hatcurXvkisoredxxxxB
\else
\ifnum#1=52 %
\hatcurXvkisoredxxxxC
\else
\ifnum#1=53 %
\hatcurXvkisoredxxxxD
\else
??????\fi
\fi
\fi
\fi
}

\newcommand{\hatcurhtreccenxxxxA}{HATS580-019}                      
\newcommand{\hatcurfieldeccenxxxxA}{\ensuremath{string}}            
\newcommand{\hatcurCCraeccenxxxxA}{\ensuremath{20^{\mathrm h}01^{\mathrm m}42.60{\mathrm s}}}                     
\newcommand{\hatcurCCdececcenxxxxA}{\ensuremath{-26{\arcdeg}04{\arcmin}39.3{\arcsec}}}                    
\newcommand{\hatcurCCmageccenxxxxA}{14.033}                         
\newcommand{\hatcurCCtwomasseccenxxxxA}{2MASS~20014273-2604392}     
\newcommand{\hatcurCCgsceccenxxxxA}{GSC~6896-01012}                 
\newcommand{\hatcurCCtassmveccenxxxxA}{\ensuremath{14.033\pm0.050}} 
\newcommand{\hatcurCCtassmvshorteccenxxxxA}{\ensuremath{14.0}}      
\newcommand{\hatcurCCtassmBeccenxxxxA}{\ensuremath{14.718\pm0.010}} 
\newcommand{\hatcurCCtassmBshorteccenxxxxA}{\ensuremath{14.7}}      
\newcommand{\hatcurCCtassmIeccenxxxxA}{\ensuremath{nff\pmnff}}      
\newcommand{\hatcurCCtassmIshorteccenxxxxA}{\ensuremath{0.0}}       
\newcommand{\hatcurCCtassmgeccenxxxxA}{\ensuremath{nff\pmnff}}      
\newcommand{\hatcurCCtassmgshorteccenxxxxA}{\ensuremath{0.0}}       
\newcommand{\hatcurCCtassmreccenxxxxA}{\ensuremath{nff\pmnff}}      
\newcommand{\hatcurCCtassmrshorteccenxxxxA}{\ensuremath{0.0}}       
\newcommand{\hatcurCCtassmieccenxxxxA}{\ensuremath{13.535\pm0.010}} 
\newcommand{\hatcurCCtassmishorteccenxxxxA}{\ensuremath{13.5}}      
\newcommand{\hatcurCCtwomassJmageccenxxxxA}{\ensuremath{12.643\pm0.025}} 
\newcommand{\hatcurCCtwomassHmageccenxxxxA}{\ensuremath{12.373\pm0.034}} 
\newcommand{\hatcurCCtwomassKmageccenxxxxA}{\ensuremath{12.289\pm0.027}} 
\newcommand{\hatcurCCcitJmageccenxxxxA}{\ensuremath{12.661\pm0.025}} 
\newcommand{\hatcurCCcitHmageccenxxxxA}{\ensuremath{12.367\pm0.034}} 
\newcommand{\hatcurCCcitKmageccenxxxxA}{\ensuremath{12.313\pm0.027}} 
\newcommand{\hatcurCCbbJmageccenxxxxA}{\ensuremath{12.709\pm0.027}} 
\newcommand{\hatcurCCbbHmageccenxxxxA}{\ensuremath{12.389\pm0.035}} 
\newcommand{\hatcurCCbbKmageccenxxxxA}{\ensuremath{12.333\pm0.027}} 
\newcommand{\hatcurCCesoJmageccenxxxxA}{\ensuremath{12.711\pm0.028}} 
\newcommand{\hatcurCCesoHmageccenxxxxA}{\ensuremath{12.385\pm0.039}} 
\newcommand{\hatcurCCesoKmageccenxxxxA}{\ensuremath{12.332\pm0.028}} 
\newcommand{\hatcurCCesoJHmageccenxxxxA}{\ensuremath{0.325\pm0.046}} 
\newcommand{\hatcurCCesoJKmageccenxxxxA}{\ensuremath{0.379\pm0.039}} 
\newcommand{\hatcurCCesoHKmageccenxxxxA}{\ensuremath{0.053\pm0.048}} 
\newcommand{\hatcurLCdipeccenxxxxA}{\ensuremath{11.3}}              
\newcommand{\hatcurLCrprstareccenxxxxA}{\ensuremath{0.1032\pm0.0029}} 
\newcommand{\hatcurLCbsqeccenxxxxA}{\ensuremath{0.135_{-0.083}^{+0.098}}} 
\newcommand{\hatcurLCimpeccenxxxxA}{\ensuremath{0.37_{-0.14}^{+0.12}}} 
\newcommand{\hatcurLCzetaeccenxxxxA}{\ensuremath{17.63_{-0.20}^{+0.26}}} 
\newcommand{\hatcurLCdureccenxxxxA}{\ensuremath{0.1271\pm0.0022}}   
\newcommand{\hatcurLCdurshorteccenxxxxA}{\ensuremath{0.1271}}       
\newcommand{\hatcurLCdurhreccenxxxxA}{\ensuremath{3.050\pm0.052}}   
\newcommand{\hatcurLCdurhrshorteccenxxxxA}{\ensuremath{3.050}}      
\newcommand{\hatcurLCqeccenxxxxA}{\ensuremath{0.03320\pm0.00057}}   
\newcommand{\hatcurLCqshorteccenxxxxA}{\ensuremath{0.033}}          
\newcommand{\hatcurLCingdureccenxxxxA}{\ensuremath{0.0137\pm0.0016}} 
\newcommand{\hatcurLCPeccenxxxxA}{\ensuremath{3.8297003\pm0.0000047}} 
\newcommand{\hatcurLCPprececcenxxxxA}{\ensuremath{3.8297003}}       
\newcommand{\hatcurLCPshorteccenxxxxA}{\ensuremath{3.8297}}         
\newcommand{\hatcurLCTeccenxxxxA}{\ensuremath{2456889.49658\pm0.00074}} 
\newcommand{\hatcurLCTAeccenxxxxA}{\ensuremath{2455277.1927\pm0.0021}} 
\newcommand{\hatcurLCTBeccenxxxxA}{\ensuremath{2457180.55384\pm0.00081}} 
\newcommand{\hatcurLChatnetmAeccenxxxxA}{\ensuremath{13.65038\pm0.00013}} 
\newcommand{\hatcurLCiblendAeccenxxxxA}{\ensuremath{0.810\pm0.073}} 
\newcommand{\hatcurLChatnetmBeccenxxxxA}{\ensuremath{13.65044\pm0.00012}} 
\newcommand{\hatcurLCiblendBeccenxxxxA}{\ensuremath{0.875\pm0.072}} 
\newcommand{\hatcurLCrhoeccenxxxxA}{\ensuremath{1.36_{-0.56}^{+1.11}}} 
\newcommand{\hatcurSMEiteffeccenxxxxA}{\ensuremath{5990\pm110}}     
\newcommand{\hatcurSMEizfeheccenxxxxA}{\ensuremath{0.300\pm0.056}}  
\newcommand{\hatcurSMEizfehshorteccenxxxxA}{\ensuremath{0.30}}      
\newcommand{\hatcurSMEiloggeccenxxxxA}{\ensuremath{4.51\pm0.17}}    
\newcommand{\hatcurSMEivsineccenxxxxA}{\ensuremath{3.76\pm0.54}}    
\newcommand{\hatcurSMEivmaceccenxxxxA}{\ensuremath{4.32\pm0.17}}    
\newcommand{\hatcurSMEivmiceccenxxxxA}{\ensuremath{1.225\pm0.085}}  
\newcommand{\hatcurLBizeccenxxxxA}{\ensuremath{0.1881}}             
\newcommand{\hatcurLBiizeccenxxxxA}{\ensuremath{0.3457}}            
\newcommand{\hatcurLBiieccenxxxxA}{\ensuremath{0.2495}}             
\newcommand{\hatcurLBiiieccenxxxxA}{\ensuremath{0.3488}}            
\newcommand{\hatcurLBiIeccenxxxxA}{\ensuremath{0.2285}}             
\newcommand{\hatcurLBiiIeccenxxxxA}{\ensuremath{0.3488}}            
\newcommand{\hatcurLBigeccenxxxxA}{\ensuremath{0.5282}}             
\newcommand{\hatcurLBiigeccenxxxxA}{\ensuremath{0.2582}}            
\newcommand{\hatcurLBireccenxxxxA}{\ensuremath{0.3362}}             
\newcommand{\hatcurLBiireccenxxxxA}{\ensuremath{0.3446}}            
\newcommand{\hatcurLBiReccenxxxxA}{\ensuremath{0.3120}}             
\newcommand{\hatcurLBiiReccenxxxxA}{\ensuremath{0.3470}}            
\newcommand{\hatcurLBikepeccenxxxxA}{\ensuremath{0.1000}}           
\newcommand{\hatcurLBiikepeccenxxxxA}{\ensuremath{0.1000}}          
\newcommand{\hatcurISOmeccenxxxxA}{\ensuremath{1.210_{-0.056}^{+0.090}}} 
\newcommand{\hatcurISOmshorteccenxxxxA}{\ensuremath{1.21}}          
\newcommand{\hatcurISOmlongeccenxxxxA}{\ensuremath{1.210_{-0.056}^{+0.090}}} 
\newcommand{\hatcurISOreccenxxxxA}{\ensuremath{1.26_{-0.13}^{+0.30}}} 
\newcommand{\hatcurISOrshorteccenxxxxA}{\ensuremath{1.26}}          
\newcommand{\hatcurISOrlongeccenxxxxA}{\ensuremath{1.26_{-0.13}^{+0.30}}} 
\newcommand{\hatcurISOrhoeccenxxxxA}{\ensuremath{0.85\pm0.31}}      
\newcommand{\hatcurISOrholongeccenxxxxA}{\ensuremath{0.85\pm0.31}}  
\newcommand{\hatcurISOloggeccenxxxxA}{\ensuremath{4.32\pm0.12}}     
\newcommand{\hatcurISOlumeccenxxxxA}{\ensuremath{1.82_{-0.44}^{+0.99}}} 
\newcommand{\hatcurISOlumshorteccenxxxxA}{\ensuremath{1.82}}        
\newcommand{\hatcurISOmveccenxxxxA}{\ensuremath{4.13\pm0.40}}       
\newcommand{\hatcurISOvieccenxxxxA}{\ensuremath{0.643\pm0.033}}     
\newcommand{\hatcurISOageeccenxxxxA}{\ensuremath{2.70_{-1.59}^{+0.94}}} 
\newcommand{\hatcurISOsigmaeccenxxxxA}{\ensuremath{0.00050\pm0.00033}} 
\newcommand{\hatcurISOMJeccenxxxxA}{\ensuremath{3.08\pm0.38}}       
\newcommand{\hatcurISOMHeccenxxxxA}{\ensuremath{2.79\pm0.37}}       
\newcommand{\hatcurISOMKeccenxxxxA}{\ensuremath{2.74\pm0.37}}       
\newcommand{\hatcurISOJKeccenxxxxA}{\ensuremath{0.350\pm0.040}}     
\newcommand{\hatcurISOspececcenxxxxA}{G}                            
\newcommand{\hatcurRVKeccenxxxxA}{\ensuremath{45\pm14}}             
\newcommand{\hatcurRVrkeccenxxxxA}{\ensuremath{0.01\pm0.28}}        
\newcommand{\hatcurRVrheccenxxxxA}{\ensuremath{0.30\pm0.21}}        
\newcommand{\hatcurRVkeccenxxxxA}{\ensuremath{0.00\pm0.16}}         
\newcommand{\hatcurRVheccenxxxxA}{\ensuremath{0.113_{-0.098}^{+0.175}}} 
\newcommand{\hatcurRVtroneeccenxxxxA}{\ensuremath{0\pm0}}           
\newcommand{\hatcurRVtrtwoeccenxxxxA}{\ensuremath{0\pm0}}           
\newcommand{\hatcurRVgammaAeccenxxxxA}{\ensuremath{-20245\pm14}}    
\newcommand{\hatcurRVjitterAeccenxxxxA}{\ensuremath{68.8\pm9.8}}    
\newcommand{\hatcurRVjittertwosiglimAeccenxxxxA}{\ensuremath{<86.0}} 
\newcommand{\hatcurRVfitrmsAeccenxxxxA}{\ensuremath{0.0}}           
\newcommand{\hatcurRVgammaBeccenxxxxA}{\ensuremath{-20173\pm62}}    
\newcommand{\hatcurRVjitterBeccenxxxxA}{\ensuremath{22\pm96}}       
\newcommand{\hatcurRVjittertwosiglimBeccenxxxxA}{\ensuremath{<237.9}} 
\newcommand{\hatcurRVfitrmsBeccenxxxxA}{\ensuremath{0.0}}           
\newcommand{\hatcurRVgammaCeccenxxxxA}{\ensuremath{9\pm12}}         
\newcommand{\hatcurRVjitterCeccenxxxxA}{\ensuremath{29\pm11}}       
\newcommand{\hatcurRVjittertwosiglimCeccenxxxxA}{\ensuremath{<51.1}} 
\newcommand{\hatcurRVfitrmsCeccenxxxxA}{\ensuremath{0.0}}           
\newcommand{\hatcurRVecceneccenxxxxA}{\ensuremath{0.16\pm0.16}}     
\newcommand{\hatcurRVeccentwosiglimeccenxxxxA}{\ensuremath{<0.516}} 
\newcommand{\hatcurRVomegaeccenxxxxA}{\ensuremath{98\pm75}}         
\newcommand{\hatcurPPieccenxxxxA}{\ensuremath{87.1_{-1.6}^{+1.1}}}  
\newcommand{\hatcurPPgeccenxxxxA}{\ensuremath{5.2_{-1.9}^{+2.6}}}   
\newcommand{\hatcurPPloggeccenxxxxA}{\ensuremath{2.72\pm0.22}}      
\newcommand{\hatcurPPareccenxxxxA}{\ensuremath{8.70_{-1.48}^{+0.98}}} 
\newcommand{\hatcurPPareleccenxxxxA}{\ensuremath{0.05105_{-0.00080}^{+0.00124}}} 
\newcommand{\hatcurPPrhoeccenxxxxA}{\ensuremath{0.21_{-0.10}^{+0.14}}} 
\newcommand{\hatcurPPmeccenxxxxA}{\ensuremath{0.37\pm0.11}}         
\newcommand{\hatcurPPmshorteccenxxxxA}{\ensuremath{0.37}}           
\newcommand{\hatcurPPmlongeccenxxxxA}{\ensuremath{0.37\pm0.11}}     
\newcommand{\hatcurPPmeeccenxxxxA}{\ensuremath{116\pm34}}           
\newcommand{\hatcurPPmeshorteccenxxxxA}{\ensuremath{116.3}}         
\newcommand{\hatcurPPmelongeccenxxxxA}{\ensuremath{116\pm34}}       
\newcommand{\hatcurPPreccenxxxxA}{\ensuremath{1.26_{-0.13}^{+0.31}}} 
\newcommand{\hatcurPPrshorteccenxxxxA}{\ensuremath{1.26}}           
\newcommand{\hatcurPPrlongeccenxxxxA}{\ensuremath{1.26_{-0.13}^{+0.31}}} 
\newcommand{\hatcurPPreeccenxxxxA}{\ensuremath{14.1_{-1.5}^{+3.4}}} 
\newcommand{\hatcurPPreshorteccenxxxxA}{\ensuremath{14.1}}          
\newcommand{\hatcurPPrelongeccenxxxxA}{\ensuremath{14.1_{-1.5}^{+3.4}}} 
\newcommand{\hatcurPPmrcorreccenxxxxA}{\ensuremath{0.08}}           
\newcommand{\hatcurPPteffeccenxxxxA}{\ensuremath{1441_{-90}^{+167}}} 
\newcommand{\hatcurPPthetaeccenxxxxA}{\ensuremath{0.0243\pm0.0080}} 
\newcommand{\hatcurPPfluxperieccenxxxxA}{\ensuremath{1.33_{-0.46}^{+2.11}}} 
\newcommand{\hatcurPPfluxperidimeccenxxxxA}{\ensuremath{9}}         
\newcommand{\hatcurPPfluxapeccenxxxxA}{\ensuremath{7.1\pm1.3}}      
\newcommand{\hatcurPPfluxapdimeccenxxxxA}{\ensuremath{8}}           
\newcommand{\hatcurPPfluxavgeccenxxxxA}{\ensuremath{9.7_{-2.2}^{+5.4}}} 
\newcommand{\hatcurPPfluxavgdimeccenxxxxA}{\ensuremath{8}}          
\newcommand{\hatcurPPfluxavglogeccenxxxxA}{\ensuremath{8.99_{-0.11}^{+0.19}}} 
\newcommand{\hatcurXsecphaseeccenxxxxA}{\ensuremath{0.50\pm0.10}}   
\newcommand{\hatcurXsecondaryeccenxxxxA}{\ensuremath{2456891.42\pm0.40}} 
\newcommand{\hatcurXsecdureccenxxxxA}{\ensuremath{0.153\pm0.040}}   
\newcommand{\hatcurXsecingdureccenxxxxA}{\ensuremath{0.019\pm0.020}} 
\newcommand{\hatcurPPphiconjeccenxxxxA}{\ensuremath{0.002_{-0.102}^{+0.076}}} 
\newcommand{\hatcurPPperieccenxxxxA}{\ensuremath{2456889.49\pm0.50}} 
\newcommand{\hatcurPPaequiveccenxxxxA}{\ensuremath{0.0378_{-0.0067}^{+0.0050}}} 
\newcommand{\hatcurPPtcirceccenxxxxA}{\ensuremath{102_{-87}^{+127}}} 
\newcommand{\hatcurPPtinfalleccenxxxxA}{\ensuremath{10000\pm310000}} 
\newcommand{\hatcurXdisteccenxxxxA}{\ensuremath{830_{-95}^{+199}}}  
\newcommand{\hatcurXAveccenxxxxA}{\ensuremath{0.33\pm0.10}}         
\newcommand{\hatcurXdistredeccenxxxxA}{\ensuremath{809_{-91}^{+192}}} 
\newcommand{\hatcurXEBVeccenxxxxA}{\ensuremath{0.108\pm0.033}}      
\newcommand{\hatcurXmvisoredeccenxxxxA}{\ensuremath{14.022\pm0.048}} 
\newcommand{\hatcurXmiisoredeccenxxxxA}{\ensuremath{13.205\pm0.022}} 
\newcommand{\hatcurXmjisoredeccenxxxxA}{\ensuremath{12.725\pm0.017}} 
\newcommand{\hatcurXmhisoredeccenxxxxA}{\ensuremath{12.393\pm0.019}} 
\newcommand{\hatcurXmkisoredeccenxxxxA}{\ensuremath{12.317\pm0.020}} 
\newcommand{\hatcurXviisoredeccenxxxxA}{\ensuremath{0.817\pm0.034}} 
\newcommand{\hatcurXvkisoredeccenxxxxA}{\ensuremath{1.705\pm0.056}} 
\newcommand{\hatcurXjhisoredeccenxxxxA}{\ensuremath{0.332\pm0.012}} 
\newcommand{\hatcurXjkisoredeccenxxxxA}{\ensuremath{0.408\pm0.012}} 
\newcommand{\hatcurCCpmraeccenxxxxA}{\ensuremath{3.2\pm1.6}}        
\newcommand{\hatcurCCpmdececcenxxxxA}{\ensuremath{1.6\pm1.6}}       
\newcommand{\hatcurCCpmeccenxxxxA}{\ensuremath{3.6\pm2.3}}          

\newcommand{\hatcurhtreccenxxxxB}{HATS601-050}                      
\newcommand{\hatcurfieldeccenxxxxB}{\ensuremath{string}}            
\newcommand{\hatcurCCraeccenxxxxB}{\ensuremath{06^{\mathrm h}51^{\mathrm m}23.40{\mathrm s}}}                     
\newcommand{\hatcurCCdececcenxxxxB}{\ensuremath{-29{\arcdeg}03{\arcmin}31.0{\arcsec}}}                    
\newcommand{\hatcurCCmageccenxxxxB}{12.471}                         
\newcommand{\hatcurCCtwomasseccenxxxxB}{2MASS~06512340-2903309}     
\newcommand{\hatcurCCgsceccenxxxxB}{GSC~6534-00607}                 
\newcommand{\hatcurCCtassmveccenxxxxB}{\ensuremath{12.471\pm0.030}} 
\newcommand{\hatcurCCtassmvshorteccenxxxxB}{\ensuremath{12.5}}      
\newcommand{\hatcurCCtassmBeccenxxxxB}{\ensuremath{13.190\pm0.030}} 
\newcommand{\hatcurCCtassmBshorteccenxxxxB}{\ensuremath{13.2}}      
\newcommand{\hatcurCCtassmIeccenxxxxB}{\ensuremath{nff\pmnff}}      
\newcommand{\hatcurCCtassmIshorteccenxxxxB}{\ensuremath{0.0}}       
\newcommand{\hatcurCCtassmgeccenxxxxB}{\ensuremath{12.766\pm0.030}} 
\newcommand{\hatcurCCtassmgshorteccenxxxxB}{\ensuremath{12.8}}      
\newcommand{\hatcurCCtassmreccenxxxxB}{\ensuremath{12.269\pm0.040}} 
\newcommand{\hatcurCCtassmrshorteccenxxxxB}{\ensuremath{12.3}}      
\newcommand{\hatcurCCtassmieccenxxxxB}{\ensuremath{12.115\pm0.040}} 
\newcommand{\hatcurCCtassmishorteccenxxxxB}{\ensuremath{12.1}}      
\newcommand{\hatcurCCtwomassJmageccenxxxxB}{\ensuremath{11.241\pm0.023}} 
\newcommand{\hatcurCCtwomassHmageccenxxxxB}{\ensuremath{10.955\pm0.024}} 
\newcommand{\hatcurCCtwomassKmageccenxxxxB}{\ensuremath{10.867\pm0.021}} 
\newcommand{\hatcurCCcitJmageccenxxxxB}{\ensuremath{11.258\pm0.023}} 
\newcommand{\hatcurCCcitHmageccenxxxxB}{\ensuremath{10.949\pm0.024}} 
\newcommand{\hatcurCCcitKmageccenxxxxB}{\ensuremath{10.891\pm0.021}} 
\newcommand{\hatcurCCbbJmageccenxxxxB}{\ensuremath{11.307\pm0.025}} 
\newcommand{\hatcurCCbbHmageccenxxxxB}{\ensuremath{10.971\pm0.025}} 
\newcommand{\hatcurCCbbKmageccenxxxxB}{\ensuremath{10.911\pm0.021}} 
\newcommand{\hatcurCCesoJmageccenxxxxB}{\ensuremath{11.309\pm0.026}} 
\newcommand{\hatcurCCesoHmageccenxxxxB}{\ensuremath{10.967\pm0.029}} 
\newcommand{\hatcurCCesoKmageccenxxxxB}{\ensuremath{10.910\pm0.022}} 
\newcommand{\hatcurCCesoJHmageccenxxxxB}{\ensuremath{0.342\pm0.037}} 
\newcommand{\hatcurCCesoJKmageccenxxxxB}{\ensuremath{0.400\pm0.034}} 
\newcommand{\hatcurCCesoHKmageccenxxxxB}{\ensuremath{0.057\pm0.036}} 
\newcommand{\hatcurLCdipeccenxxxxB}{\ensuremath{11.1}}              
\newcommand{\hatcurLCrprstareccenxxxxB}{\ensuremath{0.1010\pm0.0038}} 
\newcommand{\hatcurLCbsqeccenxxxxB}{\ensuremath{0.093_{-0.066}^{+0.095}}} 
\newcommand{\hatcurLCimpeccenxxxxB}{\ensuremath{0.30_{-0.14}^{+0.13}}} 
\newcommand{\hatcurLCzetaeccenxxxxB}{\ensuremath{16.08\pm0.16}}     
\newcommand{\hatcurLCdureccenxxxxB}{\ensuremath{0.1384\pm0.0020}}   
\newcommand{\hatcurLCdurshorteccenxxxxB}{\ensuremath{0.1384}}       
\newcommand{\hatcurLCdurhreccenxxxxB}{\ensuremath{3.321\pm0.047}}   
\newcommand{\hatcurLCdurhrshorteccenxxxxB}{\ensuremath{3.321}}      
\newcommand{\hatcurLCqeccenxxxxB}{\ensuremath{0.04130\pm0.00059}}   
\newcommand{\hatcurLCqshorteccenxxxxB}{\ensuremath{0.041}}          
\newcommand{\hatcurLCingdureccenxxxxB}{\ensuremath{0.0138\pm0.0017}} 
\newcommand{\hatcurLCPeccenxxxxB}{\ensuremath{3.3488702\pm0.0000039}} 
\newcommand{\hatcurLCPprececcenxxxxB}{\ensuremath{3.3488702}}       
\newcommand{\hatcurLCPshorteccenxxxxB}{\ensuremath{3.3489}}         
\newcommand{\hatcurLCTeccenxxxxB}{\ensuremath{2457042.00405\pm0.00058}} 
\newcommand{\hatcurLCTAeccenxxxxB}{\ensuremath{2455796.2243\pm0.0015}} 
\newcommand{\hatcurLCTBeccenxxxxB}{\ensuremath{2457299.86705\pm0.00070}} 
\newcommand{\hatcurLChatnetmeccenxxxxB}{\ensuremath{12.321600\pm0.000061}} 
\newcommand{\hatcurLCiblendeccenxxxxB}{\ensuremath{0.827\pm0.063}}  
\newcommand{\hatcurSMEiteffeccenxxxxB}{\ensuremath{5895\pm71}}      
\newcommand{\hatcurSMEizfeheccenxxxxB}{\ensuremath{0.340\pm0.030}}  
\newcommand{\hatcurSMEizfehshorteccenxxxxB}{\ensuremath{0.34}}      
\newcommand{\hatcurSMEiloggeccenxxxxB}{\ensuremath{4.51\pm0.13}}    
\newcommand{\hatcurSMEivsineccenxxxxB}{\ensuremath{3.76\pm0.37}}    
\newcommand{\hatcurSMEivmaceccenxxxxB}{\ensuremath{0.0}}            
\newcommand{\hatcurSMEivmiceccenxxxxB}{\ensuremath{0.0}}            
\newcommand{\hatcurSMEiiteffeccenxxxxB}{\ensuremath{5758\pm58}}     
\newcommand{\hatcurSMEiizfeheccenxxxxB}{\ensuremath{0.300\pm0.030}} 
\newcommand{\hatcurSMEiizfehshorteccenxxxxB}{\ensuremath{0.30}}     
\newcommand{\hatcurSMEiiloggeccenxxxxB}{\ensuremath{4.324\pm0.042}} 
\newcommand{\hatcurSMEiivsineccenxxxxB}{\ensuremath{3.98\pm0.26}}   
\newcommand{\hatcurLBizeccenxxxxB}{\ensuremath{0.2162}}             
\newcommand{\hatcurLBiizeccenxxxxB}{\ensuremath{0.3324}}            
\newcommand{\hatcurLBiieccenxxxxB}{\ensuremath{0.2833}}             
\newcommand{\hatcurLBiiieccenxxxxB}{\ensuremath{0.3306}}            
\newcommand{\hatcurLBiIeccenxxxxB}{\ensuremath{0.2606}}             
\newcommand{\hatcurLBiiIeccenxxxxB}{\ensuremath{0.3322}}            
\newcommand{\hatcurLBigeccenxxxxB}{\ensuremath{0.5877}}             
\newcommand{\hatcurLBiigeccenxxxxB}{\ensuremath{0.2128}}            
\newcommand{\hatcurLBireccenxxxxB}{\ensuremath{0.3801}}             
\newcommand{\hatcurLBiireccenxxxxB}{\ensuremath{0.3173}}            
\newcommand{\hatcurLBiReccenxxxxB}{\ensuremath{0.3531}}             
\newcommand{\hatcurLBiiReccenxxxxB}{\ensuremath{0.3221}}            
\newcommand{\hatcurLBikepeccenxxxxB}{\ensuremath{0.1000}}           
\newcommand{\hatcurLBiikepeccenxxxxB}{\ensuremath{0.1000}}          
\newcommand{\hatcurISOmeccenxxxxB}{\ensuremath{1.187\pm0.060}}      
\newcommand{\hatcurISOmshorteccenxxxxB}{\ensuremath{1.19}}          
\newcommand{\hatcurISOmlongeccenxxxxB}{\ensuremath{1.187\pm0.060}}  
\newcommand{\hatcurISOreccenxxxxB}{\ensuremath{1.44\pm0.18}}        
\newcommand{\hatcurISOrshorteccenxxxxB}{\ensuremath{1.44}}          
\newcommand{\hatcurISOrlongeccenxxxxB}{\ensuremath{1.44\pm0.18}}    
\newcommand{\hatcurISOrhoeccenxxxxB}{\ensuremath{0.56_{-0.16}^{+0.22}}} 
\newcommand{\hatcurISOrholongeccenxxxxB}{\ensuremath{0.56_{-0.16}^{+0.22}}} 
\newcommand{\hatcurISOloggeccenxxxxB}{\ensuremath{4.198\pm0.088}}   
\newcommand{\hatcurISOlumeccenxxxxB}{\ensuremath{2.04\pm0.54}}      
\newcommand{\hatcurISOlumshorteccenxxxxB}{\ensuremath{2.04}}        
\newcommand{\hatcurISOmveccenxxxxB}{\ensuremath{4.05\pm0.27}}       
\newcommand{\hatcurISOvieccenxxxxB}{\ensuremath{0.709\pm0.019}}     
\newcommand{\hatcurISOageeccenxxxxB}{\ensuremath{4.74_{-0.51}^{+0.70}}} 
\newcommand{\hatcurISOsigmaeccenxxxxB}{\ensuremath{0.00050\pm0.00012}} 
\newcommand{\hatcurISOMJeccenxxxxB}{\ensuremath{2.90\pm0.26}}       
\newcommand{\hatcurISOMHeccenxxxxB}{\ensuremath{2.56\pm0.26}}       
\newcommand{\hatcurISOMKeccenxxxxB}{\ensuremath{2.51\pm0.26}}       
\newcommand{\hatcurISOJKeccenxxxxB}{\ensuremath{0.390\pm0.010}}     
\newcommand{\hatcurISOspececcenxxxxB}{G}                            
\newcommand{\hatcurRVKeccenxxxxB}{\ensuremath{94.9\pm5.1}}          
\newcommand{\hatcurRVrkeccenxxxxB}{\ensuremath{0.24\pm0.13}}        
\newcommand{\hatcurRVrheccenxxxxB}{\ensuremath{0.34_{-0.22}^{+0.14}}} 
\newcommand{\hatcurRVkeccenxxxxB}{\ensuremath{0.100\pm0.060}}       
\newcommand{\hatcurRVheccenxxxxB}{\ensuremath{0.144\pm0.100}}       
\newcommand{\hatcurRVtroneeccenxxxxB}{\ensuremath{0\pm0}}           
\newcommand{\hatcurRVtrtwoeccenxxxxB}{\ensuremath{0\pm0}}           
\newcommand{\hatcurRVgammaAeccenxxxxB}{\ensuremath{3093\pm15}}      
\newcommand{\hatcurRVjitterAeccenxxxxB}{\ensuremath{49\pm11}}       
\newcommand{\hatcurRVfitrmsAeccenxxxxB}{\ensuremath{0.0}}           
\newcommand{\hatcurRVgammaBeccenxxxxB}{\ensuremath{3086\pm14}}      
\newcommand{\hatcurRVjitterBeccenxxxxB}{\ensuremath{58\pm12}}       
\newcommand{\hatcurRVfitrmsBeccenxxxxB}{\ensuremath{0.0}}           
\newcommand{\hatcurRVgammaCeccenxxxxB}{\ensuremath{3097.8\pm6.6}}   
\newcommand{\hatcurRVjitterCeccenxxxxB}{\ensuremath{25.2\pm7.8}}    
\newcommand{\hatcurRVfitrmsCeccenxxxxB}{\ensuremath{0.0}}           
\newcommand{\hatcurRVecceneccenxxxxB}{\ensuremath{0.190\pm0.080}}   
\newcommand{\hatcurRVeccentwosiglimeccenxxxxB}{\ensuremath{<0.330}} 
\newcommand{\hatcurRVomegaeccenxxxxB}{\ensuremath{60\pm76}}         
\newcommand{\hatcurPPieccenxxxxB}{\ensuremath{87.1\pm1.6}}          
\newcommand{\hatcurPPgeccenxxxxB}{\ensuremath{9.5\pm2.4}}           
\newcommand{\hatcurPPloggeccenxxxxB}{\ensuremath{2.98\pm0.11}}      
\newcommand{\hatcurPPareccenxxxxB}{\ensuremath{6.94\pm0.74}}        
\newcommand{\hatcurPPareleccenxxxxB}{\ensuremath{0.04639\pm0.00077}} 
\newcommand{\hatcurPPrhoeccenxxxxB}{\ensuremath{0.34_{-0.11}^{+0.16}}} 
\newcommand{\hatcurPPmeccenxxxxB}{\ensuremath{0.768\pm0.045}}       
\newcommand{\hatcurPPmshorteccenxxxxB}{\ensuremath{0.77}}           
\newcommand{\hatcurPPmlongeccenxxxxB}{\ensuremath{0.768\pm0.045}}   
\newcommand{\hatcurPPmeeccenxxxxB}{\ensuremath{244\pm14}}           
\newcommand{\hatcurPPmeshorteccenxxxxB}{\ensuremath{244.2}}         
\newcommand{\hatcurPPmelongeccenxxxxB}{\ensuremath{244\pm14}}       
\newcommand{\hatcurPPreccenxxxxB}{\ensuremath{1.41\pm0.19}}         
\newcommand{\hatcurPPrshorteccenxxxxB}{\ensuremath{1.41}}           
\newcommand{\hatcurPPrlongeccenxxxxB}{\ensuremath{1.41\pm0.19}}     
\newcommand{\hatcurPPreeccenxxxxB}{\ensuremath{15.8\pm2.1}}         
\newcommand{\hatcurPPreshorteccenxxxxB}{\ensuremath{15.8}}          
\newcommand{\hatcurPPrelongeccenxxxxB}{\ensuremath{15.8\pm2.1}}     
\newcommand{\hatcurPPmrcorreccenxxxxB}{\ensuremath{0.25}}           
\newcommand{\hatcurPPteffeccenxxxxB}{\ensuremath{1553\pm92}}        
\newcommand{\hatcurPPthetaeccenxxxxB}{\ensuremath{0.0421\pm0.0064}} 
\newcommand{\hatcurPPfluxperieccenxxxxB}{\ensuremath{1.95_{-0.62}^{+1.06}}} 
\newcommand{\hatcurPPfluxperidimeccenxxxxB}{\ensuremath{9}}         
\newcommand{\hatcurPPfluxapeccenxxxxB}{\ensuremath{9.1\pm1.3}}      
\newcommand{\hatcurPPfluxapdimeccenxxxxB}{\ensuremath{8}}           
\newcommand{\hatcurPPfluxavgeccenxxxxB}{\ensuremath{1.31_{-0.27}^{+0.36}}} 
\newcommand{\hatcurPPfluxavgdimeccenxxxxB}{\ensuremath{9}}          
\newcommand{\hatcurPPfluxavglogeccenxxxxB}{\ensuremath{9.12\pm0.10}} 
\newcommand{\hatcurXsecphaseeccenxxxxB}{\ensuremath{0.564\pm0.039}} 
\newcommand{\hatcurXsecondaryeccenxxxxB}{\ensuremath{2457043.89\pm0.13}} 
\newcommand{\hatcurXsecdureccenxxxxB}{\ensuremath{0.175\pm0.031}}   
\newcommand{\hatcurXsecingdureccenxxxxB}{\ensuremath{0.020\pm0.011}} 
\newcommand{\hatcurPPphiconjeccenxxxxB}{\ensuremath{0.062_{-0.042}^{+0.087}}} 
\newcommand{\hatcurPPperieccenxxxxB}{\ensuremath{2457041.80\pm0.28}} 
\newcommand{\hatcurPPaequiveccenxxxxB}{\ensuremath{0.0326\pm0.0035}} 
\newcommand{\hatcurPPtcirceccenxxxxB}{\ensuremath{69_{-40}^{+76}}}  
\newcommand{\hatcurPPtinfalleccenxxxxB}{\ensuremath{790_{-330}^{+550}}} 
\newcommand{\hatcurXdisteccenxxxxB}{\ensuremath{479\pm59}}          
\newcommand{\hatcurXAveccenxxxxB}{\ensuremath{0.024_{-0.024}^{+0.059}}} 
\newcommand{\hatcurXdistredeccenxxxxB}{\ensuremath{478\pm59}}       
\newcommand{\hatcurXEBVeccenxxxxB}{\ensuremath{0.0080_{-0.0080}^{+0.0190}}} 
\newcommand{\hatcurXmvisoredeccenxxxxB}{\ensuremath{12.481\pm0.027}} 
\newcommand{\hatcurXmiisoredeccenxxxxB}{\ensuremath{11.753\pm0.015}} 
\newcommand{\hatcurXmjisoredeccenxxxxB}{\ensuremath{11.303\pm0.013}} 
\newcommand{\hatcurXmhisoredeccenxxxxB}{\ensuremath{10.967\pm0.015}} 
\newcommand{\hatcurXmkisoredeccenxxxxB}{\ensuremath{10.909\pm0.016}} 
\newcommand{\hatcurXviisoredeccenxxxxB}{\ensuremath{0.728\pm0.017}} 
\newcommand{\hatcurXvkisoredeccenxxxxB}{\ensuremath{1.572\pm0.032}} 
\newcommand{\hatcurXjhisoredeccenxxxxB}{\ensuremath{0.3350\pm0.0080}} 
\newcommand{\hatcurXjkisoredeccenxxxxB}{\ensuremath{0.3930\pm0.0089}} 
\newcommand{\hatcurCCpmraeccenxxxxB}{\ensuremath{-15.5\pm1.2}}      
\newcommand{\hatcurCCpmdececcenxxxxB}{\ensuremath{-6.9\pm1.1}}      
\newcommand{\hatcurCCpmeccenxxxxB}{\ensuremath{17.0\pm1.6}}         

\newcommand{\hatcurhtreccenxxxxC}{HATS606-028}                      
\newcommand{\hatcurfieldeccenxxxxC}{\ensuremath{string}}            
\newcommand{\hatcurCCraeccenxxxxC}{\ensuremath{09^{\mathrm h}20^{\mathrm m}21.05{\mathrm s}}}                     
\newcommand{\hatcurCCdececcenxxxxC}{\ensuremath{-31{\arcdeg}16{\arcmin}09.6{\arcsec}}}                    
\newcommand{\hatcurCCmageccenxxxxC}{13.669}                         
\newcommand{\hatcurCCtwomasseccenxxxxC}{2MASS~09202105-3116095}     
\newcommand{\hatcurCCgsceccenxxxxC}{GSC~7153-01785}                 
\newcommand{\hatcurCCtassmveccenxxxxC}{\ensuremath{13.669\pm0.040}} 
\newcommand{\hatcurCCtassmvshorteccenxxxxC}{\ensuremath{13.7}}      
\newcommand{\hatcurCCtassmBeccenxxxxC}{\ensuremath{14.316\pm0.030}} 
\newcommand{\hatcurCCtassmBshorteccenxxxxC}{\ensuremath{14.3}}      
\newcommand{\hatcurCCtassmIeccenxxxxC}{\ensuremath{nff\pmnff}}      
\newcommand{\hatcurCCtassmIshorteccenxxxxC}{\ensuremath{0.0}}       
\newcommand{\hatcurCCtassmgeccenxxxxC}{\ensuremath{13.962\pm0.020}} 
\newcommand{\hatcurCCtassmgshorteccenxxxxC}{\ensuremath{14.0}}      
\newcommand{\hatcurCCtassmreccenxxxxC}{\ensuremath{13.490\pm0.060}} 
\newcommand{\hatcurCCtassmrshorteccenxxxxC}{\ensuremath{13.5}}      
\newcommand{\hatcurCCtassmieccenxxxxC}{\ensuremath{13.409\pm0.070}} 
\newcommand{\hatcurCCtassmishorteccenxxxxC}{\ensuremath{13.4}}      
\newcommand{\hatcurCCtwomassJmageccenxxxxC}{\ensuremath{12.523\pm0.034}} 
\newcommand{\hatcurCCtwomassHmageccenxxxxC}{\ensuremath{12.218\pm0.035}} 
\newcommand{\hatcurCCtwomassKmageccenxxxxC}{\ensuremath{12.114\pm0.030}} 
\newcommand{\hatcurCCcitJmageccenxxxxC}{\ensuremath{12.538\pm0.033}} 
\newcommand{\hatcurCCcitHmageccenxxxxC}{\ensuremath{12.212\pm0.035}} 
\newcommand{\hatcurCCcitKmageccenxxxxC}{\ensuremath{12.138\pm0.030}} 
\newcommand{\hatcurCCbbJmageccenxxxxC}{\ensuremath{12.590\pm0.036}} 
\newcommand{\hatcurCCbbHmageccenxxxxC}{\ensuremath{12.234\pm0.036}} 
\newcommand{\hatcurCCbbKmageccenxxxxC}{\ensuremath{12.158\pm0.030}} 
\newcommand{\hatcurCCesoJmageccenxxxxC}{\ensuremath{12.593\pm0.037}} 
\newcommand{\hatcurCCesoHmageccenxxxxC}{\ensuremath{12.230\pm0.041}} 
\newcommand{\hatcurCCesoKmageccenxxxxC}{\ensuremath{12.157\pm0.031}} 
\newcommand{\hatcurCCesoJHmageccenxxxxC}{\ensuremath{0.362\pm0.053}} 
\newcommand{\hatcurCCesoJKmageccenxxxxC}{\ensuremath{0.436\pm0.048}} 
\newcommand{\hatcurCCesoHKmageccenxxxxC}{\ensuremath{0.075\pm0.052}} 
\newcommand{\hatcurLCdipeccenxxxxC}{\ensuremath{20.0}}              
\newcommand{\hatcurLCrprstareccenxxxxC}{\ensuremath{0.1342\pm0.0029}} 
\newcommand{\hatcurLCbsqeccenxxxxC}{\ensuremath{0.175_{-0.091}^{+0.084}}} 
\newcommand{\hatcurLCimpeccenxxxxC}{\ensuremath{0.418_{-0.129}^{+0.091}}} 
\newcommand{\hatcurLCzetaeccenxxxxC}{\ensuremath{26.96\pm0.25}}     
\newcommand{\hatcurLCdureccenxxxxC}{\ensuremath{0.0862\pm0.0013}}   
\newcommand{\hatcurLCdurshorteccenxxxxC}{\ensuremath{0.0862}}       
\newcommand{\hatcurLCdurhreccenxxxxC}{\ensuremath{2.069\pm0.031}}   
\newcommand{\hatcurLCdurhrshorteccenxxxxC}{\ensuremath{2.069}}      
\newcommand{\hatcurLCqeccenxxxxC}{\ensuremath{0.06310\pm0.00096}}   
\newcommand{\hatcurLCqshorteccenxxxxC}{\ensuremath{0.063}}          
\newcommand{\hatcurLCingdureccenxxxxC}{\ensuremath{0.0122\pm0.0013}} 
\newcommand{\hatcurLCPeccenxxxxC}{\ensuremath{1.36665454\pm0.00000090}} 
\newcommand{\hatcurLCPprececcenxxxxC}{\ensuremath{1.3666545}}       
\newcommand{\hatcurLCPshorteccenxxxxC}{\ensuremath{1.3667}}         
\newcommand{\hatcurLCTeccenxxxxC}{\ensuremath{2456922.19715\pm0.00034}} 
\newcommand{\hatcurLCTAeccenxxxxC}{\ensuremath{2455972.37222\pm0.00067}} 
\newcommand{\hatcurLCTBeccenxxxxC}{\ensuremath{2457318.52697\pm0.00045}} 
\newcommand{\hatcurLChatnetmeccenxxxxC}{\ensuremath{13.62597\pm0.00012}} 
\newcommand{\hatcurLCiblendeccenxxxxC}{\ensuremath{0.911\pm0.041}}  
\newcommand{\hatcurLCrhoeccenxxxxC}{\ensuremath{1.58\pm0.43}}       
\newcommand{\hatcurSMEiteffeccenxxxxC}{\ensuremath{5760\pm130}}     
\newcommand{\hatcurSMEizfeheccenxxxxC}{\ensuremath{0.000\pm0.076}}  
\newcommand{\hatcurSMEizfehshorteccenxxxxC}{\ensuremath{0.00}}      
\newcommand{\hatcurSMEiloggeccenxxxxC}{\ensuremath{4.04\pm0.24}}    
\newcommand{\hatcurSMEivsineccenxxxxC}{\ensuremath{5.02\pm0.94}}    
\newcommand{\hatcurSMEivmaceccenxxxxC}{\ensuremath{3.97\pm0.19}}    
\newcommand{\hatcurSMEivmiceccenxxxxC}{\ensuremath{1.071\pm0.075}}  
\newcommand{\hatcurSMEiiteffeccenxxxxC}{\ensuremath{6010\pm150}}    
\newcommand{\hatcurSMEiizfeheccenxxxxC}{\ensuremath{0.22\pm0.10}}   
\newcommand{\hatcurSMEiizfehshorteccenxxxxC}{\ensuremath{0.22}}     
\newcommand{\hatcurSMEiiloggeccenxxxxC}{\ensuremath{4.448\pm0.066}} 
\newcommand{\hatcurSMEiivsineccenxxxxC}{\ensuremath{4.59\pm0.64}}   
\newcommand{\hatcurSMEiivmaceccenxxxxC}{\ensuremath{4.35\pm0.23}}   
\newcommand{\hatcurSMEiivmiceccenxxxxC}{\ensuremath{1.24\pm0.12}}   
\newcommand{\hatcurLBizeccenxxxxC}{\ensuremath{0.1828}}             
\newcommand{\hatcurLBiizeccenxxxxC}{\ensuremath{0.3458}}            
\newcommand{\hatcurLBiieccenxxxxC}{\ensuremath{0.2419}}             
\newcommand{\hatcurLBiiieccenxxxxC}{\ensuremath{0.3502}}            
\newcommand{\hatcurLBiIeccenxxxxC}{\ensuremath{0.2216}}             
\newcommand{\hatcurLBiiIeccenxxxxC}{\ensuremath{0.3497}}            
\newcommand{\hatcurLBigeccenxxxxC}{\ensuremath{0.5139}}             
\newcommand{\hatcurLBiigeccenxxxxC}{\ensuremath{0.2679}}            
\newcommand{\hatcurLBireccenxxxxC}{\ensuremath{0.3262}}             
\newcommand{\hatcurLBiireccenxxxxC}{\ensuremath{0.3487}}            
\newcommand{\hatcurLBiReccenxxxxC}{\ensuremath{0.3027}}             
\newcommand{\hatcurLBiiReccenxxxxC}{\ensuremath{0.3503}}            
\newcommand{\hatcurLBikepeccenxxxxC}{\ensuremath{0.1000}}           
\newcommand{\hatcurLBiikepeccenxxxxC}{\ensuremath{0.1000}}          
\newcommand{\hatcurISOmeccenxxxxC}{\ensuremath{1.140\pm0.062}}      
\newcommand{\hatcurISOmshorteccenxxxxC}{\ensuremath{1.14}}          
\newcommand{\hatcurISOmlongeccenxxxxC}{\ensuremath{1.140\pm0.062}}  
\newcommand{\hatcurISOreccenxxxxC}{\ensuremath{1.104_{-0.077}^{+0.129}}} 
\newcommand{\hatcurISOrshorteccenxxxxC}{\ensuremath{1.10}}          
\newcommand{\hatcurISOrlongeccenxxxxC}{\ensuremath{1.104_{-0.077}^{+0.129}}} 
\newcommand{\hatcurISOrhoeccenxxxxC}{\ensuremath{1.21_{-0.32}^{+0.23}}} 
\newcommand{\hatcurISOrholongeccenxxxxC}{\ensuremath{1.21_{-0.32}^{+0.23}}} 
\newcommand{\hatcurISOloggeccenxxxxC}{\ensuremath{4.411\pm0.061}}   
\newcommand{\hatcurISOlumeccenxxxxC}{\ensuremath{1.36_{-0.28}^{+0.42}}} 
\newcommand{\hatcurISOlumshorteccenxxxxC}{\ensuremath{1.36}}        
\newcommand{\hatcurISOmveccenxxxxC}{\ensuremath{4.46\pm0.28}}       
\newcommand{\hatcurISOvieccenxxxxC}{\ensuremath{0.654\pm0.042}}     
\newcommand{\hatcurISOageeccenxxxxC}{\ensuremath{1.9\pm1.5}}        
\newcommand{\hatcurISOsigmaeccenxxxxC}{\ensuremath{0.00300\pm0.00061}} 
\newcommand{\hatcurISOMJeccenxxxxC}{\ensuremath{3.39\pm0.23}}       
\newcommand{\hatcurISOMHeccenxxxxC}{\ensuremath{3.08\pm0.21}}       
\newcommand{\hatcurISOMKeccenxxxxC}{\ensuremath{3.03\pm0.21}}       
\newcommand{\hatcurISOJKeccenxxxxC}{\ensuremath{0.360\pm0.030}}     
\newcommand{\hatcurISOspececcenxxxxC}{G}                            
\newcommand{\hatcurRVKeccenxxxxC}{\ensuremath{391\pm24}}            
\newcommand{\hatcurRVrkeccenxxxxC}{\ensuremath{-0.01\pm0.11}}       
\newcommand{\hatcurRVrheccenxxxxC}{\ensuremath{0.27_{-0.25}^{+0.16}}} 
\newcommand{\hatcurRVkeccenxxxxC}{\ensuremath{-0.002_{-0.026}^{+0.043}}} 
\newcommand{\hatcurRVheccenxxxxC}{\ensuremath{0.079_{-0.076}^{+0.116}}} 
\newcommand{\hatcurRVtroneeccenxxxxC}{\ensuremath{0\pm0}}           
\newcommand{\hatcurRVtrtwoeccenxxxxC}{\ensuremath{0\pm0}}           
\newcommand{\hatcurRVgammaAeccenxxxxC}{\ensuremath{13442\pm46}}     
\newcommand{\hatcurRVjitterAeccenxxxxC}{\ensuremath{141\pm35}}      
\newcommand{\hatcurRVjittertwosiglimAeccenxxxxC}{\ensuremath{<207.9}} 
\newcommand{\hatcurRVfitrmsAeccenxxxxC}{\ensuremath{0.0}}           
\newcommand{\hatcurRVgammaBeccenxxxxC}{\ensuremath{13624\pm0}}      
\newcommand{\hatcurRVjitterBeccenxxxxC}{\ensuremath{0\pm14}}        
\newcommand{\hatcurRVjittertwosiglimBeccenxxxxC}{\ensuremath{<39.4}} 
\newcommand{\hatcurRVfitrmsBeccenxxxxC}{\ensuremath{0.0}}           
\newcommand{\hatcurRVgammaCeccenxxxxC}{\ensuremath{13395\pm27}}     
\newcommand{\hatcurRVjitterCeccenxxxxC}{\ensuremath{0\pm32}}        
\newcommand{\hatcurRVjittertwosiglimCeccenxxxxC}{\ensuremath{<51.7}} 
\newcommand{\hatcurRVfitrmsCeccenxxxxC}{\ensuremath{0.0}}           
\newcommand{\hatcurRVecceneccenxxxxC}{\ensuremath{0.084\pm0.079}}   
\newcommand{\hatcurRVeccentwosiglimeccenxxxxC}{\ensuremath{<0.246}} 
\newcommand{\hatcurRVomegaeccenxxxxC}{\ensuremath{93\pm62}}         
\newcommand{\hatcurPPieccenxxxxC}{\ensuremath{84.8\pm1.6}}          
\newcommand{\hatcurPPgeccenxxxxC}{\ensuremath{27.3_{-5.2}^{+3.9}}}  
\newcommand{\hatcurPPloggeccenxxxxC}{\ensuremath{3.436_{-0.093}^{+0.057}}} 
\newcommand{\hatcurPPareccenxxxxC}{\ensuremath{4.92_{-0.48}^{+0.29}}} 
\newcommand{\hatcurPPareleccenxxxxC}{\ensuremath{0.02519\pm0.00046}} 
\newcommand{\hatcurPPrhoeccenxxxxC}{\ensuremath{0.94\pm0.23}}       
\newcommand{\hatcurPPmeccenxxxxC}{\ensuremath{2.30\pm0.16}}         
\newcommand{\hatcurPPmshorteccenxxxxC}{\ensuremath{2.30}}           
\newcommand{\hatcurPPmlongeccenxxxxC}{\ensuremath{2.30\pm0.16}}     
\newcommand{\hatcurPPmeeccenxxxxC}{\ensuremath{731\pm51}}           
\newcommand{\hatcurPPmeshorteccenxxxxC}{\ensuremath{731.5}}         
\newcommand{\hatcurPPmelongeccenxxxxC}{\ensuremath{731\pm51}}       
\newcommand{\hatcurPPreccenxxxxC}{\ensuremath{1.44_{-0.11}^{+0.17}}} 
\newcommand{\hatcurPPrshorteccenxxxxC}{\ensuremath{1.44}}           
\newcommand{\hatcurPPrlongeccenxxxxC}{\ensuremath{1.44_{-0.11}^{+0.17}}} 
\newcommand{\hatcurPPreeccenxxxxC}{\ensuremath{16.2_{-1.2}^{+2.0}}} 
\newcommand{\hatcurPPreshorteccenxxxxC}{\ensuremath{16.2}}          
\newcommand{\hatcurPPrelongeccenxxxxC}{\ensuremath{16.2_{-1.2}^{+2.0}}} 
\newcommand{\hatcurPPmrcorreccenxxxxC}{\ensuremath{0.44}}           
\newcommand{\hatcurPPteffeccenxxxxC}{\ensuremath{1901_{-93}^{+124}}} 
\newcommand{\hatcurPPthetaeccenxxxxC}{\ensuremath{0.0698\pm0.0068}} 
\newcommand{\hatcurPPfluxperieccenxxxxC}{\ensuremath{3.48_{-0.86}^{+2.27}}} 
\newcommand{\hatcurPPfluxperidimeccenxxxxC}{\ensuremath{9}}         
\newcommand{\hatcurPPfluxapeccenxxxxC}{\ensuremath{2.46\pm0.35}}    
\newcommand{\hatcurPPfluxapdimeccenxxxxC}{\ensuremath{9}}           
\newcommand{\hatcurPPfluxavgeccenxxxxC}{\ensuremath{2.95_{-0.54}^{+0.84}}} 
\newcommand{\hatcurPPfluxavgdimeccenxxxxC}{\ensuremath{9}}          
\newcommand{\hatcurPPfluxavglogeccenxxxxC}{\ensuremath{9.470\pm0.095}} 
\newcommand{\hatcurXsecphaseeccenxxxxC}{\ensuremath{0.499\pm0.026}} 
\newcommand{\hatcurXsecondaryeccenxxxxC}{\ensuremath{2456922.879\pm0.036}} 
\newcommand{\hatcurXsecdureccenxxxxC}{\ensuremath{0.098\pm0.013}}   
\newcommand{\hatcurXsecingdureccenxxxxC}{\ensuremath{0.0145\pm0.0046}} 
\newcommand{\hatcurPPphiconjeccenxxxxC}{\ensuremath{-0.006_{-0.132}^{+0.049}}} 
\newcommand{\hatcurPPperieccenxxxxC}{\ensuremath{2456922.21\pm0.21}} 
\newcommand{\hatcurPPaequiveccenxxxxC}{\ensuremath{0.0216\pm0.0022}} 
\newcommand{\hatcurPPtcirceccenxxxxC}{\ensuremath{4.4\pm2.0}}       
\newcommand{\hatcurPPtinfalleccenxxxxC}{\ensuremath{18.5\pm6.2}}    
\newcommand{\hatcurXdisteccenxxxxC}{\ensuremath{669_{-57}^{+83}}}   
\newcommand{\hatcurXAveccenxxxxC}{\ensuremath{0.068_{-0.068}^{+0.113}}} 
\newcommand{\hatcurXdistredeccenxxxxC}{\ensuremath{671_{-57}^{+81}}} 
\newcommand{\hatcurXEBVeccenxxxxC}{\ensuremath{0.022_{-0.022}^{+0.037}}} 
\newcommand{\hatcurXmvisoredeccenxxxxC}{\ensuremath{13.689\pm0.041}} 
\newcommand{\hatcurXmiisoredeccenxxxxC}{\ensuremath{12.989\pm0.024}} 
\newcommand{\hatcurXmjisoredeccenxxxxC}{\ensuremath{12.552\pm0.019}} 
\newcommand{\hatcurXmhisoredeccenxxxxC}{\ensuremath{12.231\pm0.023}} 
\newcommand{\hatcurXmkisoredeccenxxxxC}{\ensuremath{12.173\pm0.023}} 
\newcommand{\hatcurXviisoredeccenxxxxC}{\ensuremath{0.700\pm0.027}} 
\newcommand{\hatcurXvkisoredeccenxxxxC}{\ensuremath{1.514\pm0.052}} 
\newcommand{\hatcurXjhisoredeccenxxxxC}{\ensuremath{0.320\pm0.017}} 
\newcommand{\hatcurXjkisoredeccenxxxxC}{\ensuremath{0.377\pm0.016}} 
\newcommand{\hatcurCCpmraeccenxxxxC}{\ensuremath{-27.7\pm3.7}}      
\newcommand{\hatcurCCpmdececcenxxxxC}{\ensuremath{16.0\pm3.7}}      
\newcommand{\hatcurCCpmeccenxxxxC}{\ensuremath{32.0\pm5.2}}         

\newcommand{\hatcurhtreccenxxxxD}{HATS655-001}                      
\newcommand{\hatcurfieldeccenxxxxD}{\ensuremath{string}}            
\newcommand{\hatcurCCraeccenxxxxD}{\ensuremath{11^{\mathrm h}46^{\mathrm m}30.72{\mathrm s}}}                     
\newcommand{\hatcurCCdececcenxxxxD}{\ensuremath{-33{\arcdeg}51{\arcmin}36.2{\arcsec}}}                    
\newcommand{\hatcurCCmageccenxxxxD}{13.790}                         
\newcommand{\hatcurCCtwomasseccenxxxxD}{2MASS~11463084-3351361}     
\newcommand{\hatcurCCgsceccenxxxxD}{GSC~7225-00413}                 
\newcommand{\hatcurCCtassmveccenxxxxD}{\ensuremath{13.790\pm0.030}} 
\newcommand{\hatcurCCtassmvshorteccenxxxxD}{\ensuremath{13.8}}      
\newcommand{\hatcurCCtassmBeccenxxxxD}{\ensuremath{14.548\pm0.040}} 
\newcommand{\hatcurCCtassmBshorteccenxxxxD}{\ensuremath{14.5}}      
\newcommand{\hatcurCCtassmIeccenxxxxD}{\ensuremath{nff\pmnff}}      
\newcommand{\hatcurCCtassmIshorteccenxxxxD}{\ensuremath{0.0}}       
\newcommand{\hatcurCCtassmgeccenxxxxD}{\ensuremath{14.137\pm0.020}} 
\newcommand{\hatcurCCtassmgshorteccenxxxxD}{\ensuremath{14.1}}      
\newcommand{\hatcurCCtassmreccenxxxxD}{\ensuremath{13.579\pm0.030}} 
\newcommand{\hatcurCCtassmrshorteccenxxxxD}{\ensuremath{13.6}}      
\newcommand{\hatcurCCtassmieccenxxxxD}{\ensuremath{13.30\pm0.28}}   
\newcommand{\hatcurCCtassmishorteccenxxxxD}{\ensuremath{13.3}}      
\newcommand{\hatcurCCtwomassJmageccenxxxxD}{\ensuremath{12.458\pm0.025}} 
\newcommand{\hatcurCCtwomassHmageccenxxxxD}{\ensuremath{12.088\pm0.027}} 
\newcommand{\hatcurCCtwomassKmageccenxxxxD}{\ensuremath{12.046\pm0.027}} 
\newcommand{\hatcurCCcitJmageccenxxxxD}{\ensuremath{12.473\pm0.025}} 
\newcommand{\hatcurCCcitHmageccenxxxxD}{\ensuremath{12.083\pm0.028}} 
\newcommand{\hatcurCCcitKmageccenxxxxD}{\ensuremath{12.070\pm0.027}} 
\newcommand{\hatcurCCbbJmageccenxxxxD}{\ensuremath{12.525\pm0.027}} 
\newcommand{\hatcurCCbbHmageccenxxxxD}{\ensuremath{12.104\pm0.028}} 
\newcommand{\hatcurCCbbKmageccenxxxxD}{\ensuremath{12.090\pm0.027}} 
\newcommand{\hatcurCCesoJmageccenxxxxD}{\ensuremath{12.528\pm0.028}} 
\newcommand{\hatcurCCesoHmageccenxxxxD}{\ensuremath{12.097\pm0.029}} 
\newcommand{\hatcurCCesoKmageccenxxxxD}{\ensuremath{12.089\pm0.028}} 
\newcommand{\hatcurCCesoJHmageccenxxxxD}{\ensuremath{0.431\pm0.039}} 
\newcommand{\hatcurCCesoJKmageccenxxxxD}{\ensuremath{0.439\pm0.040}} 
\newcommand{\hatcurCCesoHKmageccenxxxxD}{\ensuremath{0.0080\pm0.0070}} 
\newcommand{\hatcurLCdipeccenxxxxD}{\ensuremath{19.0}}              
\newcommand{\hatcurLCrprstareccenxxxxD}{\ensuremath{0.1260\pm0.0030}} 
\newcommand{\hatcurLCbsqeccenxxxxD}{\ensuremath{0.038_{-0.028}^{+0.055}}} 
\newcommand{\hatcurLCimpeccenxxxxD}{\ensuremath{0.195_{-0.093}^{+0.111}}} 
\newcommand{\hatcurLCzetaeccenxxxxD}{\ensuremath{15.454\pm0.097}}   
\newcommand{\hatcurLCdureccenxxxxD}{\ensuremath{0.1465\pm0.0016}}   
\newcommand{\hatcurLCdurshorteccenxxxxD}{\ensuremath{0.1465}}       
\newcommand{\hatcurLCdurhreccenxxxxD}{\ensuremath{3.516\pm0.040}}   
\newcommand{\hatcurLCdurhrshorteccenxxxxD}{\ensuremath{3.516}}      
\newcommand{\hatcurLCqeccenxxxxD}{\ensuremath{0.03800\pm0.00043}}   
\newcommand{\hatcurLCqshorteccenxxxxD}{\ensuremath{0.038}}          
\newcommand{\hatcurLCingdureccenxxxxD}{\ensuremath{0.0171\pm0.0013}} 
\newcommand{\hatcurLCPeccenxxxxD}{\ensuremath{3.8537777\pm0.0000042}} 
\newcommand{\hatcurLCPprececcenxxxxD}{\ensuremath{3.8537777}}       
\newcommand{\hatcurLCPshorteccenxxxxD}{\ensuremath{3.8538}}         
\newcommand{\hatcurLCTeccenxxxxD}{\ensuremath{2457283.00191\pm0.00046}} 
\newcommand{\hatcurLCTAeccenxxxxD}{\ensuremath{2455672.1229\pm0.0017}} 
\newcommand{\hatcurLCTBeccenxxxxD}{\ensuremath{2457433.29925\pm0.00052}} 
\newcommand{\hatcurLChatnetmeccenxxxxD}{\ensuremath{13.62311\pm0.00010}} 
\newcommand{\hatcurLCiblendeccenxxxxD}{\ensuremath{0.904\pm0.046}}  
\newcommand{\hatcurLCrhoeccenxxxxD}{\ensuremath{0.82_{-0.31}^{+0.24}}} 
\newcommand{\hatcurSMEiteffeccenxxxxD}{\ensuremath{5700\pm110}}     
\newcommand{\hatcurSMEizfeheccenxxxxD}{\ensuremath{-0.040\pm0.063}} 
\newcommand{\hatcurSMEizfehshorteccenxxxxD}{\ensuremath{-0.04}}     
\newcommand{\hatcurSMEiloggeccenxxxxD}{\ensuremath{4.49\pm0.14}}    
\newcommand{\hatcurSMEivsineccenxxxxD}{\ensuremath{2.04\pm0.98}}    
\newcommand{\hatcurSMEivmaceccenxxxxD}{\ensuremath{3.87\pm0.17}}    
\newcommand{\hatcurSMEivmiceccenxxxxD}{\ensuremath{1.037\pm0.063}}  
\newcommand{\hatcurSMEiiteffeccenxxxxD}{\ensuremath{5644\pm94}}     
\newcommand{\hatcurSMEiizfeheccenxxxxD}{\ensuremath{0.010\pm0.066}} 
\newcommand{\hatcurSMEiizfehshorteccenxxxxD}{\ensuremath{0.01}}     
\newcommand{\hatcurSMEiiloggeccenxxxxD}{\ensuremath{4.338\pm0.022}} 
\newcommand{\hatcurSMEiivsineccenxxxxD}{\ensuremath{2.50\pm0.76}}   
\newcommand{\hatcurSMEiivmaceccenxxxxD}{\ensuremath{3.79\pm0.14}}   
\newcommand{\hatcurSMEiivmiceccenxxxxD}{\ensuremath{1.006\pm0.049}} 
\newcommand{\hatcurLBizeccenxxxxD}{\ensuremath{0.2257}}             
\newcommand{\hatcurLBiizeccenxxxxD}{\ensuremath{0.3193}}            
\newcommand{\hatcurLBiieccenxxxxD}{\ensuremath{0.2888}}             
\newcommand{\hatcurLBiiieccenxxxxD}{\ensuremath{0.3179}}            
\newcommand{\hatcurLBiIeccenxxxxD}{\ensuremath{0.2676}}             
\newcommand{\hatcurLBiiIeccenxxxxD}{\ensuremath{0.3192}}            
\newcommand{\hatcurLBigeccenxxxxD}{\ensuremath{0.5830}}             
\newcommand{\hatcurLBiigeccenxxxxD}{\ensuremath{0.2134}}            
\newcommand{\hatcurLBireccenxxxxD}{\ensuremath{0.3816}}             
\newcommand{\hatcurLBiireccenxxxxD}{\ensuremath{0.3101}}            
\newcommand{\hatcurLBiReccenxxxxD}{\ensuremath{0.3559}}             
\newcommand{\hatcurLBiiReccenxxxxD}{\ensuremath{0.3132}}            
\newcommand{\hatcurLBikepeccenxxxxD}{\ensuremath{0.1000}}           
\newcommand{\hatcurLBiikepeccenxxxxD}{\ensuremath{0.1000}}          
\newcommand{\hatcurISOmeccenxxxxD}{\ensuremath{0.981_{-0.040}^{+0.055}}} 
\newcommand{\hatcurISOmshorteccenxxxxD}{\ensuremath{0.98}}          
\newcommand{\hatcurISOmlongeccenxxxxD}{\ensuremath{0.981_{-0.040}^{+0.055}}} 
\newcommand{\hatcurISOreccenxxxxD}{\ensuremath{1.19_{-0.10}^{+0.22}}} 
\newcommand{\hatcurISOrshorteccenxxxxD}{\ensuremath{1.19}}          
\newcommand{\hatcurISOrlongeccenxxxxD}{\ensuremath{1.19_{-0.10}^{+0.22}}} 
\newcommand{\hatcurISOrhoeccenxxxxD}{\ensuremath{0.82\pm0.27}}      
\newcommand{\hatcurISOrholongeccenxxxxD}{\ensuremath{0.82\pm0.27}}  
\newcommand{\hatcurISOloggeccenxxxxD}{\ensuremath{4.28\pm0.11}}     
\newcommand{\hatcurISOlumeccenxxxxD}{\ensuremath{1.30_{-0.24}^{+0.53}}} 
\newcommand{\hatcurISOlumshorteccenxxxxD}{\ensuremath{1.30}}        
\newcommand{\hatcurISOmveccenxxxxD}{\ensuremath{4.57\pm0.32}}       
\newcommand{\hatcurISOvieccenxxxxD}{\ensuremath{0.736\pm0.027}}     
\newcommand{\hatcurISOageeccenxxxxD}{\ensuremath{9.4\pm1.9}}        
\newcommand{\hatcurISOsigmaeccenxxxxD}{\ensuremath{0.00060\pm0.00017}} 
\newcommand{\hatcurISOMJeccenxxxxD}{\ensuremath{3.38\pm0.31}}       
\newcommand{\hatcurISOMHeccenxxxxD}{\ensuremath{3.02\pm0.31}}       
\newcommand{\hatcurISOMKeccenxxxxD}{\ensuremath{2.96\pm0.31}}       
\newcommand{\hatcurISOJKeccenxxxxD}{\ensuremath{0.420\pm0.020}}     
\newcommand{\hatcurISOspececcenxxxxD}{G}                            
\newcommand{\hatcurRVKeccenxxxxD}{\ensuremath{80\pm12}}             
\newcommand{\hatcurRVrkeccenxxxxD}{\ensuremath{0.00\pm0.17}}        
\newcommand{\hatcurRVrheccenxxxxD}{\ensuremath{0.22\pm0.23}}        
\newcommand{\hatcurRVkeccenxxxxD}{\ensuremath{0.002\pm0.067}}       
\newcommand{\hatcurRVheccenxxxxD}{\ensuremath{0.061_{-0.069}^{+0.151}}} 
\newcommand{\hatcurRVtroneeccenxxxxD}{\ensuremath{0\pm0}}           
\newcommand{\hatcurRVtrtwoeccenxxxxD}{\ensuremath{0\pm0}}           
\newcommand{\hatcurRVgammaAeccenxxxxD}{\ensuremath{71950.2\pm7.3}}  
\newcommand{\hatcurRVjitterAeccenxxxxD}{\ensuremath{16\pm14}}       
\newcommand{\hatcurRVjittertwosiglimAeccenxxxxD}{\ensuremath{<38.1}} 
\newcommand{\hatcurRVfitrmsAeccenxxxxD}{\ensuremath{0.0}}           
\newcommand{\hatcurRVgammaBeccenxxxxD}{\ensuremath{71941\pm25}}     
\newcommand{\hatcurRVjitterBeccenxxxxD}{\ensuremath{0\pm25}}        
\newcommand{\hatcurRVjittertwosiglimBeccenxxxxD}{\ensuremath{<56.4}} 
\newcommand{\hatcurRVfitrmsBeccenxxxxD}{\ensuremath{0.0}}           
\newcommand{\hatcurRVecceneccenxxxxD}{\ensuremath{0.10\pm0.10}}     
\newcommand{\hatcurRVeccentwosiglimeccenxxxxD}{\ensuremath{<0.330}} 
\newcommand{\hatcurRVomegaeccenxxxxD}{\ensuremath{100\pm100}}       
\newcommand{\hatcurPPieccenxxxxD}{\ensuremath{88.56_{-1.08}^{+0.72}}} 
\newcommand{\hatcurPPgeccenxxxxD}{\ensuremath{6.9\pm2.0}}           
\newcommand{\hatcurPPloggeccenxxxxD}{\ensuremath{2.84_{-0.17}^{+0.10}}} 
\newcommand{\hatcurPPareccenxxxxD}{\ensuremath{8.63_{-1.27}^{+0.77}}} 
\newcommand{\hatcurPPareleccenxxxxD}{\ensuremath{0.04780_{-0.00067}^{+0.00088}}} 
\newcommand{\hatcurPPrhoeccenxxxxD}{\ensuremath{0.236\pm0.094}}     
\newcommand{\hatcurPPmeccenxxxxD}{\ensuremath{0.600\pm0.087}}       
\newcommand{\hatcurPPmshorteccenxxxxD}{\ensuremath{0.60}}           
\newcommand{\hatcurPPmlongeccenxxxxD}{\ensuremath{0.600\pm0.087}}   
\newcommand{\hatcurPPmeeccenxxxxD}{\ensuremath{191\pm28}}           
\newcommand{\hatcurPPmeshorteccenxxxxD}{\ensuremath{190.8}}         
\newcommand{\hatcurPPmelongeccenxxxxD}{\ensuremath{191\pm28}}       
\newcommand{\hatcurPPreccenxxxxD}{\ensuremath{1.46_{-0.13}^{+0.28}}} 
\newcommand{\hatcurPPrshorteccenxxxxD}{\ensuremath{1.46}}           
\newcommand{\hatcurPPrlongeccenxxxxD}{\ensuremath{1.46_{-0.13}^{+0.28}}} 
\newcommand{\hatcurPPreeccenxxxxD}{\ensuremath{16.3_{-1.5}^{+3.2}}} 
\newcommand{\hatcurPPreshorteccenxxxxD}{\ensuremath{16.3}}          
\newcommand{\hatcurPPrelongeccenxxxxD}{\ensuremath{16.3_{-1.5}^{+3.2}}} 
\newcommand{\hatcurPPmrcorreccenxxxxD}{\ensuremath{0.03}}           
\newcommand{\hatcurPPteffeccenxxxxD}{\ensuremath{1363_{-63}^{+121}}} 
\newcommand{\hatcurPPthetaeccenxxxxD}{\ensuremath{0.0397\pm0.0080}} 
\newcommand{\hatcurPPfluxperieccenxxxxD}{\ensuremath{9.3_{-2.3}^{+8.4}}} 
\newcommand{\hatcurPPfluxperidimeccenxxxxD}{\ensuremath{8}}         
\newcommand{\hatcurPPfluxapeccenxxxxD}{\ensuremath{6.59\pm0.97}}    
\newcommand{\hatcurPPfluxapdimeccenxxxxD}{\ensuremath{8}}           
\newcommand{\hatcurPPfluxavgeccenxxxxD}{\ensuremath{7.8_{-1.3}^{+3.2}}} 
\newcommand{\hatcurPPfluxavgdimeccenxxxxD}{\ensuremath{8}}          
\newcommand{\hatcurPPfluxavglogeccenxxxxD}{\ensuremath{8.891_{-0.083}^{+0.148}}} 
\newcommand{\hatcurXsecphaseeccenxxxxD}{\ensuremath{0.501\pm0.043}} 
\newcommand{\hatcurXsecondaryeccenxxxxD}{\ensuremath{2457284.93\pm0.17}} 
\newcommand{\hatcurXsecdureccenxxxxD}{\ensuremath{0.165\pm0.039}}   
\newcommand{\hatcurXsecingdureccenxxxxD}{\ensuremath{0.020\pm0.015}} 
\newcommand{\hatcurPPphiconjeccenxxxxD}{\ensuremath{0.002_{-0.085}^{+0.234}}} 
\newcommand{\hatcurPPperieccenxxxxD}{\ensuremath{2457282.99\pm0.71}} 
\newcommand{\hatcurPPaequiveccenxxxxD}{\ensuremath{0.0420_{-0.0062}^{+0.0041}}} 
\newcommand{\hatcurPPtcirceccenxxxxD}{\ensuremath{89\pm57}}         
\newcommand{\hatcurPPtinfalleccenxxxxD}{\ensuremath{2900\pm1700}}   
\newcommand{\hatcurXdisteccenxxxxD}{\ensuremath{671_{-59}^{+125}}}  
\newcommand{\hatcurXAveccenxxxxD}{\ensuremath{0.113\pm0.078}}       
\newcommand{\hatcurXdistredeccenxxxxD}{\ensuremath{660_{-57}^{+123}}} 
\newcommand{\hatcurXEBVeccenxxxxD}{\ensuremath{0.036\pm0.025}}      
\newcommand{\hatcurXmvisoredeccenxxxxD}{\ensuremath{13.792\pm0.029}} 
\newcommand{\hatcurXmiisoredeccenxxxxD}{\ensuremath{12.997\pm0.019}} 
\newcommand{\hatcurXmjisoredeccenxxxxD}{\ensuremath{12.510\pm0.016}} 
\newcommand{\hatcurXmhisoredeccenxxxxD}{\ensuremath{12.136\pm0.017}} 
\newcommand{\hatcurXmkisoredeccenxxxxD}{\ensuremath{12.068\pm0.018}} 
\newcommand{\hatcurXviisoredeccenxxxxD}{\ensuremath{0.793\pm0.022}} 
\newcommand{\hatcurXvkisoredeccenxxxxD}{\ensuremath{1.724\pm0.036}} 
\newcommand{\hatcurXjhisoredeccenxxxxD}{\ensuremath{0.374\pm0.012}} 
\newcommand{\hatcurXjkisoredeccenxxxxD}{\ensuremath{0.442\pm0.012}} 
\newcommand{\hatcurCCpmraeccenxxxxD}{\ensuremath{-35.0\pm2.0}}      
\newcommand{\hatcurCCpmdececcenxxxxD}{\ensuremath{-5.0\pm2.1}}      
\newcommand{\hatcurCCpmeccenxxxxD}{\ensuremath{35.4\pm2.9}}         

\newcommand{\hatcurCCbbHmageccen}[1]{\ifnum#1=50 %
\hatcurCCbbHmageccenxxxxA
\else
\ifnum#1=51 %
\hatcurCCbbHmageccenxxxxB
\else
\ifnum#1=52 %
\hatcurCCbbHmageccenxxxxC
\else
\ifnum#1=53 %
\hatcurCCbbHmageccenxxxxD
\else
??????\fi
\fi
\fi
\fi
}
\newcommand{\hatcurCCbbJmageccen}[1]{\ifnum#1=50 %
\hatcurCCbbJmageccenxxxxA
\else
\ifnum#1=51 %
\hatcurCCbbJmageccenxxxxB
\else
\ifnum#1=52 %
\hatcurCCbbJmageccenxxxxC
\else
\ifnum#1=53 %
\hatcurCCbbJmageccenxxxxD
\else
??????\fi
\fi
\fi
\fi
}
\newcommand{\hatcurCCbbKmageccen}[1]{\ifnum#1=50 %
\hatcurCCbbKmageccenxxxxA
\else
\ifnum#1=51 %
\hatcurCCbbKmageccenxxxxB
\else
\ifnum#1=52 %
\hatcurCCbbKmageccenxxxxC
\else
\ifnum#1=53 %
\hatcurCCbbKmageccenxxxxD
\else
??????\fi
\fi
\fi
\fi
}
\newcommand{\hatcurCCcitHmageccen}[1]{\ifnum#1=50 %
\hatcurCCcitHmageccenxxxxA
\else
\ifnum#1=51 %
\hatcurCCcitHmageccenxxxxB
\else
\ifnum#1=52 %
\hatcurCCcitHmageccenxxxxC
\else
\ifnum#1=53 %
\hatcurCCcitHmageccenxxxxD
\else
??????\fi
\fi
\fi
\fi
}
\newcommand{\hatcurCCcitJmageccen}[1]{\ifnum#1=50 %
\hatcurCCcitJmageccenxxxxA
\else
\ifnum#1=51 %
\hatcurCCcitJmageccenxxxxB
\else
\ifnum#1=52 %
\hatcurCCcitJmageccenxxxxC
\else
\ifnum#1=53 %
\hatcurCCcitJmageccenxxxxD
\else
??????\fi
\fi
\fi
\fi
}
\newcommand{\hatcurCCcitKmageccen}[1]{\ifnum#1=50 %
\hatcurCCcitKmageccenxxxxA
\else
\ifnum#1=51 %
\hatcurCCcitKmageccenxxxxB
\else
\ifnum#1=52 %
\hatcurCCcitKmageccenxxxxC
\else
\ifnum#1=53 %
\hatcurCCcitKmageccenxxxxD
\else
??????\fi
\fi
\fi
\fi
}
\newcommand{\hatcurCCdececcen}[1]{\ifnum#1=50 %
\hatcurCCdececcenxxxxA
\else
\ifnum#1=51 %
\hatcurCCdececcenxxxxB
\else
\ifnum#1=52 %
\hatcurCCdececcenxxxxC
\else
\ifnum#1=53 %
\hatcurCCdececcenxxxxD
\else
??????\fi
\fi
\fi
\fi
}
\newcommand{\hatcurCCesoHKmageccen}[1]{\ifnum#1=50 %
\hatcurCCesoHKmageccenxxxxA
\else
\ifnum#1=51 %
\hatcurCCesoHKmageccenxxxxB
\else
\ifnum#1=52 %
\hatcurCCesoHKmageccenxxxxC
\else
\ifnum#1=53 %
\hatcurCCesoHKmageccenxxxxD
\else
??????\fi
\fi
\fi
\fi
}
\newcommand{\hatcurCCesoHmageccen}[1]{\ifnum#1=50 %
\hatcurCCesoHmageccenxxxxA
\else
\ifnum#1=51 %
\hatcurCCesoHmageccenxxxxB
\else
\ifnum#1=52 %
\hatcurCCesoHmageccenxxxxC
\else
\ifnum#1=53 %
\hatcurCCesoHmageccenxxxxD
\else
??????\fi
\fi
\fi
\fi
}
\newcommand{\hatcurCCesoJHmageccen}[1]{\ifnum#1=50 %
\hatcurCCesoJHmageccenxxxxA
\else
\ifnum#1=51 %
\hatcurCCesoJHmageccenxxxxB
\else
\ifnum#1=52 %
\hatcurCCesoJHmageccenxxxxC
\else
\ifnum#1=53 %
\hatcurCCesoJHmageccenxxxxD
\else
??????\fi
\fi
\fi
\fi
}
\newcommand{\hatcurCCesoJKmageccen}[1]{\ifnum#1=50 %
\hatcurCCesoJKmageccenxxxxA
\else
\ifnum#1=51 %
\hatcurCCesoJKmageccenxxxxB
\else
\ifnum#1=52 %
\hatcurCCesoJKmageccenxxxxC
\else
\ifnum#1=53 %
\hatcurCCesoJKmageccenxxxxD
\else
??????\fi
\fi
\fi
\fi
}
\newcommand{\hatcurCCesoJmageccen}[1]{\ifnum#1=50 %
\hatcurCCesoJmageccenxxxxA
\else
\ifnum#1=51 %
\hatcurCCesoJmageccenxxxxB
\else
\ifnum#1=52 %
\hatcurCCesoJmageccenxxxxC
\else
\ifnum#1=53 %
\hatcurCCesoJmageccenxxxxD
\else
??????\fi
\fi
\fi
\fi
}
\newcommand{\hatcurCCesoKmageccen}[1]{\ifnum#1=50 %
\hatcurCCesoKmageccenxxxxA
\else
\ifnum#1=51 %
\hatcurCCesoKmageccenxxxxB
\else
\ifnum#1=52 %
\hatcurCCesoKmageccenxxxxC
\else
\ifnum#1=53 %
\hatcurCCesoKmageccenxxxxD
\else
??????\fi
\fi
\fi
\fi
}
\newcommand{\hatcurCCgsceccen}[1]{\ifnum#1=50 %
\hatcurCCgsceccenxxxxA
\else
\ifnum#1=51 %
\hatcurCCgsceccenxxxxB
\else
\ifnum#1=52 %
\hatcurCCgsceccenxxxxC
\else
\ifnum#1=53 %
\hatcurCCgsceccenxxxxD
\else
??????\fi
\fi
\fi
\fi
}
\newcommand{\hatcurCCmageccen}[1]{\ifnum#1=50 %
\hatcurCCmageccenxxxxA
\else
\ifnum#1=51 %
\hatcurCCmageccenxxxxB
\else
\ifnum#1=52 %
\hatcurCCmageccenxxxxC
\else
\ifnum#1=53 %
\hatcurCCmageccenxxxxD
\else
??????\fi
\fi
\fi
\fi
}
\newcommand{\hatcurCCpmdececcen}[1]{\ifnum#1=50 %
\hatcurCCpmdececcenxxxxA
\else
\ifnum#1=51 %
\hatcurCCpmdececcenxxxxB
\else
\ifnum#1=52 %
\hatcurCCpmdececcenxxxxC
\else
\ifnum#1=53 %
\hatcurCCpmdececcenxxxxD
\else
??????\fi
\fi
\fi
\fi
}
\newcommand{\hatcurCCpmeccen}[1]{\ifnum#1=50 %
\hatcurCCpmeccenxxxxA
\else
\ifnum#1=51 %
\hatcurCCpmeccenxxxxB
\else
\ifnum#1=52 %
\hatcurCCpmeccenxxxxC
\else
\ifnum#1=53 %
\hatcurCCpmeccenxxxxD
\else
??????\fi
\fi
\fi
\fi
}
\newcommand{\hatcurCCpmraeccen}[1]{\ifnum#1=50 %
\hatcurCCpmraeccenxxxxA
\else
\ifnum#1=51 %
\hatcurCCpmraeccenxxxxB
\else
\ifnum#1=52 %
\hatcurCCpmraeccenxxxxC
\else
\ifnum#1=53 %
\hatcurCCpmraeccenxxxxD
\else
??????\fi
\fi
\fi
\fi
}
\newcommand{\hatcurCCraeccen}[1]{\ifnum#1=50 %
\hatcurCCraeccenxxxxA
\else
\ifnum#1=51 %
\hatcurCCraeccenxxxxB
\else
\ifnum#1=52 %
\hatcurCCraeccenxxxxC
\else
\ifnum#1=53 %
\hatcurCCraeccenxxxxD
\else
??????\fi
\fi
\fi
\fi
}
\newcommand{\hatcurCCtassmBeccen}[1]{\ifnum#1=50 %
\hatcurCCtassmBeccenxxxxA
\else
\ifnum#1=51 %
\hatcurCCtassmBeccenxxxxB
\else
\ifnum#1=52 %
\hatcurCCtassmBeccenxxxxC
\else
\ifnum#1=53 %
\hatcurCCtassmBeccenxxxxD
\else
??????\fi
\fi
\fi
\fi
}
\newcommand{\hatcurCCtassmBshorteccen}[1]{\ifnum#1=50 %
\hatcurCCtassmBshorteccenxxxxA
\else
\ifnum#1=51 %
\hatcurCCtassmBshorteccenxxxxB
\else
\ifnum#1=52 %
\hatcurCCtassmBshorteccenxxxxC
\else
\ifnum#1=53 %
\hatcurCCtassmBshorteccenxxxxD
\else
??????\fi
\fi
\fi
\fi
}
\newcommand{\hatcurCCtassmgeccen}[1]{\ifnum#1=50 %
\hatcurCCtassmgeccenxxxxA
\else
\ifnum#1=51 %
\hatcurCCtassmgeccenxxxxB
\else
\ifnum#1=52 %
\hatcurCCtassmgeccenxxxxC
\else
\ifnum#1=53 %
\hatcurCCtassmgeccenxxxxD
\else
??????\fi
\fi
\fi
\fi
}
\newcommand{\hatcurCCtassmgshorteccen}[1]{\ifnum#1=50 %
\hatcurCCtassmgshorteccenxxxxA
\else
\ifnum#1=51 %
\hatcurCCtassmgshorteccenxxxxB
\else
\ifnum#1=52 %
\hatcurCCtassmgshorteccenxxxxC
\else
\ifnum#1=53 %
\hatcurCCtassmgshorteccenxxxxD
\else
??????\fi
\fi
\fi
\fi
}
\newcommand{\hatcurCCtassmieccen}[1]{\ifnum#1=50 %
\hatcurCCtassmieccenxxxxA
\else
\ifnum#1=51 %
\hatcurCCtassmieccenxxxxB
\else
\ifnum#1=52 %
\hatcurCCtassmieccenxxxxC
\else
\ifnum#1=53 %
\hatcurCCtassmieccenxxxxD
\else
??????\fi
\fi
\fi
\fi
}
\newcommand{\hatcurCCtassmIeccen}[1]{\ifnum#1=50 %
\hatcurCCtassmIeccenxxxxA
\else
\ifnum#1=51 %
\hatcurCCtassmIeccenxxxxB
\else
\ifnum#1=52 %
\hatcurCCtassmIeccenxxxxC
\else
\ifnum#1=53 %
\hatcurCCtassmIeccenxxxxD
\else
??????\fi
\fi
\fi
\fi
}
\newcommand{\hatcurCCtassmishorteccen}[1]{\ifnum#1=50 %
\hatcurCCtassmishorteccenxxxxA
\else
\ifnum#1=51 %
\hatcurCCtassmishorteccenxxxxB
\else
\ifnum#1=52 %
\hatcurCCtassmishorteccenxxxxC
\else
\ifnum#1=53 %
\hatcurCCtassmishorteccenxxxxD
\else
??????\fi
\fi
\fi
\fi
}
\newcommand{\hatcurCCtassmIshorteccen}[1]{\ifnum#1=50 %
\hatcurCCtassmIshorteccenxxxxA
\else
\ifnum#1=51 %
\hatcurCCtassmIshorteccenxxxxB
\else
\ifnum#1=52 %
\hatcurCCtassmIshorteccenxxxxC
\else
\ifnum#1=53 %
\hatcurCCtassmIshorteccenxxxxD
\else
??????\fi
\fi
\fi
\fi
}
\newcommand{\hatcurCCtassmreccen}[1]{\ifnum#1=50 %
\hatcurCCtassmreccenxxxxA
\else
\ifnum#1=51 %
\hatcurCCtassmreccenxxxxB
\else
\ifnum#1=52 %
\hatcurCCtassmreccenxxxxC
\else
\ifnum#1=53 %
\hatcurCCtassmreccenxxxxD
\else
??????\fi
\fi
\fi
\fi
}
\newcommand{\hatcurCCtassmrshorteccen}[1]{\ifnum#1=50 %
\hatcurCCtassmrshorteccenxxxxA
\else
\ifnum#1=51 %
\hatcurCCtassmrshorteccenxxxxB
\else
\ifnum#1=52 %
\hatcurCCtassmrshorteccenxxxxC
\else
\ifnum#1=53 %
\hatcurCCtassmrshorteccenxxxxD
\else
??????\fi
\fi
\fi
\fi
}
\newcommand{\hatcurCCtassmveccen}[1]{\ifnum#1=50 %
\hatcurCCtassmveccenxxxxA
\else
\ifnum#1=51 %
\hatcurCCtassmveccenxxxxB
\else
\ifnum#1=52 %
\hatcurCCtassmveccenxxxxC
\else
\ifnum#1=53 %
\hatcurCCtassmveccenxxxxD
\else
??????\fi
\fi
\fi
\fi
}
\newcommand{\hatcurCCtassmvshorteccen}[1]{\ifnum#1=50 %
\hatcurCCtassmvshorteccenxxxxA
\else
\ifnum#1=51 %
\hatcurCCtassmvshorteccenxxxxB
\else
\ifnum#1=52 %
\hatcurCCtassmvshorteccenxxxxC
\else
\ifnum#1=53 %
\hatcurCCtassmvshorteccenxxxxD
\else
??????\fi
\fi
\fi
\fi
}
\newcommand{\hatcurCCtwomasseccen}[1]{\ifnum#1=50 %
\hatcurCCtwomasseccenxxxxA
\else
\ifnum#1=51 %
\hatcurCCtwomasseccenxxxxB
\else
\ifnum#1=52 %
\hatcurCCtwomasseccenxxxxC
\else
\ifnum#1=53 %
\hatcurCCtwomasseccenxxxxD
\else
??????\fi
\fi
\fi
\fi
}
\newcommand{\hatcurCCtwomassHmageccen}[1]{\ifnum#1=50 %
\hatcurCCtwomassHmageccenxxxxA
\else
\ifnum#1=51 %
\hatcurCCtwomassHmageccenxxxxB
\else
\ifnum#1=52 %
\hatcurCCtwomassHmageccenxxxxC
\else
\ifnum#1=53 %
\hatcurCCtwomassHmageccenxxxxD
\else
??????\fi
\fi
\fi
\fi
}
\newcommand{\hatcurCCtwomassJmageccen}[1]{\ifnum#1=50 %
\hatcurCCtwomassJmageccenxxxxA
\else
\ifnum#1=51 %
\hatcurCCtwomassJmageccenxxxxB
\else
\ifnum#1=52 %
\hatcurCCtwomassJmageccenxxxxC
\else
\ifnum#1=53 %
\hatcurCCtwomassJmageccenxxxxD
\else
??????\fi
\fi
\fi
\fi
}
\newcommand{\hatcurCCtwomassKmageccen}[1]{\ifnum#1=50 %
\hatcurCCtwomassKmageccenxxxxA
\else
\ifnum#1=51 %
\hatcurCCtwomassKmageccenxxxxB
\else
\ifnum#1=52 %
\hatcurCCtwomassKmageccenxxxxC
\else
\ifnum#1=53 %
\hatcurCCtwomassKmageccenxxxxD
\else
??????\fi
\fi
\fi
\fi
}
\newcommand{\hatcurfieldeccen}[1]{\ifnum#1=50 %
\hatcurfieldeccenxxxxA
\else
\ifnum#1=51 %
\hatcurfieldeccenxxxxB
\else
\ifnum#1=52 %
\hatcurfieldeccenxxxxC
\else
\ifnum#1=53 %
\hatcurfieldeccenxxxxD
\else
??????\fi
\fi
\fi
\fi
}
\newcommand{\hatcurhtreccen}[1]{\ifnum#1=50 %
\hatcurhtreccenxxxxA
\else
\ifnum#1=51 %
\hatcurhtreccenxxxxB
\else
\ifnum#1=52 %
\hatcurhtreccenxxxxC
\else
\ifnum#1=53 %
\hatcurhtreccenxxxxD
\else
??????\fi
\fi
\fi
\fi
}
\newcommand{\hatcurISOageeccen}[1]{\ifnum#1=50 %
\hatcurISOageeccenxxxxA
\else
\ifnum#1=51 %
\hatcurISOageeccenxxxxB
\else
\ifnum#1=52 %
\hatcurISOageeccenxxxxC
\else
\ifnum#1=53 %
\hatcurISOageeccenxxxxD
\else
??????\fi
\fi
\fi
\fi
}
\newcommand{\hatcurISOJKeccen}[1]{\ifnum#1=50 %
\hatcurISOJKeccenxxxxA
\else
\ifnum#1=51 %
\hatcurISOJKeccenxxxxB
\else
\ifnum#1=52 %
\hatcurISOJKeccenxxxxC
\else
\ifnum#1=53 %
\hatcurISOJKeccenxxxxD
\else
??????\fi
\fi
\fi
\fi
}
\newcommand{\hatcurISOloggeccen}[1]{\ifnum#1=50 %
\hatcurISOloggeccenxxxxA
\else
\ifnum#1=51 %
\hatcurISOloggeccenxxxxB
\else
\ifnum#1=52 %
\hatcurISOloggeccenxxxxC
\else
\ifnum#1=53 %
\hatcurISOloggeccenxxxxD
\else
??????\fi
\fi
\fi
\fi
}
\newcommand{\hatcurISOlumeccen}[1]{\ifnum#1=50 %
\hatcurISOlumeccenxxxxA
\else
\ifnum#1=51 %
\hatcurISOlumeccenxxxxB
\else
\ifnum#1=52 %
\hatcurISOlumeccenxxxxC
\else
\ifnum#1=53 %
\hatcurISOlumeccenxxxxD
\else
??????\fi
\fi
\fi
\fi
}
\newcommand{\hatcurISOlumshorteccen}[1]{\ifnum#1=50 %
\hatcurISOlumshorteccenxxxxA
\else
\ifnum#1=51 %
\hatcurISOlumshorteccenxxxxB
\else
\ifnum#1=52 %
\hatcurISOlumshorteccenxxxxC
\else
\ifnum#1=53 %
\hatcurISOlumshorteccenxxxxD
\else
??????\fi
\fi
\fi
\fi
}
\newcommand{\hatcurISOmeccen}[1]{\ifnum#1=50 %
\hatcurISOmeccenxxxxA
\else
\ifnum#1=51 %
\hatcurISOmeccenxxxxB
\else
\ifnum#1=52 %
\hatcurISOmeccenxxxxC
\else
\ifnum#1=53 %
\hatcurISOmeccenxxxxD
\else
??????\fi
\fi
\fi
\fi
}
\newcommand{\hatcurISOMHeccen}[1]{\ifnum#1=50 %
\hatcurISOMHeccenxxxxA
\else
\ifnum#1=51 %
\hatcurISOMHeccenxxxxB
\else
\ifnum#1=52 %
\hatcurISOMHeccenxxxxC
\else
\ifnum#1=53 %
\hatcurISOMHeccenxxxxD
\else
??????\fi
\fi
\fi
\fi
}
\newcommand{\hatcurISOMJeccen}[1]{\ifnum#1=50 %
\hatcurISOMJeccenxxxxA
\else
\ifnum#1=51 %
\hatcurISOMJeccenxxxxB
\else
\ifnum#1=52 %
\hatcurISOMJeccenxxxxC
\else
\ifnum#1=53 %
\hatcurISOMJeccenxxxxD
\else
??????\fi
\fi
\fi
\fi
}
\newcommand{\hatcurISOMKeccen}[1]{\ifnum#1=50 %
\hatcurISOMKeccenxxxxA
\else
\ifnum#1=51 %
\hatcurISOMKeccenxxxxB
\else
\ifnum#1=52 %
\hatcurISOMKeccenxxxxC
\else
\ifnum#1=53 %
\hatcurISOMKeccenxxxxD
\else
??????\fi
\fi
\fi
\fi
}
\newcommand{\hatcurISOmlongeccen}[1]{\ifnum#1=50 %
\hatcurISOmlongeccenxxxxA
\else
\ifnum#1=51 %
\hatcurISOmlongeccenxxxxB
\else
\ifnum#1=52 %
\hatcurISOmlongeccenxxxxC
\else
\ifnum#1=53 %
\hatcurISOmlongeccenxxxxD
\else
??????\fi
\fi
\fi
\fi
}
\newcommand{\hatcurISOmshorteccen}[1]{\ifnum#1=50 %
\hatcurISOmshorteccenxxxxA
\else
\ifnum#1=51 %
\hatcurISOmshorteccenxxxxB
\else
\ifnum#1=52 %
\hatcurISOmshorteccenxxxxC
\else
\ifnum#1=53 %
\hatcurISOmshorteccenxxxxD
\else
??????\fi
\fi
\fi
\fi
}
\newcommand{\hatcurISOmveccen}[1]{\ifnum#1=50 %
\hatcurISOmveccenxxxxA
\else
\ifnum#1=51 %
\hatcurISOmveccenxxxxB
\else
\ifnum#1=52 %
\hatcurISOmveccenxxxxC
\else
\ifnum#1=53 %
\hatcurISOmveccenxxxxD
\else
??????\fi
\fi
\fi
\fi
}
\newcommand{\hatcurISOreccen}[1]{\ifnum#1=50 %
\hatcurISOreccenxxxxA
\else
\ifnum#1=51 %
\hatcurISOreccenxxxxB
\else
\ifnum#1=52 %
\hatcurISOreccenxxxxC
\else
\ifnum#1=53 %
\hatcurISOreccenxxxxD
\else
??????\fi
\fi
\fi
\fi
}
\newcommand{\hatcurISOrhoeccen}[1]{\ifnum#1=50 %
\hatcurISOrhoeccenxxxxA
\else
\ifnum#1=51 %
\hatcurISOrhoeccenxxxxB
\else
\ifnum#1=52 %
\hatcurISOrhoeccenxxxxC
\else
\ifnum#1=53 %
\hatcurISOrhoeccenxxxxD
\else
??????\fi
\fi
\fi
\fi
}
\newcommand{\hatcurISOrholongeccen}[1]{\ifnum#1=50 %
\hatcurISOrholongeccenxxxxA
\else
\ifnum#1=51 %
\hatcurISOrholongeccenxxxxB
\else
\ifnum#1=52 %
\hatcurISOrholongeccenxxxxC
\else
\ifnum#1=53 %
\hatcurISOrholongeccenxxxxD
\else
??????\fi
\fi
\fi
\fi
}
\newcommand{\hatcurISOrlongeccen}[1]{\ifnum#1=50 %
\hatcurISOrlongeccenxxxxA
\else
\ifnum#1=51 %
\hatcurISOrlongeccenxxxxB
\else
\ifnum#1=52 %
\hatcurISOrlongeccenxxxxC
\else
\ifnum#1=53 %
\hatcurISOrlongeccenxxxxD
\else
??????\fi
\fi
\fi
\fi
}
\newcommand{\hatcurISOrshorteccen}[1]{\ifnum#1=50 %
\hatcurISOrshorteccenxxxxA
\else
\ifnum#1=51 %
\hatcurISOrshorteccenxxxxB
\else
\ifnum#1=52 %
\hatcurISOrshorteccenxxxxC
\else
\ifnum#1=53 %
\hatcurISOrshorteccenxxxxD
\else
??????\fi
\fi
\fi
\fi
}
\newcommand{\hatcurISOsigmaeccen}[1]{\ifnum#1=50 %
\hatcurISOsigmaeccenxxxxA
\else
\ifnum#1=51 %
\hatcurISOsigmaeccenxxxxB
\else
\ifnum#1=52 %
\hatcurISOsigmaeccenxxxxC
\else
\ifnum#1=53 %
\hatcurISOsigmaeccenxxxxD
\else
??????\fi
\fi
\fi
\fi
}
\newcommand{\hatcurISOspececcen}[1]{\ifnum#1=50 %
\hatcurISOspececcenxxxxA
\else
\ifnum#1=51 %
\hatcurISOspececcenxxxxB
\else
\ifnum#1=52 %
\hatcurISOspececcenxxxxC
\else
\ifnum#1=53 %
\hatcurISOspececcenxxxxD
\else
??????\fi
\fi
\fi
\fi
}
\newcommand{\hatcurISOvieccen}[1]{\ifnum#1=50 %
\hatcurISOvieccenxxxxA
\else
\ifnum#1=51 %
\hatcurISOvieccenxxxxB
\else
\ifnum#1=52 %
\hatcurISOvieccenxxxxC
\else
\ifnum#1=53 %
\hatcurISOvieccenxxxxD
\else
??????\fi
\fi
\fi
\fi
}
\newcommand{\hatcurLBigeccen}[1]{\ifnum#1=50 %
\hatcurLBigeccenxxxxA
\else
\ifnum#1=51 %
\hatcurLBigeccenxxxxB
\else
\ifnum#1=52 %
\hatcurLBigeccenxxxxC
\else
\ifnum#1=53 %
\hatcurLBigeccenxxxxD
\else
??????\fi
\fi
\fi
\fi
}
\newcommand{\hatcurLBiieccen}[1]{\ifnum#1=50 %
\hatcurLBiieccenxxxxA
\else
\ifnum#1=51 %
\hatcurLBiieccenxxxxB
\else
\ifnum#1=52 %
\hatcurLBiieccenxxxxC
\else
\ifnum#1=53 %
\hatcurLBiieccenxxxxD
\else
??????\fi
\fi
\fi
\fi
}
\newcommand{\hatcurLBiIeccen}[1]{\ifnum#1=50 %
\hatcurLBiIeccenxxxxA
\else
\ifnum#1=51 %
\hatcurLBiIeccenxxxxB
\else
\ifnum#1=52 %
\hatcurLBiIeccenxxxxC
\else
\ifnum#1=53 %
\hatcurLBiIeccenxxxxD
\else
??????\fi
\fi
\fi
\fi
}
\newcommand{\hatcurLBiigeccen}[1]{\ifnum#1=50 %
\hatcurLBiigeccenxxxxA
\else
\ifnum#1=51 %
\hatcurLBiigeccenxxxxB
\else
\ifnum#1=52 %
\hatcurLBiigeccenxxxxC
\else
\ifnum#1=53 %
\hatcurLBiigeccenxxxxD
\else
??????\fi
\fi
\fi
\fi
}
\newcommand{\hatcurLBiiieccen}[1]{\ifnum#1=50 %
\hatcurLBiiieccenxxxxA
\else
\ifnum#1=51 %
\hatcurLBiiieccenxxxxB
\else
\ifnum#1=52 %
\hatcurLBiiieccenxxxxC
\else
\ifnum#1=53 %
\hatcurLBiiieccenxxxxD
\else
??????\fi
\fi
\fi
\fi
}
\newcommand{\hatcurLBiiIeccen}[1]{\ifnum#1=50 %
\hatcurLBiiIeccenxxxxA
\else
\ifnum#1=51 %
\hatcurLBiiIeccenxxxxB
\else
\ifnum#1=52 %
\hatcurLBiiIeccenxxxxC
\else
\ifnum#1=53 %
\hatcurLBiiIeccenxxxxD
\else
??????\fi
\fi
\fi
\fi
}
\newcommand{\hatcurLBiikepeccen}[1]{\ifnum#1=50 %
\hatcurLBiikepeccenxxxxA
\else
\ifnum#1=51 %
\hatcurLBiikepeccenxxxxB
\else
\ifnum#1=52 %
\hatcurLBiikepeccenxxxxC
\else
\ifnum#1=53 %
\hatcurLBiikepeccenxxxxD
\else
??????\fi
\fi
\fi
\fi
}
\newcommand{\hatcurLBiireccen}[1]{\ifnum#1=50 %
\hatcurLBiireccenxxxxA
\else
\ifnum#1=51 %
\hatcurLBiireccenxxxxB
\else
\ifnum#1=52 %
\hatcurLBiireccenxxxxC
\else
\ifnum#1=53 %
\hatcurLBiireccenxxxxD
\else
??????\fi
\fi
\fi
\fi
}
\newcommand{\hatcurLBiiReccen}[1]{\ifnum#1=50 %
\hatcurLBiiReccenxxxxA
\else
\ifnum#1=51 %
\hatcurLBiiReccenxxxxB
\else
\ifnum#1=52 %
\hatcurLBiiReccenxxxxC
\else
\ifnum#1=53 %
\hatcurLBiiReccenxxxxD
\else
??????\fi
\fi
\fi
\fi
}
\newcommand{\hatcurLBiizeccen}[1]{\ifnum#1=50 %
\hatcurLBiizeccenxxxxA
\else
\ifnum#1=51 %
\hatcurLBiizeccenxxxxB
\else
\ifnum#1=52 %
\hatcurLBiizeccenxxxxC
\else
\ifnum#1=53 %
\hatcurLBiizeccenxxxxD
\else
??????\fi
\fi
\fi
\fi
}
\newcommand{\hatcurLBikepeccen}[1]{\ifnum#1=50 %
\hatcurLBikepeccenxxxxA
\else
\ifnum#1=51 %
\hatcurLBikepeccenxxxxB
\else
\ifnum#1=52 %
\hatcurLBikepeccenxxxxC
\else
\ifnum#1=53 %
\hatcurLBikepeccenxxxxD
\else
??????\fi
\fi
\fi
\fi
}
\newcommand{\hatcurLBireccen}[1]{\ifnum#1=50 %
\hatcurLBireccenxxxxA
\else
\ifnum#1=51 %
\hatcurLBireccenxxxxB
\else
\ifnum#1=52 %
\hatcurLBireccenxxxxC
\else
\ifnum#1=53 %
\hatcurLBireccenxxxxD
\else
??????\fi
\fi
\fi
\fi
}
\newcommand{\hatcurLBiReccen}[1]{\ifnum#1=50 %
\hatcurLBiReccenxxxxA
\else
\ifnum#1=51 %
\hatcurLBiReccenxxxxB
\else
\ifnum#1=52 %
\hatcurLBiReccenxxxxC
\else
\ifnum#1=53 %
\hatcurLBiReccenxxxxD
\else
??????\fi
\fi
\fi
\fi
}
\newcommand{\hatcurLBizeccen}[1]{\ifnum#1=50 %
\hatcurLBizeccenxxxxA
\else
\ifnum#1=51 %
\hatcurLBizeccenxxxxB
\else
\ifnum#1=52 %
\hatcurLBizeccenxxxxC
\else
\ifnum#1=53 %
\hatcurLBizeccenxxxxD
\else
??????\fi
\fi
\fi
\fi
}
\newcommand{\hatcurLCbsqeccen}[1]{\ifnum#1=50 %
\hatcurLCbsqeccenxxxxA
\else
\ifnum#1=51 %
\hatcurLCbsqeccenxxxxB
\else
\ifnum#1=52 %
\hatcurLCbsqeccenxxxxC
\else
\ifnum#1=53 %
\hatcurLCbsqeccenxxxxD
\else
??????\fi
\fi
\fi
\fi
}
\newcommand{\hatcurLCdipeccen}[1]{\ifnum#1=50 %
\hatcurLCdipeccenxxxxA
\else
\ifnum#1=51 %
\hatcurLCdipeccenxxxxB
\else
\ifnum#1=52 %
\hatcurLCdipeccenxxxxC
\else
\ifnum#1=53 %
\hatcurLCdipeccenxxxxD
\else
??????\fi
\fi
\fi
\fi
}
\newcommand{\hatcurLCdureccen}[1]{\ifnum#1=50 %
\hatcurLCdureccenxxxxA
\else
\ifnum#1=51 %
\hatcurLCdureccenxxxxB
\else
\ifnum#1=52 %
\hatcurLCdureccenxxxxC
\else
\ifnum#1=53 %
\hatcurLCdureccenxxxxD
\else
??????\fi
\fi
\fi
\fi
}
\newcommand{\hatcurLCdurhreccen}[1]{\ifnum#1=50 %
\hatcurLCdurhreccenxxxxA
\else
\ifnum#1=51 %
\hatcurLCdurhreccenxxxxB
\else
\ifnum#1=52 %
\hatcurLCdurhreccenxxxxC
\else
\ifnum#1=53 %
\hatcurLCdurhreccenxxxxD
\else
??????\fi
\fi
\fi
\fi
}
\newcommand{\hatcurLCdurhrshorteccen}[1]{\ifnum#1=50 %
\hatcurLCdurhrshorteccenxxxxA
\else
\ifnum#1=51 %
\hatcurLCdurhrshorteccenxxxxB
\else
\ifnum#1=52 %
\hatcurLCdurhrshorteccenxxxxC
\else
\ifnum#1=53 %
\hatcurLCdurhrshorteccenxxxxD
\else
??????\fi
\fi
\fi
\fi
}
\newcommand{\hatcurLCdurshorteccen}[1]{\ifnum#1=50 %
\hatcurLCdurshorteccenxxxxA
\else
\ifnum#1=51 %
\hatcurLCdurshorteccenxxxxB
\else
\ifnum#1=52 %
\hatcurLCdurshorteccenxxxxC
\else
\ifnum#1=53 %
\hatcurLCdurshorteccenxxxxD
\else
??????\fi
\fi
\fi
\fi
}
\newcommand{\hatcurLChatnetmAeccen}[1]{\ifnum#1=50 %
\hatcurLChatnetmAeccenxxxxA
\else
??????\fi
}
\newcommand{\hatcurLChatnetmBeccen}[1]{\ifnum#1=50 %
\hatcurLChatnetmBeccenxxxxA
\else
??????\fi
}
\newcommand{\hatcurLChatnetmeccen}[1]{\ifnum#1=51 %
\hatcurLChatnetmeccenxxxxB
\else
\ifnum#1=52 %
\hatcurLChatnetmeccenxxxxC
\else
\ifnum#1=53 %
\hatcurLChatnetmeccenxxxxD
\else
??????\fi
\fi
\fi
}
\newcommand{\hatcurLCiblendAeccen}[1]{\ifnum#1=50 %
\hatcurLCiblendAeccenxxxxA
\else
??????\fi
}
\newcommand{\hatcurLCiblendBeccen}[1]{\ifnum#1=50 %
\hatcurLCiblendBeccenxxxxA
\else
??????\fi
}
\newcommand{\hatcurLCiblendeccen}[1]{\ifnum#1=51 %
\hatcurLCiblendeccenxxxxB
\else
\ifnum#1=52 %
\hatcurLCiblendeccenxxxxC
\else
\ifnum#1=53 %
\hatcurLCiblendeccenxxxxD
\else
??????\fi
\fi
\fi
}
\newcommand{\hatcurLCimpeccen}[1]{\ifnum#1=50 %
\hatcurLCimpeccenxxxxA
\else
\ifnum#1=51 %
\hatcurLCimpeccenxxxxB
\else
\ifnum#1=52 %
\hatcurLCimpeccenxxxxC
\else
\ifnum#1=53 %
\hatcurLCimpeccenxxxxD
\else
??????\fi
\fi
\fi
\fi
}
\newcommand{\hatcurLCingdureccen}[1]{\ifnum#1=50 %
\hatcurLCingdureccenxxxxA
\else
\ifnum#1=51 %
\hatcurLCingdureccenxxxxB
\else
\ifnum#1=52 %
\hatcurLCingdureccenxxxxC
\else
\ifnum#1=53 %
\hatcurLCingdureccenxxxxD
\else
??????\fi
\fi
\fi
\fi
}
\newcommand{\hatcurLCPeccen}[1]{\ifnum#1=50 %
\hatcurLCPeccenxxxxA
\else
\ifnum#1=51 %
\hatcurLCPeccenxxxxB
\else
\ifnum#1=52 %
\hatcurLCPeccenxxxxC
\else
\ifnum#1=53 %
\hatcurLCPeccenxxxxD
\else
??????\fi
\fi
\fi
\fi
}
\newcommand{\hatcurLCPprececcen}[1]{\ifnum#1=50 %
\hatcurLCPprececcenxxxxA
\else
\ifnum#1=51 %
\hatcurLCPprececcenxxxxB
\else
\ifnum#1=52 %
\hatcurLCPprececcenxxxxC
\else
\ifnum#1=53 %
\hatcurLCPprececcenxxxxD
\else
??????\fi
\fi
\fi
\fi
}
\newcommand{\hatcurLCPshorteccen}[1]{\ifnum#1=50 %
\hatcurLCPshorteccenxxxxA
\else
\ifnum#1=51 %
\hatcurLCPshorteccenxxxxB
\else
\ifnum#1=52 %
\hatcurLCPshorteccenxxxxC
\else
\ifnum#1=53 %
\hatcurLCPshorteccenxxxxD
\else
??????\fi
\fi
\fi
\fi
}
\newcommand{\hatcurLCqeccen}[1]{\ifnum#1=50 %
\hatcurLCqeccenxxxxA
\else
\ifnum#1=51 %
\hatcurLCqeccenxxxxB
\else
\ifnum#1=52 %
\hatcurLCqeccenxxxxC
\else
\ifnum#1=53 %
\hatcurLCqeccenxxxxD
\else
??????\fi
\fi
\fi
\fi
}
\newcommand{\hatcurLCqshorteccen}[1]{\ifnum#1=50 %
\hatcurLCqshorteccenxxxxA
\else
\ifnum#1=51 %
\hatcurLCqshorteccenxxxxB
\else
\ifnum#1=52 %
\hatcurLCqshorteccenxxxxC
\else
\ifnum#1=53 %
\hatcurLCqshorteccenxxxxD
\else
??????\fi
\fi
\fi
\fi
}
\newcommand{\hatcurLCrhoeccen}[1]{\ifnum#1=50 %
\hatcurLCrhoeccenxxxxA
\else
\ifnum#1=52 %
\hatcurLCrhoeccenxxxxC
\else
\ifnum#1=53 %
\hatcurLCrhoeccenxxxxD
\else
??????\fi
\fi
\fi
}
\newcommand{\hatcurLCrprstareccen}[1]{\ifnum#1=50 %
\hatcurLCrprstareccenxxxxA
\else
\ifnum#1=51 %
\hatcurLCrprstareccenxxxxB
\else
\ifnum#1=52 %
\hatcurLCrprstareccenxxxxC
\else
\ifnum#1=53 %
\hatcurLCrprstareccenxxxxD
\else
??????\fi
\fi
\fi
\fi
}
\newcommand{\hatcurLCTAeccen}[1]{\ifnum#1=50 %
\hatcurLCTAeccenxxxxA
\else
\ifnum#1=51 %
\hatcurLCTAeccenxxxxB
\else
\ifnum#1=52 %
\hatcurLCTAeccenxxxxC
\else
\ifnum#1=53 %
\hatcurLCTAeccenxxxxD
\else
??????\fi
\fi
\fi
\fi
}
\newcommand{\hatcurLCTBeccen}[1]{\ifnum#1=50 %
\hatcurLCTBeccenxxxxA
\else
\ifnum#1=51 %
\hatcurLCTBeccenxxxxB
\else
\ifnum#1=52 %
\hatcurLCTBeccenxxxxC
\else
\ifnum#1=53 %
\hatcurLCTBeccenxxxxD
\else
??????\fi
\fi
\fi
\fi
}
\newcommand{\hatcurLCTeccen}[1]{\ifnum#1=50 %
\hatcurLCTeccenxxxxA
\else
\ifnum#1=51 %
\hatcurLCTeccenxxxxB
\else
\ifnum#1=52 %
\hatcurLCTeccenxxxxC
\else
\ifnum#1=53 %
\hatcurLCTeccenxxxxD
\else
??????\fi
\fi
\fi
\fi
}
\newcommand{\hatcurLCzetaeccen}[1]{\ifnum#1=50 %
\hatcurLCzetaeccenxxxxA
\else
\ifnum#1=51 %
\hatcurLCzetaeccenxxxxB
\else
\ifnum#1=52 %
\hatcurLCzetaeccenxxxxC
\else
\ifnum#1=53 %
\hatcurLCzetaeccenxxxxD
\else
??????\fi
\fi
\fi
\fi
}
\newcommand{\hatcurPPaequiveccen}[1]{\ifnum#1=50 %
\hatcurPPaequiveccenxxxxA
\else
\ifnum#1=51 %
\hatcurPPaequiveccenxxxxB
\else
\ifnum#1=52 %
\hatcurPPaequiveccenxxxxC
\else
\ifnum#1=53 %
\hatcurPPaequiveccenxxxxD
\else
??????\fi
\fi
\fi
\fi
}
\newcommand{\hatcurPPareccen}[1]{\ifnum#1=50 %
\hatcurPPareccenxxxxA
\else
\ifnum#1=51 %
\hatcurPPareccenxxxxB
\else
\ifnum#1=52 %
\hatcurPPareccenxxxxC
\else
\ifnum#1=53 %
\hatcurPPareccenxxxxD
\else
??????\fi
\fi
\fi
\fi
}
\newcommand{\hatcurPPareleccen}[1]{\ifnum#1=50 %
\hatcurPPareleccenxxxxA
\else
\ifnum#1=51 %
\hatcurPPareleccenxxxxB
\else
\ifnum#1=52 %
\hatcurPPareleccenxxxxC
\else
\ifnum#1=53 %
\hatcurPPareleccenxxxxD
\else
??????\fi
\fi
\fi
\fi
}
\newcommand{\hatcurPPfluxapdimeccen}[1]{\ifnum#1=50 %
\hatcurPPfluxapdimeccenxxxxA
\else
\ifnum#1=51 %
\hatcurPPfluxapdimeccenxxxxB
\else
\ifnum#1=52 %
\hatcurPPfluxapdimeccenxxxxC
\else
\ifnum#1=53 %
\hatcurPPfluxapdimeccenxxxxD
\else
??????\fi
\fi
\fi
\fi
}
\newcommand{\hatcurPPfluxapeccen}[1]{\ifnum#1=50 %
\hatcurPPfluxapeccenxxxxA
\else
\ifnum#1=51 %
\hatcurPPfluxapeccenxxxxB
\else
\ifnum#1=52 %
\hatcurPPfluxapeccenxxxxC
\else
\ifnum#1=53 %
\hatcurPPfluxapeccenxxxxD
\else
??????\fi
\fi
\fi
\fi
}
\newcommand{\hatcurPPfluxavgdimeccen}[1]{\ifnum#1=50 %
\hatcurPPfluxavgdimeccenxxxxA
\else
\ifnum#1=51 %
\hatcurPPfluxavgdimeccenxxxxB
\else
\ifnum#1=52 %
\hatcurPPfluxavgdimeccenxxxxC
\else
\ifnum#1=53 %
\hatcurPPfluxavgdimeccenxxxxD
\else
??????\fi
\fi
\fi
\fi
}
\newcommand{\hatcurPPfluxavgeccen}[1]{\ifnum#1=50 %
\hatcurPPfluxavgeccenxxxxA
\else
\ifnum#1=51 %
\hatcurPPfluxavgeccenxxxxB
\else
\ifnum#1=52 %
\hatcurPPfluxavgeccenxxxxC
\else
\ifnum#1=53 %
\hatcurPPfluxavgeccenxxxxD
\else
??????\fi
\fi
\fi
\fi
}
\newcommand{\hatcurPPfluxavglogeccen}[1]{\ifnum#1=50 %
\hatcurPPfluxavglogeccenxxxxA
\else
\ifnum#1=51 %
\hatcurPPfluxavglogeccenxxxxB
\else
\ifnum#1=52 %
\hatcurPPfluxavglogeccenxxxxC
\else
\ifnum#1=53 %
\hatcurPPfluxavglogeccenxxxxD
\else
??????\fi
\fi
\fi
\fi
}
\newcommand{\hatcurPPfluxperidimeccen}[1]{\ifnum#1=50 %
\hatcurPPfluxperidimeccenxxxxA
\else
\ifnum#1=51 %
\hatcurPPfluxperidimeccenxxxxB
\else
\ifnum#1=52 %
\hatcurPPfluxperidimeccenxxxxC
\else
\ifnum#1=53 %
\hatcurPPfluxperidimeccenxxxxD
\else
??????\fi
\fi
\fi
\fi
}
\newcommand{\hatcurPPfluxperieccen}[1]{\ifnum#1=50 %
\hatcurPPfluxperieccenxxxxA
\else
\ifnum#1=51 %
\hatcurPPfluxperieccenxxxxB
\else
\ifnum#1=52 %
\hatcurPPfluxperieccenxxxxC
\else
\ifnum#1=53 %
\hatcurPPfluxperieccenxxxxD
\else
??????\fi
\fi
\fi
\fi
}
\newcommand{\hatcurPPgeccen}[1]{\ifnum#1=50 %
\hatcurPPgeccenxxxxA
\else
\ifnum#1=51 %
\hatcurPPgeccenxxxxB
\else
\ifnum#1=52 %
\hatcurPPgeccenxxxxC
\else
\ifnum#1=53 %
\hatcurPPgeccenxxxxD
\else
??????\fi
\fi
\fi
\fi
}
\newcommand{\hatcurPPieccen}[1]{\ifnum#1=50 %
\hatcurPPieccenxxxxA
\else
\ifnum#1=51 %
\hatcurPPieccenxxxxB
\else
\ifnum#1=52 %
\hatcurPPieccenxxxxC
\else
\ifnum#1=53 %
\hatcurPPieccenxxxxD
\else
??????\fi
\fi
\fi
\fi
}
\newcommand{\hatcurPPloggeccen}[1]{\ifnum#1=50 %
\hatcurPPloggeccenxxxxA
\else
\ifnum#1=51 %
\hatcurPPloggeccenxxxxB
\else
\ifnum#1=52 %
\hatcurPPloggeccenxxxxC
\else
\ifnum#1=53 %
\hatcurPPloggeccenxxxxD
\else
??????\fi
\fi
\fi
\fi
}
\newcommand{\hatcurPPmeccen}[1]{\ifnum#1=50 %
\hatcurPPmeccenxxxxA
\else
\ifnum#1=51 %
\hatcurPPmeccenxxxxB
\else
\ifnum#1=52 %
\hatcurPPmeccenxxxxC
\else
\ifnum#1=53 %
\hatcurPPmeccenxxxxD
\else
??????\fi
\fi
\fi
\fi
}
\newcommand{\hatcurPPmeeccen}[1]{\ifnum#1=50 %
\hatcurPPmeeccenxxxxA
\else
\ifnum#1=51 %
\hatcurPPmeeccenxxxxB
\else
\ifnum#1=52 %
\hatcurPPmeeccenxxxxC
\else
\ifnum#1=53 %
\hatcurPPmeeccenxxxxD
\else
??????\fi
\fi
\fi
\fi
}
\newcommand{\hatcurPPmelongeccen}[1]{\ifnum#1=50 %
\hatcurPPmelongeccenxxxxA
\else
\ifnum#1=51 %
\hatcurPPmelongeccenxxxxB
\else
\ifnum#1=52 %
\hatcurPPmelongeccenxxxxC
\else
\ifnum#1=53 %
\hatcurPPmelongeccenxxxxD
\else
??????\fi
\fi
\fi
\fi
}
\newcommand{\hatcurPPmeshorteccen}[1]{\ifnum#1=50 %
\hatcurPPmeshorteccenxxxxA
\else
\ifnum#1=51 %
\hatcurPPmeshorteccenxxxxB
\else
\ifnum#1=52 %
\hatcurPPmeshorteccenxxxxC
\else
\ifnum#1=53 %
\hatcurPPmeshorteccenxxxxD
\else
??????\fi
\fi
\fi
\fi
}
\newcommand{\hatcurPPmlongeccen}[1]{\ifnum#1=50 %
\hatcurPPmlongeccenxxxxA
\else
\ifnum#1=51 %
\hatcurPPmlongeccenxxxxB
\else
\ifnum#1=52 %
\hatcurPPmlongeccenxxxxC
\else
\ifnum#1=53 %
\hatcurPPmlongeccenxxxxD
\else
??????\fi
\fi
\fi
\fi
}
\newcommand{\hatcurPPmrcorreccen}[1]{\ifnum#1=50 %
\hatcurPPmrcorreccenxxxxA
\else
\ifnum#1=51 %
\hatcurPPmrcorreccenxxxxB
\else
\ifnum#1=52 %
\hatcurPPmrcorreccenxxxxC
\else
\ifnum#1=53 %
\hatcurPPmrcorreccenxxxxD
\else
??????\fi
\fi
\fi
\fi
}
\newcommand{\hatcurPPmshorteccen}[1]{\ifnum#1=50 %
\hatcurPPmshorteccenxxxxA
\else
\ifnum#1=51 %
\hatcurPPmshorteccenxxxxB
\else
\ifnum#1=52 %
\hatcurPPmshorteccenxxxxC
\else
\ifnum#1=53 %
\hatcurPPmshorteccenxxxxD
\else
??????\fi
\fi
\fi
\fi
}
\newcommand{\hatcurPPperieccen}[1]{\ifnum#1=50 %
\hatcurPPperieccenxxxxA
\else
\ifnum#1=51 %
\hatcurPPperieccenxxxxB
\else
\ifnum#1=52 %
\hatcurPPperieccenxxxxC
\else
\ifnum#1=53 %
\hatcurPPperieccenxxxxD
\else
??????\fi
\fi
\fi
\fi
}
\newcommand{\hatcurPPphiconjeccen}[1]{\ifnum#1=50 %
\hatcurPPphiconjeccenxxxxA
\else
\ifnum#1=51 %
\hatcurPPphiconjeccenxxxxB
\else
\ifnum#1=52 %
\hatcurPPphiconjeccenxxxxC
\else
\ifnum#1=53 %
\hatcurPPphiconjeccenxxxxD
\else
??????\fi
\fi
\fi
\fi
}
\newcommand{\hatcurPPreccen}[1]{\ifnum#1=50 %
\hatcurPPreccenxxxxA
\else
\ifnum#1=51 %
\hatcurPPreccenxxxxB
\else
\ifnum#1=52 %
\hatcurPPreccenxxxxC
\else
\ifnum#1=53 %
\hatcurPPreccenxxxxD
\else
??????\fi
\fi
\fi
\fi
}
\newcommand{\hatcurPPreeccen}[1]{\ifnum#1=50 %
\hatcurPPreeccenxxxxA
\else
\ifnum#1=51 %
\hatcurPPreeccenxxxxB
\else
\ifnum#1=52 %
\hatcurPPreeccenxxxxC
\else
\ifnum#1=53 %
\hatcurPPreeccenxxxxD
\else
??????\fi
\fi
\fi
\fi
}
\newcommand{\hatcurPPrelongeccen}[1]{\ifnum#1=50 %
\hatcurPPrelongeccenxxxxA
\else
\ifnum#1=51 %
\hatcurPPrelongeccenxxxxB
\else
\ifnum#1=52 %
\hatcurPPrelongeccenxxxxC
\else
\ifnum#1=53 %
\hatcurPPrelongeccenxxxxD
\else
??????\fi
\fi
\fi
\fi
}
\newcommand{\hatcurPPreshorteccen}[1]{\ifnum#1=50 %
\hatcurPPreshorteccenxxxxA
\else
\ifnum#1=51 %
\hatcurPPreshorteccenxxxxB
\else
\ifnum#1=52 %
\hatcurPPreshorteccenxxxxC
\else
\ifnum#1=53 %
\hatcurPPreshorteccenxxxxD
\else
??????\fi
\fi
\fi
\fi
}
\newcommand{\hatcurPPrhoeccen}[1]{\ifnum#1=50 %
\hatcurPPrhoeccenxxxxA
\else
\ifnum#1=51 %
\hatcurPPrhoeccenxxxxB
\else
\ifnum#1=52 %
\hatcurPPrhoeccenxxxxC
\else
\ifnum#1=53 %
\hatcurPPrhoeccenxxxxD
\else
??????\fi
\fi
\fi
\fi
}
\newcommand{\hatcurPPrlongeccen}[1]{\ifnum#1=50 %
\hatcurPPrlongeccenxxxxA
\else
\ifnum#1=51 %
\hatcurPPrlongeccenxxxxB
\else
\ifnum#1=52 %
\hatcurPPrlongeccenxxxxC
\else
\ifnum#1=53 %
\hatcurPPrlongeccenxxxxD
\else
??????\fi
\fi
\fi
\fi
}
\newcommand{\hatcurPPrshorteccen}[1]{\ifnum#1=50 %
\hatcurPPrshorteccenxxxxA
\else
\ifnum#1=51 %
\hatcurPPrshorteccenxxxxB
\else
\ifnum#1=52 %
\hatcurPPrshorteccenxxxxC
\else
\ifnum#1=53 %
\hatcurPPrshorteccenxxxxD
\else
??????\fi
\fi
\fi
\fi
}
\newcommand{\hatcurPPtcirceccen}[1]{\ifnum#1=50 %
\hatcurPPtcirceccenxxxxA
\else
\ifnum#1=51 %
\hatcurPPtcirceccenxxxxB
\else
\ifnum#1=52 %
\hatcurPPtcirceccenxxxxC
\else
\ifnum#1=53 %
\hatcurPPtcirceccenxxxxD
\else
??????\fi
\fi
\fi
\fi
}
\newcommand{\hatcurPPteffeccen}[1]{\ifnum#1=50 %
\hatcurPPteffeccenxxxxA
\else
\ifnum#1=51 %
\hatcurPPteffeccenxxxxB
\else
\ifnum#1=52 %
\hatcurPPteffeccenxxxxC
\else
\ifnum#1=53 %
\hatcurPPteffeccenxxxxD
\else
??????\fi
\fi
\fi
\fi
}
\newcommand{\hatcurPPthetaeccen}[1]{\ifnum#1=50 %
\hatcurPPthetaeccenxxxxA
\else
\ifnum#1=51 %
\hatcurPPthetaeccenxxxxB
\else
\ifnum#1=52 %
\hatcurPPthetaeccenxxxxC
\else
\ifnum#1=53 %
\hatcurPPthetaeccenxxxxD
\else
??????\fi
\fi
\fi
\fi
}
\newcommand{\hatcurPPtinfalleccen}[1]{\ifnum#1=50 %
\hatcurPPtinfalleccenxxxxA
\else
\ifnum#1=51 %
\hatcurPPtinfalleccenxxxxB
\else
\ifnum#1=52 %
\hatcurPPtinfalleccenxxxxC
\else
\ifnum#1=53 %
\hatcurPPtinfalleccenxxxxD
\else
??????\fi
\fi
\fi
\fi
}
\newcommand{\hatcurRVecceneccen}[1]{\ifnum#1=50 %
\hatcurRVecceneccenxxxxA
\else
\ifnum#1=51 %
\hatcurRVecceneccenxxxxB
\else
\ifnum#1=52 %
\hatcurRVecceneccenxxxxC
\else
\ifnum#1=53 %
\hatcurRVecceneccenxxxxD
\else
??????\fi
\fi
\fi
\fi
}
\newcommand{\hatcurRVeccentwosiglimeccen}[1]{\ifnum#1=50 %
\hatcurRVeccentwosiglimeccenxxxxA
\else
\ifnum#1=51 %
\hatcurRVeccentwosiglimeccenxxxxB
\else
\ifnum#1=52 %
\hatcurRVeccentwosiglimeccenxxxxC
\else
\ifnum#1=53 %
\hatcurRVeccentwosiglimeccenxxxxD
\else
??????\fi
\fi
\fi
\fi
}
\newcommand{\hatcurRVfitrmsAeccen}[1]{\ifnum#1=50 %
\hatcurRVfitrmsAeccenxxxxA
\else
\ifnum#1=51 %
\hatcurRVfitrmsAeccenxxxxB
\else
\ifnum#1=52 %
\hatcurRVfitrmsAeccenxxxxC
\else
\ifnum#1=53 %
\hatcurRVfitrmsAeccenxxxxD
\else
??????\fi
\fi
\fi
\fi
}
\newcommand{\hatcurRVfitrmsBeccen}[1]{\ifnum#1=50 %
\hatcurRVfitrmsBeccenxxxxA
\else
\ifnum#1=51 %
\hatcurRVfitrmsBeccenxxxxB
\else
\ifnum#1=52 %
\hatcurRVfitrmsBeccenxxxxC
\else
\ifnum#1=53 %
\hatcurRVfitrmsBeccenxxxxD
\else
??????\fi
\fi
\fi
\fi
}
\newcommand{\hatcurRVfitrmsCeccen}[1]{\ifnum#1=50 %
\hatcurRVfitrmsCeccenxxxxA
\else
\ifnum#1=51 %
\hatcurRVfitrmsCeccenxxxxB
\else
\ifnum#1=52 %
\hatcurRVfitrmsCeccenxxxxC
\else
??????\fi
\fi
\fi
}
\newcommand{\hatcurRVgammaAeccen}[1]{\ifnum#1=50 %
\hatcurRVgammaAeccenxxxxA
\else
\ifnum#1=51 %
\hatcurRVgammaAeccenxxxxB
\else
\ifnum#1=52 %
\hatcurRVgammaAeccenxxxxC
\else
\ifnum#1=53 %
\hatcurRVgammaAeccenxxxxD
\else
??????\fi
\fi
\fi
\fi
}
\newcommand{\hatcurRVgammaBeccen}[1]{\ifnum#1=50 %
\hatcurRVgammaBeccenxxxxA
\else
\ifnum#1=51 %
\hatcurRVgammaBeccenxxxxB
\else
\ifnum#1=52 %
\hatcurRVgammaBeccenxxxxC
\else
\ifnum#1=53 %
\hatcurRVgammaBeccenxxxxD
\else
??????\fi
\fi
\fi
\fi
}
\newcommand{\hatcurRVgammaCeccen}[1]{\ifnum#1=50 %
\hatcurRVgammaCeccenxxxxA
\else
\ifnum#1=51 %
\hatcurRVgammaCeccenxxxxB
\else
\ifnum#1=52 %
\hatcurRVgammaCeccenxxxxC
\else
??????\fi
\fi
\fi
}
\newcommand{\hatcurRVheccen}[1]{\ifnum#1=50 %
\hatcurRVheccenxxxxA
\else
\ifnum#1=51 %
\hatcurRVheccenxxxxB
\else
\ifnum#1=52 %
\hatcurRVheccenxxxxC
\else
\ifnum#1=53 %
\hatcurRVheccenxxxxD
\else
??????\fi
\fi
\fi
\fi
}
\newcommand{\hatcurRVjitterAeccen}[1]{\ifnum#1=50 %
\hatcurRVjitterAeccenxxxxA
\else
\ifnum#1=51 %
\hatcurRVjitterAeccenxxxxB
\else
\ifnum#1=52 %
\hatcurRVjitterAeccenxxxxC
\else
\ifnum#1=53 %
\hatcurRVjitterAeccenxxxxD
\else
??????\fi
\fi
\fi
\fi
}
\newcommand{\hatcurRVjitterBeccen}[1]{\ifnum#1=50 %
\hatcurRVjitterBeccenxxxxA
\else
\ifnum#1=51 %
\hatcurRVjitterBeccenxxxxB
\else
\ifnum#1=52 %
\hatcurRVjitterBeccenxxxxC
\else
\ifnum#1=53 %
\hatcurRVjitterBeccenxxxxD
\else
??????\fi
\fi
\fi
\fi
}
\newcommand{\hatcurRVjitterCeccen}[1]{\ifnum#1=50 %
\hatcurRVjitterCeccenxxxxA
\else
\ifnum#1=51 %
\hatcurRVjitterCeccenxxxxB
\else
\ifnum#1=52 %
\hatcurRVjitterCeccenxxxxC
\else
??????\fi
\fi
\fi
}
\newcommand{\hatcurRVjittertwosiglimAeccen}[1]{\ifnum#1=50 %
\hatcurRVjittertwosiglimAeccenxxxxA
\else
\ifnum#1=52 %
\hatcurRVjittertwosiglimAeccenxxxxC
\else
\ifnum#1=53 %
\hatcurRVjittertwosiglimAeccenxxxxD
\else
??????\fi
\fi
\fi
}
\newcommand{\hatcurRVjittertwosiglimBeccen}[1]{\ifnum#1=50 %
\hatcurRVjittertwosiglimBeccenxxxxA
\else
\ifnum#1=52 %
\hatcurRVjittertwosiglimBeccenxxxxC
\else
\ifnum#1=53 %
\hatcurRVjittertwosiglimBeccenxxxxD
\else
??????\fi
\fi
\fi
}
\newcommand{\hatcurRVjittertwosiglimCeccen}[1]{\ifnum#1=50 %
\hatcurRVjittertwosiglimCeccenxxxxA
\else
\ifnum#1=52 %
\hatcurRVjittertwosiglimCeccenxxxxC
\else
??????\fi
\fi
}
\newcommand{\hatcurRVkeccen}[1]{\ifnum#1=50 %
\hatcurRVkeccenxxxxA
\else
\ifnum#1=51 %
\hatcurRVkeccenxxxxB
\else
\ifnum#1=52 %
\hatcurRVkeccenxxxxC
\else
\ifnum#1=53 %
\hatcurRVkeccenxxxxD
\else
??????\fi
\fi
\fi
\fi
}
\newcommand{\hatcurRVKeccen}[1]{\ifnum#1=50 %
\hatcurRVKeccenxxxxA
\else
\ifnum#1=51 %
\hatcurRVKeccenxxxxB
\else
\ifnum#1=52 %
\hatcurRVKeccenxxxxC
\else
\ifnum#1=53 %
\hatcurRVKeccenxxxxD
\else
??????\fi
\fi
\fi
\fi
}
\newcommand{\hatcurRVomegaeccen}[1]{\ifnum#1=50 %
\hatcurRVomegaeccenxxxxA
\else
\ifnum#1=51 %
\hatcurRVomegaeccenxxxxB
\else
\ifnum#1=52 %
\hatcurRVomegaeccenxxxxC
\else
\ifnum#1=53 %
\hatcurRVomegaeccenxxxxD
\else
??????\fi
\fi
\fi
\fi
}
\newcommand{\hatcurRVrheccen}[1]{\ifnum#1=50 %
\hatcurRVrheccenxxxxA
\else
\ifnum#1=51 %
\hatcurRVrheccenxxxxB
\else
\ifnum#1=52 %
\hatcurRVrheccenxxxxC
\else
\ifnum#1=53 %
\hatcurRVrheccenxxxxD
\else
??????\fi
\fi
\fi
\fi
}
\newcommand{\hatcurRVrkeccen}[1]{\ifnum#1=50 %
\hatcurRVrkeccenxxxxA
\else
\ifnum#1=51 %
\hatcurRVrkeccenxxxxB
\else
\ifnum#1=52 %
\hatcurRVrkeccenxxxxC
\else
\ifnum#1=53 %
\hatcurRVrkeccenxxxxD
\else
??????\fi
\fi
\fi
\fi
}
\newcommand{\hatcurRVtroneeccen}[1]{\ifnum#1=50 %
\hatcurRVtroneeccenxxxxA
\else
\ifnum#1=51 %
\hatcurRVtroneeccenxxxxB
\else
\ifnum#1=52 %
\hatcurRVtroneeccenxxxxC
\else
\ifnum#1=53 %
\hatcurRVtroneeccenxxxxD
\else
??????\fi
\fi
\fi
\fi
}
\newcommand{\hatcurRVtrtwoeccen}[1]{\ifnum#1=50 %
\hatcurRVtrtwoeccenxxxxA
\else
\ifnum#1=51 %
\hatcurRVtrtwoeccenxxxxB
\else
\ifnum#1=52 %
\hatcurRVtrtwoeccenxxxxC
\else
\ifnum#1=53 %
\hatcurRVtrtwoeccenxxxxD
\else
??????\fi
\fi
\fi
\fi
}
\newcommand{\hatcurSMEiiloggeccen}[1]{\ifnum#1=51 %
\hatcurSMEiiloggeccenxxxxB
\else
\ifnum#1=52 %
\hatcurSMEiiloggeccenxxxxC
\else
\ifnum#1=53 %
\hatcurSMEiiloggeccenxxxxD
\else
??????\fi
\fi
\fi
}
\newcommand{\hatcurSMEiiteffeccen}[1]{\ifnum#1=51 %
\hatcurSMEiiteffeccenxxxxB
\else
\ifnum#1=52 %
\hatcurSMEiiteffeccenxxxxC
\else
\ifnum#1=53 %
\hatcurSMEiiteffeccenxxxxD
\else
??????\fi
\fi
\fi
}
\newcommand{\hatcurSMEiivmaceccen}[1]{\ifnum#1=52 %
\hatcurSMEiivmaceccenxxxxC
\else
\ifnum#1=53 %
\hatcurSMEiivmaceccenxxxxD
\else
??????\fi
\fi
}
\newcommand{\hatcurSMEiivmiceccen}[1]{\ifnum#1=52 %
\hatcurSMEiivmiceccenxxxxC
\else
\ifnum#1=53 %
\hatcurSMEiivmiceccenxxxxD
\else
??????\fi
\fi
}
\newcommand{\hatcurSMEiivsineccen}[1]{\ifnum#1=51 %
\hatcurSMEiivsineccenxxxxB
\else
\ifnum#1=52 %
\hatcurSMEiivsineccenxxxxC
\else
\ifnum#1=53 %
\hatcurSMEiivsineccenxxxxD
\else
??????\fi
\fi
\fi
}
\newcommand{\hatcurSMEiizfeheccen}[1]{\ifnum#1=51 %
\hatcurSMEiizfeheccenxxxxB
\else
\ifnum#1=52 %
\hatcurSMEiizfeheccenxxxxC
\else
\ifnum#1=53 %
\hatcurSMEiizfeheccenxxxxD
\else
??????\fi
\fi
\fi
}
\newcommand{\hatcurSMEiizfehshorteccen}[1]{\ifnum#1=51 %
\hatcurSMEiizfehshorteccenxxxxB
\else
\ifnum#1=52 %
\hatcurSMEiizfehshorteccenxxxxC
\else
\ifnum#1=53 %
\hatcurSMEiizfehshorteccenxxxxD
\else
??????\fi
\fi
\fi
}
\newcommand{\hatcurSMEiloggeccen}[1]{\ifnum#1=50 %
\hatcurSMEiloggeccenxxxxA
\else
\ifnum#1=51 %
\hatcurSMEiloggeccenxxxxB
\else
\ifnum#1=52 %
\hatcurSMEiloggeccenxxxxC
\else
\ifnum#1=53 %
\hatcurSMEiloggeccenxxxxD
\else
??????\fi
\fi
\fi
\fi
}
\newcommand{\hatcurSMEiteffeccen}[1]{\ifnum#1=50 %
\hatcurSMEiteffeccenxxxxA
\else
\ifnum#1=51 %
\hatcurSMEiteffeccenxxxxB
\else
\ifnum#1=52 %
\hatcurSMEiteffeccenxxxxC
\else
\ifnum#1=53 %
\hatcurSMEiteffeccenxxxxD
\else
??????\fi
\fi
\fi
\fi
}
\newcommand{\hatcurSMEivmaceccen}[1]{\ifnum#1=50 %
\hatcurSMEivmaceccenxxxxA
\else
\ifnum#1=51 %
\hatcurSMEivmaceccenxxxxB
\else
\ifnum#1=52 %
\hatcurSMEivmaceccenxxxxC
\else
\ifnum#1=53 %
\hatcurSMEivmaceccenxxxxD
\else
??????\fi
\fi
\fi
\fi
}
\newcommand{\hatcurSMEivmiceccen}[1]{\ifnum#1=50 %
\hatcurSMEivmiceccenxxxxA
\else
\ifnum#1=51 %
\hatcurSMEivmiceccenxxxxB
\else
\ifnum#1=52 %
\hatcurSMEivmiceccenxxxxC
\else
\ifnum#1=53 %
\hatcurSMEivmiceccenxxxxD
\else
??????\fi
\fi
\fi
\fi
}
\newcommand{\hatcurSMEivsineccen}[1]{\ifnum#1=50 %
\hatcurSMEivsineccenxxxxA
\else
\ifnum#1=51 %
\hatcurSMEivsineccenxxxxB
\else
\ifnum#1=52 %
\hatcurSMEivsineccenxxxxC
\else
\ifnum#1=53 %
\hatcurSMEivsineccenxxxxD
\else
??????\fi
\fi
\fi
\fi
}
\newcommand{\hatcurSMEizfeheccen}[1]{\ifnum#1=50 %
\hatcurSMEizfeheccenxxxxA
\else
\ifnum#1=51 %
\hatcurSMEizfeheccenxxxxB
\else
\ifnum#1=52 %
\hatcurSMEizfeheccenxxxxC
\else
\ifnum#1=53 %
\hatcurSMEizfeheccenxxxxD
\else
??????\fi
\fi
\fi
\fi
}
\newcommand{\hatcurSMEizfehshorteccen}[1]{\ifnum#1=50 %
\hatcurSMEizfehshorteccenxxxxA
\else
\ifnum#1=51 %
\hatcurSMEizfehshorteccenxxxxB
\else
\ifnum#1=52 %
\hatcurSMEizfehshorteccenxxxxC
\else
\ifnum#1=53 %
\hatcurSMEizfehshorteccenxxxxD
\else
??????\fi
\fi
\fi
\fi
}
\newcommand{\hatcurXAveccen}[1]{\ifnum#1=50 %
\hatcurXAveccenxxxxA
\else
\ifnum#1=51 %
\hatcurXAveccenxxxxB
\else
\ifnum#1=52 %
\hatcurXAveccenxxxxC
\else
\ifnum#1=53 %
\hatcurXAveccenxxxxD
\else
??????\fi
\fi
\fi
\fi
}
\newcommand{\hatcurXdisteccen}[1]{\ifnum#1=50 %
\hatcurXdisteccenxxxxA
\else
\ifnum#1=51 %
\hatcurXdisteccenxxxxB
\else
\ifnum#1=52 %
\hatcurXdisteccenxxxxC
\else
\ifnum#1=53 %
\hatcurXdisteccenxxxxD
\else
??????\fi
\fi
\fi
\fi
}
\newcommand{\hatcurXdistredeccen}[1]{\ifnum#1=50 %
\hatcurXdistredeccenxxxxA
\else
\ifnum#1=51 %
\hatcurXdistredeccenxxxxB
\else
\ifnum#1=52 %
\hatcurXdistredeccenxxxxC
\else
\ifnum#1=53 %
\hatcurXdistredeccenxxxxD
\else
??????\fi
\fi
\fi
\fi
}
\newcommand{\hatcurXEBVeccen}[1]{\ifnum#1=50 %
\hatcurXEBVeccenxxxxA
\else
\ifnum#1=51 %
\hatcurXEBVeccenxxxxB
\else
\ifnum#1=52 %
\hatcurXEBVeccenxxxxC
\else
\ifnum#1=53 %
\hatcurXEBVeccenxxxxD
\else
??????\fi
\fi
\fi
\fi
}
\newcommand{\hatcurXjhisoredeccen}[1]{\ifnum#1=50 %
\hatcurXjhisoredeccenxxxxA
\else
\ifnum#1=51 %
\hatcurXjhisoredeccenxxxxB
\else
\ifnum#1=52 %
\hatcurXjhisoredeccenxxxxC
\else
\ifnum#1=53 %
\hatcurXjhisoredeccenxxxxD
\else
??????\fi
\fi
\fi
\fi
}
\newcommand{\hatcurXjkisoredeccen}[1]{\ifnum#1=50 %
\hatcurXjkisoredeccenxxxxA
\else
\ifnum#1=51 %
\hatcurXjkisoredeccenxxxxB
\else
\ifnum#1=52 %
\hatcurXjkisoredeccenxxxxC
\else
\ifnum#1=53 %
\hatcurXjkisoredeccenxxxxD
\else
??????\fi
\fi
\fi
\fi
}
\newcommand{\hatcurXmhisoredeccen}[1]{\ifnum#1=50 %
\hatcurXmhisoredeccenxxxxA
\else
\ifnum#1=51 %
\hatcurXmhisoredeccenxxxxB
\else
\ifnum#1=52 %
\hatcurXmhisoredeccenxxxxC
\else
\ifnum#1=53 %
\hatcurXmhisoredeccenxxxxD
\else
??????\fi
\fi
\fi
\fi
}
\newcommand{\hatcurXmiisoredeccen}[1]{\ifnum#1=50 %
\hatcurXmiisoredeccenxxxxA
\else
\ifnum#1=51 %
\hatcurXmiisoredeccenxxxxB
\else
\ifnum#1=52 %
\hatcurXmiisoredeccenxxxxC
\else
\ifnum#1=53 %
\hatcurXmiisoredeccenxxxxD
\else
??????\fi
\fi
\fi
\fi
}
\newcommand{\hatcurXmjisoredeccen}[1]{\ifnum#1=50 %
\hatcurXmjisoredeccenxxxxA
\else
\ifnum#1=51 %
\hatcurXmjisoredeccenxxxxB
\else
\ifnum#1=52 %
\hatcurXmjisoredeccenxxxxC
\else
\ifnum#1=53 %
\hatcurXmjisoredeccenxxxxD
\else
??????\fi
\fi
\fi
\fi
}
\newcommand{\hatcurXmkisoredeccen}[1]{\ifnum#1=50 %
\hatcurXmkisoredeccenxxxxA
\else
\ifnum#1=51 %
\hatcurXmkisoredeccenxxxxB
\else
\ifnum#1=52 %
\hatcurXmkisoredeccenxxxxC
\else
\ifnum#1=53 %
\hatcurXmkisoredeccenxxxxD
\else
??????\fi
\fi
\fi
\fi
}
\newcommand{\hatcurXmvisoredeccen}[1]{\ifnum#1=50 %
\hatcurXmvisoredeccenxxxxA
\else
\ifnum#1=51 %
\hatcurXmvisoredeccenxxxxB
\else
\ifnum#1=52 %
\hatcurXmvisoredeccenxxxxC
\else
\ifnum#1=53 %
\hatcurXmvisoredeccenxxxxD
\else
??????\fi
\fi
\fi
\fi
}
\newcommand{\hatcurXsecdureccen}[1]{\ifnum#1=50 %
\hatcurXsecdureccenxxxxA
\else
\ifnum#1=51 %
\hatcurXsecdureccenxxxxB
\else
\ifnum#1=52 %
\hatcurXsecdureccenxxxxC
\else
\ifnum#1=53 %
\hatcurXsecdureccenxxxxD
\else
??????\fi
\fi
\fi
\fi
}
\newcommand{\hatcurXsecingdureccen}[1]{\ifnum#1=50 %
\hatcurXsecingdureccenxxxxA
\else
\ifnum#1=51 %
\hatcurXsecingdureccenxxxxB
\else
\ifnum#1=52 %
\hatcurXsecingdureccenxxxxC
\else
\ifnum#1=53 %
\hatcurXsecingdureccenxxxxD
\else
??????\fi
\fi
\fi
\fi
}
\newcommand{\hatcurXsecondaryeccen}[1]{\ifnum#1=50 %
\hatcurXsecondaryeccenxxxxA
\else
\ifnum#1=51 %
\hatcurXsecondaryeccenxxxxB
\else
\ifnum#1=52 %
\hatcurXsecondaryeccenxxxxC
\else
\ifnum#1=53 %
\hatcurXsecondaryeccenxxxxD
\else
??????\fi
\fi
\fi
\fi
}
\newcommand{\hatcurXsecphaseeccen}[1]{\ifnum#1=50 %
\hatcurXsecphaseeccenxxxxA
\else
\ifnum#1=51 %
\hatcurXsecphaseeccenxxxxB
\else
\ifnum#1=52 %
\hatcurXsecphaseeccenxxxxC
\else
\ifnum#1=53 %
\hatcurXsecphaseeccenxxxxD
\else
??????\fi
\fi
\fi
\fi
}
\newcommand{\hatcurXviisoredeccen}[1]{\ifnum#1=50 %
\hatcurXviisoredeccenxxxxA
\else
\ifnum#1=51 %
\hatcurXviisoredeccenxxxxB
\else
\ifnum#1=52 %
\hatcurXviisoredeccenxxxxC
\else
\ifnum#1=53 %
\hatcurXviisoredeccenxxxxD
\else
??????\fi
\fi
\fi
\fi
}
\newcommand{\hatcurXvkisoredeccen}[1]{\ifnum#1=50 %
\hatcurXvkisoredeccenxxxxA
\else
\ifnum#1=51 %
\hatcurXvkisoredeccenxxxxB
\else
\ifnum#1=52 %
\hatcurXvkisoredeccenxxxxC
\else
\ifnum#1=53 %
\hatcurXvkisoredeccenxxxxD
\else
??????\fi
\fi
\fi
\fi
}

\newcommand{\hatcurxxxxA}{HATS-50}
\newcommand{\hatcurbxxxxA}{HATS-50b}
\newcommand{\hatcurcxxxxA}{HATS-50c}

\newcommand{\hatcurplanetnumxxxxA}{50}

\newcommand{\hatcurCCtwomassshortxxxxA}{20014273-2604392}

\newcommand{\hatcurRVgammaabsxxxxA}{\hatcurRVgammaA{\hatcurplanetnumxxxxA}}                           

\newcommand{\hatcurRVgammarelxxxxA}{\hatcurRVgammaA{\hatcurplanetnumxxxxA}}                           

\newcommand{\hatcurCCtassvixxxxA}{\ensuremath{NULL\pm NULL}}                  

\newcommand{\hatcurSMEversionxxxxA}{i}                                       

\newcommand{\hatcurisoshortxxxxA}{YY}
\newcommand{\hatcurisofullxxxxA}{Yonsei-Yale (YY)}
\newcommand{\hatcurisocitexxxxA}{yi:2001}

\newcommand{\hatcurlumindxxxxA}{\arstar}

\newcommand{\hatcurjhkfilsetxxxxA}{ESO}

\newcommand{\hatcurSMEteffxxxxA}{\ifthenelse{\equal{\hatcurSMEversionxxxxA}{i}}{\hatcurSMEiteff{\hatcurplanetnumxxxxA}}{\hatcurSMEiiteff{\hatcurplanetnumxxxxA}}}
\newcommand{\hatcurSMEzfehxxxxA}{\ifthenelse{\equal{\hatcurSMEversionxxxxA}{i}}{\hatcurSMEizfeh{\hatcurplanetnumxxxxA}}{\hatcurSMEiizfeh{\hatcurplanetnumxxxxA}}}
\newcommand{\hatcurSMEzfehshortxxxxA}{\ifthenelse{\equal{\hatcurSMEversionxxxxA}{i}}{\hatcurSMEizfehshort{\hatcurplanetnumxxxxA}}{\hatcurSMEiizfehshort{\hatcurplanetnumxxxxA}}}
\newcommand{\hatcurSMEloggxxxxA}{\ifthenelse{\equal{\hatcurSMEversionxxxxA}{i}}{\hatcurSMEilogg{\hatcurplanetnumxxxxA}}{\hatcurSMEiilogg{\hatcurplanetnumxxxxA}}}
\newcommand{\hatcurSMEvsinxxxxA}{\ifthenelse{\equal{\hatcurSMEversionxxxxA}{i}}{\hatcurSMEivsin{\hatcurplanetnumxxxxA}}{\hatcurSMEiivsin{\hatcurplanetnumxxxxA}}}
\newcommand{\hatcurSMEvmacxxxxA}{\ifthenelse{\equal{\hatcurSMEversionxxxxA}{i}}{\hatcurSMEivmac{\hatcurplanetnumxxxxA}}{\hatcurSMEiivmac{\hatcurplanetnumxxxxA}}}
\newcommand{\hatcurSMEvmicxxxxA}{\ifthenelse{\equal{\hatcurSMEversionxxxxA}{i}}{\hatcurSMEivmic{\hatcurplanetnumxxxxA}}{\hatcurSMEiivmic{\hatcurplanetnumxxxxA}}}

\newcommand{\hatcurxxxxB}{HATS-51}
\newcommand{\hatcurbxxxxB}{HATS-51b}
\newcommand{\hatcurcxxxxB}{HATS-51c}

\newcommand{\hatcurplanetnumxxxxB}{51}

\newcommand{\hatcurCCtwomassshortxxxxB}{06512340-2903309}

\newcommand{\hatcurRVgammaabsxxxxB}{\hatcurRVgammaA{\hatcurplanetnumxxxxB}}                           

\newcommand{\hatcurRVgammarelxxxxB}{\hatcurRVgammaA{\hatcurplanetnumxxxxB}}                           

\newcommand{\hatcurCCtassvixxxxB}{\ensuremath{NULL\pm NULL}}                  

\newcommand{\hatcurSMEversionxxxxB}{ii}                                       

\newcommand{\hatcurisoshortxxxxB}{YY}
\newcommand{\hatcurisofullxxxxB}{Yonsei-Yale (YY)}
\newcommand{\hatcurisocitexxxxB}{yi:2001}

\newcommand{\hatcurlumindxxxxB}{\arstar}

\newcommand{\hatcurjhkfilsetxxxxB}{ESO}

\newcommand{\hatcurSMEteffxxxxB}{\ifthenelse{\equal{\hatcurSMEversionxxxxB}{i}}{\hatcurSMEiteff{\hatcurplanetnumxxxxB}}{\hatcurSMEiiteff{\hatcurplanetnumxxxxB}}}
\newcommand{\hatcurSMEzfehxxxxB}{\ifthenelse{\equal{\hatcurSMEversionxxxxB}{i}}{\hatcurSMEizfeh{\hatcurplanetnumxxxxB}}{\hatcurSMEiizfeh{\hatcurplanetnumxxxxB}}}
\newcommand{\hatcurSMEzfehshortxxxxB}{\ifthenelse{\equal{\hatcurSMEversionxxxxB}{i}}{\hatcurSMEizfehshort{\hatcurplanetnumxxxxB}}{\hatcurSMEiizfehshort{\hatcurplanetnumxxxxB}}}
\newcommand{\hatcurSMEloggxxxxB}{\ifthenelse{\equal{\hatcurSMEversionxxxxB}{i}}{\hatcurSMEilogg{\hatcurplanetnumxxxxB}}{\hatcurSMEiilogg{\hatcurplanetnumxxxxB}}}
\newcommand{\hatcurSMEvsinxxxxB}{\ifthenelse{\equal{\hatcurSMEversionxxxxB}{i}}{\hatcurSMEivsin{\hatcurplanetnumxxxxB}}{\hatcurSMEiivsin{\hatcurplanetnumxxxxB}}}
\newcommand{\hatcurSMEvmacxxxxB}{\ifthenelse{\equal{\hatcurSMEversionxxxxB}{i}}{\hatcurSMEivmac{\hatcurplanetnumxxxxB}}{\hatcurSMEiivmac{\hatcurplanetnumxxxxB}}}
\newcommand{\hatcurSMEvmicxxxxB}{\ifthenelse{\equal{\hatcurSMEversionxxxxB}{i}}{\hatcurSMEivmic{\hatcurplanetnumxxxxB}}{\hatcurSMEiivmic{\hatcurplanetnumxxxxB}}}

\newcommand{\hatcurxxxxC}{HATS-52}
\newcommand{\hatcurbxxxxC}{HATS-52b}
\newcommand{\hatcurcxxxxC}{HATS-52c}

\newcommand{\hatcurplanetnumxxxxC}{52}

\newcommand{\hatcurCCtwomassshortxxxxC}{09202105-3116095}

\newcommand{\hatcurRVgammaabsxxxxC}{\hatcurRVgammaA{\hatcurplanetnumxxxxC}}                           

\newcommand{\hatcurRVgammarelxxxxC}{\hatcurRVgammaA{\hatcurplanetnumxxxxC}}                           

\newcommand{\hatcurCCtassvixxxxC}{\ensuremath{NULL\pm NULL}}                  

\newcommand{\hatcurSMEversionxxxxC}{ii}                                       

\newcommand{\hatcurisoshortxxxxC}{YY}
\newcommand{\hatcurisofullxxxxC}{Yonsei-Yale (YY)}
\newcommand{\hatcurisocitexxxxC}{yi:2001}

\newcommand{\hatcurlumindxxxxC}{\arstar}

\newcommand{\hatcurjhkfilsetxxxxC}{ESO}

\newcommand{\hatcurSMEteffxxxxC}{\ifthenelse{\equal{\hatcurSMEversionxxxxC}{i}}{\hatcurSMEiteff{\hatcurplanetnumxxxxC}}{\hatcurSMEiiteff{\hatcurplanetnumxxxxC}}}
\newcommand{\hatcurSMEzfehxxxxC}{\ifthenelse{\equal{\hatcurSMEversionxxxxC}{i}}{\hatcurSMEizfeh{\hatcurplanetnumxxxxC}}{\hatcurSMEiizfeh{\hatcurplanetnumxxxxC}}}
\newcommand{\hatcurSMEzfehshortxxxxC}{\ifthenelse{\equal{\hatcurSMEversionxxxxC}{i}}{\hatcurSMEizfehshort{\hatcurplanetnumxxxxC}}{\hatcurSMEiizfehshort{\hatcurplanetnumxxxxC}}}
\newcommand{\hatcurSMEloggxxxxC}{\ifthenelse{\equal{\hatcurSMEversionxxxxC}{i}}{\hatcurSMEilogg{\hatcurplanetnumxxxxC}}{\hatcurSMEiilogg{\hatcurplanetnumxxxxC}}}
\newcommand{\hatcurSMEvsinxxxxC}{\ifthenelse{\equal{\hatcurSMEversionxxxxC}{i}}{\hatcurSMEivsin{\hatcurplanetnumxxxxC}}{\hatcurSMEiivsin{\hatcurplanetnumxxxxC}}}
\newcommand{\hatcurSMEvmacxxxxC}{\ifthenelse{\equal{\hatcurSMEversionxxxxC}{i}}{\hatcurSMEivmac{\hatcurplanetnumxxxxC}}{\hatcurSMEiivmac{\hatcurplanetnumxxxxC}}}
\newcommand{\hatcurSMEvmicxxxxC}{\ifthenelse{\equal{\hatcurSMEversionxxxxC}{i}}{\hatcurSMEivmic{\hatcurplanetnumxxxxC}}{\hatcurSMEiivmic{\hatcurplanetnumxxxxC}}}

\newcommand{\hatcurxxxxD}{HATS-53}
\newcommand{\hatcurbxxxxD}{HATS-53b}
\newcommand{\hatcurcxxxxD}{HATS-53c}

\newcommand{\hatcurplanetnumxxxxD}{53}

\newcommand{\hatcurCCtwomassshortxxxxD}{11463084-3351361}

\newcommand{\hatcurRVgammaabsxxxxD}{\hatcurRVgammaA{\hatcurplanetnumxxxxD}}                           

\newcommand{\hatcurRVgammarelxxxxD}{\hatcurRVgammaA{\hatcurplanetnumxxxxD}}                           

\newcommand{\hatcurCCtassvixxxxD}{\ensuremath{NULL\pm NULL}}                  

\newcommand{\hatcurSMEversionxxxxD}{ii}                                       

\newcommand{\hatcurisoshortxxxxD}{YY}
\newcommand{\hatcurisofullxxxxD}{Yonsei-Yale (YY)}
\newcommand{\hatcurisocitexxxxD}{yi:2001}

\newcommand{\hatcurlumindxxxxD}{\arstar}

\newcommand{\hatcurjhkfilsetxxxxD}{ESO}

\newcommand{\hatcurSMEteffxxxxD}{\ifthenelse{\equal{\hatcurSMEversionxxxxD}{i}}{\hatcurSMEiteff{\hatcurplanetnumxxxxD}}{\hatcurSMEiiteff{\hatcurplanetnumxxxxD}}}
\newcommand{\hatcurSMEzfehxxxxD}{\ifthenelse{\equal{\hatcurSMEversionxxxxD}{i}}{\hatcurSMEizfeh{\hatcurplanetnumxxxxD}}{\hatcurSMEiizfeh{\hatcurplanetnumxxxxD}}}
\newcommand{\hatcurSMEzfehshortxxxxD}{\ifthenelse{\equal{\hatcurSMEversionxxxxD}{i}}{\hatcurSMEizfehshort{\hatcurplanetnumxxxxD}}{\hatcurSMEiizfehshort{\hatcurplanetnumxxxxD}}}
\newcommand{\hatcurSMEloggxxxxD}{\ifthenelse{\equal{\hatcurSMEversionxxxxD}{i}}{\hatcurSMEilogg{\hatcurplanetnumxxxxD}}{\hatcurSMEiilogg{\hatcurplanetnumxxxxD}}}
\newcommand{\hatcurSMEvsinxxxxD}{\ifthenelse{\equal{\hatcurSMEversionxxxxD}{i}}{\hatcurSMEivsin{\hatcurplanetnumxxxxD}}{\hatcurSMEiivsin{\hatcurplanetnumxxxxD}}}
\newcommand{\hatcurSMEvmacxxxxD}{\ifthenelse{\equal{\hatcurSMEversionxxxxD}{i}}{\hatcurSMEivmac{\hatcurplanetnumxxxxD}}{\hatcurSMEiivmac{\hatcurplanetnumxxxxD}}}
\newcommand{\hatcurSMEvmicxxxxD}{\ifthenelse{\equal{\hatcurSMEversionxxxxD}{i}}{\hatcurSMEivmic{\hatcurplanetnumxxxxD}}{\hatcurSMEiivmic{\hatcurplanetnumxxxxD}}}

\newcommand{\hatcur}[1]{\ifnum#1=50 %
\hatcurxxxxA
\else
\ifnum#1=51 %
\hatcurxxxxB
\else
\ifnum#1=52 %
\hatcurxxxxC
\else
\ifnum#1=53 %
\hatcurxxxxD
\else
??????\fi
\fi
\fi
\fi
}
\newcommand{\hatcurb}[1]{\ifnum#1=50 %
\hatcurbxxxxA
\else
\ifnum#1=51 %
\hatcurbxxxxB
\else
\ifnum#1=52 %
\hatcurbxxxxC
\else
\ifnum#1=53 %
\hatcurbxxxxD
\else
??????\fi
\fi
\fi
\fi
}
\newcommand{\hatcurc}[1]{\ifnum#1=50 %
\hatcurcxxxxA
\else
\ifnum#1=51 %
\hatcurcxxxxB
\else
\ifnum#1=52 %
\hatcurcxxxxC
\else
\ifnum#1=53 %
\hatcurcxxxxD
\else
??????\fi
\fi
\fi
\fi
}
\newcommand{\hatcurCCtassvi}[1]{\ifnum#1=50 %
\hatcurCCtassvixxxxA
\else
\ifnum#1=51 %
\hatcurCCtassvixxxxB
\else
\ifnum#1=52 %
\hatcurCCtassvixxxxC
\else
\ifnum#1=53 %
\hatcurCCtassvixxxxD
\else
??????\fi
\fi
\fi
\fi
}
\newcommand{\hatcurCCtwomassshort}[1]{\ifnum#1=50 %
\hatcurCCtwomassshortxxxxA
\else
\ifnum#1=51 %
\hatcurCCtwomassshortxxxxB
\else
\ifnum#1=52 %
\hatcurCCtwomassshortxxxxC
\else
\ifnum#1=53 %
\hatcurCCtwomassshortxxxxD
\else
??????\fi
\fi
\fi
\fi
}
\newcommand{\hatcurisocite}[1]{\ifnum#1=50 %
\hatcurisocitexxxxA
\else
\ifnum#1=51 %
\hatcurisocitexxxxB
\else
\ifnum#1=52 %
\hatcurisocitexxxxC
\else
\ifnum#1=53 %
\hatcurisocitexxxxD
\else
??????\fi
\fi
\fi
\fi
}
\newcommand{\hatcurisofull}[1]{\ifnum#1=50 %
\hatcurisofullxxxxA
\else
\ifnum#1=51 %
\hatcurisofullxxxxB
\else
\ifnum#1=52 %
\hatcurisofullxxxxC
\else
\ifnum#1=53 %
\hatcurisofullxxxxD
\else
??????\fi
\fi
\fi
\fi
}
\newcommand{\hatcurisoshort}[1]{\ifnum#1=50 %
\hatcurisoshortxxxxA
\else
\ifnum#1=51 %
\hatcurisoshortxxxxB
\else
\ifnum#1=52 %
\hatcurisoshortxxxxC
\else
\ifnum#1=53 %
\hatcurisoshortxxxxD
\else
??????\fi
\fi
\fi
\fi
}
\newcommand{\hatcurjhkfilset}[1]{\ifnum#1=50 %
\hatcurjhkfilsetxxxxA
\else
\ifnum#1=51 %
\hatcurjhkfilsetxxxxB
\else
\ifnum#1=52 %
\hatcurjhkfilsetxxxxC
\else
\ifnum#1=53 %
\hatcurjhkfilsetxxxxD
\else
??????\fi
\fi
\fi
\fi
}
\newcommand{\hatcurlumind}[1]{\ifnum#1=50 %
\hatcurlumindxxxxA
\else
\ifnum#1=51 %
\hatcurlumindxxxxB
\else
\ifnum#1=52 %
\hatcurlumindxxxxC
\else
\ifnum#1=53 %
\hatcurlumindxxxxD
\else
??????\fi
\fi
\fi
\fi
}
\newcommand{\hatcurplanetnum}[1]{\ifnum#1=50 %
\hatcurplanetnumxxxxA
\else
\ifnum#1=51 %
\hatcurplanetnumxxxxB
\else
\ifnum#1=52 %
\hatcurplanetnumxxxxC
\else
\ifnum#1=53 %
\hatcurplanetnumxxxxD
\else
??????\fi
\fi
\fi
\fi
}
\newcommand{\hatcurRVgammaabs}[1]{\ifnum#1=50 %
\hatcurRVgammaabsxxxxA
\else
\ifnum#1=51 %
\hatcurRVgammaabsxxxxB
\else
\ifnum#1=52 %
\hatcurRVgammaabsxxxxC
\else
\ifnum#1=53 %
\hatcurRVgammaabsxxxxD
\else
??????\fi
\fi
\fi
\fi
}
\newcommand{\hatcurRVgammarel}[1]{\ifnum#1=50 %
\hatcurRVgammarelxxxxA
\else
\ifnum#1=51 %
\hatcurRVgammarelxxxxB
\else
\ifnum#1=52 %
\hatcurRVgammarelxxxxC
\else
\ifnum#1=53 %
\hatcurRVgammarelxxxxD
\else
??????\fi
\fi
\fi
\fi
}
\newcommand{\hatcurSMElogg}[1]{\ifnum#1=50 %
\hatcurSMEloggxxxxA
\else
\ifnum#1=51 %
\hatcurSMEloggxxxxB
\else
\ifnum#1=52 %
\hatcurSMEloggxxxxC
\else
\ifnum#1=53 %
\hatcurSMEloggxxxxD
\else
??????\fi
\fi
\fi
\fi
}
\newcommand{\hatcurSMEteff}[1]{\ifnum#1=50 %
\hatcurSMEteffxxxxA
\else
\ifnum#1=51 %
\hatcurSMEteffxxxxB
\else
\ifnum#1=52 %
\hatcurSMEteffxxxxC
\else
\ifnum#1=53 %
\hatcurSMEteffxxxxD
\else
??????\fi
\fi
\fi
\fi
}
\newcommand{\hatcurSMEversion}[1]{\ifnum#1=50 %
\hatcurSMEversionxxxxA
\else
\ifnum#1=51 %
\hatcurSMEversionxxxxB
\else
\ifnum#1=52 %
\hatcurSMEversionxxxxC
\else
\ifnum#1=53 %
\hatcurSMEversionxxxxD
\else
??????\fi
\fi
\fi
\fi
}
\newcommand{\hatcurSMEvmac}[1]{\ifnum#1=50 %
\hatcurSMEvmacxxxxA
\else
\ifnum#1=51 %
\hatcurSMEvmacxxxxB
\else
\ifnum#1=52 %
\hatcurSMEvmacxxxxC
\else
\ifnum#1=53 %
\hatcurSMEvmacxxxxD
\else
??????\fi
\fi
\fi
\fi
}
\newcommand{\hatcurSMEvmic}[1]{\ifnum#1=50 %
\hatcurSMEvmicxxxxA
\else
\ifnum#1=51 %
\hatcurSMEvmicxxxxB
\else
\ifnum#1=52 %
\hatcurSMEvmicxxxxC
\else
\ifnum#1=53 %
\hatcurSMEvmicxxxxD
\else
??????\fi
\fi
\fi
\fi
}
\newcommand{\hatcurSMEvsin}[1]{\ifnum#1=50 %
\hatcurSMEvsinxxxxA
\else
\ifnum#1=51 %
\hatcurSMEvsinxxxxB
\else
\ifnum#1=52 %
\hatcurSMEvsinxxxxC
\else
\ifnum#1=53 %
\hatcurSMEvsinxxxxD
\else
??????\fi
\fi
\fi
\fi
}
\newcommand{\hatcurSMEzfeh}[1]{\ifnum#1=50 %
\hatcurSMEzfehxxxxA
\else
\ifnum#1=51 %
\hatcurSMEzfehxxxxB
\else
\ifnum#1=52 %
\hatcurSMEzfehxxxxC
\else
\ifnum#1=53 %
\hatcurSMEzfehxxxxD
\else
??????\fi
\fi
\fi
\fi
}
\newcommand{\hatcurSMEzfehshort}[1]{\ifnum#1=50 %
\hatcurSMEzfehshortxxxxA
\else
\ifnum#1=51 %
\hatcurSMEzfehshortxxxxB
\else
\ifnum#1=52 %
\hatcurSMEzfehshortxxxxC
\else
\ifnum#1=53 %
\hatcurSMEzfehshortxxxxD
\else
??????\fi
\fi
\fi
\fi
}

\newcounter{planetcounter}

\newboolean{emulateapj}
\setboolean{emulateapj}{true}

\newboolean{rvtablelong}
\setboolean{rvtablelong}{true}

\newboolean{astroph}
\setboolean{astroph}{true}

\shortauthors{Henning et al.}
\shorttitle{\hatcur{50}\lowercase{b}--\hatcur{53}\lowercase{b}}
\ifthenelse{\boolean{emulateapj}}{
    
}{
    
}

\begin{document}

\title{
\hatcurb{50} Through \hatcurb{53}: Four Transiting Hot Jupiters orbiting G-type stars Discovered by the HATSouth Survey\footnote{
The HATSouth network is operated by a collaboration consisting of
Princeton University (PU), the Max Planck Institute f\"ur Astronomie
(MPIA), the Australian National University (ANU), and the Pontificia
Universidad Cat\'olica de Chile (PUC).  The station at Las Campanas
Observatory (LCO) of the Carnegie Institute is operated by PU in
conjunction with PUC, the station at the High Energy Spectroscopic
Survey (H.E.S.S.) site is operated in conjunction with MPIA, and the
station at Siding Spring Observatory (SSO) is operated jointly with
ANU.
Based in part on observations made with the ESO\,3.6\,m, the NTT, the MPG\,2.2\,m and Euler\,1.2 m Telescopes at the ESO Observatory in La Silla.
Based in part on observations made with the 3.9\,m Anglo-Australian Telescope and the ANU\,2.3\,m Telescope, both at SSO.
Based in part on observations made with the Keck-I Telescope at Mauna Kea Observatory in Hawaii.
Based in part on observations obtained with the facilities of the Las Cumbres Observatory Global Telescope and with the Perth Exoplanet Survey Telescope.
}
}

\correspondingauthor{Thomas Henning}
\email{henning@mpia.de}

\author{Th. Henning}
\affil{Max Planck Institute for Astronomy, K\"{o}nigstuhl 17, 69117 -- Heidelberg, Germany}

\author{L. Mancini}
\affil{Department of Physics, University of Rome Tor Vergata, Via della Ricerca Scientifica 1, 00133 -- Rome, Italy}
\affil{Max Planck Institute for Astronomy, K\"{o}nigstuhl 17, 69117 -- Heidelberg, Germany}
\affil{INAF -- Astrophysical Observatory of Turin, via Osservatorio 20, 10025 -- Pino Torinese, Italy}

\author{P. Sarkis}
\affil{Max Planck Institute for Astronomy, K\"{o}nigstuhl 17, 69117 -- Heidelberg, Germany}

\author{G.~\'{A}. Bakos}
\altaffiliation{Packard Fellow}
\affil{Department of Astrophysical Sciences, Princeton University, NJ 08544, USA}

\author{J.~D. Hartman}
\affil{Department of Astrophysical Sciences, Princeton University, NJ 08544, USA}

\author{D. Bayliss}
\affil{Department of Physics, University of Warwick, Coventry CV4 7AL, UK}

\author{J. Bento}
\affil{Research School of Astronomy and Astrophysics, Australian National University, Canberra, ACT 2611, Australia}

\author{W. Bhatti}
\affil{Department of Astrophysical Sciences, Princeton University, NJ 08544, USA}

\author{R. Brahm}
\affil{Millennium Institute of Astrophysics, Av. Vicu\~{n}a Mackenna 4860, 7820436 Macul, Santiago, Chile}\affil{Instituto de Astrof\'{i}sica, Pontificia Universidad Cat\'{o}lica de Chile, Av. Vicu\~{n}a Mackenna 4860, 7820436 Macul, Santiago, Chile}

\author{S. Ciceri}
\affil{Department of Astronomy, Stockholm University, SE-106 91 Stockholm, Sweden}

\author{Z. Csubry}
\affil{Department of Astrophysical Sciences, Princeton University, NJ 08544, USA}

\author{M. de Val-Borro}
\affil{Department of Astrophysical Sciences, Princeton University, NJ 08544, USA}

\author{N. Espinoza}
\affil{Instituto de Astrof\'{i}sica, Pontificia Universidad Cat\'{o}lica de Chile, Av. Vicu\~{n}a Mackenna 4860, 7820436 Macul, Santiago, Chile}
\affil{Millennium Institute of Astrophysics, Av. Vicu\~{n}a Mackenna 4860, 7820436 Macul, Santiago, Chile}

\author{B.~J. Fulton}
\affil{California Institute of Technology, Pasadena, CA 91125, USA}

\author{A.~W. Howard}
\affil{California Institute of Technology, Pasadena, CA 91125, USA}

\author{H.~T. Isaacson}
\affil{Astronomy Department, University of California, Berkeley, CA, 94720, USA}

\author{A. Jord\'an}
\affil{Instituto de Astrof\'{i}sica, Pontificia Universidad Cat\'{o}lica de Chile, Av. Vicu\~{n}a Mackenna 4860, 7820436 Macul, Santiago, Chile}
\affil{Millennium Institute of Astrophysics, Av. Vicu\~{n}a Mackenna 4860, 7820436 Macul, Santiago, Chile}
\affil{Max Planck Institute for Astronomy, K\"{o}nigstuhl 17, 69117 -- Heidelberg, Germany}

\author{G.~W. Marcy}
\affil{Astronomy Department, University of California, Berkeley, CA, 94720, USA}

\author{K. Penev}
\affil{Department of Astrophysical Sciences, Princeton University, NJ 08544, USA}

\author{M. Rabus}
\affil{Instituto de Astrof\'{i}sica, Pontificia Universidad Cat\'{o}lica de Chile, Av. Vicu\~{n}a Mackenna 4860, 7820436 Macul, Santiago, Chile}
\affil{Max Planck Institute for Astronomy, K\"{o}nigstuhl 17, 69117 -- Heidelberg, Germany}

\author{V. Suc}
\affil{Instituto de Astrof\'{i}sica, Pontificia Universidad Cat\'{o}lica de Chile, Av. Vicu\~{n}a Mackenna 4860, 7820436 Macul, Santiago, Chile}

\author{T.~G. Tan}
\affil{Perth Exoplanet Survey Telescope, Perth, Australia}

\author{C.~G. Tinney}
\affil{Australian Centre for Astrobiology, School of Physics, University of New South Wales, NSW 2052, Australia}
\affil{Exoplanetary Science at UNSW, School of Physics, University of New South Wales, NSW 2052, Australia}

\author{D.~J. Wright}
\affil{Australian Centre for Astrobiology, School of Physics, University of New South Wales, NSW 2052, Australia}
\affil{Exoplanetary Science at UNSW, School of Physics, University of New South Wales, NSW 2052, Australia}

\author{G. Zhou}
\affil{Harvard-Smithsonian Center for Astrophysics, 60 Garden St., Cambridge, MA 02138, USA}

\author{S. Durkan}
\affil{Astrophysics Research Centre, Queens University, Belfast, Belfast, Northern Ireland, UK}

\author{J. Lazar}
\affil{Hungarian Astronomical Association, 1451 Budapest, Hungary}

\author{I. Papp}
\affil{Hungarian Astronomical Association, 1451 Budapest, Hungary}

\author{P. Sari}
\affil{Hungarian Astronomical Association, 1451 Budapest, Hungary}

\begin{abstract}

\setcounter{footnote}{10}
We report the discovery of four close-in transiting exoplanets (\hatcurb{50} through \hatcurb{53}), discovered using the HATSouth three-continent network of homogeneous and automated telescopes. These new exoplanets belong to the class of hot Jupiters and orbit G-type dwarf stars, with brightness in the range $V=12.5-14.0$\,mag. While \hatcur{53} has many physical characteristics similar to the Sun, the other three stars appear to be metal rich ($\feh=0.2-0.3$), larger and more massive. Three of the new exoplanets, namely \hatcurb{50}, \hatcurb{51} and \hatcurb{53}, have low density (\hatcurb{50}: \hatcurPPmlong{50}\,\mjup, \hatcurPPrlong{50}\,\rjup; \hatcurb{51}: \hatcurPPmlong{51}\,\mjup, \hatcurPPrlong{51}\,\rjup; \hatcurb{53}: \hatcurPPmlong{53}\,\mjup, \hatcurPPrlong{53}\,\rjup) and similar orbital period (\hatcurLCPshort{50}\,d, \hatcurLCPshort{51}\,d, \hatcurLCPshort{53}\,d, respectively). Instead, \hatcurb{52} is more dense (mass \hatcurPPmlong{52}\,\mjup\ and radius \hatcurPPrlong{52}\,\rjup) and has a shorter orbital period (\hatcurLCPshort{52}\,d). It also receives an intensive radiation from its parent star and, consequently, presents a high equilibrium temperature ($T_{\rm eq}=\hatcurPPteff{52}$\,K). \hatcur{50} shows a marginal additional transit feature consistent with an ultra-short period hot super Neptune (upper mass limit $0.16$\,\mjup), which will be able to be confirmed with TESS photometry.
\setcounter{footnote}{0}
\end{abstract}

\keywords{
    planetary systems ---
    stars: individual (
\setcounter{planetcounter}{1}
\hatcur{50},
\hatcurCCgsc{50}\loopcommanoperiod
\setcounter{planetcounter}{2}
\hatcur{51},
\hatcurCCgsc{51}\loopcommanoperiod
\setcounter{planetcounter}{3}
\hatcur{52},
\hatcurCCgsc{52}\loopcommanoperiod
\setcounter{planetcounter}{3}
\hatcur{53},
\hatcurCCgsc{53}\loopcommanoperiod
\setcounter{planetcounter}{4}
) 
    techniques: spectroscopic, photometric
}


\section{Introduction}
\label{sec:introduction}
Ground-based transit surveys, based on small robotic telescopes, are a versatile tool for the detection of transiting exoplanets and the precise measurement of planetary radii and masses. They have provided key contributions to exoplanetary science by discovering extremely interesting objects (e.g. WASP-12b: \citealp{hebb:2009}; GJ\,1124b: \citealp{charbonneau:2009}, HAT-P-11b: \citealp{bakos:2010:hat11}), and are still revealing astonishing planetary systems (some of the most recent ones are, for example, GJ\,1132: \citealp{berta:2015}; XO-2: \citealp{burke:2007,desidera:2014,damasso:2015}; WASP-47: \citealp{hellier:2012,becker:2015}; Trappist-1: \citealp{gillon:2016}; KELT-9: \citealp{gaudi:2017}). 

Due to observational and instrumental limitations, these surveys are particularly sensitive for detecting hot Jupiters, which are a class of exoplanets formed by gas giant planets, similar to Jupiter in terms of size, mass and composition, but having  shorter orbital periods ($P_{\rm orb}<10$\,days). Considering the proximity to their parent stars and since they are more massive and larger than ice and rocky planets, hot Jupiters are often excellent targets for the follow-up characterization of their physical properties and atmospheres.

Thanks to the efforts of various teams (e.g. HATNet: \citealp{bakos:2004:hatnet}; WASP: \citealp{pollacco:2006}; KELT: \citealp{pepper:2007,pepper:2012}; MEarth: \citealp{charbonneau:2009}; QES: \citealp{alsubai:2013}; NGTS: \citealp{wheatley:2017}), who set up and ran ground-based surveys for many years, we currently know roughly 300 hot Jupiters, whose physical and orbital parameters have been well determined. However they represent less than 10\% of $\approx 3500$ confirmed exoplanets\footnote{Data taken from the NASA Exoplanet Archive: https://exoplanetarchive.opac.caltech.edu}. In fact, one of the greatest achievements obtained by the {\it Kepler} space-telescope survey \citep{borucki:2011} was to establish the statistical abundance of the different classes of exoplanets in the Galaxy, revealing that giant planets are rarer than small-size rocky and Neptunian-type planets \citep{fressin:2013,dressing:2013}. However, even though hot Jupiters are relatively rare, there are many open questions which make these bodies extremely interesting to study.
 
Debated are the theories that have been proposed to explain their formation and evolution, including in-situ scenarios \citep{bodenheimer:2000,boley:2016,batygin:2016} and physical mechanisms that reasonably forced them to migrate, from the snowline, so close to their parent star \citep{lin:1996,rasio:1996,fabrycky:2007,chatterjee:2008,marzari:2009,bitsch:2011}. It remains to be fully understood why giant exoplanets with similar masses present such a wide range of radii (see \citealp{thorngren:2017}). 
Particularly intriguing is, finally, the fact that the most recent studies of hot-Jupiters atmospheres have shown a wide range of different results, including Rayleigh scattering, Na and K absorption, detection of molecules, like H$_{2}$O and titanium oxide, and flat transmission spectra probably caused by the presence of thick clouds or hazes \citep{sing:2016,sedaghati:2017}.

In order to give the right answers to these and other theoretical and phenomenological questions concerning hot Jupiters, it is mandatory to have a large enough sample for statistical studies. Ground-based surveys have been conceived for this purpose and the current challenge is to try to fill all the parameter space of exoplanet properties, in particular those zones where the investigation is particularly hard due to observational biases.

In this context, we are undertaking the HATSouth project with the aim to detect new transiting exoplanet systems. The HATSouth survey consists of a network of 24 homogeneous telescopes, which are mounted on six automated units distributed in pairs over three continents (South America, Africa, and Australia). The large number of telescopes and the wide separation between the HATSouth stations increases the sensitivity to exoplanets orbiting faint stars ($12\,{\rm mag}<V<16\,{\rm mag}$) and having long orbital periods ($>10$\,days) (Bakos et al. 2013).

In this work, we present four new transiting extrasolar planets: \hatcurb{50}, \hatcurb{51}, \hatcurb{52} and \hatcurb{53}. The paper is organized as follows: in Sect.~\ref{sec:obs} we describe the detection of the photometric transit signal by the HATSouth survey and the spectroscopic and photometric follow-up observations performed to confirm the exoplanetary nature of the candidates. Then, in Sect.~\ref{sec:analysis}, we jointly analyze the data to determine the stellar and planetary parameters, ruling out false positive scenarios. Our results are finally summarized and discussed in Sect.~\ref{sec:discussion}.

\section{Observations}
\label{sec:obs}

\subsection{The HATSouth survey}
\label{sec:survey}
The four new exoplanets reported in this work have been detected thanks to the HATSouth survey\footnote{http://hatsouth.org/} \citep{bakos:2013:hatsouth}. This is a network of robotic wide-field telescopes, composed of six identical units located in three stations. The stations are distributed over three continents in the southern hemisphere, i.e. Las Campanas Observatory (LCO) in Chile, the H.E.S.S. site in Namibia, and Siding Spring Observatory (SSO) in Australia. Each unit consists of a single mount with four 18\,cm Takahashi astrographs with a focal length of 500\,mm, and four Apogee U16M Alta CCD cameras, which have 4k$\times$4k pixels of size $9.0\, \mu$m. With a plate scale of $3.7$\,arcsec\,pixel$^{-1}$, the total mosaic field-of-view on the sky is $8^{\circ} \times 8^{\circ}$. The survey operates in the visual, through Sloan-$r$ filters, and the scientific images are obtained using an exposure time of 4\,minutes. They are then automatically calibrated with bias, dark and flat images and are stored in the HATSouth archive at Princeton University. Each stellar field is monitored for roughly $2-3$ months from each station, in order to get a 24\,h coverage, thus exploiting the great advantage coming from their large separation in longitude. Once a long time-series sequence ($>7\,000$ images) for a single field is collected, then aperture photometry is performed to get light curves for each star with $9 \lesssim r \lesssim 16$\,mag in the field. The resulting light curves are treated with decorrelation and detrending algorithms\footnote{External Parameter Decorrelation (EPD; \citealp{bakos:2010:hat11}); Trend Filtering Algorithm (TFA; \citealp{kovacs:2005:TFA}).} and, finally, we look for possible transiting-planet periodic signals by running the BLS (Box-fitting Least Squares; \citealp{kovacs:2002:BLS}) code for each of them. Planet candidates detected from the survey undergo spectroscopic characterization and, finally, their planetary nature is confirmed or excluded by precise radial-velocity measurements and photometric follow-up observations \citep{penev:2013:hats1}.

Since first light in 2009, the HATSouth survey has so far produced 6.25 million light curves for 5.07 million stars from observations covering 2609 square degrees. This is due to the overlap between the pointing of the cameras on a single mount, and between the different pointing positions we use to tile the sky into target fields. As a matter of fact, some stars have multiple light curves from different cameras and pointing positions (e.g. HATS-4: \citealp{jordan:2014:hats4}). Based on these observations, we have so far identified 1883 candidate transiting planets of which 1120 of them have undergone follow-up observations. This leads so far to the determination that 636 of the candidates are false alarms or false positives, while we confirmed and published 44 planets. 

Notable cases are: the two super-Neptunes HATS-7b \citep{bakos:2015:hats7} and HATS-8b \citep{bayliss:2015:hats8}; HATS-6b, a warm Saturn-mass exoplanet orbiting a M star \citep{hartman:2015:hats6}; HATS-17b, the longest period transiting exoplanet discovered so far by a wide-field ground-based photometric survey \citep{brahm:2016:hats17}; HATS-18b, an extremely short-period planet spinning-up its host star \citep{penev:2016:hats18}; the detection of several very low mass stars ($0.1-0.2 \, M_{\sun}$) in eclipsing binary systems \citep{zhou:2014:mebs,zhou:2015}. 

Currently, dozens of other exoplanets have been confirmed by the HATSouth team and are undergoing analysis and preparation for publication.

\subsection{Photometric detection}
\label{sec:detection}
This study presents the discovery of four new transiting planetary systems, which were detected following the procedure described above, and confirmed based on follow-up observations as described in the next sections; the new systems are \hatcur{50}, \hatcur{51}, \hatcur{52} and \hatcur{53}. Each of them are composed of a moderately bright G-type star and a hot-Jupiter-type planet. The orbital periods are $\hatcurLCPshort{50}$\,d, $\hatcurLCPshort{51}$\,d, $\hatcurLCPshort{52}$\,d and $\hatcurLCPshort{53}$\,d for \hatcurb{50}, \hatcurb{51}, \hatcurb{52} and \hatcurb{53}, respectively, implying that we are dealing with new close-in hot Jupiters. Stellar coordinates, magnitudes and cross-identifications are shown in \reftabl{stellar}. In particular, the magnitudes of the fours stars in the optical bands were taken from APASS \citep{henden:2009}, as listed in the UCAC\,4 catalogue \citep{zacharias:2012:ucac4}, while those in the NIR bands are from the 2MASS catalogue.

A summary of the HATSouth photometric observations for these objects is reported in Table~\ref{tab:photobs}. In particular, the four stars were observed thousands of times by the HATSouth telescopes between March 2010 and July 2013; the corresponding phase-folded light curves are plotted in Figure~\ref{fig:hatsouth}, clearly showing typical transiting-planet signals with transit depths around $1-2\%$. 

\subsection{Searching for additional periodic signals in the time-series survey data}
\label{sec:periodic_signals}
%
\ifthenelse{\boolean{emulateapj}}{
    \begin{figure*}[!ht]
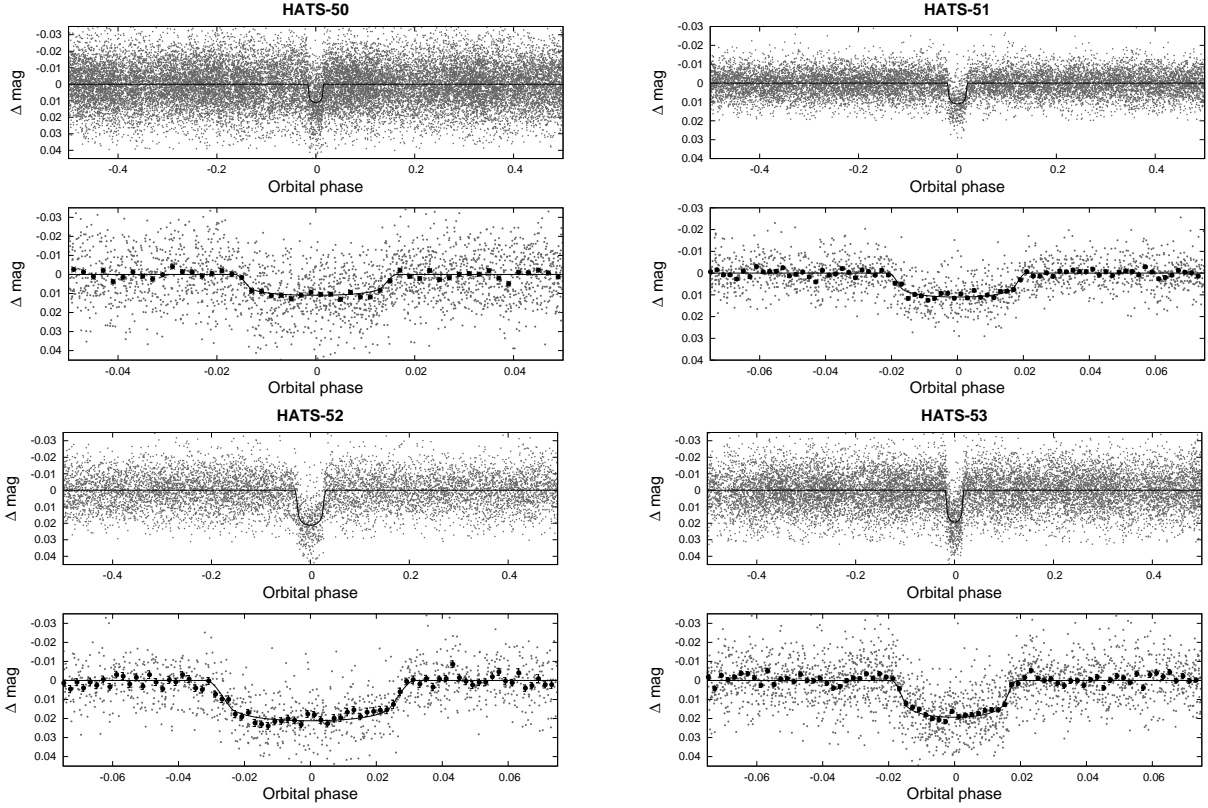

}{
    \begin{figure}[!ht]
}
\plottwo{\hatcurhtr{50}-hs.eps}{\hatcurhtr{51}-hs.eps}
\plottwo{\hatcurhtr{52}-hs.eps}{\hatcurhtr{53}-hs.eps}
\caption{
Phase-folded unbinned HATSouth light curves for \hatcur{50} (upper left), \hatcur{51} (upper right), \hatcur{52} (lower left) and \hatcur{53} (lower right). In each case we show two panels. The top panel shows the full light curve, while the bottom panel shows the light curve zoomed-in on the transit. The solid lines show the model fits to the light curves. The dark filled circles in the bottom panels show the light curves binned in phase with a bin size of 0.002.
\label{fig:hatsouth}}
\ifthenelse{\boolean{emulateapj}}{
    \end{figure*}
}{
    \end{figure}
}

\begin{figure}
\plotone{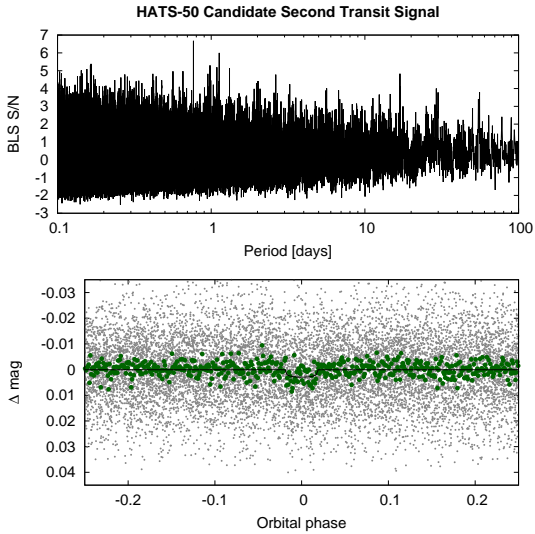}
\caption{
Search for additional periodic signals in the light curve of \hatcur{50} due to other transiting planets.
{\it Top panel}: BLS spectrum. {\it Bottom panel}: Unbinned light curve of \hatcur{50} (grey points) phase folded with the $0.76624822$\,days transit signal. The green points are the phase-binned values (bin size of 0.001). The line is a \citet{mandel:2002} transit-model fit to the light curve.
}
\label{fig:hats50c_candidate}
\end{figure}

After having detected the planetary signals, we further analyze each of the four data sets to search for potential stellar variability/activity and additional periodic signals due to other transiting planets. This analysis was carried out by running BLS on the residuals of each HATSouth light curve, and studying the Generalized Lomb-Scargle periodogram (GLS; \citealp{zechmeister:2009}). The results of these additional checks are the following.
\begin{itemize}
\item By running BLS, we did not detected any significant periodic transit signal in the residuals of the \hatcur{51}, \hatcur{52} and \hatcur{53} light curves. For the case of \hatcur{50}, we noticed a marginal transit signal with a period of $0.76624822$\,days, $T_{\rm C} = 2455274.38586$, depth of $3.2$\,mmag, and duration of 46\,minutes (bottom panel of Fig.~\ref{fig:hats50c_candidate}). The candidate transit signal has a S/N of 7.5 in the phase-folded light curve and a BLS Signal-Detection-Efficiency (SDE) value of 7.38.
Even though this signal is below our threshold for selecting candidates to follow-up (top panel of Fig.~\ref{fig:hats50c_candidate}), it is worth noting it given the presence of the confirmed hot Jupiter and the possibility of close companions to hot-Jupiters \citep{becker:2015}. Therefore, \hatcur{50} could be an interesting target for a further short-cadence, time-series photometric monitoring with a more precise instrument, like TESS \citep{ricker:2017}. However, given the density of the host star \hatcur{50}, inferred from modelling the transits of \hatcurb{50} (see Table~\ref{tab:stellar}), and the duration of the transits for the candidate signal, the candidate planet would have an orbital inclination that differs by more than 10$^{\circ}$ from that of the hot Jupiter. We finally note that a \citet{mandel:2002} model fit indicates an implied radius in the super-Neptune regime. So, if this signal was really caused by an additional transiting planet, it would be firmly in the Neptune desert. \\
\item By running GLS, we did not detect any significant periodic signal in the GLS spectrum of the light curve obtained for \hatcur{50}, \hatcur{51} and \hatcur{53}. For these three systems, we placed a $95\%$ confidence upper limit of $0.95$\,mmag on the amplitude of any periodic signal between 0.01\,days and 100\,days. For \hatcur{52}, we detected a sinusoidal signal with a periodicity of $P = 15.63703 \pm 0.00066$\,days and an amplitude of $1.23 \pm 0.25$\,mmag. The peak has a signal-to-noise ratio of $20.3$ in the spectrum, a periodogram value of $\Delta \chi^2 / \chi^2_{0} = 0.0054$, and a false alarm probability (FAP), assuming Gaussian white noise, of $2 \times 10^{-5}$. 
We also assessed the FAP of GLS by performing a bootstrap analysis and obtaining a distribution of peak signals. From this, we estimated a more accurate false alarm probability of $9 \times 10^{-5}$.

This sinusoidal periodic signal can be related to the stellar activity and, therefore, presumably indicates its rotation period. 
However, considering that \hatcur{52} has a radius of $\rstar=\hatcurISOrlong{52}\,\rsun$ (see Table~\ref{tab:stellar}), this value of $P$ implies that $v_{\rm eq} = 3.60 \pm 0.59$\,\kms\ for \hatcur{52}, which is $1.5\sigma$ below the spectroscopically determined value of $\vsini = \hatcurSMEvsin{52}$\,\kms. So, there is a slight tension between these measurements if we identify the photometric periodicity of $P = 15.6$\,days with the rotation period of the star.
\end{itemize}

\startlongtable
\ifthenelse{\boolean{emulateapj}}{
    \begin{deluxetable*}{llrrrr}
}{
    \begin{deluxetable}{llrrrr}
}
\tablewidth{0pc}
\tabletypesize{\scriptsize}
\tablecaption{
    Summary of photometric observations
    \label{tab:photobs}
}
\tablehead{
    \multicolumn{1}{c}{Instrument/Field\tablenotemark{a}} &
    \multicolumn{1}{c}{Date(s)} &
    \multicolumn{1}{c}{\# Images} &
    \multicolumn{1}{c}{Cadence\tablenotemark{b}} &
    \multicolumn{1}{c}{Filter} &
    \multicolumn{1}{c}{Precision\tablenotemark{c}} \\
    \multicolumn{1}{c}{} &
    \multicolumn{1}{c}{} &
    \multicolumn{1}{c}{} &
    \multicolumn{1}{c}{(sec)} &
    \multicolumn{1}{c}{} &
    \multicolumn{1}{c}{(mmag)}
}
\startdata
\sidehead{\textbf{\hatcur{50}}}
~~~~HS-2.4/G580 & 2010 Mar--2011 Aug & 6072 & 294 & $r$ & 12.6 \\
~~~~HS-4.4/G580 & 2010 Mar--2011 Aug & 3082 & 298 & $r$ & 13.2 \\
~~~~HS-6.4/G580 & 2010 Mar--2011 May & 742 & 297 & $r$ & 14.1 \\
~~~~HS-1.3/G625 & 2012 Jun--2012 Oct & 4662 & 291 & $r$ & 13.2 \\
~~~~HS-3.3/G625 & 2012 Jun--2012 Oct & 5357 & 293 & $r$ & 13.0 \\
~~~~HS-5.3/G625 & 2012 Jun--2012 Oct & 1724 & 293 & $r$ & 12.7 \\
~~~~PEST~0.3\,m & 2014 Aug 04 & 202 & 133 & $R_{c}$ & 4.8 \\
~~~~LCOGT~1\,m+CTIO/sinistro & 2015 May 11 & 57 & 226 & $i$ & 1.3 \\
~~~~LCOGT~1\,m+SAAO/SBIG & 2015 Jun 06 & 136 & 150 & $i$ & 4.1 \\
\sidehead{\textbf{\hatcur{51}}}
~~~~HS-1.2/G601 & 2011 Aug--2012 Jan & 4806 & 296 & $r$ & 6.2 \\
~~~~HS-3.2/G601 & 2011 Aug--2012 Jan & 4062 & 296 & $r$ & 6.6 \\
~~~~HS-5.2/G601 & 2011 Aug--2012 Jan & 3083 & 290 & $r$ & 6.8 \\
~~~~LCOGT~1\,m+CTIO/sinistro & 2014 Oct 31 & 36 & 228 & $i$ & 0.8 \\
~~~~LCOGT~1\,m+SSO/SBIG & 2015 Mar 07 & 172 & 76 & $i$ & 2.9 \\
~~~~LCOGT~1\,m+SAAO/SBIG & 2015 Mar 10 & 92 & 141 & $i$ & 1.7 \\
~~~~LCOGT~1\,m+CTIO/sinistro & 2015 Oct 03 & 71 & 159 & $i$ & 1.8 \\
\sidehead{\textbf{\hatcur{52}}}
~~~~HS-2.1/G606 & 2012 Feb--2012 Jun & 3753 & 291 & $r$ & 9.1 \\
~~~~HS-4.1/G606 & 2012 Feb--2012 Jun & 2778 & 300 & $r$ & 11.8 \\
~~~~HS-6.1/G606 & 2012 Feb--2012 Jun & 1184 & 299 & $r$ & 9.8 \\
~~~~PEST~0.3\,m & 2015 Feb 06 & 193 & 132 & $R_{C}$ & 12.8 \\
~~~~LCOGT~1\,m+CTIO/sinistro & 2015 May 12 & 38 & 226 & $i$ & 4.1 \\
~~~~LCOGT~1\,m+SSO/SBIG & 2015 May 13 & 53 & 195 & $i$ & 1.4 \\
~~~~LCOGT~1\,m+CTIO/sinistro & 2015 May 16 & 42 & 226 & $i$ & 1.9 \\
~~~~LCOGT~1\,m+CTIO/sinistro & 2015 Oct 23 & 100 & 54 & $i$ & 6.4 \\
\sidehead{\textbf{\hatcur{53}}}
~~~~HS-2.4/G610 & 2011 Apr--2013 Jul & 5496 & 280 & $r$ & 10.8 \\
~~~~HS-4.4/G610 & 2013 Jan--2013 Jul & 3739 & 323 & $r$ & 10.8 \\
~~~~HS-6.4/G610 & 2011 Apr--2013 Jul & 3578 & 282 & $r$ & 11.8 \\
~~~~LCOGT~1\,m+CTIO/sinistro & 2016 Feb 02 & 89 & 219 & $i$ & 1.6 \\
~~~~LCOGT~1\,m+SAAO/SBIG & 2016 Feb 10 & 48 & 192 & $i$ & 2.0 \\
~~~~PEST~0.3\,m & 2016 Feb 14 & 156 & 132 & $R_{C}$ & 11.9 \\
\enddata
\tablenotetext{a}{
    For HATSouth data we list the HATSouth unit, CCD and field name
    from which the observations are taken. HS-1 and -2 are located at
    Las Campanas Observatory in Chile, HS-3 and -4 are located at the
    H.E.S.S. site in Namibia, and HS-5 and -6 are located at Siding
    Spring Observatory in Australia. Each unit has 4 CCDs. Each field
    corresponds to one of 838 fixed pointings used to cover the full
    4$\pi$ celestial sphere. All data from a given HATSouth field and
    CCD number are reduced together, while detrending through External
    Parameter Decorrelation (EPD) is done independently for each
    unique unit+CCD+field combination.
}
\tablenotetext{b}{
    The median time between consecutive images rounded to the nearest
    second. Due to factors such as weather, the day--night cycle,
    guiding and focus corrections the cadence is only approximately
    uniform over short timescales.
}
\tablenotetext{c}{
    The RMS of the residuals from the best-fit model.
} \ifthenelse{\boolean{emulateapj}}{
    \end{deluxetable*}
}{
    \end{deluxetable}
}

\subsection{Spectroscopic Observations}
\label{sec:obsspec}

\ifthenelse{\boolean{emulateapj}}{
    \begin{deluxetable*}{llrrrrr}
}{
    \begin{deluxetable}{llrrrrrrrr}
}
\tablewidth{0pc}
\tabletypesize{\scriptsize}
\tablecaption{
    Summary of spectroscopy observations 
    \label{tab:specobs}
}
\tablehead{
    \multicolumn{1}{c}{Instrument}          &
    \multicolumn{1}{c}{UT Date(s)}             &
    \multicolumn{1}{c}{\# Spec.}   &
    \multicolumn{1}{c}{Res.}          &
    \multicolumn{1}{c}{S/N Range\tablenotemark{a}}           &
    \multicolumn{1}{c}{$\gamma_{\rm RV}$\tablenotemark{b}} &
    \multicolumn{1}{c}{RV Precision\tablenotemark{c}} \\
    &
    &
    &
    \multicolumn{1}{c}{$\Delta \lambda$/$\lambda$/1000} &
    &
    \multicolumn{1}{c}{(\kms)}              &
    \multicolumn{1}{c}{(\ms)}
}
\startdata
\sidehead{\textbf{\hatcur{50}}}\\ [-12pt]
ANU~2.3\,m/WiFeS & 2014 Jun 3--5 & 3 & 7 & 21--90 & -23.4 & 4000 \\
ANU~2.3\,m/WiFeS & 2014 Jun 4 & 1 & 3 & 76 & $\cdots$ & $\cdots$ \\
Euler~1.2\,m/CORALIE & 2014 Jun--Sep & 4\tablenotemark{d} & 60 & 7--13 & -20.176 & 73 \\
MPG~2.2\,m/FEROS & 2014 Jul--2016 Sep & 32\tablenotemark{d} & 48 & 14--55 & -20.250 & 72 \\
Keck-I~10\,m/HIRES+I$_{2}$ & 2014 Sep--2015 Jul & 7 & 48 & 110--155 & $\cdots$ & 29 \\
Keck-I~10\,m/HIRES & 2015 Jul 5 & 1 & 48 & 70 & $\cdots$ & $\cdots$ \\ [4pt]
\sidehead{\textbf{\hatcur{51}}}\\ [-12pt]
ANU~2.3\,m/WiFeS & 2014 Oct 7 & 1 & 3 & 46 & $\cdots$ & $\cdots$ \\
ANU~2.3\,m/WiFeS & 2014 Oct 8--12 & 4 & 7 & 33--67 & 2.0 & 4000 \\
Euler~1.2\,m/CORALIE & 2014 Oct--2016 Nov & 24\tablenotemark{d} & 60 & 8--30 & 3.086 & 55 \\
MPG~2.2\,m/FEROS & 2014 Dec--2015 Feb & 14 & 48 & 60--97 & 3.087 & 55 \\
AAT~3.9\,m/CYCLOPS & 2015 Feb--May & 13 & 70 & $\cdots$ & 3.087 & 30 \\ [4pt]
\sidehead{\textbf{\hatcur{52}}}\\ [-12pt]
ANU~2.3\,m/WiFeS & 2014 Jul 3 & 1 & 3 & 58 & $\cdots$ & $\cdots$ \\
ANU~2.3\,m/WiFeS & 2014 Jul--2015 Jan & 4 & 7 & 14--50 & 13.8 & 4000 \\
Euler~1.2\,m/CORALIE & 2015 Mar 28 & 1 & 60 & 11 & 13.30 & $\cdots$ \\
ESO~3.6\,m/HARPS & 2015 Apr 6--8 & 3 & 115 & 10--13 & 13.384 & 15 \\
MPG~2.2\,m/FEROS & 2015 Jun--2016 Feb & 11 & 48 & 25--48 & 13.456 & 127 \\ [4pt]
\sidehead{\textbf{\hatcur{53}}}\\ [-12pt]
ANU~2.3\,m/WiFeS & 2015 Mar 30 & 1 & 3 & 30 & $\cdots$ & $\cdots$ \\
ANU~2.3\,m/WiFeS & 2015 Mar--Apr & 2 & 7 & 6--33 & 68.5 & 4000 \\
ESO~3.6\,m/HARPS & 2015 Apr 7--8 & 2 & 115 & 7--12 & 71.945 & 23 \\
MPG~2.2\,m/FEROS & 2015 Jun 6--20 & 11 & 48 & 21--40 & 71.950 & 28 \\ [4pt]
\enddata 
\tablenotetext{a}{
    S/N per resolution element near 5180\,\AA.
}
\tablenotetext{b}{
    For high-precision RV observations included in the orbit determination this is the zero-point RV from the best-fit orbit. For other instruments it is the mean value. We do not provide this quantity for the lower resolution WiFeS observations which were only used to measure stellar atmospheric parameters, or for the Keck-I/HIRES spectra of \hatcur{50} from which only relative velocities have been measured.
}
\tablenotetext{c}{
    For high-precision RV observations included in the orbit
    determination this is the scatter in the RV residuals from the
    best-fit orbit (which may include astrophysical jitter), for other
    instruments this is either an estimate of the precision (not
    including jitter), or the measured standard deviation. We do not
    provide this quantity for low-resolution observations from the
    ANU~2.3\,m/WiFeS.
}
\tablenotetext{d}{
    We list here the total number of spectra collected for each instrument, including observations that were excluded from the analysis due to very low S/N or substantial sky contamination. For \hatcur{50} we excluded one CORALIE spectrum and 5 FEROS spectra from the analysis. For \hatcur{51} we excluded 3 CORALIE spectra.
}
\ifthenelse{\boolean{emulateapj}}{
    \end{deluxetable*}
}{
    \end{deluxetable}
}
%
\startlongtable
\tabletypesize{\scriptsize}
\ifthenelse{\boolean{emulateapj}}{
    \begin{deluxetable*}{lrrrrrl}
}{
    \begin{deluxetable}{lrrrrrl}
}
\tablewidth{0pc}
\tablecaption{
    Relative radial velocities and bisector spans for \hatcur{50}--\hatcur{53}.
    \label{tab:rvs}
}
\tablehead{
    \colhead{BJD} &
    \colhead{RV\tablenotemark{a}} &
    \colhead{\ensuremath{\sigma_{\rm RV}}\tablenotemark{b}} &
    \colhead{BS} &
    \colhead{\ensuremath{\sigma_{\rm BS}}} &
    \colhead{Phase} &
    \colhead{Instrument}\\
    \colhead{\hbox{(2,450,000$+$)}} &
    \colhead{(\ms)} &
    \colhead{(\ms)} &
    \colhead{(\ms)} &
    \colhead{(\ms)} &
    \colhead{} &
    \colhead{}
}
\startdata
\multicolumn{7}{c}{\bf HATS-50} \\
\hline\\
    \input{\hatcurhtr{50}_rvtable.tex}
\cutinhead{\bf HATS-51}
    \input{\hatcurhtr{51}_rvtable.tex}
\cutinhead{\bf HATS-52}
    \input{\hatcurhtr{52}_rvtable.tex}
\cutinhead{\bf HATS-53}
    \input{\hatcurhtr{53}_rvtable.tex}
\enddata
\tablenotetext{a}{
    The zero-point of these velocities is arbitrary. An overall offset
    $\gamma_{\rm rel}$ fitted independently to the velocities from
    each instrument has been subtracted.
}
\tablenotetext{b}{
    Internal errors excluding the component of astrophysical jitter
    considered in \refsecl{globmod}.
}
\ifthenelse{\boolean{rvtablelong}}{
}{
    \tablecomments{
    }
} 
\ifthenelse{\boolean{emulateapj}}{
    \end{deluxetable*}
}{
    \end{deluxetable}
}

\setcounter{planetcounter}{1}
%
\ifthenelse{\boolean{emulateapj}}{
    \begin{figure*} [ht]
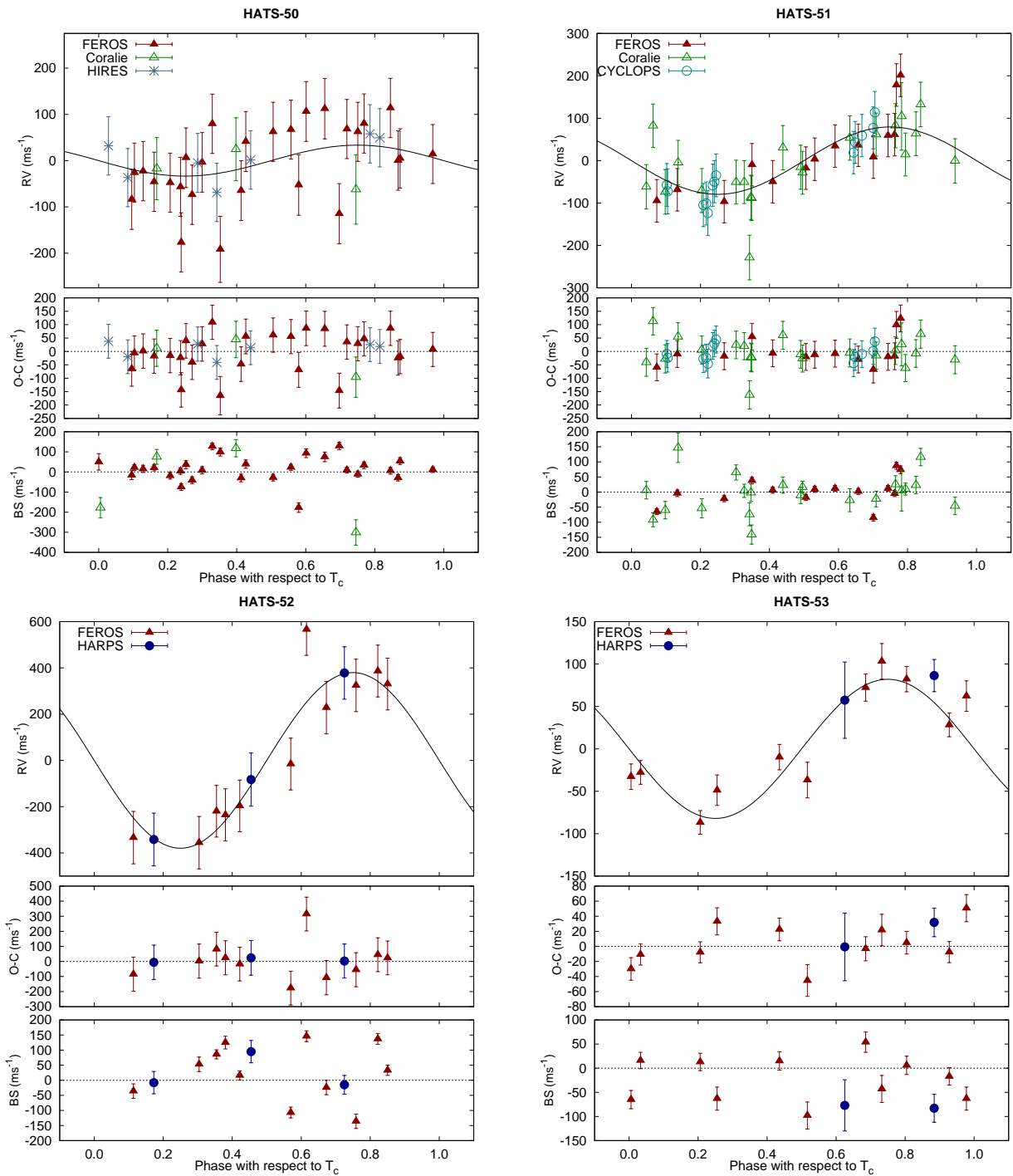

}{
    \begin{figure}[ht]
}
\plottwo{\hatcurhtr{50}-rv.eps}{\hatcurhtr{51}-rv.eps}
\plottwo{\hatcurhtr{52}-rv.eps}{\hatcurhtr{53}-rv.eps}
\caption{
    Phased high-precision RV measurements for \hbox{\hatcur{50}{}} (upper left), \hbox{\hatcur{51}{}} (upper right), \hbox{\hatcur{52}{}} (lower left) and \hbox{\hatcur{53}{}} (lower right). The instruments used are labelled in the plots. In each case we show three panels. The top panel shows the phased measurements together with our best-fit model (see \reftabl{planetparam}) for each system. Zero-phase corresponds to the time of mid-transit. The center-of-mass velocity has been subtracted. The second panel shows the velocity $O\!-\!C$ residuals from the best fit. The error bars include the jitter terms listed in \reftabl{planetparam} added in quadrature to the formal errors for each instrument. The third panel shows the bisector spans (BS). Note the different vertical scales of the panels.
}
\label{fig:rvbis}
\ifthenelse{\boolean{emulateapj}}{
    \end{figure*}
}{
    \end{figure}
}

The first step that was undertaken in confirming the planetary nature of the four planetary candidates was to obtain a spectral reconnaissance of their host stars. This allows us to rule out the usual false positive cases (giant stars, binary systems, and blending with faint eclipsing-binary systems). For this purpose, we used the Wide Field Spectrograph (WiFeS; \citealp{dopita:2007}), mounted on the ANU 2.3\,m telescope at SSO. The spectroscopic parameters were estimated by taking low-resolution spectra ($R=3000$). All the four targets were identified as dwarf stars. We also took medium-resolution spectra ($R=7000$), with the aim to search for possible RV variations at the $\sim 2$\,km\,s$^{-1}$ level, which are useful to rule out possible stellar companions. Details about the data reduction and the processing of the WiFeS spectra are summarized in \citet{bayliss:2013:hats3}.

Precise RV measurements of the targets were then acquired by using several medium- and large-class telescopes, equipped with high-resolution spectrographs and working on wide ranges of optical wavelengths. They are summarized in Table~\ref{tab:specobs}, together with their main characteristics. With these instruments, it was possible to measure periodic RV variation of the stars, which is compatible with the presence of planet-type objects orbiting around them. In particular, we mainly used the FEROS spectrograph \citep{kaufer:1998}, which is mounted on the MPG\,2.2\,m telescope at the ESO Observatory in La Silla, for monitoring the four targets. Other spectra were collected thanks to CORALIE \citep{queloz:2001} on the Euler 1.2\,m telescope, HARPS \citep{mayor:2003} on the ESO\,3.6\,m telescope, which are also located at the La Silla observatory, and CYCLOPS mounted on the 3.9\,m Anglo-Australian Telescope at SSO. For the case of \hatcur{50}, which is the faintest star of the four ($V=\hatcurCCtassmvshort{50}$\,mag), we needed higher signal-to-noise (S/N) measurements. These were obtained by taking seven spectra with the HIRES spectrograph \citep{vogt:1994} on the Keck-I\,10\,m telescope at Mauna Kea Observatory in Hawaii.

More details about the instruments, the data reduction and the computation of the corresponding RVs can be found in the previous works of the HATSouth team, e.g. \citet{penev:2013:hats1,mohlerfischer:2013:hats2,bayliss:2013:hats3}. In particular, HARPS, FEROS and CORALIE spectra were analysed with the method described in \citet{jordan:2014:hats4} and \citet{ceres:2017}, while those coming from CYCLOPS in \citet{addison:2013}. Finally, we refer the reader to \citet{bakos:2015:hats7} and \citet{howard:2010} for the analysis of the Keck/HIRES spectra. 

The values of the RV measurements are reported in Table~\ref{tab:rvs}, while the phased RVs and bisector spans (BS) are plotted for each system in \reffigl{rvbis}. We also used the FEROS high-resolution spectra for an accurate determination of the stellar spectroscopic parameters (effective temperature, metal abundance and projected rotational velocity) by applying the ZASPE (Zonal Atmospherical Stellar Parameter Estimator) routine \citep{brahm:2017:zaspe}. This analysis is discussed in Sect.~\ref{sec:stelparam}. 


\subsection{Photometric follow-up observations}
\label{sec:phot}

%
\setcounter{planetcounter}{1}
%
\begin{figure*}[!ht]
\plotone{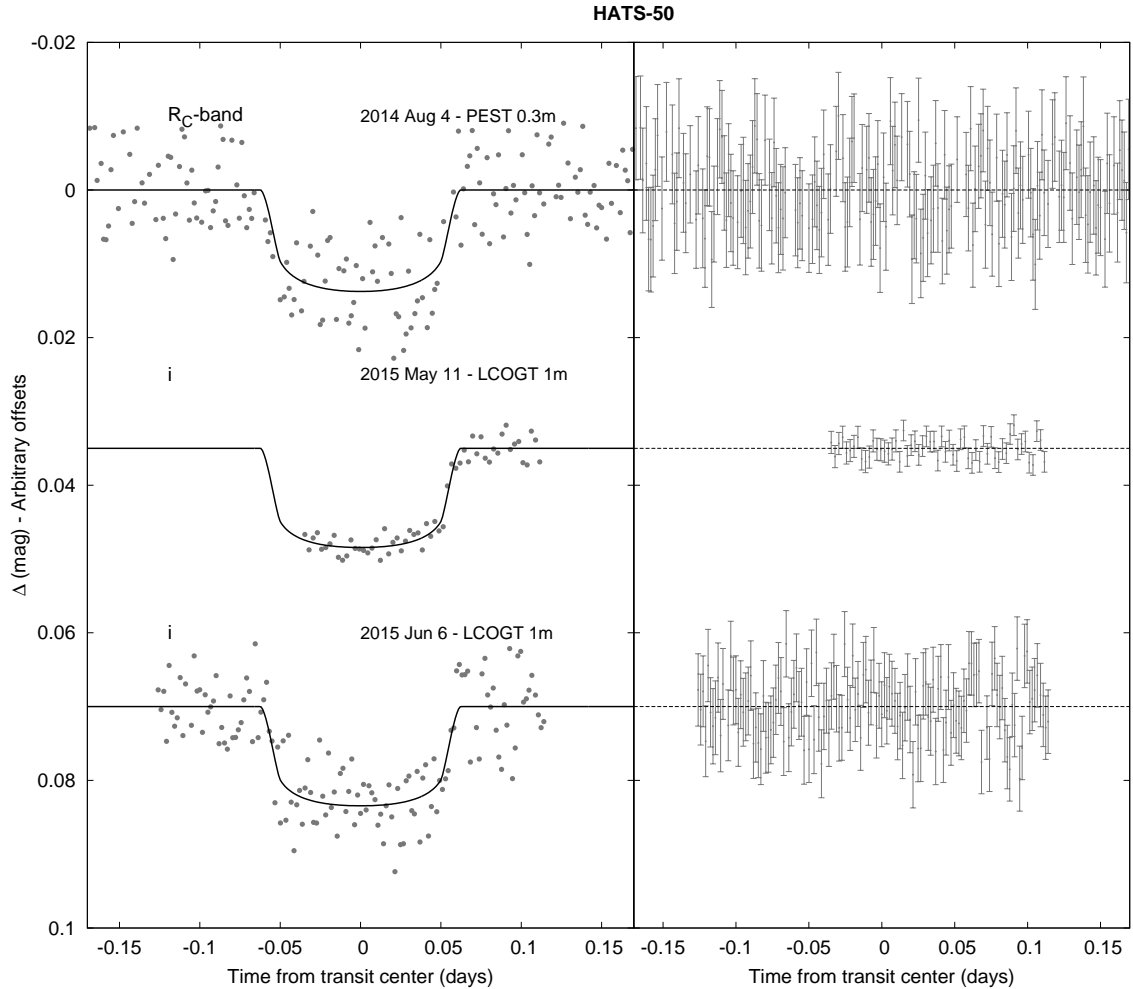}
\caption{
    Unbinned transit \lcs{} for \hatcur{50}.  The light curves have been
    corrected for quadratic trends in time, and linear trends with up
    to three parameters characterizing the shape of the PSF, fitted
    simultaneously with the transit model.
    The dates of the events, filters and instruments used are
    indicated.  Light curves following the first are displaced
    vertically for clarity.  Our best fit from the global modeling
    described in \refsecl{globmod} is shown by the solid lines. The
    residuals from the best-fit model are shown on the right-hand-side in the same
    order as the original light curves.  The error bars represent the
    photon and background shot noise, plus the readout noise.
}
\label{fig:lc:50}
\end{figure*}

\begin{figure*}[!ht]
\plotone{\hatcurhtr{51}-lc.eps}
\caption{
    Same as Fig.~\ref{fig:lc:50}, here we show \lcs{} for \hatcur{51}.
}
\label{fig:lc:51}
\end{figure*}

\begin{figure*}[!ht]
\plotone{\hatcurhtr{52}-lc.eps}
\caption{
    Same as Fig.~\ref{fig:lc:50}, here we show \lcs{} for \hatcur{52}.
}
\label{fig:lc:52}
\end{figure*}

\begin{figure*}[!ht]
\plotone{\hatcurhtr{53}-lc.eps}
\caption{
    Same as Fig.~\ref{fig:lc:50}, here we show \lcs{} for \hatcur{53}.
}
\label{fig:lc:53}
\end{figure*}

Another important step for confirming and characterizing a transiting exoplanetary system consists of performing photometric follow-up observations of transit events. We can thus derive more precise measurements, with respect to the survey data, of the transit depth, duration, mid-transit time and contact points of the corresponding light curves. An accurate knowledge of these photometric parameters are vital for robustly constraining the orbital parameters of the system and the physical parameters of both the star and the planet.

One complete transit event was observed with the PEST 0.3\,m telescope\footnote{http://pestobservatory.com} for \hatcurb{50}, \hatcurb{52}, and \hatcurb{53} through a $R$-band filter. Details of this telescope and the method used for reducing the data are described in \citet{zhou:2014:mebs}. Other eleven transit light curves of the four targets were recorded using the 1-m telescopes (CTIO, SAAO, SSO) in the Las Cumbres Observatory Global Telescope network (LCOGT: \citealp{brown:2013:lcogt}) and Sloan-$i^{\prime}$ filters. The LCOGT telescopes and the corresponding data reduction are described in \citet{hartman:2015:hats6}. An excerpt of these observations is reported in \reftabl{photobs}. The light curves are plotted in Figures~\ref{fig:lc:50}, \ref{fig:lc:51}, \ref{fig:lc:52}, and \ref{fig:lc:53}, for \hatcur{50}, \hatcur{51}, \hatcur{52}, and \hatcur{53}, respectively, and 
are compared to our best-fit models.

\ifthenelse{\boolean{emulateapj}}{
    \begin{deluxetable*}{llrrrrl}
}{
    \begin{deluxetable}{llrrrrl}
}
\tablewidth{0pc}
\tablecaption{
    Light curve data for \hatcur{50}, \hatcur{51}, \hatcur{52} and \hatcur{53}\label{tab:phfu}.
}
\tablehead{
    \colhead{Object\tablenotemark{a}} &
    \colhead{BJD\tablenotemark{b}} & 
    \colhead{Mag\tablenotemark{c}} & 
    \colhead{\ensuremath{\sigma_{\rm Mag}}} &
    \colhead{Mag(orig)\tablenotemark{d}} & 
    \colhead{Filter} &
    \colhead{Instrument} \\
    \colhead{} &
    \colhead{\hbox{~~~~(2,400,000$+$)~~~~}} & 
    \colhead{} & 
    \colhead{} &
    \colhead{} & 
    \colhead{} &
    \colhead{}
}
\startdata
   HATS-50 & $ 55451.44267 $ & $   0.00058 $ & $   0.00718 $ & $ \cdots $ & $ r$ &  HS/G580.4\\
   HATS-50 & $ 55765.47934 $ & $   0.00588 $ & $   0.00716 $ & $ \cdots $ & $ r$ &  HS/G580.4\\
   HATS-50 & $ 55788.45820 $ & $   0.00832 $ & $   0.00928 $ & $ \cdots $ & $ r$ &  HS/G580.4\\
   HATS-50 & $ 55516.54955 $ & $  -0.00368 $ & $   0.01095 $ & $ \cdots $ & $ r$ &  HS/G580.4\\
   HATS-50 & $ 55478.25269 $ & $  -0.01940 $ & $   0.00810 $ & $ \cdots $ & $ r$ &  HS/G580.4\\
   HATS-50 & $ 55451.44609 $ & $  -0.01299 $ & $   0.00728 $ & $ \cdots $ & $ r$ &  HS/G580.4\\
   HATS-50 & $ 55738.67412 $ & $  -0.00224 $ & $   0.00857 $ & $ \cdots $ & $ r$ &  HS/G580.4\\
   HATS-50 & $ 55470.59492 $ & $  -0.00388 $ & $   0.00781 $ & $ \cdots $ & $ r$ &  HS/G580.4\\
   HATS-50 & $ 55765.48280 $ & $  -0.00187 $ & $   0.00699 $ & $ \cdots $ & $ r$ &  HS/G580.4\\
   HATS-50 & $ 55516.55298 $ & $   0.01526 $ & $   0.01119 $ & $ \cdots $ & $ r$ &  HS/G580.4\\

\enddata
\tablenotetext{a}{
    Either \hatcur{50}, \hatcur{51}, \hatcur{52} or \hatcur{53}.
}
\tablenotetext{b}{
    Barycentric Julian Date is computed directly from the UTC time
    without correction for leap seconds.
}
\tablenotetext{c}{
    The out-of-transit level has been subtracted. For observations
    made with the HATSouth instruments (identified by ``HS'' in the
    ``Instrument'' column) these magnitudes have been corrected for
    trends using the EPD and TFA procedures applied {\em prior} to
    fitting the transit model. This procedure may lead to an
    artificial dilution in the transit depths. The blend factors for
    the HATSouth light curves are listed in
    Table~\ref{tab:planetparam}. For
    observations made with follow-up instruments (anything other than
    ``HS'' in the ``Instrument'' column), the magnitudes have been
    corrected for a quadratic trend in time, and for variations
    correlated with up to three PSF shape parameters, fit simultaneously
    with the transit.
}
\tablenotetext{d}{
    Raw magnitude values without correction for the quadratic trend in
    time, or for trends correlated with the seeing. These are only
    reported for the follow-up observations.
}
\tablecomments{
    This table is available in a machine-readable form in the online
    journal.  A portion is shown here for guidance regarding its form
    and content.
}
\ifthenelse{\boolean{emulateapj}}{
    \end{deluxetable*}
}{
    \end{deluxetable}
}

\subsection{Lucky Imaging}
\label{sec:luckyimaging}

\ifthenelse{\boolean{emulateapj}}{
    \begin{figure*}[!ht]
}{
    \begin{figure}[!ht]
}
\plottwo{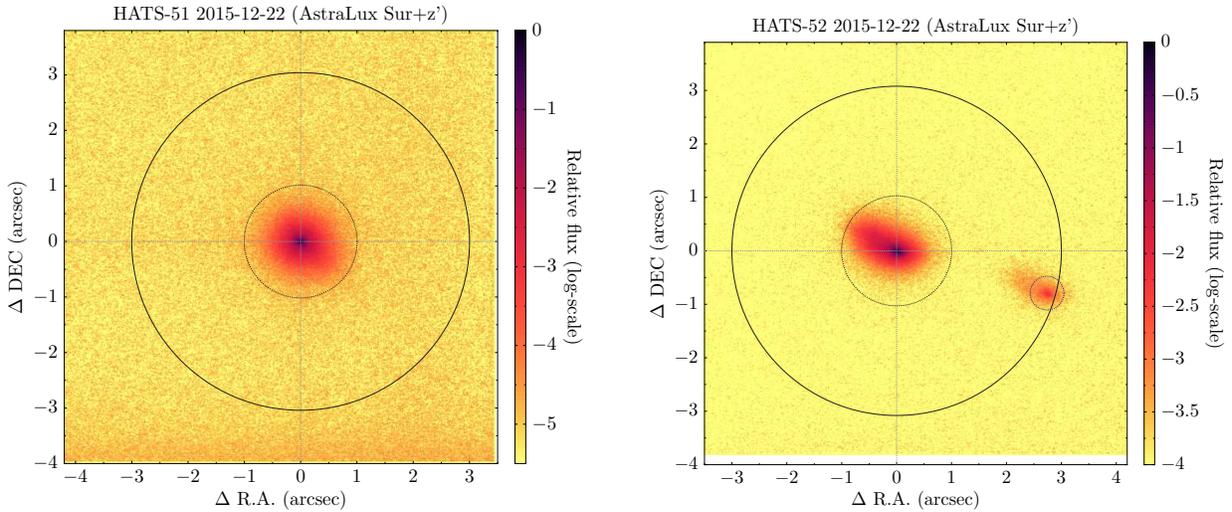}{\hatcur{52}_astralux.eps}
\caption{
    Astralux lucky images of \hatcur{51} ({\em left}) and \hatcur{52}
    ({\em right}). No neighboring source is detected for
    \hatcur{51}. For \hatcur{52} a neighbor is clearly detected at
    $\Delta {\rm R.A.} \approx 3^{\arcsec}$, $\Delta {\rm Dec.}
    \approx -1^{\arcsec}$ and with $\Delta z^{\prime} = 2.457 \pm
    0.013$\,mag.
\label{fig:luckyimages}}
\ifthenelse{\boolean{emulateapj}}{
    \end{figure*}
}{
    \end{figure}
}

\ifthenelse{\boolean{emulateapj}}{
    \begin{figure*}[!ht]
}{
    \begin{figure}[!ht]
}
\plottwo{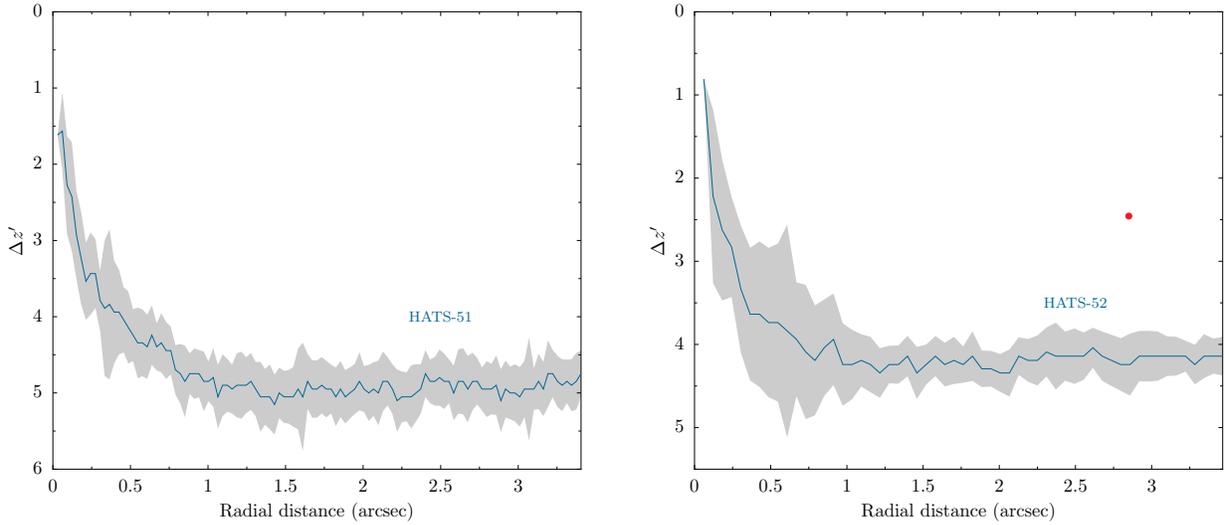}{contrast_curve_\hatcur{52}.eps}
\caption{
 Contrast curves for \hatcur{51} ({\em left}), and \hatcur{52} ({\em right}) based on our AstraLux Sur $z^{\prime}-band$ observations. Gray bands show the uncertainty given by the scatter in the contrast in the azimuthal direction at a given radius. The neighbor to \hatcur{52} is marked with the red filled circle.
\label{fig:luckyimagecontrastcurves}}
\ifthenelse{\boolean{emulateapj}}{
    \end{figure*}
}{
    \end{figure}
}

Lucky imaging observations were obtained through a $z^{\prime}$ filter for the \hatcur{51} and \hatcur{52} systems using the Astralux Sur camera \citep{hippler:2009} on the New Technology Telescope (NTT), at La Silla Observatory in Chile, on the nights of 2015 December 22 and 23. Observations with this facility were carried out and reduced following 
\citet{espinoza:2016:hats25hats30}, but a plate scale of 15.20\,mas\,pixel$^{-1}$ (derived in the work of \citealp{janson:2017}) was used. Figure~\ref{fig:luckyimages} shows the reduced final images for each system, while Figure~\ref{fig:luckyimagecontrastcurves} shows the contrast curves based on these images produced using the technique and software described in \citet{espinoza:2016:hats25hats30}. 

For \hatcur{52} a neighboring source is clearly detected at a distance of $2.74 \pm 0.03$\,arcsec to the east and $0.79 \pm 0.03$\,arcsec to the south from the target (i.e., at a distance of $2.85 \pm 0.03$\,arcsec from the target), with $\Delta z^{\prime}= 2.457 \pm 0.013$\,mag, relative to the target. Based on the photometric follow-up observations of this system that were carried out with the LCOGT~1\,m telescope network (Section~\ref{sec:phot}), we were able to determine that the transits occur around the star \hatcur{52}, and not the neighbor. The final combined image has an effective FWHM of $0\farcs0722 \pm 0\farcs0050$. The same source was detected by the GAIA space observatory (GAIA Data Release 1; \citealp{lindegren:2016}) at $\approx 2\farcs8$ separation to the East from \hatcur{52}, with $\Delta G_{\rm GAIA}=2.26$. Closer sources were not detected.

For \hatcur{51} no neighbors were detected with the Astralux Sur camera (effective FWHM of $0\farcs0431 \pm 0\farcs0053$), neither with GAIA within $10^{\prime \prime}$. Instead, based on GAIA data, we report that \hatcur{50} has a neighbor at $2\farcs1$ to the east ($\Delta\,G=2.97$), while \hatcur{53} has no neighbors within $8\farcs$.

\section{Analysis}
\label{sec:analysis}

Here we describe the analysis of the observational data, which were presented in the previous section, with the aim to get complete physical characterizations of the \hatcur{50}, \hatcur{51}, \hatcur{52} and \hatcur{53} planetary systems.

\subsection{Properties of the parent star}
\label{sec:stelparam}

As anticipated before, we used high-resolution FEROS spectra for determining the atmospheric properties (metallicity, effective temperature and surface gravity) of the four stars. The spectra were analysed with the ZASPE routine, which is comprehensively described in \citet{brahm:2017:zaspe}. 

Then, we followed the methodology of \citet{sozzetti:2007} for determining other stellar parameters (mass, radius, luminosity, age, etc.) together with their uncertainties. In brief, we performed a Markov chain Monte Carlo (MCMC) global analysis of our photometric and spectroscopic data, based on ($i$) the stellar effective temperature, $\teffstar$, and stellar metal abundance, [Fe/H], which were both determined with ZASPE, ($ii$) the stellar mean density, $\rhostar$, estimated by modelling the photometric transit light curves, and ($iii$) using the Yonsei-Yale \citep[YY;][]{yi:2001} evolutionary tracks.

We determined the YY isochrones for each of the four systems over a wide range of ages. The values of the stellar parameters were obtained from the best agreement between the resulting values of $\rhostar$ and $\teffstar$ and those estimated from the data. \reffigl{iso} shows the locations of each star on an $\teffstar$--$\rhostar$ diagram.
From this analysis, we kept the values of the stellar logarithmic surface gravities, $\loggstar$, and used them as fixed parameters for a second iteration with ZASPE, which returned the final values of the parameters of the four stars. They are reported in \reftabl{stellar} and all our objects are G-type stars. 

The most likely values for \teffstar\ and \rhostar\ for \hatcur{52} fall at a higher density than the lowest age isochrone tabulated in the models (see bottom-left panel in \reffigl{iso}). The models and observations are consistent within $2\sigma$. For determining the physical parameters of this star we exclude any \teffstar\--\rhostar\--\feh\ point in the Markov chain which does not match to a stellar model. In \reftabl{stellar} we list for each star the median stellar density based both on the full Mark chain (i.e., without enforcing a match to the stellar models) and on the chain after excluding points that do not match to a model.

We note that, while \hatcur{53} has physical characteristics similar to the Sun ($\teffstar=\hatcurSMEteff{53}$, $\feh=\hatcurSMEzfeh{53}$, $\mstar=\hatcurISOmlong{53}\,\msun$, $\rstar=\hatcurISOrlong{53}\,\rsun$), \hatcur{50}, \hatcur{51} and \hatcur{52} are more massive, larger, and metal richer ($\feh=\hatcurSMEzfeh{50}$, $\feh=\hatcurSMEzfeh{51}$ and $\feh=\hatcurSMEzfeh{52}$ for \hatcur{50}, \hatcur{51} and \hatcur{52}, respectively).  We note that \hatcur{51} is much less dense ($\rhostar=\hatcurISOrho{51}\,$g\,cm$^{-3}$) than the other three stars due to its large radius ($\rstar=\hatcurISOrlong{51}\,\rsun$). Moreover, our analysis indicates that \hatcur{50} and \hatcur{52} are quite young, i.e. \hatcurISOage{50}\,Gyr and \hatcurISOage{52}\,Gyr, respectively; these estimates are both consistent with their Zero Age Main Sequence implying a
$95\%$ confidence upper limit on the age of $t<3.9$\,Gyr and $t<3.8$\,Gyr, for \hatcur{50} and \hatcur{52} respectively. \hatcur{51} has an intermediate age (\hatcurISOage{51}\,Gyr), whereas \hatcur{53} is quite old (\hatcurISOage{53}\,Gyr).

We also estimated the distance of the four stars by comparing their broad-multi-band photometry taken from public astronomical archives (see \reftabl{stellar}) with the predicted magnitudes in each filter from the isochrones. The extinction was determined assuming a $R_{V} = 3.1$ law from \citet{cardelli:1989}. 
For a consistency check, we used the NED online extinction calculator, based on Galactic extinction maps, for estimating the expected total line of site extinction for each source. Three of them (\hatcur{51}, \hatcur{52} and \hatcur{53}) passed this check, as we found values greater than the inferred $A_{V}$. In the case of \hatcur{50}, our estimated $A_{V}$ is very close to the value determined from the dust maps.

\ifthenelse{\boolean{emulateapj}}{
    \begin{figure*}[!ht]
}{
    \begin{figure}[!ht]
}
\plottwo{\hatcurhtr{50}-iso-rho.eps}{\hatcurhtr{51}-iso-rho.eps}
\plottwo{\hatcurhtr{52}-iso-rho.eps}{\hatcurhtr{53}-iso-rho.eps}
\caption{
    Model isochrones from \cite{\hatcurisocite{50}} for the measured
    metallicities of \hatcur{50} (upper left), \hatcur{51} (upper right), \hatcur{52} (lower left) and \hatcur{53} (lower right). We show models for ages of 0.2\,Gyr and 1.0 to 14.0\,Gyr in 1.0\,Gyr increments (ages increasing from left to right). The
    adopted values of $\teffstar$ and \rhostar\ are shown together with
    their 1$\sigma$ and 2$\sigma$ confidence ellipsoids.  The initial
    values of \teffstar\ and \rhostar\ from the first ZASPE and \lc\
    analyses of \hatcur{50} and \hatcur{53} are represented with open triangles.
}
\label{fig:iso}
\ifthenelse{\boolean{emulateapj}}{
    \end{figure*}
}{
    \end{figure}
}

\ifthenelse{\boolean{emulateapj}}{
    \begin{deluxetable*}{lccccl}
}{
    \begin{deluxetable}{lccccl}
}
\tablewidth{0pc}
\tabletypesize{\scriptsize}
\tablecaption{
    Stellar parameters for \hatcur{50}--\hatcur{53}
    \label{tab:stellar}
}
\tablehead{
    \multicolumn{1}{c}{} &
    \multicolumn{1}{c}{\bf HATS-50} &
    \multicolumn{1}{c}{\bf HATS-51} &
    \multicolumn{1}{c}{\bf HATS-52} &
    \multicolumn{1}{c}{\bf HATS-53} &
    \multicolumn{1}{c}{} \\
    \multicolumn{1}{c}{~~~~~~~~Parameter~~~~~~~~} &
    \multicolumn{1}{c}{Value}                     &
    \multicolumn{1}{c}{Value}                     &
    \multicolumn{1}{c}{Value}                     &
    \multicolumn{1}{c}{Value}                     &
    \multicolumn{1}{c}{Source}
}
\startdata
\noalign{\vskip -3pt}
\sidehead{Astrometric properties and cross-identifications}
~~~~2MASS-ID\dotfill               & \hatcurCCtwomassshort{50}  & \hatcurCCtwomassshort{51} & \hatcurCCtwomassshort{52} & \hatcurCCtwomassshort{53} & \\
~~~~GSC-ID\dotfill                 & \hatcurCCgsc{50}      & \hatcurCCgsc{51}     & \hatcurCCgsc{52}     & \hatcurCCgsc{53}     & \\
~~~~R.A. (J2000)\dotfill            & \hatcurCCra{50}       & \hatcurCCra{51}    & \hatcurCCra{52}    & \hatcurCCra{53}    & 2MASS\\
~~~~Dec. (J2000)\dotfill            & \hatcurCCdec{50}      & \hatcurCCdec{51}   & \hatcurCCdec{52}   & \hatcurCCdec{53}   & 2MASS\\
~~~~$\mu_{\rm R.A.}$ (\masy)              & \hatcurCCpmra{50}     & \hatcurCCpmra{51} & \hatcurCCpmra{52} & \hatcurCCpmra{53} & UCAC4\\
~~~~$\mu_{\rm Dec.}$ (\masy)              & \hatcurCCpmdec{50}    & \hatcurCCpmdec{51} & \hatcurCCpmdec{52} & \hatcurCCpmdec{53} & UCAC4\\
\sidehead{Spectroscopic properties}
~~~~$\teffstar$ (K)\dotfill         &  \hatcurSMEteff{50}   & \hatcurSMEteff{51} & \hatcurSMEteff{52} & \hatcurSMEteff{53} & ZASPE\tablenotemark{a}\\
~~~~$\feh$\dotfill                  &  \hatcurSMEzfeh{50}   & \hatcurSMEzfeh{51} & \hatcurSMEzfeh{52} & \hatcurSMEzfeh{53} & ZASPE               \\
~~~~$\vsini$ (\kms)\dotfill         &  \hatcurSMEvsin{50}   & \hatcurSMEvsin{51} & \hatcurSMEvsin{52} & \hatcurSMEvsin{53} & ZASPE                \\
~~~~$\vmac$ (\kms)\dotfill          &  $\hatcurSMEvmac{50}$   & $\hatcurSMEvmac{51}$ & $\hatcurSMEvmac{52}$ & $\hatcurSMEvmac{53}$ & Assumed              \\
~~~~$\vmic$ (\kms)\dotfill          &  $\hatcurSMEvmic{50}$   & $\hatcurSMEvmic{51}$ & $\hatcurSMEvmic{52}$ & $\hatcurSMEvmic{53}$ & Assumed              \\
~~~~$\gamma_{\rm RV}$ (\ms)\dotfill&  \hatcurRVgammaabs{50}  & \hatcurRVgammaabs{51} & \hatcurRVgammaabs{52} & \hatcurRVgammaabs{53} & FEROS\tablenotemark{b}  \\
\sidehead{Photometric properties}
~~~~$G$ (mag)\dotfill               &  13.8  & 12.24 & 13.54 & 13.57 & GAIA DR1\tablenotemark{c} \\
~~~~$B$ (mag)\dotfill               &  \hatcurCCtassmB{50}  & \hatcurCCtassmB{51} & \hatcurCCtassmB{52} & \hatcurCCtassmB{53} & APASS\tablenotemark{d} \\
~~~~$V$ (mag)\dotfill               &  \hatcurCCtassmv{50}  & \hatcurCCtassmv{51} & \hatcurCCtassmv{52} & \hatcurCCtassmv{53} & APASS\tablenotemark{d} \\
~~~~$g$ (mag)\dotfill               &  $\cdots$  & \hatcurCCtassmg{51} & \hatcurCCtassmg{52} & \hatcurCCtassmg{53} & APASS\tablenotemark{d} \\
~~~~$r$ (mag)\dotfill               &  $\cdots$  & \hatcurCCtassmr{51} & \hatcurCCtassmr{52} & \hatcurCCtassmr{53} & APASS\tablenotemark{d} \\
~~~~$i$ (mag)\dotfill               &  \hatcurCCtassmi{50}  & \hatcurCCtassmi{51} & \hatcurCCtassmi{52} & \hatcurCCtassmi{53} & APASS\tablenotemark{d} \\
~~~~$J$ (mag)\dotfill               &  \hatcurCCtwomassJmag{50} & \hatcurCCtwomassJmag{51} & \hatcurCCtwomassJmag{52} & \hatcurCCtwomassJmag{53} & 2MASS           \\
~~~~$H$ (mag)\dotfill               &  \hatcurCCtwomassHmag{50} & \hatcurCCtwomassHmag{51} & \hatcurCCtwomassHmag{52} & \hatcurCCtwomassHmag{53} & 2MASS           \\
~~~~$K_s$ (mag)\dotfill             &  \hatcurCCtwomassKmag{50} & \hatcurCCtwomassKmag{51} & \hatcurCCtwomassKmag{52} & \hatcurCCtwomassKmag{53} & 2MASS           \\
\sidehead{Derived properties}
~~~~$\mstar$ ($\msun$)\dotfill      &  \hatcurISOmlong{50}   & \hatcurISOmlong{51} & \hatcurISOmlong{52} & \hatcurISOmlong{53} & YY+$\rhostar$+ZASPE \tablenotemark{e}\\
~~~~$\rstar$ ($\rsun$)\dotfill      &  \hatcurISOrlong{50}   & \hatcurISOrlong{51} & \hatcurISOrlong{52} & \hatcurISOrlong{53} & YY+$\rhostar$+ZASPE         \\
~~~~$\loggstar$ (cgs)\dotfill       &  \hatcurISOlogg{50}    & \hatcurISOlogg{51} & \hatcurISOlogg{52} & \hatcurISOlogg{53} & YY+$\rhostar$+ZASPE         \\
~~~~$\rhostar$ (\gcmc) \tablenotemark{f}\dotfill       &  \hatcurLCrho{50}    & \hatcurLCrho{51} & \hatcurLCrho{52} & \hatcurLCrho{53} & Light curves         \\
~~~~$\rhostar$ (\gcmc) \tablenotemark{f}\dotfill       &  \hatcurISOrho{50}    & \hatcurISOrho{51} & \hatcurISOrho{52} & \hatcurISOrho{53} & YY+Light curves+ZASPE         \\
~~~~$\lstar$ ($\lsun$)\dotfill      &  \hatcurISOlum{50}     & \hatcurISOlum{51} & \hatcurISOlum{52} & \hatcurISOlum{53} & YY+$\rhostar$+ZASPE         \\
~~~~$M_V$ (mag)\dotfill             &  \hatcurISOmv{50}      & \hatcurISOmv{51} & \hatcurISOmv{52} & \hatcurISOmv{53} & YY+$\rhostar$+ZASPE         \\
~~~~$M_K$ (mag,\hatcurjhkfilset{50})\dotfill &  \hatcurISOMK{50} & \hatcurISOMK{51} & \hatcurISOMK{52} & \hatcurISOMK{53} & YY+$\rhostar$+ZASPE         \\
~~~~Age (Gyr)\dotfill               &  \hatcurISOage{50}     & \hatcurISOage{51} & \hatcurISOage{52} & \hatcurISOage{53} & YY+$\rhostar$+ZASPE         \\
~~~~$A_{V}$ (mag)\dotfill               &  \hatcurXAv{50}     & \hatcurXAv{51} & \hatcurXAv{52} & \hatcurXAv{53} & YY+$\rhostar$+ZASPE         \\
~~~~Distance (pc)\dotfill           &  \hatcurXdistred{50}\phn  & \hatcurXdistred{51} & \hatcurXdistred{52} & \hatcurXdistred{53} & YY+$\rhostar$+ZASPE\\
\enddata
\tablecomments{
For all four systems we adopt a model in which the orbit is assumed to be circular. See the discussion in Section~\ref{sec:globmod}.
}
\tablenotetext{a}{
    ZASPE = Zonal Atmospherical Stellar Parameter Estimator routine
    for the analysis of high-resolution spectra
    \citep{brahm:2017:zaspe}, applied to the FEROS spectra of each system. These parameters rely primarily on ZASPE, but have a small
    dependence also on the iterative analysis incorporating the
    isochrone search and global modeling of the data.
}
\tablenotetext{b}{
    The error on $\gamma_{\rm RV}$ is determined from the orbital fit
    to the RV measurements, and does not include the systematic
    uncertainty in transforming the velocities to the IAU standard
    system. The velocities have not been corrected for gravitational
    redshifts.
} \tablenotetext{c}{
    From GAIA Data Release 1 \citep{lindegren:2016}. \hatcur{50} has a neighbour at $2.1^{\prime \prime}$ to the east ($\Delta\,G=2.97$); \hatcur{51} has no neighbour within $10^{\prime \prime}$; \hatcur{52} has a neighbour at $2.8^{\prime \prime}$ to east ($\Delta\,G=2.26$). \hatcur{53} has no neighbour within $8^{\prime \prime}$.
}
\tablenotetext{d}{
From APASS DR6 for as listed in the UCAC 4 catalog \citep{zacharias:2012:ucac4}.      
}
\tablenotetext{e}{
    \hatcurisoshort{50}+\rhostar+ZASPE = Based on the \hatcurisoshort{50}
    isochrones \citep{\hatcurisocite{50}}, \rhostar\ as a luminosity
    indicator, and the ZASPE results.
}
\tablenotetext{f}{
    In the case of $\rhostar$ we list two values. The first value is
    determined from the global fit to the light curves and RV data,
    without imposing a constraint that the parameters match the
    stellar evolution models. The second value results from
    restricting the posterior distribution to combinations of
    $\rhostar$+$\teffstar$+$\feh$ that match to a \hatcurisoshort{50}
    stellar model.
}
\ifthenelse{\boolean{emulateapj}}{
    \end{deluxetable*}
}{
    \end{deluxetable}
}

\subsection{Excluding blend scenarios}
\label{sec:blend}

In order to exclude blend scenarios we carried out an analysis following \citet{hartman:2012:hat39hat41}. We attempt to model the available photometric data (including light curves and catalog broad-band photometric measurements) for each object as a blend between an eclipsing binary star system and a third star along the line of sight. The physical properties of the stars are constrained using the Padova isochrones \citep{girardi:2000}, while we also require that the brightest of the three stars in the blend has atmospheric parameters consistent with those measured with ZASPE. We also simulate composite cross-correlation functions (CCFs) and use them to predict RVs and BSs for each blend scenario considered. The results for each system are as follows:
\begin{itemize}
\item {\em \hatcur{50}} -- all blend models tested for this system can be rejected in favor of a model of a single star with a planet with greater than $3\sigma$ confidence based solely on the photometry. Moreover the blend models that come closest to fitting the photometric data (those rejected with less than $5\sigma$ confidence) would yield large bisector span variations in excess of 1\,\kms, whereas the measured scatter in the BS values is only 62\,\ms\ based on FEROS. Based on this we conclude that \hatcur{50} is not a blended stellar eclipsing binary system.
\item {\em \hatcur{51}} -- we find that the best-fit blend models are indistinguishable from the best-fit planet model based on the photometry. However, all blend models tested which fit the photometry (i.e., those which cannot be rejected in favor of the best-fit single-star plus planet model with at least $5\sigma$ confidence) would have been easily identified as composite systems based on the spectroscopy, with BS and/or RV variations in excess of 1\,\kms. For comparison the measured FEROS BSs have a scatter of 46\,\ms. Based on this we rule out stellar eclipsing binary blend scenarios.
\item {\em \hatcur{52}} -- similar to \hatcur{50}, all blend models tested for this system can be rejected in favor of a model of a single star with a planet with greater than $3\sigma$ confidence based solely on the photometry, while blend models that come closest to fitting the photometric data (those rejected with less than $5\sigma$ confidence) would yield large bisector span variations in excess of 1\,\kms. In this case measured scatter in the BS values is 96\,\ms\ based on FEROS. Based on this we conclude that \hatcur{52} is not a blended stellar eclipsing binary system.
\item {\em \hatcur{53}} -- similar to \hatcur{50}, all blend models tested for this system can be rejected in favor of a model of a single star with a planet with greater than $4\sigma$ confidence based solely on the photometry, while blend models that come closest to fitting the photometric data (those rejected with less than $5\sigma$ confidence) would yield large bisector span variations in excess of 1\,\kms. In this case measured scatter in the BS values is 47\,\ms\ based on FEROS. Based on this we conclude that \hatcur{53} is not a blended stellar eclipsing binary system.
\end{itemize}

\subsection{Global modeling of the data}
\label{sec:globmod}
The physical parameters of the four planetary systems were estimated by modelling the HATSouth photometry, the follow-up photometry, and the high-precision RV measurements. For this task, we followed the robust procedures developed by the HAT team, which are exhaustively described in several of their exoplanet-discovery papers (e.g. \citealp{pal:2008:hat7,bakos:2010:hat11,hartman:2012:hat39hat41,hartman:2015:hats6}). Here we give a brief summary.

The transit light curves taken by the HATSouth telescopes (Figure~\ref{fig:hatsouth}) were fitted by using \citet{mandel:2002} models and considering possible dilution of the transit depth; this was done for taking care possible ($i$) blending from neighboring stars ($ii$) or over-correction made when the light curves were detrended during the reduction phase. 

Concerning the photometric follow-up observations (Figures~\ref{fig:lc:50}, \ref{fig:lc:51}, \ref{fig:lc:52}, and \ref{fig:lc:53}), the systematic noise of each data set was corrected during the modeling of the corresponding light curve, by including a quadratic trend in time. We also included linear trends with three parameters describing the measured shape of the PSF. These were included to account for possible variations in the photometry resulting from PSF-shape changes that can happen during the transit observation due to poor guiding or non-photometric conditions.

The RV curves, which we presented in Sect.~\ref{sec:obsspec}, are composed of points that were measured with different spectrographs, which can present different zero-points and can be affected by RV jitter as well. Therefore, we modelled the RV curves (Figure~\ref{fig:rvbis}) with Keplerian orbits considering the zero-point and the RV jitter for each instrument as free parameters. 

Finally, the values of the physical parameters of the exoplanetary systems were obtained by exploring their parameter spaces by means of a Differential Evolution Markov Chain Monte Carlo procedure. This allowed to identify the most likely
values for the parameters together with their $1\sigma$ confidence interval.

We also investigated the possibility that the orbits of the four planets are eccentric. This was done by performing the joint fit of each of the four data sets with both fixed circular orbits and free-eccentricity models. Then we estimated the Bayesian evidence for each scenario following the method of Weinberg et al. (2013). This method involves using the Markov Chains produced in modelling the observations to identify a region of high posterior probability which dominates the Bayesian evidence, and then carrying out a Monte Carlo integration over this small domain to estimate the evidence. We find that in all cases a model with a fixed circular orbit has a higher Bayesian evidence than a model where the eccentricity is allowed to vary, and we adopt the fixed circular orbit model for each system.

The resulting parameters for each system are reported in \reftabl{planetparam} and indicate that three of the planets are puffy, low-density, hot giants (\hatcurb{50}, \hatcurb{51} and \hatcurb{53}) with $3$\,d$<P_{\rm orb}<4$\,d, while the fourth (\hatcurb{52}) is a high-density, massive ($M_{\rm p}\approx 2.2\,\mjup$), close-in ($P_{\rm orb}=1.37$\,d; $a\approx0.025$\,au; $T_{\rm eq}\approx 1830$\,K) hot Jupiter. The fours planets have a radius larger than Jupiter, and the least massive of the four is \hatcurb{50} with a mass of $\approx 0.4\,\mjup$. 

\startlongtable
%
\ifthenelse{\boolean{emulateapj}}{
    \begin{deluxetable*}{lcccc}
}{
    \begin{deluxetable}{lcccc}
}
\tabletypesize{\scriptsize}
\tablecaption{Orbital and planetary parameters for \hatcurb{50}--\hatcurb{53}\label{tab:planetparam}}
\tablehead{
    \multicolumn{1}{c}{} &
    \multicolumn{1}{c}{\bf HATS-50b} &
    \multicolumn{1}{c}{\bf HATS-51b} &
    \multicolumn{1}{c}{\bf HATS-52b} &
    \multicolumn{1}{c}{\bf HATS-53b} \\ 
    \multicolumn{1}{c}{~~~~~~~~~~~~~~~Parameter~~~~~~~~~~~~~~~} &
    \multicolumn{1}{c}{Value} &
    \multicolumn{1}{c}{Value} &
    \multicolumn{1}{c}{Value} &
    \multicolumn{1}{c}{Value}
}
\startdata
\noalign{\vskip -3pt}
\sidehead{\Lc{} parameters}
~~~$P$ (days)             \dotfill    & $\hatcurLCP{50}$ & $\hatcurLCP{51}$ & $\hatcurLCP{52}$ & $\hatcurLCP{53}$ \\
~~~$T_c$ (${\rm BJD}$)    
      \tablenotemark{a}   \dotfill    & $\hatcurLCT{50}$ & $\hatcurLCT{51}$ & $\hatcurLCT{52}$ & $\hatcurLCT{53}$ \\
~~~$T_{14}$ (days)
      \tablenotemark{a}   \dotfill    & $\hatcurLCdur{50}$ & $\hatcurLCdur{51}$ & $\hatcurLCdur{52}$ & $\hatcurLCdur{53}$ \\
~~~$T_{12} = T_{34}$ (days)
      \tablenotemark{a}   \dotfill    & $\hatcurLCingdur{50}$ & $\hatcurLCingdur{51}$ & $\hatcurLCingdur{52}$ & $\hatcurLCingdur{53}$ \\
~~~$\arstar$              \dotfill    & $\hatcurPPar{50}$ & $\hatcurPPar{51}$ & $\hatcurPPar{52}$ & $\hatcurPPar{53}$ \\
~~~$\zrstar$ \tablenotemark{b}             \dotfill    & $\hatcurLCzeta{50}$\phn & $\hatcurLCzeta{51}$\phn & $\hatcurLCzeta{52}$\phn & $\hatcurLCzeta{53}$\phn \\
~~~$\rpl/\rstar$          \dotfill    & $\hatcurLCrprstar{50}$ & $\hatcurLCrprstar{51}$ & $\hatcurLCrprstar{52}$ & $\hatcurLCrprstar{53}$ \\
~~~$b^2$                  \dotfill    & $\hatcurLCbsq{50}$ & $\hatcurLCbsq{51}$ & $\hatcurLCbsq{52}$ & $\hatcurLCbsq{53}$ \\
~~~$b \equiv a \cos i/\rstar$
                          \dotfill    & $\hatcurLCimp{50}$ & $\hatcurLCimp{51}$ & $\hatcurLCimp{52}$ & $\hatcurLCimp{53}$ \\
~~~$i$ (deg)              \dotfill    & $\hatcurPPi{50}$\phn & $\hatcurPPi{51}$\phn & $\hatcurPPi{52}$\phn & $\hatcurPPi{53}$\phn \\
\sidehead{HATSouth dilution factors \tablenotemark{c}}
~~~Dilution factor 1 \dotfill & \hatcurLCiblendA{50} & \hatcurLCiblend{51} & $\hatcurLCiblend{52}$ & \hatcurLCiblend{53} \\
~~~Dilution factor 2 \dotfill & \hatcurLCiblendB{50} & $\cdots$ & $\cdots$ & $\cdots$ \\
\sidehead{Limb-darkening coefficients \tablenotemark{d}}
~~~$c_1,r$                  \dotfill    & $\hatcurLBir{50}$ & $\hatcurLBir{51}$ & $\hatcurLBir{52}$ & $\hatcurLBir{53}$ \\
~~~$c_2,r$                  \dotfill    & $\hatcurLBiir{50}$ & $\hatcurLBiir{51}$ & $\hatcurLBiir{52}$ & $\hatcurLBiir{53}$ \\
~~~$\hatcurLBiR{53}$                  \dotfill    & $\hatcurLBiR{50}$ & $\cdots$ & $\cdots$ & $\hatcurLBiiR{53}$ \\
~~~$c_2,R$                  \dotfill    & $\hatcurLBiiR{50}$ & $\cdots$ & $\cdots$ & $\cdots$ \\
~~~$c_1,i$                  \dotfill    & $\hatcurLBii{50}$ & $\hatcurLBii{51}$ & $\hatcurLBii{52}$ & $\hatcurLBii{53}$ \\
~~~$c_2,i$                  \dotfill    & $\hatcurLBiii{50}$ & $\hatcurLBiii{51}$ & $\hatcurLBiii{52}$ & $\hatcurLBiii{53}$ \\
\sidehead{RV parameters}
~~~$K$ (\ms)              \dotfill    & $\hatcurRVK{50}$\phn\phn & $\hatcurRVK{51}$\phn\phn & $\hatcurRVK{52}$\phn\phn & $\hatcurRVK{53}$\phn\phn \\
~~~$e$ \tablenotemark{e}               \dotfill    & $\hatcurRVeccentwosiglimeccen{50}$ & $\hatcurRVeccentwosiglimeccen{51}$ & $\hatcurRVeccentwosiglimeccen{52}$ & $\hatcurRVeccentwosiglimeccen{53}$ \\
~~~RV jitter FEROS (\ms) \tablenotemark{f}       \dotfill    & $\hatcurRVjitterA{50}$ & $\hatcurRVjitterA{51}$ & $\hatcurRVjitterA{52}$ & \hatcurRVjitterA{53} \\
~~~RV jitter HARPS (\ms)        \dotfill    & $\cdots$ & $\cdots$ & \hatcurRVjittertwosiglimC{52} & \hatcurRVjittertwosiglimB{53} \\
~~~RV jitter CYCLOPS (\ms)        \dotfill    & $\cdots$ & $\hatcurRVjitterC{51}$ & $\cdots$ & $\cdots$ \\
~~~RV jitter CORALIE (\ms)        \dotfill    & $\hatcurRVjittertwosiglimB{50}$ & $\hatcurRVjitterB{51}$ & $\cdots$ & $\cdots$ \\
~~~RV jitter HIRES (\ms)        \dotfill    & $\hatcurRVjitterC{50}$ & $\cdots$ & $\cdots$ & $\cdots$ \\
\sidehead{Planetary parameters}
~~~$\mpl$ ($\mjup$)       \dotfill    & $\hatcurPPmlong{50}$ & $\hatcurPPmlong{51}$ & $\hatcurPPmlong{52}$ & $\hatcurPPmlong{53}$ \\
~~~$\rpl$ ($\rjup$)       \dotfill    & $\hatcurPPrlong{50}$ & $\hatcurPPrlong{51}$ & $\hatcurPPrlong{52}$ & $\hatcurPPrlong{53}$ \\
~~~$C(\mpl,\rpl)$
    \tablenotemark{g}     \dotfill    & $\hatcurPPmrcorr{50}$ & $\hatcurPPmrcorr{51}$ & $\hatcurPPmrcorr{52}$ & $\hatcurPPmrcorr{53}$ \\
~~~$\rhopl$ (\gcmc)       \dotfill    & $\hatcurPPrho{50}$ & $\hatcurPPrho{51}$ & $\hatcurPPrho{52}$ & $\hatcurPPrho{53}$ \\
~~~$\log g_p$ (cgs)       \dotfill    & $\hatcurPPlogg{50}$ & $\hatcurPPlogg{51}$ & $\hatcurPPlogg{52}$ & $\hatcurPPlogg{53}$ \\
~~~$a$ (AU)               \dotfill    & $\hatcurPParel{50}$ & $\hatcurPParel{51}$ & $\hatcurPParel{52}$ & $\hatcurPParel{53}$ \\
~~~$T_{\rm eq}$ (K)        \dotfill   & $\hatcurPPteff{50}$ & $\hatcurPPteff{51}$ & $\hatcurPPteff{52}$ & $\hatcurPPteff{53}$ \\
~~~$\Theta$ \tablenotemark{h} \dotfill & $\hatcurPPtheta{50}$ & $\hatcurPPtheta{51}$ & $\hatcurPPtheta{52}$ & $\hatcurPPtheta{53}$ \\
~~~$\log_{10}\langle F \rangle$ (cgs) \tablenotemark{i}
                          \dotfill    & $\hatcurPPfluxavglog{50}$ & $\hatcurPPfluxavglog{51}$ & $\hatcurPPfluxavglog{52}$ & $\hatcurPPfluxavglog{53}$ \\
\enddata
\tablecomments{
For all four systems we adopt a model in which the orbit is assumed to be circular. See the discussion in Section~\ref{sec:globmod}.
}
\tablenotetext{a}{
    Times are in Barycentric Julian Date calculated directly from UTC {\em without} correction for leap seconds.
    \ensuremath{T_c}: Reference epoch of
    mid transit that minimizes the correlation with the orbital
    period.
    \ensuremath{T_{12}}: total transit duration, time
    between first to last contact;
    \ensuremath{T_{12}=T_{34}}: ingress/egress time, time between first
    and second, or third and fourth contact.
}
\tablenotetext{b}{
   Reciprocal of the half duration of the transit used as a jump parameter in our MCMC analysis in place of $\arstar$. It is related to $\arstar$ by the expression $\zrstar = \arstar(2\pi(1+e\sin\omega))/(P\sqrt{1-b^2}\sqrt{1-e^2})$ \citep{bakos:2010:hat11}.
}
\tablenotetext{c}{
    Scaling factor applied to the model transit that is fit to the HATSouth light curves. This factor accounts for dilution of the transit due to blending from neighboring stars and over-filtering of the light curve.  These factors are varied in the fit, with independent values adopted for each HATSouth light curve. The factors listed for \hatcur{50} are for the G580.4 and G625.3 light curves, respectively. For \hatcur{51}, we list the factor for 601.2. For \hatcur{52} the listed factor is for G606.1. For \hatcur{53}, the listed factor is for G610.4.
}
\tablenotetext{d}{
    Values for a quadratic law, adopted from the tabulations by
    \cite{claret:2004} according to the spectroscopic (ZASPE) parameters
    listed in \reftabl{stellar}.
}
\tablenotetext{e}{
    The 95\% confidence upper limit on the eccentricity determined
    when $\sqrt{e}\cos\omega$ and $\sqrt{e}\sin\omega$ are allowed to
    vary in the fit.
}
\tablenotetext{f}{
    Term added in quadrature to the formal RV uncertainties for each
    instrument. This is treated as a free parameter in the fitting
    routine. In cases where the jitter is consistent with zero, we
    list its 95\% confidence upper limit.
}
\tablenotetext{g}{
    Correlation coefficient between the planetary mass \mpl\ and radius
    \rpl\ estimated from the posterior parameter distribution.
}
\tablenotetext{h}{
    The Safronov number is given by $\Theta = \frac{1}{2}(V_{\rm
    esc}/V_{\rm orb})^2 = (a/\rpl)(\mpl / \mstar )$
    \citep[see][]{hansen:2007}.
}
\tablenotetext{i}{
    Incoming flux per unit surface area, averaged over the orbit.
}
\ifthenelse{\boolean{emulateapj}}{
    \end{deluxetable*}
}{
    \end{deluxetable}
}

\subsection{Mass upper limit for HATS-50c}
\label{sec:mass_HATS-50c}
\begin{figure}
\plotone{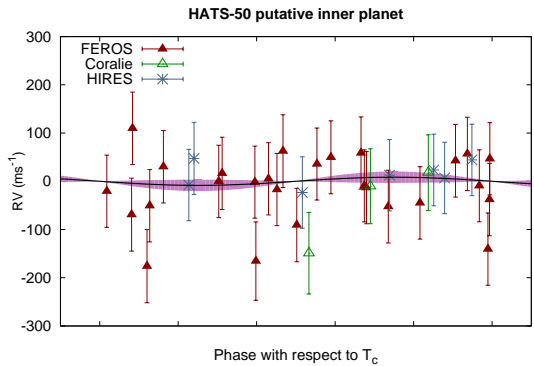}
\caption{Phased high-precision RV measurements for HATS-50, after subtracting the orbital variation due to the confirmed transiting hot Jupiter, \hatcurb{50}, and phase-folded at the period (0.77\,days) of the candidate inner transiting planet, \hatcur{50}c. The line shows the best-fit circular orbit at this period, while the shaded region shows the 1-sigma uncertainty bounds on this model.
The instruments used are labelled in the plots.
}
\label{fig:rvresid-secondplanet}
\end{figure}

In sect.~\ref{sec:periodic_signals} we have discussed the possibility that \hatcur{50} may host another planet (\hatcur{50}c) with a shorter orbital period (0.77\,days). This putative planet could be the cause of the substantial residuals of the RV measurements of \hatcur{50}, which are showed in Figure~\ref{fig:rvbis} (top-left panel) after removing the model for planet b.
We have therefore deeply investigated this possibility. Figure~\ref{fig:rvresid-secondplanet} shows the RVs for HATS-50, after subtracting the orbital variation due to the confirmed transiting hot Jupiter, and phase-folded at the period of the candidate inner transiting planet. The line shows the best-fit circular orbit at this period, while the shaded region shows the 1-$\sigma$ uncertainty bounds on this model.  We find the RVs are consistent with no variation at this period, with a best-fit RV semi-amplitude of $K = 8.4 \pm 11.8$\,\ms. The $95\%$ confidence upper limit on the mass of the candidate inner transiting planet is thus $M_{\rm pl,c} < 0.16$\,\mjup.

\section{Summary and discussion}
\label{sec:discussion}

\begin{figure*}
\plotone{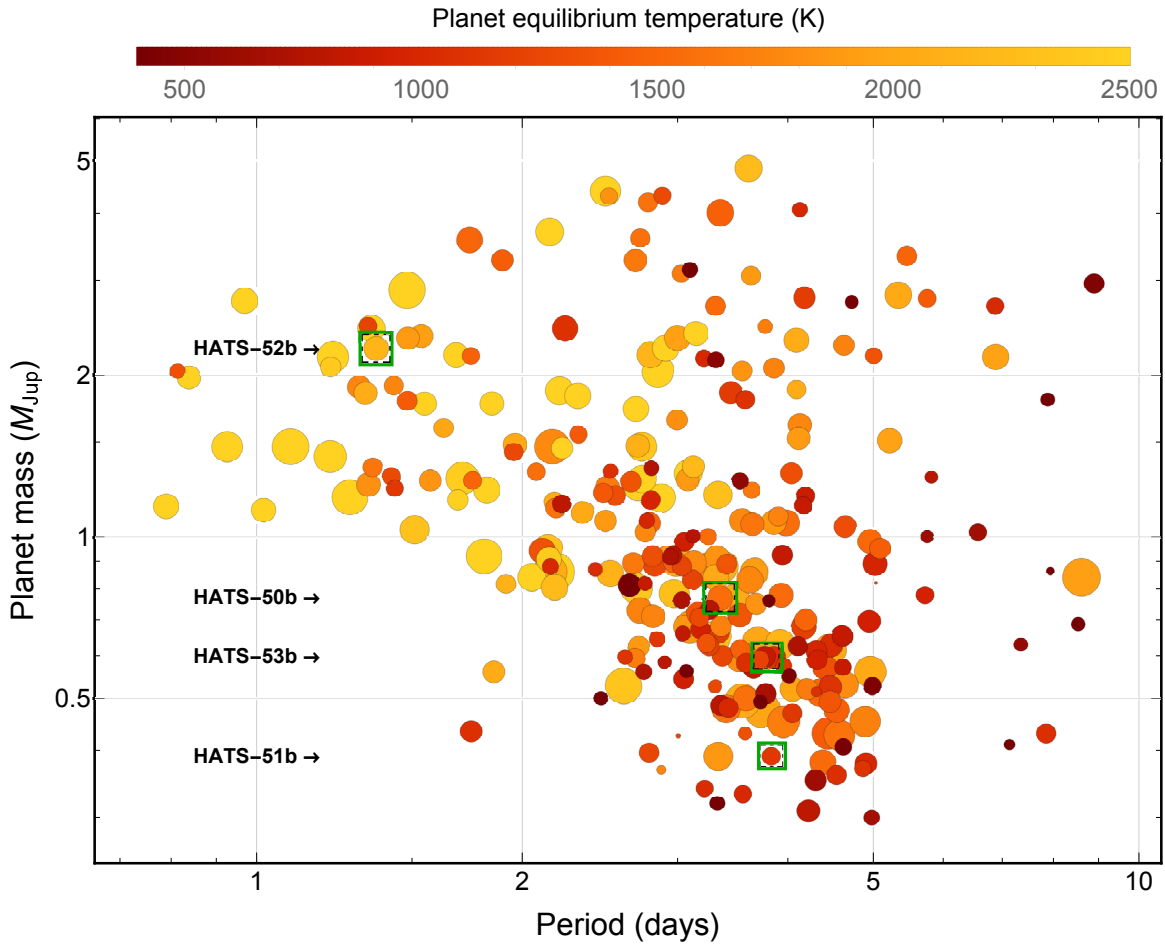}
\caption{
Mass-period diagram of all known transiting hot Jupiters, i.e. transiting exoplanets in the mass range $0.3\,M_{\rm J} < M_{\rm p} < 5\,M_{\rm J}$ and with an orbital period less than 10 days. The planets are represented by circles, whose size is proportional to their radius. Color indicates equilibrium temperature. The positions of \hatcurb{50}, \hatcurb{51}, \hatcurb{52}, \hatcurb{53} are highlighted with green boxes. The error bars have been suppressed for clarity. Data taken from the Transiting Extrasolar Planet Catalogue (TEPCat).
}
\label{fig:diagram}
\end{figure*}
\begin{figure}
\plotone{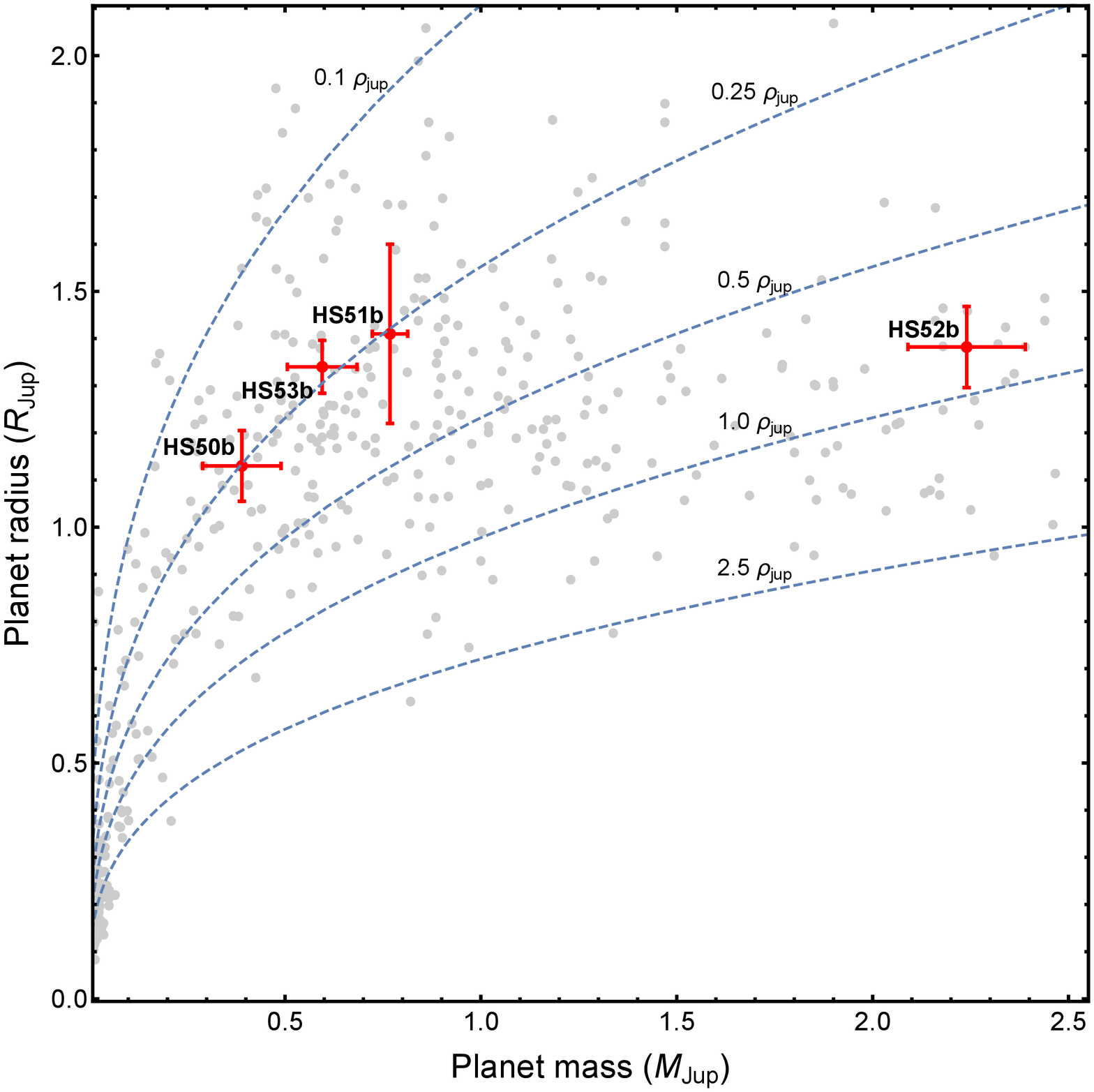}
\caption{
The masses and radii of the known transiting extrasolar planets. The plot is restricted to exoplanets with values of the mass until $2.5\,M_{\rm J}$ and radius until $2.0\,R_{\rm J}$. Grey points denote values taken from TEPCat. Their error bars
have been suppressed for clarity. The new HATS exoplanets, \hatcurb{50}, \hatcurb{51}, \hatcurb{52} and \hatcurb{53}, are shown in red points with error bars. Dotted lines show where density is 2.5, 1.0, 0.5, 0.25 and 0.1 $\rho_{\rm J}$.
}
\label{fig:diagram2}
\end{figure}
\begin{figure*}
\plotone{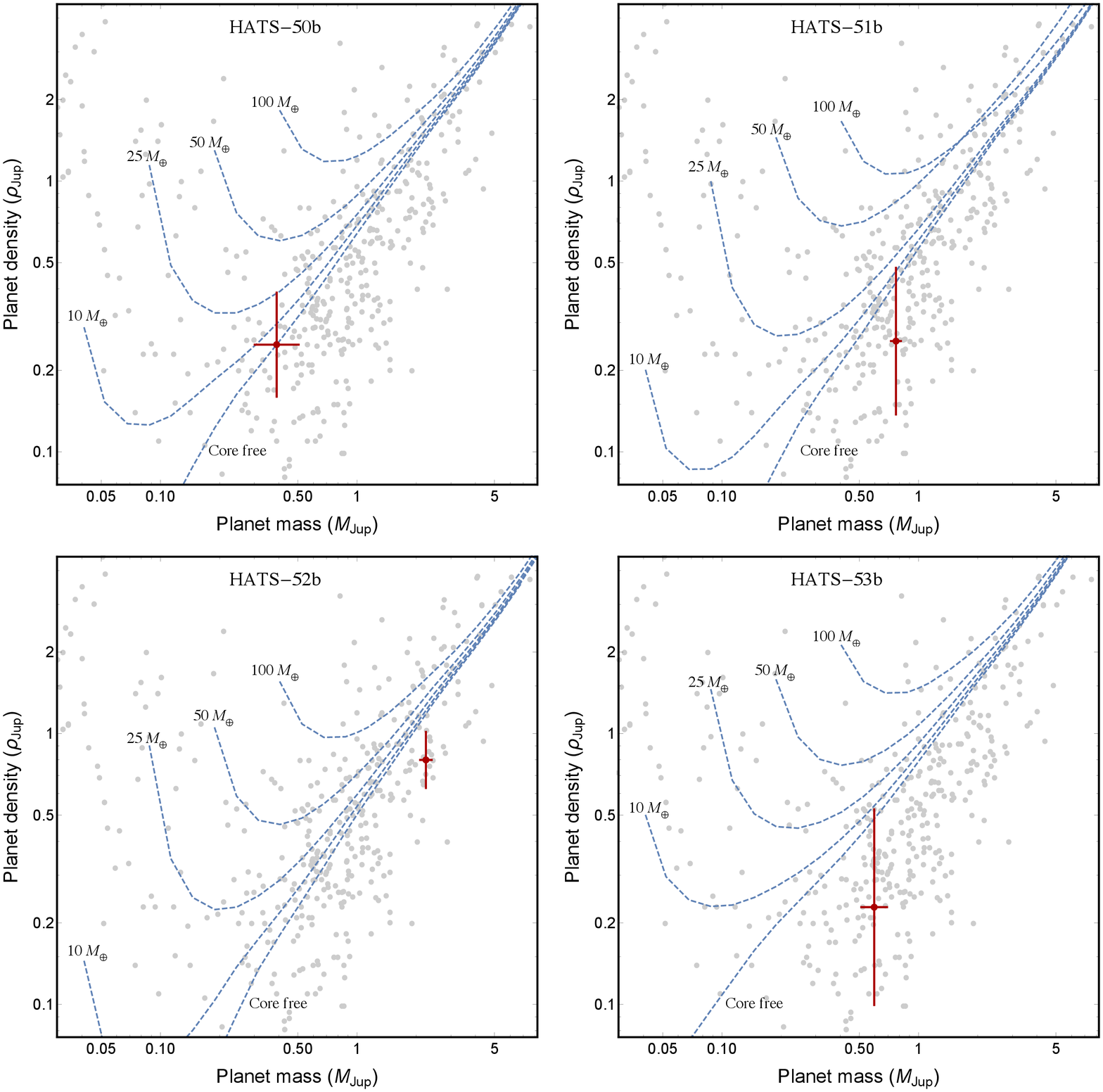}
\caption{
The mass-density diagram of the currently known transiting exoplanets. The grey points denote values taken from TEPCat. Their error bars have been suppressed for clarity. The position of \hatcurb{50}, \hatcurb{51}, \hatcurb{52}, \hatcurb{53} are shown in red with error bars in the top-left, top-right, bottom-left and bottom-right panel, respectively. Four planetary models, with various heavy-element core masses (10, 25, 50, and 100 Earth mass) and another without a core \citep{fortney:2007} are plotted for comparison. They were estimated for a planet at 0.045\,au from a parent star with an age of 1.0\,Gyr (top-left panel), 0.045\,au and 3.16\,Gyr (top-right panel), 0.02\,au and 1.0\,Gyr (bottom-left panel), and 0.045\,au and 10\,Gyr (bottom-right panel).
}
\label{fig:diagram3}
\end{figure*}

Having now exceeded 50 discoveries\footnote{The papers describing the discovery of the HATS exoplanets from HATS-36 to HATS-49 are under review or close to being submitted.}, HATSouth turns out to be one of the most efficient ground-based survey for detecting transiting exoplanets. Thanks to systematic photometric observations of southern-sky regions with the HATSouth robotic telescopes, we have presented the discovery of four new hot Jupiters, namely \hatcurb{50}, \hatcurb{51}, \hatcurb{52} and \hatcurb{53}. After their detection with the survey facilities, their planetary nature was robustly confirmed through photometric follow-up observations, and extensive RV measurements, as described in the previous sections. 

All the photometric and spectroscopic data that we have collected were used for fully characterizing these new exoplanetary systems. From the analysis of the parents stars we found that they are G-type main-sequence stars and have very different ages. While \hatcur{50} and \hatcur{52} are young ($\approx 1.2$\,Gyr), \hatcur{51} has an age similar to the Sun ($\approx 4.7$\,Gyr), and \hatcur{51} appears to be very old ($\approx 9.0$\,Gyr). Three of them resulted to be metal rich (\hatcur{50}: $\feh=\hatcurSMEzfeh{50}$; \hatcur{51}: $\feh=\hatcurSMEzfeh{51}$; \hatcur{52}: $\feh=\hatcurSMEzfeh{52}$;), whereas \hatcur{53} presents a metal abundance similar to the Sun, $\feh=\hatcurSMEzfeh{53}$.

Figure~\ref{fig:diagram} shows the positions of the four new HATS planets in the current planet period-mass diagram. They are plotted together with all the other known transiting hot Jupiters, i.e. exoplanets having a mass in the range $0.3 \, M_{\rm J} < M_{\rm p} < 5\,M_{\rm J}$ and an orbital period less than 10 days (data taken from the TEPCat catalogue\footnote{The Transiting Extrasolar Planet Catalogue (TEPCat) is available at http://www.astro.keele.ac.uk/jkt/tepcat/ \citep{southworth:2011}.} on October 30, 2017). While \hatcurb{51} and \hatcurb{53} have a similar Safranov number (see Table~\ref{tab:planetparam}) and are located in regions of the diagram where the hot Jupiters are very packed, \hatcurb{50} and \hatcurb{52} are in less-populated regions of the diagram, highlighting the well-known desert of low-mass Jupiters and Neptunes at low orbital periods (e.g. \citealp{mazeh:2005,benitez:2011,mazeh:2016}). 

The inflated size of \hatcurb{50}, \hatcurb{51} and \hatcurb{53} is evident from Figure~\ref{fig:diagram2}, in which the mass-radius diagram of known transiting exoplanets (with mass and radius up to $2.5\,M_{\rm J}$ and $2.0\,R_{\rm J}$, respectively) is shown. The three planets exhibit a similar density. Instead, due to its mass, \hatcurb{52} occupies a zone a slightly apart from the other three and from the crowd of giant exoplanets, similar to the physical characteristics of WASP-36b \citep{mancini:2016} and Kepler-17b \citep{desert:2011}. Moreover, since the stellar radiation that it receives from its star is $\approx 2.6\times10^9$\,erg~sec$^{-1}$, \hatcurb{52} is very hot ($T_{\rm eq}=\hatcurPPteff{52}$\,K) and belongs to the pM class of hot Jupiters, according to the terminology of \citet{fortney:2008}\footnote{The hypothesis proposed by \citet{fortney:2008} is to divide hot Jupiters into two classes (pM- and pL-class planets, analogous to the M- and L-type dwarfs), depending on the presence, in their atmospheres, of strong absorbers such as gaseous TiO and VO.}. 

The panels of Figure~\ref{fig:diagram3} show the position of the four planets in the mass-density diagram of the currently known transiting exoplanets. Each planet is compared with five different theoretical models estimated by \citet{fortney:2007}. Each model has a different core of heavy-elements, i.e. 0, 10, 25, 50, and 100 Earth mass and each panel shows models that were estimated for different values of planet-star separation and stellar age, as explained in the caption of the figure. 
The four planets have densities comparable with models of core-free planets.
One pontential explanation could be that the planets are bloated which would provide the incorrect impression of a too small core mass \citep{thorngren:2017}. An alternative explanation would be that relatively low opacities would allow gas runaway accretion also for lower core masses \citep{mordasini:2014,ormel:2014}. In a recent investigation based on results of the Juno mission the core mass of Jupiter was estimated to be in the range between $7-25$ Earth mass \citep{wahl:2017} which points to a relatively small core mass.

Finally, we would like to remark the possible existence of an inner planet in the \hatcur{50} planetary system. The analysis of the photometric data of the HATSouth survey has actually revealed a small transit signal with duration of 46 minutes (see Fig.~\ref{fig:hats50c_candidate}), yet with a SDE below our threshold for selecting it as a planet candidate. The radius, estimated from the best-fitting model of the HATS photometry and the upper limit of its mass, as coming from the RV measurements, suggest that this putative planet c has physical characteristics of a super Neptune. Its short periodicity (0.77\,days) place it in the Neptune desert, thus making it an interesting candidate to possibly confirm or invalidate with more performing astronomical facilities, as the next space telescope TESS will be.

\acknowledgements 
Development of the HATSouth
project was funded by NSF MRI grant NSF/AST-0723074, operations have
been supported by NASA grants NNX09AB29G, NNX12AH91H, and NNX17AB61G, and follow-up
observations receive partial support from grant NSF/AST-1108686.
J.H.\ acknowledges support from NASA grant NNX14AE87G.
A.J.\ acknowledges support from FONDECYT project 1171208, BASAL CATA
PFB-06, and project IC120009 ``Millennium Institute of Astrophysics
(MAS)'' of the Millenium Science Initiative, Chilean Ministry of
Economy. N.E.\ is supported by CONICYT-PCHA/Doctorado
Nacional. R.B.\ and N.E.\ acknowledge support from project
IC120009 ``Millenium Institute of Astrophysics (MAS)'' of the
Millennium Science Initiative, Chilean Ministry of Economy.
V.S.\ acknowledges support form BASAL CATA PFB-06.  
A.V. is supported by the NSF Graduate Research Fellowship, Grant No. DGE 1144152.
This work also uses observations obtained with facilities of the Las
Cumbres Observatory Global Telescope (LCOGT).
This work is based on observations collected with HARPS at the European Organisation for Astronomical Research in the Southern Hemisphere under ESO programme 095.C-0367.
This work has made use of data from the European Space Agency (ESA) mission {\it Gaia} (\url{https://www.cosmos.esa.int/gaia}), processed by the {\it Gaia} Data Processing and Analysis Consortium (DPAC,
\url{https://www.cosmos.esa.int/web/gaia/dpac/consortium}).
Funding for the DPAC has been provided by national institutions, in particular the institutions participating in the {\it Gaia} Multilateral Agreement.
We acknowledge the use of the AAVSO Photometric All-Sky Survey (APASS),
funded by the Robert Martin Ayers Sciences Fund, and the SIMBAD
database, operated at CDS, Strasbourg, France.
Operations at the MPG~2.2\,m Telescope are jointly performed by the Max Planck Gesellschaft and the European Southern Observatory in La Silla. We thank the MPG\,2.2\,m telescope support team for their technical assistance during observations."
This work is based in part on observations carried out with the Keck-I telescope at Mauna Kea Observatory in Hawaii. Time on this facility was awarded through the Australian community access. Australian community access to the Keck Observatory was supported through the Australian Government's National Collaborative Research Infrastructure Strategy, via the Department of Education and Training, and an Australian Government astronomy research infrastructure grant, via the Department of Industry and Science.
The authors wish to thank the anonymous referee for its useful comments.
\clearpage
\bibliographystyle{aasjournal}
\bibliography{hatsbib}

\clearpage

\end{document}